\documentclass[3p, 12pt]{elsarticle}

\makeatletter
\def\ps@pprintTitle{%
    \let\@oddhead\@empty
    \let\@evenhead\@empty
    \let\@evenfoot\@oddfoot
    }
\makeatother

\usepackage{tabularx}
\usepackage{amsmath}
\usepackage{amssymb}
\usepackage{amsthm}
\usepackage{mathrsfs}
\usepackage{bm}
\usepackage[usenames,dvipsnames]{color}
\usepackage{graphicx}
\usepackage{float}
\usepackage[ruled,vlined,algo2e]{algorithm2e}
\usepackage{enumitem}
\usepackage{array}
\usepackage{multirow}
\usepackage{makecell}
\usepackage{empheq}
\usepackage{cases}
\usepackage{textcomp}
\usepackage{gensymb}
\usepackage{subcaption}
\usepackage{hyperref}
\hypersetup{
	colorlinks=true,       
	linkcolor=blue,        
	citecolor=blue,        
	filecolor=magenta,     
	urlcolor=blue    
}
\usepackage[capitalise,noabbrev]{cleveref}
\usepackage{booktabs}
\usepackage{diagbox}

\usepackage{placeins}

\newcommand{\R}{\mathbb{R}}

\newcommand{\Z}{\mathbb{Z}}

\DeclareMathOperator*{\argmax}{arg\,max}
\newtheorem{remark}{Remark}

\newcommand{\dd}{\,\textrm{d}}

\newcommand{\calA}{{\mathcal{A}}}

\newcommand{\calD}{{\mathcal{D}}}

\newcommand{\calF}{{\mathcal{F}}}
\newcommand{\calG}{{\mathcal{G}}}
\newcommand{\calH}{{\mathcal{H}}}

\newcommand{\calO}{{\mathcal{O}}}

\newcommand{\calS}{{\mathcal{S}}}

\newcommand{\calU}{{\mathcal{U}}}

\crefname{algocf}{Algorithm}{Algorithms}
\Crefname{algocf}{Algorithm}{Algorithms}
\crefname{appendix}{}{}
\Crefname{appendix}{}{}
\crefname{figure}{Figure}{Figures}
\crefname{equation}{}{}
\Crefname{equation}{}{}
\Crefname{figure}{Figure}{Figures}

\newtheorem{lemma}{Lemma}[section]

\newcommand{\iid}{\stackrel{\text{i.i.d.}}{\sim}}

\begin{document}

\begin{frontmatter}
\title{Optimal Experimental Design for Reliable Learning of History-Dependent Constitutive Laws}
\author{Kaushik Bhattacharya}
\ead{bhatta@caltech.edu}
\author{Lianghao Cao\corref{cor1}}
\ead{lianghao@caltech.edu}
\author{Andrew Stuart}
\ead{astuart@caltech.edu}

\address{California Institute of Technology, 1200 E. California Blvd., Pasadena, CA 91125, USA.}

\cortext[cor1]{Corresponding author}
\begin{abstract}
History-dependent constitutive models serve as macroscopic closures for the aggregated effects of micromechanics. Their parameters are typically learned from experimental data. With a limited experimental budget, eliciting the full range of responses needed to characterize the constitutive relation can be difficult. As a result, the data can be well explained by a range of parameter choices, leading to parameter estimates that are uncertain or unreliable. To address this issue, we propose a Bayesian optimal experimental design framework to quantify, interpret, and maximize the utility of experimental designs for reliable learning of history-dependent constitutive models.  In this framework, the design utility is defined as the expected reduction in parametric uncertainty or the expected information gain. This enables \emph{in silico} design optimization using simulated data and reduces the cost of physical experiments for reliable parameter identification.

We introduce two approximations that make this framework practical for advanced material testing with expensive forward models and high-dimensional data: (i) a Gaussian approximation of the expected information gain, and (ii) a surrogate approximation of the Fisher information matrix. The former enables efficient design optimization and interpretation, while the latter extends this approach to batched design optimization by amortizing the cost of repeated utility evaluations. Our numerical studies of uniaxial tests for viscoelastic solids show that optimized specimen geometries and loading paths yield image and force data that significantly improve parameter identifiability relative to random designs, especially for parameters associated with memory effects.

\end{abstract}

\begin{keyword}
optimal experimental design, Bayesian inverse problems, machine learning, solid mechanics, material testing
\end{keyword}
\end{frontmatter}

\section{Introduction}

\subsection{Background and Motivation}

Constitutive relations close the balance equations of continuum mechanics by mapping the kinematic histories of continua to their internal response. To determine the constitutive relations for a system of interest, one first selects a constitutive model with unspecified parameters based on physical considerations. Then, the system-specific parameter values are identified from experiments conducted under controlled and simplified scenarios. The calibrated constitutive models are then deployed to predict quantities of interest in more complex, realistic scenarios, informing critical decision-making for the system. The quality of these predictions and the subsequent decisions hinges on the quality of the parameter identification. Consequently, it is of great interest to develop techniques that reliably identify constitutive parameters from experiments.

The parameters defining constitutive relations are often not directly measurable. Instead, the parameters are identified by solving the inverse problem: finding parameter values that minimize the discrepancy between experimental data and model predictions. These predictions are generated by a forward model that consists of two parts: (i) the balance equations, adapted to the specific experimental setup, such as geometry and boundary conditions; and (ii) an observation model that predicts the experimental data from the solutions of the balance equations. Three major challenges arise when solving this inverse problem in practice:
\begin{enumerate}[label=(\roman*)]
    \item \textit{Computational Complexity.} Repeated solution of the balance equations, required to solve the inverse problem, can render the computations intractable under time and computing resource constraints.
    \item \textit{Robustness.} There are inevitable discrepancies between the forward model and the reality of experiments, such as data noise and model misspecification. If the inverse problem is unstable, these discrepancies can be amplified, leading to inaccurate or unphysical parameter estimates.
    \item \textit{Data Quality.} Experimental data is often limited and noisy and, as a consequence, may not contain sufficient information to constrain the parameters. This leads to uncertain or unreliable parameter identification, thereby limiting the predictive power of the calibrated constitutive model.
\end{enumerate}
These three challenges interact. Lack of robustness amplifies the downstream effects of uncertain or unreliable parameter estimates \cite{kirsch2021introduction, ghattas2021learning}. This can be addressed by the Bayesian approach to inversion \cite{stuart2010inverse}, which quantifies uncertainty and improves robustness, but it places a heavy computational burden. Finally, we note that experimental design \cite{chaloner1995bayesian, Huan2024Optimal} can mitigate issues with data quality but adds to the computational burden.

In the context of solid-like materials, various experimental and computational techniques have been developed to address the challenges above; however, data-quality limitations remain largely unresolved. For example, conventional material testing, such as uniaxial loading and shear rheometry, prioritizes simplicity and robustness; however, data quality is often limited due to highly constrained experimental setups \cite{pierron2021towards}. Emerging techniques that use full-field observations \cite{avril2008overview} capture detailed, heterogeneous displacement fields from images, thereby increasing the information capacity of the data for parameter identification. However, exploiting this capacity to improve parameter identifiability remains challenging, as it increases the complexity of the experimental design---see Material Testing 2.0 \cite{pierron2021towards} and a review of related work on material testing for parameter identification in \cref{ssec:related_works}.

This paper addresses data-quality limitations in constitutive parameter identification using optimal experimental design. A specific focus is on history-dependent constitutive models in which stress depends on deformation history through a memory representation, such as hereditary integrals, fractional time derivatives, or internal state variables. In inelastic materials, the history dependence reflects the combined effects of dissipative microscale mechanisms, such as dislocation glide, chain segment relaxation, and microvoid growth. Identifying their constitutive parameters requires experimental data characterizing the material response under a diverse set of deformation histories. Designing a small number of experiments that generate such information-rich data is challenging for two primary reasons: (i) it is often unclear what type of deformation histories are most relevant to parameter identifiability; and (ii) most experiments cannot directly impose heterogeneous deformation histories pointwise to enrich the resulting full-field observations. For certain constitutive models, we have insights into experimental designs that can isolate or amplify signals of operative mechanisms in experimental data, such as stress relaxation in viscoelastic solids. However, gaining such insight is not straightforward for general, nonlinear constitutive models, such as neural network models \cite{liu2023learning, bhattacharya2025learning, fuhg2025review}, which have high expressivity but low interpretability. This motivates our focus and the contributions we now overview.

\subsection{Main Contributions and Outline of Paper}

We introduce a Bayesian framework to quantify, interpret, and maximize the utility of experimental designs for reliable parameter identification. Departing from intuition-driven or heuristic strategies, we employ the expected information gain to directly quantify parameter identifiability for a given design. Furthermore, we address the limitations of data-driven \cite{lookman2019active} and sensitivity-based approaches \cite{fayad2025, pukelsheim2006optimal}, which typically require exploratory physical testing with suboptimal designs to inform the optimization process. In contrast, our framework explores the joint design-parameter space using simulated experimental outcomes, thereby eliminating the reliance on exploratory testing and reducing total testing time and costs. Our specific contributions are as follows: 

\begin{enumerate}[label=(C\Roman*)]
\item We formulate a Bayesian optimal experimental design framework to improve the parameter identifiability of history-dependent constitutive models from experimental data. It uses expected information gain to assess data quality and to enable \textit{in silico} design optimization. \label{CI}

\item We employ a Bayesian D-optimal design utility based on a Gaussian approximation of the expected information gain for efficient design optimization and interpretation. \label{CII}

\item We propose and numerically validate a method that amortizes the cost of batched design optimization by surrogate modeling of the Fisher information matrix. \label{CIII}

\item We study the proposed framework for designing uniaxial tests for viscoelastic solids. The results show that the optimized specimen shape and loading path yield high-quality image and force data, resulting in more reliable parameter estimates than those obtained from randomly selected tests. The implementation is available through the following link:
\begin{center}
    \url{https://github.com/lcao11/constitutive_oed/}.
\end{center}
\label{CIV}
\end{enumerate}

The paper is organized as follows. We conclude this section with a review of related work in \cref{ssec:related_works}. \cref{sec:constitutive_laws} introduces history-dependent constitutive laws, particularly with memory representation based on internal state variables. \cref{sec:BIP} presents the Bayesian formulation of constitutive parameter identification from experimental data. \cref{sec:oed} introduces Bayesian optimal experimental design, where the EIG design utility (CI) and its approximations (CII and CIII) are derived. \cref{sec:results_linear,sec:results_nonlinear} present our numerical study on the design of uniaxial tests to learn linear and nonlinear viscoelasticity (CIV), respectively. \cref{sec:conclusion} contains concluding remarks. The appendices contain details of the numerical implementation.

\subsection{Related Work}
\label{ssec:related_works}

We provide further details concerning each of the
contributions in this paper and review related work. The purpose of this review is to set our contributions in the context of the literature rather than to serve as an exhaustive survey. We first address Bayesian optimal experimental design for material testing (CI) in \cref{sssec:review_CI} and Bayesian D-optimal design (CII) in \cref{sssec:review_CII}. We then discuss surrogate modeling for experimental design (CIII) in \cref{sssec:review_CIII}, followed by advanced material testing for parameter identification (CIV) in \cref{sssec:review_CIV}. 

\subsubsection{Bayesian Optimal Experimental Design for Material Testing}\label{sssec:review_CI}

\paragraph{Details of \labelcref{CI}} We use the expected information gain (EIG) as a baseline measure of the utility of experimental design for reliable parameter identification. The EIG is based on a Bayesian formulation of parameter identification that yields a posterior distribution rather than a point estimate from the experimental data. The EIG quantifies the expected reduction in parametric uncertainty, measured by information entropy, from model-predicted data at the candidate design. It can also be interpreted as the mutual information between the parameters and data.\hfill$\diamond$

\paragraph{Expected Information Gain}
 Lindley \cite{lindley1956information} first proposed using the EIG to measure experiment utility, a concept now standard in Bayesian experimental design; see reviews on this topic in \cite{chaloner1995bayesian, alexanderian2021optimal, Huan2024Optimal, Rainforth2024Modern}. For material testing, the EIG has been shown to be effective for designing loading steps for biaxial testing \cite{ricciardi2024bayesian}, Weber number and bubble radius for inertial microcavitation rheometry \cite{chu2025bayesian}, and grain orientation for spherical indentation \cite{castillo2019bayesian}. These works employ the incremental EIG for myopic sequential design, in which the posterior from the previous experiment serves as the prior for generating the next design. They use simplified but approximate forward models that circumvent the balance equations. In contrast, our forward models incorporate them to fully assimilate image data that capture spatially heterogeneous material responses. Additionally, our focus is on non-adaptive batched design, which can serve as an incremental step in sequential design. Finally, the batched and myopic sequential designs are special cases of sequential designs formulated as Markov decision processes \cite{huan2016sequential, villarreal2023design}.\hfill$\diamond$

\paragraph{Stress State Entropy} Recently, information-theoretic metrics have been proposed to design specimen geometries for uniaxial and biaxial loading that maximize the heterogeneity of induced stress fields. Prominently, the stress state entropy framework \cite{ihuaenyi2024seeking,ihuaenyi2025mechanics} employs information entropy to quantify the diversity of stress invariants, such as triaxiality and Lode angle, within a single experiment. This framework assumes that maximizing stress state diversity is equivalent to maximizing the information content for parameter identification. In contrast, we consider the EIG, which directly assesses parameter identifiability and quantifies the mutual information between the data and the parameters. \hfill$\diamond$

\subsubsection{Bayesian D-Optimal Design Based on Gaussian Posterior Approximation} \label{sssec:review_CII}

\paragraph{Details of \labelcref{CII}} The EIG is expensive to estimate and can be difficult to interpret. We therefore adopt the Bayesian D-optimal design utility derived from the EIG using a Gaussian posterior approximation. This replaces the nested loop of forward model evaluations for EIG estimation with a single loop of Fisher information matrix (FIM) evaluations, thereby substantially reducing the cost of utility evaluations. Moreover, because Gaussian marginals are available in closed form, the utility can be evaluated for any parameter subset at negligible additional cost, enabling quantitative diagnosis of how design choices affect recovery of parameters governing specific behaviors such as anisotropy and stress relaxation. \hfill$\diamond$

\paragraph{Sensitivity-Based Design} Our approach is connected to sensitivity-based optimal design of specimen geometry \cite{bertin2016optimization, chamoin2020coupling} and loading path \cite{fayad2025} for uniaxial or biaxial testing. Their design formulations are equivalent to classical D-optimal designs \cite{pukelsheim2006optimal, Huan2024Optimal}, which maximize the logarithm of the determinant of the FIM for a linear or linearized forward model; this connection was recently noted by Fayad et al.~\cite{fayad2025}. For nonlinear models, design optimization involves an iterative process of experimentation, parameter identification, and optimization using the linearized model at the parameter point estimate. This procedure faces a dilemma: reliable estimates are required for good design, yet good design requires reliable estimates. In contrast, our approach does not require a point estimate of the parameter to initiate the optimization. We formulate a risk-neutral optimization under prior uncertainty, eliminating the need for exploratory physical testing with suboptimal designs.\hfill$\diamond$

\paragraph{Gaussian Posterior Approximation} There are two commonly-used approximations derived from a quadratic expansion of the log-posterior centered at different parameter estimates: (i) the Laplace approximation \cite{Beck2018Fast, long2013fast, goda2020multilevel, wu2023fast, Overstall2018Appraoch}, centered at the maximum a posteriori (MAP) estimator, and (ii) a data-independent local approximation \cite{Englezou2022Approximate, long2013fast, wu2023fast}, centered at the data-generating parameters. Although (i) typically provides higher fidelity, it requires parameter identification for each data realization. By contrast, (ii) is cheaper because it avoids this optimization procedure and exploits the known data-generating parameters. When the forward model evaluation requires solving the balance equations, the repeated parameter identification for each utility evaluation creates a computational bottleneck in design optimization. This issue is exacerbated when history-dependent constitutive models are used. This motivates our use of (ii).\hfill$\diamond$

\paragraph{Bayesian D-optimal design} The connection between Bayesian D-optimal design utility and the EIG under a Gaussian posterior approximation is well established \cite{chaloner1995bayesian}. Such approximation is widely used to mitigate the prohibitive cost of nested Monte Carlo estimators of EIG \cite{Ryan2003Estimating, Rainforth2018Nested}. In practice, the Gaussian approximation is employed either as (i) an analytic replacement of the posterior density in EIG approximation \cite{long2013fast, Long2015Laplace, wu2023fast, Overstall2018Appraoch}  or (ii) an importance sampling proposal to accelerate EIG estimation \cite{Beck2018Fast, goda2020multilevel, Englezou2022Approximate}. The Bayesian D-optimal utility emerges in case (i), which is adopted in this work.\hfill$\diamond$

\subsubsection{Surrogate Modeling For Optimal Experimental Design} \label{sssec:review_CIII}

\paragraph{Details of \labelcref{CIII}} Material testing often involves repeated experiments, where a batch of experiments with random designs may yield more useful information than a single optimized experiment. This motivates batched experimental design optimization. However, its computational cost increases linearly with the batch size. To amortize this cost, we train a surrogate model of the FIM, incurring a one-time cost to generate training samples that is independent of the batch size. Then the trained surrogate model predicts the FIMs during batched design optimization, making it tractable even with large batch sizes.\hfill$\diamond$

\paragraph{Surrogate Utility and Surrogate Forward Model} Most surrogate modeling approaches for experimental design either learn the utility function \cite{muller1995optimal, overstall2017bayesian} or the forward model \cite{huan2013simulation, piyush2021surrogate, wu2023large, go2025accurate}. Surrogate utility emulates the design utility at a fixed batch size, and its training cost depends on that batch size. In our numerical studies, we also use a Gaussian process to emulate the utility for design optimization (i.e., Bayesian optimization). In contrast, the training costs of surrogate forward models are independent of batch size, and the surrogate can be used to evaluate likelihood across multiple experiments. However, training surrogate forward models can be challenging when the dimensions of the parameters, design variables, or data are high. For advanced material testing, snapshots of image data or full-field observations contain millions of data points, making it difficult to construct surrogate forward models without resorting to additional data compression techniques.\hfill$\diamond$

\paragraph{Surrogate Fisher Information Matrix}
The FIM size is independent of the data dimension, yet scales quadratically with the parameter dimension. When the parameter dimension is much smaller than the data dimension, it is easier to construct surrogate FIMs than surrogate forward models. We therefore adopt this approach in our work, training a neural network to emulate the FIM as a function of the parameter and design variables. Similarly, Lan et al.~\cite{lan2016emulation} use a Gaussian process to emulate the FIM for fast metric tensor predictions in MCMC. Recent works explore using the derivative-informed neural operator \cite{oleary2024dino} to emulate both the forward model and its Jacobian, which is then used for FIM predictions in geometric MCMC \cite{cao2025derivative} and optimal experimental design \cite{go2025accurate}.\hfill$\diamond$

\subsubsection{Advanced Material Testing for Parameter Identification} \label{sssec:review_CIV}

\paragraph{Details of \labelcref{CIV}} We provide numerical case studies on uniaxial testing of linear and nonlinear viscoelastic solids. The experiment yields reaction force measurements and image snapshots of the deformed specimen as data for parameter identification. The following experimental designs are considered simultaneously:
\begin{enumerate}[label=(\roman*)]
    \item \textit{Specimen Geometry.} We consider a thin, flat specimen with a centered elliptical hole. The hole's aspect ratio and orientation are design variables.
    \item \textit{Loading Path.} The axial loading path is prescribed by the strain values at discrete time points with interpolation. These strain values are design variables.
\end{enumerate}
We note that these design variables are chosen to demonstrate the applicability of our framework to advanced material testing using image or full-field observations. Our design optimization formulation is general and not limited to these specific settings.\hfill$\diamond$

\paragraph{Full-Field Observations} Digital image correlation (DIC) \cite{chu1985application,hild2006dic, sutton2009image} infers displacement fields by correlating images of a speckled surface before and after deformation. It has also been extended to obtain out-of-plane \cite{luo1993accurate, pan2009dic} and 3D displacement fields \cite{bay1999dvc, buljac2018dvc, yang2020augumented}. Full-field observations have high information capacity for parameter identification. For example, one can probe many different deformation histories in a single test using specimens with complex geometries \cite{bertin2016optimization}. This information capacity can be further leveraged through topology optimization of specimen geometry \cite{chamoin2020coupling, goncalves2023design, ghouli2025topology}. In our numerical case studies, we also incorporate specimen geometry into the experimental design.
\hfill$\diamond$

\paragraph{Parameter Identification from Full-Field Observations} Numerical methods such as the finite element model updating (FEMU) method \cite{kenneth1971finite, chen2025finite} and the virtual fields (VFM) method \cite{grediac2006vfm, Pierron2012vfm} can exploit full-field observations for parameter identification; see \cite{avril2008overview} for a review. Other approaches that treat parameter identification as PDE-constrained inverse problems \cite{akerson2025learning, oberai2003solution} are most relevant to this work. These inverse-problem formulations employ weighted least-squares data misfits, which are equivalent to MAP estimation problems under a Bayesian formulation with additive Gaussian noise \cite{stuart2010inverse, ghattas2021learning}.\hfill$\diamond$

\paragraph{Parameter Identification from Images}

In our numerical studies, images are used directly for parameter identification rather than the displacements inferred from them. This aligns with integrated DIC and other image-based approaches \cite{besnard2006finite, mathieu2015estimation, wihardja2025constitutive, kirchhoff2024inference}, which avoid inverting for displacement fields and directly minimize image misfit or the grayscale residual. This integrated approach has been shown to yield more accurate and reliable parameter estimates than the two-step approach of first applying DIC, then performing parameter identification \cite{ruybalid2016comparison}. The motivation for our use of image data is the simplicity of image noise models, such as additive and uncorrelated Gaussian or Poisson--Gaussian noise \cite{Janesick2007PhotonTransfer, foi2008pratical}, which leads to closed-form likelihoods and allows for FIM evaluations using the Jacobian of the forward model \cite{chao2016fisher}. In contrast, the noise models for the inferred displacement fields are nonphysical and nonlinear \cite{schreier2000systematic} with spatial correlation \cite{blaysat2016effect}.\hfill$\diamond$

\paragraph{Bayesian Formulation of Parameter Identification}

The theory and computational methods for Bayesian inverse problems have seen substantial development in the past two decades; see, e.g., \cite{stuart2010inverse, ghattas2021learning, kirsch2021introduction}. It has also been applied to parameter identification using conventional testing; see, e.g., \cite{rappel2020tutorial}. Recent works have also applied this formulation to full-field observations \cite{ bhattacharyya2023bayesian, jafari2025bayesian, touminet2025bayesian, mahmoud2025sequential} or image data \cite{gaynutdinova2023bayesian}. The Bayesian formulation considered in our work is general and not restricted to full-field observations; for instance, our numerical case studies consider both force and image data. Our formulation involving image data is related to Gaynutdinova et al.~\cite{gaynutdinova2023bayesian}, which extends integrated DIC to define a pixel-level likelihood.\hfill$\diamond$

\section{History-Dependent Constitutive Laws}\label{sec:constitutive_laws}

In this section, we present history-dependent constitutive laws based on the internal variable theory \cite{coleman1967thermodynamics} to model the isothermal behavior of materials. In this framework, the state of a material is described not only by the deformation gradient ${F}$, but also by a set of scalar or tensorial internal states, $\{\alpha_i\}_{i=1}^{N}$ (denoted $\alpha=\{\alpha_i\}$ for brevity, and similarly for other variables.) The evolution of these internal states is typically needed to account for changes in the material microstructure. However, we emphasize that the experimental design methodology introduced in this work is not restricted to internal variable theory and can also be applied to other memory representations, such as time integrals and fractional derivatives.

In \cref{ssec:CMC}, we describe some common modeling considerations arising in
history-dependent constitutive laws. \cref{subsec:viscoelastic,subsec:viscohyper} then describe, respectively, the specific cases of linear viscoelasticity and anisotropic finite-strain viscoelasticity used in our numerical examples. For a general discussion of viscoelasticity, we refer the reader to \cite{stefanie1998theory, lake2009viscoelastic}.

\subsection{Common Modeling Considerations}
\label{ssec:CMC}

\paragraph{Helmholtz Free Energy} The Helmholtz free energy $\Psi$, a 
measure of stored material energy, is assumed to depend only on $F$ and 
$\alpha$:
\begin{equation*}
    \Psi = \Psi(F, \{\alpha_i\}).
\end{equation*}
The first Piola--Kirchhoff (PK1) stress, denoted as ${P}$, is the stress measure that is work-conjugate to $F$. It is often assumed to be additively decomposed into an equilibrium component, defined through the free energy, 
and a non-equilibrium component,
\begin{equation*}
    {P} = {P}^{\textrm{eq}} + {P}^{\textrm{neq}},\quad {P}^{\textrm{eq}} = \partial_F\Psi({F}, \{\alpha_i\}).
\end{equation*}
\hfill$\diamond$

\paragraph{Dissipation Inequality and Kinetic Laws} For an isothermal process, the second law of thermodynamics requires that the rate of internal dissipation be non-negative:
\begin{equation}\label{eq:energetic_force}
    {P}^{\textrm{neq}}:\dot{{F}} + \sum_{i=1}^{n_\alpha} {A}_i\,\dot{\alpha}_i \geq 0,\quad {A}_i:= -\partial_{\alpha_i}\Psi({F}, \{\alpha_j\}).
\end{equation}
where $\dot{F}$ denotes the rate of change of the deformation gradient, and ${A}_i$ is the thermodynamic driving force conjugate to the internal state $\alpha_i$ defined as the partial derivative of free energy.

The dissipation inequality constrains the admissible forms for the dissipative stress ${P}^{\textrm{neq}}$ and the kinetic laws that govern the evolution of the internal states. We consider the following commonly adopted form of 
kinetic law, defined through function $\mathcal{G}=\{\mathcal{G}_i\}$:
\begin{align*}
    \dot{\alpha}_i = \mathcal{G}_i(F, \dot{F}, \{\alpha_j\}, \{A_j\}).
\end{align*}
\hfill$\diamond$

\paragraph{Hereditary Operator} Together, the modeling choices for the free energy potential $\Psi$, kinetic laws $\{\calG_i\}$, and the initial condition of the internal states form a constitutive model. This model can be viewed as a hereditary operator $\mathcal{M}$ that predicts the stress state in the reference frame from the history of deformation:
\begin{equation*}
    {P}(t) = \mathcal{M}\left(\{{F}(s)\}_{s\in[0,t]}, \theta\right),
\end{equation*}
where $\theta$ represents the constitutive parameters.\hfill$\diamond$

\paragraph{Material Frame Indifference}
The Helmholtz free energy $\Psi$ and the kinetic laws $\calG_i$ should satisfy the material frame indifference principle, which can be achieved by defining $\Psi$ and $\{\mathcal{G}_i\}$ through objective tensors that depend on $F$ and $\dot{F}$, such as the right Cauchy--Green tensor $C \coloneqq F^\top F$, and the rate of deformation tensor $D = \textrm{sym}(\dot{F}F^{-1})$, where $\textrm{sym}(A)=(A + A^{\top})/2$.\hfill$\diamond$

\paragraph{Dissipative Potential}

A common approach for designing the kinetic laws to meet the constraint of the dissipation inequality is to postulate a dissipation potential $\Phi$ defined as
\begin{equation*}
    \Phi = \Phi(F, \dot{{F}}, \{\alpha_i\}, \{\dot{\alpha}_i\}),
\end{equation*}
from which the dissipative quantities are then derived:
\begin{equation}\label{eq:dissipative_force}
    {P}^{\textrm{neq}} = \partial_{\dot{F}}\Phi(F, \dot{F},\{\dot{\alpha}_i\},  \{\alpha_i\}),\quad {A}_i = \partial_{\dot{\alpha}_i}\Phi(F, \dot{F},\{\alpha_i\},  \{\dot{\alpha}_j\}).
\end{equation}
The convexity of $\Phi$ with respect to the rate variables $\dot{F}$ and $\{\dot{\alpha}\}$ ensures the satisfaction of dissipation inequality. 
The kinetic law for each internal state is then found by equating the energetic driving force in \cref{eq:energetic_force} with the dissipative force in \cref{eq:dissipative_force}.\hfill$\diamond$

\subsection{Example I: Linear Viscoelasticity}\label{subsec:viscoelastic}
We consider three-dimensional linear viscoelasticity. Let ${\alpha}_i:[0,T]\to\R^{3\times 3}_{\textrm{sym}}$ for $1\leq i\leq N$ denote a set of tensorial internal variables. Let $\varepsilon = \frac{1}{2}(F + F^{\top}) - I$ denote the infinitesimal strain tensor. The Helmholtz free energy and the dissipative potential are given by
\begin{align*}
    \Psi(\varepsilon, {\alpha}; \{\mathbb{C}_i\}) &= \frac{1}{2}{\varepsilon}:\mathbb{C}_{0}:{\varepsilon} + \frac{1}{2}\sum_{i=1}^{N} \left({\varepsilon}-{\alpha}_i\right):\mathbb{C}_i:\left({\varepsilon}-{\alpha}_i\right),\\
    \Phi(\dot{{\alpha}}; \{\tau_i\}, \{\mathbb{C}_i\}) &=  \frac{1}{2}\sum_{i=1}^{N} {\tau_i}\dot{{\alpha}}_i:\mathbb{C}_i:\dot{{\alpha}}_i,
\end{align*}
where $\{\mathbb{C}_i\}$ are symmetric and coercive 4th-order tensors representing the elastic and viscous effects and $\{\tau_i\}$ are the relaxation 
times. The resulting constitutive model is the generalized Maxwell model that predicts the stress $\sigma\approx P$ in $t\in[0,T]$:
\begin{align*}
    {\sigma}(t) &= \mathbb{C}_0:{\varepsilon}(t) + \sum_{i=1}^{N} \mathbb{C}_i: ({\varepsilon}(t)-{\alpha}_i(t)),\\
    \dot{{\alpha}}_i(t) &= \tau_i^{-1}({\varepsilon}(t)-{\alpha}_i(t)), && 1\leq i\leq N,\\
    {\alpha}_i(0) &= 0, && 1\leq i\leq N.
\end{align*}
This constitutive model is equivalent to the following memory kernel model under the Prony series expansion when $\varepsilon(0)=0$:
\begin{align*}
     {\sigma}(t) &= \mathbb{C}_0:{\varepsilon}(t) +  \int_{0}^t \mathbb{K}(t-s; \{\mathbb{C}_i\}, \{\tau_i\}):{\dot{\varepsilon}}(s) \dd s,\\
     \mathbb{K}(t) &= \sum_{i=1}^{N} \mathbb{C}_i\exp(-t/\tau_i).
\end{align*}
In this linear viscoelastic constitutive model, at most $d_{\theta} = 22N + 21$ parameters need to be determined, since each 4th-order tensor contains at most 21 independent components. For the numerical example in \cref{sec:results_linear}, we use this model with $N=2$ and orthotropic plane stress, leading to 11 parameters.

\subsection{Example II: Anisotropic Finite-Strain Viscoelasticity}\label{subsec:viscohyper}
We consider a three-dimensional finite-strain, anisotropic nonlinear viscoelastic solid with $N$ parallel viscous branches \cite{holzapfel2001viscoelastic}. The equilibrium response is hyperelastic with fiber reinforcement, and the rate dependence arises from non-linear viscous branches driven by a dual dissipative potential. In this model, the internal state variables are $\{F^{\textrm{v}}_i\}_{i=1}^{N}$, each represents the viscous deformation gradient for the $i$-th viscous branch. A multiplicative decomposition of the deformation gradient is assumed for each branch: $F = F^{\textrm{e}}_iF^{\textrm{v}}_i$, where $F^{\textrm{e}}_i$ is the elastic deformation gradient for the $i$-th branch. The total stress response is the sum of an equilibrium response and $N$ non-equilibrium responses. The Helmholtz free energy is given by an equilibrium energy $\Psi^{\textrm{eq}}$ and a set of non-equilibrium energies $\{\Psi_i^{\textrm{neq}}\}$:
\begin{align*}
    \Psi(F, \{F^{\textrm{v}}_i\}) = \Psi^{\textrm{eq}}(C) + \sum_{i=1}^{N} \Psi^{\textrm{neq}}_{i}(C^{\textrm{e}}_i),
\end{align*}
where $C^{\textrm{e}}_i = (F_i^{\textrm{e}})^{\top} F^{\textrm{e}}_i$ is the strain tensor of the intermediate configuration for each viscous branch. 

Let $\overline{(\cdot)} = J^{-2/3}(\cdot)$ be the isochoric operator where $J=\det(F)$. The relevant invariants of $\overline{C}$ are $\overline{I}_1 = \textrm{tr}(\overline{C})$ and $\overline{I}_4 = \overline{C}:A_0$, where $A_0 = a_0\otimes a_0$ denotes the structure tensor and $a_0$ is the fiber orientation unit vector. The equilibrium energy is given by
\begin{equation*}
    \Psi^{\textrm{eq}}(C;\mu_{\infty}, \kappa, k_1, k_2, a_0) = \frac{\mu_\infty}{2}(\overline{I}_1 - 3) + \frac{\kappa}{2} (\ln J)^2 + \frac{k_1}{2k_2} \left( \exp\left( k_2\langle \overline{I}_4 - 1 \rangle^2 \right) - 1 \right) + \frac{\tau k_1}{2} \langle 1 - \overline{I}_4 \rangle^2,
\end{equation*}
where $\mu_{\infty}$ is the shear modulus of the matrix (i.e., the background material), $\kappa$ is the bulk modulus, $k_1$ is the initial (small-strain) fiber stiffness, $k_2$ is the stiffening rate, and $\langle \cdot \rangle$ is the Macaulay bracket defined as $\langle x\rangle\coloneqq \max\{0, x\}$. The fiber energy includes both tensile and compressive components \cite{holzapfel2015on}. The exponential term is active only when fibers are in tension ($\overline{I}_4 > 1$), while a small quadratic compressive penalty, scaled by $\tau=0.1$, is active in compression ($\overline{I}_4 < 1$) to prevent numerical instabilities (arising, for
example, from buckling). To capture the reorientation of fibers during viscous flow, the non-equilibrium fiber invariants are defined in the intermediate configuration:
\begin{align*}
\overline{I}_{4, i}^{e} = \overline{C^{\textrm{e}}_i} : {A}^{\textrm{v}}_i, \quad \textrm{with} \quad {A}^{\textrm{v}}_i = {a}^{\textrm{v}}_i \otimes {a}^{\textrm{v}}_i, \quad
{a}^{\textrm{v}}_i = \frac{F^{\textrm{v}}_i a_0}{\|F^{\textrm{v}}_i a_0\|}.
\end{align*}
Physically, ${a}^{\textrm{v}}_i$ represents the fiber direction convected by the viscous flow $F^{\textrm{v}}_i$ prior to elastic stretching \cite{nguyen2007modeling}. We define $\overline{I}_{1, i}^{e} = \textrm{tr}(\overline{C^{\textrm{e}}_i})$ and assume the viscous flow is isochoric, which implies $J_i^{\textrm{v}}=1$ and $\textrm{det}(C^{\textrm{e}}_i)=J^2$. The forms of the non-equilibrium energies are:
\begin{equation*}
    \Psi^{\textrm{neq}}_{i}(C^{\textrm{e}}_i;\mu_i, k_{1,i}, k_{2,i}, a_0) = \frac{\mu_i}{2} (\overline{I}_{1,i}^{\textrm{e}} - 3) + \frac{k_{1, i}}{2k_{2, i}}\left(\exp\left(k_{2, i}\langle \overline{I}_{4,i}^{\textrm{e}}-1\rangle^2\right)-1\right) + \frac{\tau k_{1, i}}{2} \langle 1 - \overline{I}_{4,i}^{\textrm{e}} \rangle^2,
\end{equation*}
where $\{\mu_i\}$, $\{k_{1,i}\}$, $\{k_{2,i}\}$ are parameters of the non-equilibrium energies.

The dissipation for each viscous branch is governed by the conjugate pair $(M^{\textrm{v}}_i, L^{\textrm{v}}_i)$, where $L^{\textrm{v}}_i = \dot{F}^{\textrm{v}}_i(F^{\textrm{v}}_i)^{-1}$ is the viscous velocity gradient and $M^{\textrm{v}}_i = C^{\textrm{e}}_i S^{\textrm{e}}_i$ is the Mandel stress for the $i$-th branch, with $S^{\textrm{e}}_i=2\nabla\Psi^{\textrm{neq}}_{i}(C_i^{\textrm{e}})$ being the thermodynamic driving force. The kinetic law is prescribed by dual dissipative potentials $\{\Phi_i^*\}$ \cite{ortiz1999variational, simo1998computational}:
\begin{equation*}
    \Phi_i^*(M^{\textrm{v}}_i; \gamma_i, m_i) = \frac{\gamma_i}{m_i+1} \|\textrm{dev}(\textrm{sym}(M^{\textrm{v}}_i))\| ^{m_i+1},
\end{equation*}
where $\{\gamma_i\}$ are the rate coefficients, $\{m_i\}$ are the rate-sensitivity exponents.

The explicit form of the constitutive law for $t\in[0,T]$ is given by
\begin{align*}
    P(t) &= F(t)S(t) , \\
    S(t) &= 2 \nabla\Psi^{\textrm{eq}}(C(t)) + 2\sum_{i=1}^N F^{\textrm{v}}_i(t)^{-1}  \nabla\Psi^{\textrm{neq}}_{i}(C^{\textrm{e}}_i(t)) F^{\textrm{v}}_i(t)^{-\top},\\
    M^{\textrm{v}}_i(t) &= 2 C^{\textrm{e}}_i(t)  \nabla\Psi^{\textrm{neq}}_{i}(C^{\textrm{e}}_i(t)), && 1\leq i\leq N, \\
    L^{\textrm{v}}_i(t) &= \gamma_i \|\textrm{dev}(\textrm{sym}(M^{\textrm{v}}_i))\|^{m_i-1} \textrm{dev}(\textrm{sym}(M^{\textrm{v}}_i)), && 1\leq i\leq N,\\
    \dot{F}^{\textrm{v}}_i(t) &= L^{\textrm{v}}_i(t) F^{\textrm{v}}_i(t), && 1\leq i\leq N,\\
    F^{\textrm{v}}_i(0)&= I, && 1\leq i\leq N.
\end{align*}
In this nonlinear viscoelastic model, there are in total $5N+4$ constitutive parameters to be determined from experimental data.

\section{Bayesian Inverse Problems}\label{sec:BIP}
In this section, we formulate Bayesian inverse problems (BIPs) for identifying constitutive parameters from experimental data. We first define the forward model, comprising the balance equation (\cref{subsec:material_response}) and observation operator (\cref{subsec:observation}). We then present the model-constrained BIP framework in \cref{subsec:bip} and extend it to multiple experiments in \cref{subsec:multiple_experiments}.

\subsection{The Balance Equations}\label{subsec:material_response}
We consider an open, bounded domain $\Omega_0\subset\R^{3}$ occupied by a material undergoing deformation. We assume the deformation is described by a displacement field ${u}({X},t)$, where ${X}\in\Omega_0$ denotes and $t\in[0, T]$ denotes time with a terminal time $T>0$. We are interested in the material response to external stimuli, such as body forces, boundary traction forces, boundary conditions, or initial conditions. Let $z\in\R^{d_z}$ denote the experimental design variables. The governing equation for the displacement field is given by
\begin{subequations}
\begin{align}
    \rho \ddot{{u}}({X},t) - \textrm{Div}_{{X}} ({P}({X},t)) &= {f}({X},t; z), &({X}, t)\in\Omega_0(z)\times[0, T],\\
    {P}({X},t) &= \mathcal{M}\left(\left\{{F}({X},s)\right\}_{s\in[0, t]};\theta\right), &({X}, t)\in\Omega_0(z)\times[0, T],\\
    {P}({X},t) {n_0}({X}) &= {\tau}_n({X}, t; z), &({X}, t)\in\Gamma_N(z)\times[0, T],\\
    {u}({X},t) &= {u}_D({X},t;z), &({X}, t)\in\Gamma_D(z)\times[0, T],\\
    {u}({X},0) &= {u}_0({X};z), & {X}\in\Omega_0(z),\\
    \dot{u}({X},0) &= {v}_0({X};z), & {X}\in\Omega_0(z),
\end{align}
\end{subequations}
where $\rho$ is the material density, ${f}$ is the external force, $n_0$ is the outward normal vector of the material domain, ${\tau}_n$ is the boundary traction defined over the material domain boundary $\Gamma_N\subseteq \partial\Omega_0$, ${u}_D$ is the Dirichlet boundary defined over the material domain boundary $\Gamma_D\subseteq \partial\Omega_0$, and ${u}_0$ and ${v}_0$ are the initial condition for the displacement field and its velocity. In abstract form, the solution operator of the governing equation is denoted as the map from the experimental design variables and the constitutive parameters to the trajectory of the material displacement:
\begin{equation}
\label{eq:defS}
    u = \calS(\theta, z).
\end{equation}

\paragraph{Example: Uniaxial Tests with Loading Path and Specimen Shape Design}

We consider a uniaxial test of a flat, thin specimen under the 2D plane-stress assumption. Let $\Omega_0(z_s)=([0, L_1]\times [0, L_2])\setminus \Omega_H(z_s)$ denote a rectangular-shaped material surface with holes $\Omega_H(z_s)\subset [0, L_1]\times [0, L_2]$ parametrized by the shape parameters $z_s\in\R^{d_{z_s}}$. The material is clamped at the left boundary (zero displacement) and pulled on the right boundary according to a loading path $u_l:[0,1]\to[0, u_{\textrm{max}}]$. We take $u_l(0)=0$ and $u_{\textrm{max}}>0$ as the maximum loading position. Let $z_l\in[0, u_\textrm{max}]^{d_{z_l}}$ denote the equally spaced control points of the loading path. We prescribe the loading path via interpolation, i.e., 
\begin{equation*}
    u_l(t_i; z_l) = [z_l]_i,\quad t_i = i/d_{z_l},\quad i=1,\dots, d_{z_l}.
\end{equation*}

\begin{figure}
    \centering
    \includegraphics[width=0.9\linewidth]{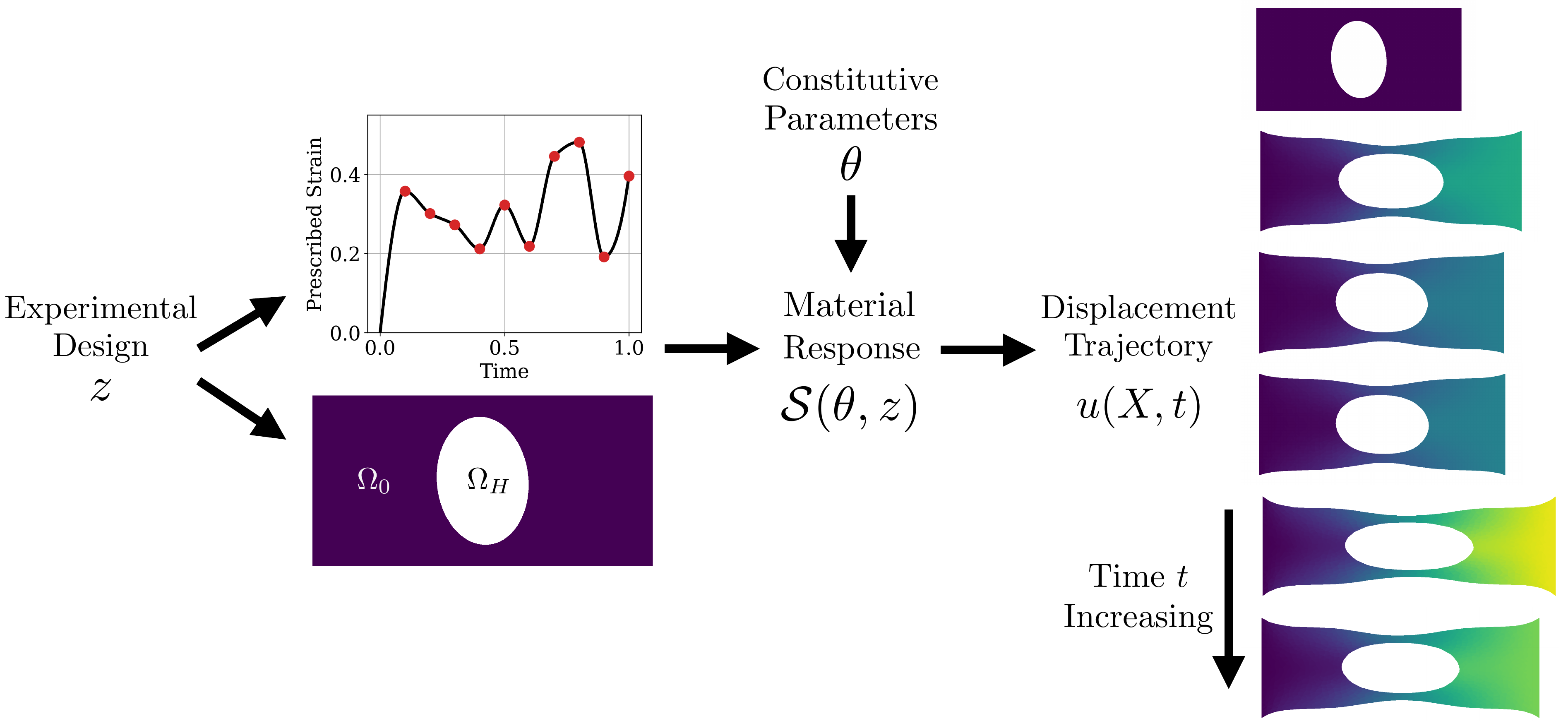}
    \caption{A visualization of a model for the described uniaxial test with loading path and specimen shape design. The dots on the loading path, i.e., the prescribed strain, are the control points. The color on the deformed material represents the magnitude of the displacement.}
    \label{fig:uniaxial_demo}
\end{figure}

Under the quasi-static assumption, this setup is obtained from the balance equations above by setting $\rho\ddot{u}=0$, $f=0$, and $\tau_n=0$, while retaining the same hereditary constitutive law. The resulting equilibrium equation is
\begin{equation*}
    - \textrm{Div}_{{X}} ({P}({X},t)) = 0, \quad ({X}, t)\in\Omega_0(z_s)\times[0, T],
\end{equation*}
with Dirichlet boundary conditions
\begin{equation*}
    u(X,t)=0 \quad \text{on } \{0\}\times[0,L_2],
    \qquad
    u(X,t)=\left(u_l(t;z_l),0\right) \quad \text{on } \{L_1\}\times[0,L_2],
\end{equation*}
a traction-free boundary condition on the remaining boundary, and zero initial displacement. \cref{fig:uniaxial_demo} shows examples for the input and output of the solution operator $u=\mathcal{S}(\theta,z)$ for this setup of uniaxial testing.\hfill $\diamond$

\subsection{The Observation Operator and the Forward Model}\label{subsec:observation}
During experiments, observations of the material's response are collected, including images and reaction forces. The observations can also be designed, for example, with a speckled pattern, and thus may depend on the design variables $z$. In an abstract form, we model the experimental data $y\in\R^{d_{y}}$ as
\begin{equation}
\label{eq:defO}
    y = \calO({u}, \theta, z, \eta),\quad \eta \sim \pi_{\eta}.
\end{equation}
Here $\calO$ is the observation operator that returns the experimental data given the displacement fields, the constitutive parameters, and the experimental design. The unknown noise $\eta\in\R^{d_{y}}$ follows the noise distribution $\pi_{\eta}$. 

By combining the solution operator of the governing equation, encapsulated in
\eqref{eq:defS}, with the observation model \eqref{eq:defO}, 
we arrive at our forward model $\calF$:
\begin{align}
    y = \calF(\theta, z, \eta),\quad \calF(\theta, z, \eta)\coloneqq\calO(\calS(\theta, z), \theta, z, \eta).
\end{align}

\paragraph{Example: Image Data}
We consider image data that captures a 2D surface of a deformed material. The surface is painted with a speckled pattern, and this data is often used in inferring displacement fields via DIC; see \cref{ssec:related_works}. Here, we define a model for a snapshot of image data produced at time $t\in[0, T]$, with a large field of view that covers the full range of deformation in the experiment. The details of this model are provided in \cref{app:dic}.

\begin{figure}
    \centering
    \includegraphics[width=0.75\linewidth]{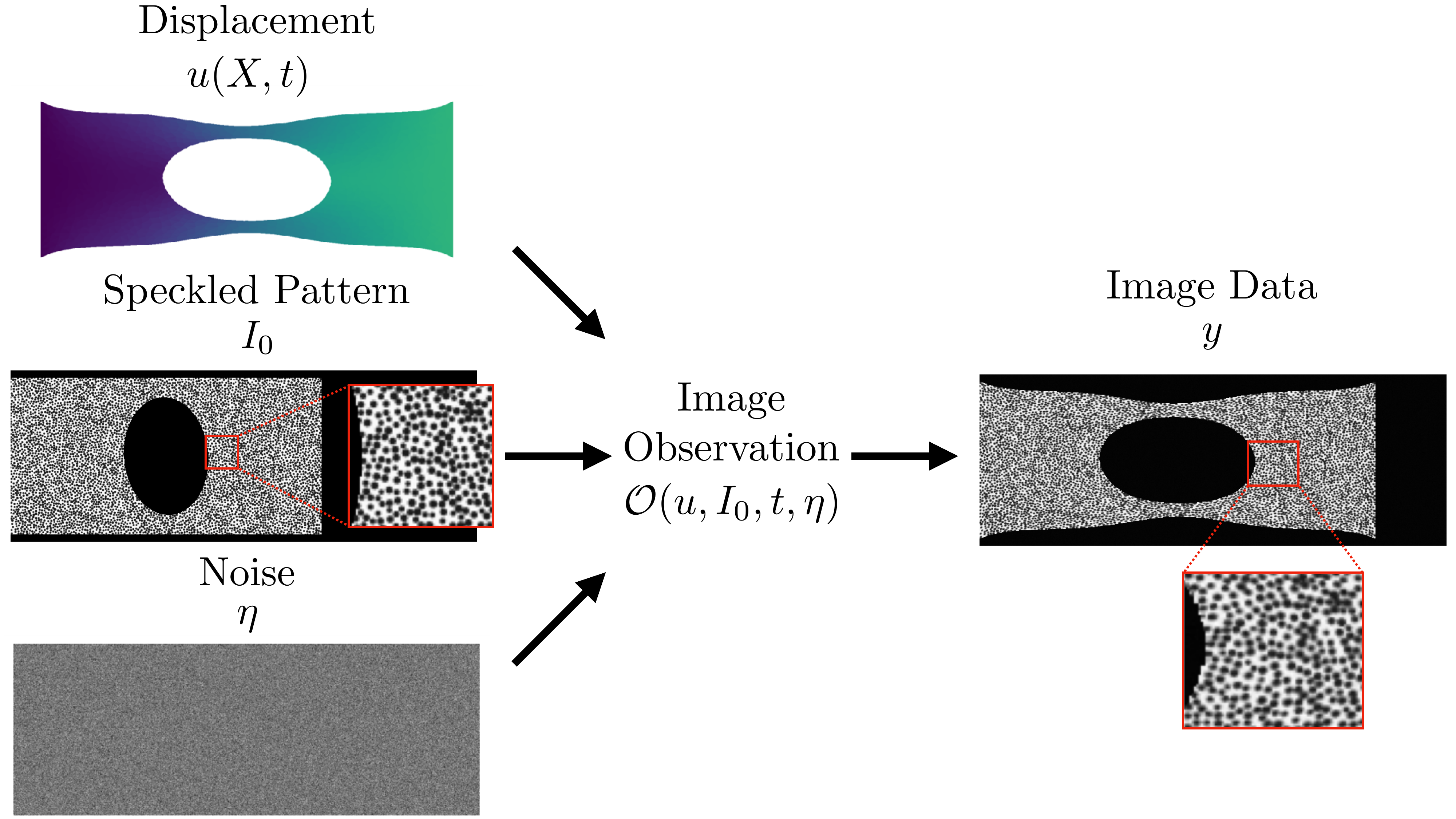}
    \caption{A visualization of the observation operator for a single snapshot of image data at time $t$ for the uniaxial testing described in \cref{subsec:material_response}.}
    \label{fig:image_observation}
\end{figure}

Let $I_0\in[0,1]^{n_W\times n_H}$ denote a reference image of the speckled pattern painted on a flat material surface $\partial\Omega_{\textrm{I}}\subset\Omega$. Pixels located outside the surface (i.e., the environment) are masked to zero intensity. First, we define an image deformation operator $\mathcal{W}$ that uses the displacement value at each pixel of the reference image and a push forward method to obtain an image of the deformed material $I_{\textrm{pred}}\in[0,1]^{n_W\times n_H}$ with masked environment, i.e.,
\begin{equation}\label{eq:deformed_image}
    I_{\textrm{pred}} = \mathcal{W}(u(\cdot, t)|_{\partial\Omega_{I}}, I_0) \in [0,1]^{n_W\times n_H},
\end{equation}
where $u(\cdot, t)|_{\partial\Omega_{I}}$ denotes the displacement field on the surface at time $t$. The deformed image is perturbed with white noise that has a variance $\sigma_{\eta}^2$, and a soft mask $\mathcal{A}(I_{\textrm{pred}})\in[0,1]^{n_W\times n_H}$ is then applied to remove the intensity of pixels corresponding to the environment:
\begin{equation*}
    y = \mathcal{A}(I_{\textrm{pred}}) \odot (I_{\textrm{pred}} + \sigma_{\eta}\eta), \quad [\eta]_{ij}\sim \mathcal{N}(0, 1),
\end{equation*}
where $\odot$ denotes pixel-wise multiplication. \cref{fig:image_observation} contains examples of the observation operator $y=\mathcal{O}(u, I_0, t, \eta)$ defined through the procedure above. Finally, we note that one may include $I_0$ and $t$ as design variables. In our numerical examples, however, we choose to fix them and design the specimen geometry and loading path. \hfill $\diamond$

\subsection{Model-Constrained Bayesian Inverse Problems}\label{subsec:bip}
    We now adopt a Bayesian probabilistic framework for inferring the constitutive parameters from the data. Let $\Theta$ denote the parameter random vector following the prior distribution $\pi_{\Theta}$. We may define a data random vector $Y$ for experimental design variables $z$ as
    \begin{equation*}
        Y(z) = \calF(\Theta, z, \eta),\quad \Theta\sim\pi_{\Theta},\quad \eta\sim\pi_{\eta}.
    \end{equation*}
    We denote the conditional distribution of $Y(z)$ given $\theta$, i.e., the likelihood, as $\pi_{Y(z)\mid\Theta}(\cdot \mid \theta)$. A common form of the likelihood can be derived from the additive noise model, which assumes that there exists a parameter-to-observable map $\calF_{\textrm{pto}}$ such that $Y(z) = \calF_{\textrm{pto}}(\Theta, z) + \eta$. In this case, we have
    \begin{equation*}
        \pi_{Y(z)\mid\Theta}(y\mid\theta)\coloneqq \pi_{\eta}(y-\calF_{\textrm{pto}}(\theta, z)).
    \end{equation*}
    The posterior distribution of the constitutive parameters, denoted by $\pi_{\Theta\mid Y(z)}$, is defined via the Bayes' rule
    \begin{equation*}
        \pi_{\Theta\mid Y(z)}(\theta|y) = \frac{\pi_{Y(z)\mid\Theta}(y\mid\theta)}{\pi_{Y(z)}(y)}\pi_{\Theta}(\theta),
    \end{equation*}
    where the data marginal distribution or marginal likelihood $\pi_{Y(z)}$ is given by
    \begin{equation}\label{eq:data_marginal}
        \pi_{Y(z)}(y) = \mathbb{E}_{\theta\sim \pi_\Theta}[\pi_{Y(z)\mid\Theta}(y\mid\theta)].
    \end{equation}
    Given the observed data $y$ produced with design $z$, the goal of the Bayesian inverse problem is to characterize the posterior $\pi_{\Theta\mid Y(z)}(\cdot\mid y)$, which encapsulates the parametric uncertainty after the experiment.

\subsection{Bayesian Inverse Problems for Multiple Experiments}\label{subsec:multiple_experiments}
Now we consider the case where $N_E$ experiments are conducted with the same setup described by the model $\calF$ but with a set of different experimental design variables $\{z_i\}$, e.g., different loading paths, test specimen shapes, and observations. In this case, we have a set of data random vectors $\{Y_i(z_i)\}$ defined by
\begin{equation*}
    Y_i(z_i) = \mathcal{F}(\Theta, z_i, \eta_i),\quad \Theta\sim\pi_{\Theta}, \quad \eta_i\iid \pi_{\eta_i},\quad 1\leq i\leq N_E.
\end{equation*}
In this case, the likelihood is given by
\begin{equation}\label{eq:batched_likelihood}
    \pi_{\{Y_i(z_i)\}|\Theta}(\{y_i\}|\theta)\coloneqq\prod_{i=1}^{N_E} \pi_{Y_i(z_i)\mid\Theta}(y_i\mid\theta).
\end{equation}
We have the following Bayes' rule for the posterior of the constitutive parameters:
\begin{equation*}
        \pi_{\Theta\mid \{Y_i(z_i)\}}(\theta|\{y_i\}) = \frac{\pi_{\{Y_i(z_i)\}\mid\Theta}(\{y_i\}\mid\theta)}{\pi_{\{Y_i(z_i)\}}(\{y_i\})}\pi_{\Theta}(\theta),
    \end{equation*}
where the data marginal $\pi_{\{Y_i(z_i)\}}$ is given by
    \begin{equation*}
        \pi_{\{Y_i(z_i)\}}(y) = \mathbb{E}_{\theta\sim \pi_\Theta}[\pi_{\{Y_i(z_i)\}\mid\Theta}(y\mid\theta)].
    \end{equation*}
While we are interested in the BIP for multiple experiments, it is a straightforward extension of the BIP for a single experiment. We will use the simplified notation for a single experiment in the subsequent analysis, except when the distinction is important.
    
\section{Bayesian Optimal Experimental Design}\label{sec:oed}
Building on the BIP formulation from \cref{sec:BIP}, \cref{subsec:eig} formulates Bayesian optimal experimental design (BOED) using the expected information gain (EIG), which covers \ref{CI}. Then, in \cref{subsec:eig_equivalence}, we present two equivalent forms of the EIG based on expected entropy reduction and mutual information. These formulations motivate the three EIG estimation approaches considered in this work: 
\begin{enumerate}[label=(\roman*)]
    \item Gaussian posterior approximation (\cref{subsec:eig_gauss}), which is efficient but biased. This covers \ref{CII}.
    \item Surrogate Fisher information matrix (\cref{subsec:ESFIM}), which improves the efficiency of the Gaussian approximation for batched design. This covers \ref{CIII}.
    \item Nested Monte Carlo (\cref{subsec:NMCE}), a baseline method that is asymptotically consistent but computationally demanding.
\end{enumerate}
Lastly, in \cref{subsec:discussion_estimators} we discuss the suitability of the first two approaches for learning history-dependent constitutive models from high-dimensional data.

\subsection{Expected Information Gain as the Design Utility}\label{subsec:eig}
We derive the EIG from a BIP perspective. Suppose that the data $y$ is observed from the experiment with design $z$, and the BIP is solved to obtain the posterior $\pi_{\Theta|Y(z)}(\cdot|y)$. The success of the experiment can be assessed via the reduction of uncertainty, or the \emph{information gain} in the constitutive parameters. A useful metric for the level of uncertainty in a probability distribution $p$ is given by the information entropy $\mathcal{H}$ defined as
\begin{equation*}
    \mathcal{H}(p) \coloneqq -\mathbb{E}_{x\sim p}\left[\log p(x)\right].
\end{equation*}
Following our Bayesian formulation of parameter identification in \cref{sec:BIP}, the information gain (IG) from an experiment with the design $z$ and observed data $y$ can be defined through the difference in the information entropy between the prior and the posterior:
\begin{equation}\label{eq:information_gain}
    \textrm{IG}(y, z)\coloneqq \mathcal{H}(\pi_{\Theta})-\mathcal{H}\left(\pi_{\Theta\mid Y(z)}(\cdot\mid y)\right).
\end{equation}
While the IG enables us to assess data quality for identifying constitutive parameters, the primary goal of BOED is to propose experiments without access to the experimental outcomes. A natural solution to this dilemma is to use the forward model $\mathcal{F}$ to simulate data. This gives rise to the expected IG (EIG) defined as an expectation of the IG over the data marginal $\pi_{Y(z)}$:
\begin{equation}\label{eq:eig_posterior}
    \textrm{EIG}(z)\coloneqq \mathbb{E}_{y\sim \pi_{Y(z)}}\left[\textrm{IG}(y, z)\right].
\end{equation}
To see where the prior and the forward model come into play in EIG, notice that sampling from the data marginal $y\sim \pi_{Y(z)}$ follows a procedure given by:
\begin{enumerate}[label=(\roman*)]
    \item Sample constitutive parameters from the prior $\theta\sim \pi_{\Theta}$.
    \item Sample noise $\eta\sim\pi_{\eta}$.
    \item Evaluate the forward model $\calF(\theta, z, \eta)$ to obtain a simulated data sample at the constitutive parameters $\theta$ and the experimental design $z$.
\end{enumerate}

The objective of BOED is to find the optimal design $z^{\dagger}$ that maximizes the EIG subject to regularization:
\begin{equation}\label{eq:eig_maximization}
    z^{\dagger} \in \argmax_{z\in\R^{d_z}} \left\{\textrm{EIG}(z) - \mathcal{P}(z)\right\},
\end{equation}
where the regularization function $\mathcal{P}:\mathbb{R}^{d_z}\to\overline{\mathbb{R}}$ imposes bound constraints and penalties on the design variables. In this section, we introduce three different estimators for the EIG. Once an estimator is chosen, the maximization in \cref{eq:eig_maximization} can be solved using an appropriate optimization algorithm. Examples include Bayesian optimization, which is applicable when the estimator's gradient is unavailable, and stochastic gradient descent, which is applicable when an unbiased gradient estimator is available; see \cite[Section 4]{Huan2024Optimal} for a detailed discussion of these algorithms.

\subsection{Equivalent Forms of the Expected Information Gain} \label{subsec:eig_equivalence}
The EIG has numerous equivalent forms and interpretations. While they have been discussed in the literature (see, e.g., \cite{lindley1956information, Huan2024Optimal, Rainforth2024Modern}), we provide the following lemma to make an explicit connection between the expected entropy reduction form in \cref{eq:eig_posterior} and the mutual information forms of the EIG. The proof can be found in \cref{app:proof}.

\begin{lemma}[The Expected Information Gain as Mutual Information]
\label{lem:eig_equivalence}
The EIG, defined as the expected reduction in information entropy from the prior to the posterior in \cref{eq:eig_posterior}, is equivalent to the mutual information, denoted by $\mathrm{MI}$, between the parameters and the data:
\begin{equation*}
    \mathrm{EIG}(z) \equiv \mathrm{MI}(\Theta;Y(z)).
\end{equation*}
Here, the mutual information has two equivalent definitions:
\begin{align*}
    \mathrm{MI}(\Theta;Y(z)) &\coloneqq \mathbb{E}_{y\sim\pi_{Y(z)}}\left[D_{\mathrm{KL}}\left( \pi_{\Theta\mid Y(z)}(\cdot\mid y) \,\|\, \pi_{\Theta}\right)\right],\\
    \mathrm{MI}(\Theta;Y(z))&\coloneqq\mathbb{E}_{\theta\sim\pi_{\Theta}}\left[D_{\mathrm{KL}}\left( \pi_{Y(z)\mid\Theta}(\cdot\mid \theta) \,\|\, \pi_{Y(z)}\right)\right],
\end{align*}
where $D_{\mathrm{KL}}$ denotes the Kullback--Leibler divergence.
\end{lemma}

While theoretically identical, the two forms of the EIG pose distinct computational challenges. The expected entropy reduction form is conceptually intuitive and directly measures uncertainty minimization, but it is computationally expensive for nonlinear models because it requires repeated posterior density evaluations and sampling across data realizations. The likelihood-based mutual information form is useful when likelihood evaluation is tractable, but estimating the marginal data density $\pi_{Y(z)}$ can be difficult. The remainder of this section is devoted to deriving estimators of the EIG starting from its two forms. Their advantages and limitations when learning history-dependent constitutive laws will be discussed.

\subsection{Estimator Based on Gaussian Approximation}\label{subsec:eig_gauss}
We consider the EIG in the form of \cref{eq:eig_posterior}. Estimating \textrm{EIG} in this form is often intractable due to the prohibitive cost of posterior sampling and density evaluations across many data realizations, each of which requires running MCMC and evaluating the data marginal density. This issue can be addressed by first constructing a Gaussian posterior approximation and then using it in place of the posterior for approximate, rapid EIG estimation. 

This subsection is organized as follows. We derive a Gaussian approximation to the posterior in \cref{subsubsec:gaussia_approx}. The formulas for the resulting EIG approximation that yield the Bayesian D-optimal design utility are shown in \cref{sssec:apply_gauss}. Lastly, we discuss the numerical evaluations as well as limitations and advantages of this approximation in \cref{ssec:numerical_evalaution,sssec:limitation}, respectively.

\subsubsection{Gaussian Posterior Approximation}\label{subsubsec:gaussia_approx} First, notice that the data realization $y$ that is used to evaluate the IG in \cref{eq:eig_posterior} is generated by forward model evaluation $y=\mathcal{F}(\theta, z, \eta)$ at the prior sample $\theta\sim\pi_{\Theta}$ and the noise sample $\eta\sim\pi_{\eta}$, as described in \cref{subsec:eig}. We consider a data-independent local Gaussian approximation, denoted as $\pi_{\textrm{loc}}(\,\cdot\,;\theta, z)$. It captures the expected posterior behavior given the data-generating parameters $\theta$, averaging out variation across data realizations due to the random noise $\eta$. Compared with the Laplace approximation, which centers the Gaussian approximation at the MAP point that depends on the data realization, this data-independent approximation circumvents the optimization step for parameter identification. This optimization procedure can be computationally intensive when evaluating the forward model requires solving a transient PDE with history-dependent constitutive models.

We make the following assumptions on the likelihood and the prior:
\begin{enumerate}[label=(\roman*)]
    \item The log-likelihood $\ell(\theta'; y, z)\coloneqq \log \pi_{Y(z)\mid\Theta}(y\mid\theta')$ is twice continuously differentiable.
    \item The likelihood satisfies regularity conditions so that differentiation may be interchanged with expectation under $y\sim\pi_{Y(z)|\Theta}(\cdot\mid\theta)$.
    \item The log-prior $\log\pi_{\Theta}$ is twice continuously differentiable on the interior of its support.
    \item The prior $\pi_{\Theta}$ is strictly log-concave on the interior of its support.
\end{enumerate}
At the data-generating parameter $\theta$, the expected score satisfies the first Bartlett identity \cite{barlett1953approximate}, and the Fisher information matrix (FIM) is defined as the negative expected Hessian of the log-likelihood:
\begin{equation}\label{eq:bartlett_identity_and_fim_genprior}
    \mathbb{E}_{y\sim \pi_{Y(z)\mid\Theta}(\cdot\mid\theta)}\!\left[ \nabla_{\theta'} \ell(\theta; y, z)\right] = 0,\quad 
    \mathrm{FIM}(\theta, z) \coloneqq -\mathbb{E}_{y\sim \pi_{Y(z)\mid\Theta}(\cdot\mid\theta)}\!\left[ \nabla^2_{\theta'} \ell(\theta; y, z)\right].
\end{equation}
We define the expected log-posterior (up to an additive constant independent of $\theta'$) as
\begin{equation*}
    \phi(\theta'; \theta, z)
    \coloneqq
    \mathbb{E}_{y\sim \pi_{Y(z)\mid\Theta}(\cdot\mid\theta)}\!\left[\ell(\theta'; y, z)\right]
    + \log \pi_{\Theta}(\theta').
\end{equation*}
Using \cref{eq:bartlett_identity_and_fim_genprior}, the gradient and Hessian of $\phi$ at $\theta$ are
\begin{equation*}
    \nabla_{\theta'}\phi(\theta;\theta, z)=\nabla \log \pi_{\Theta}(\theta),\qquad
    \nabla^2_{\theta'}\phi(\theta;\theta, z)= -\mathrm{FIM}(\theta, z)+ \nabla^2 \log \pi_{\Theta}(\theta).
\end{equation*}
Consequently, a quadratic expansion of $\phi$ around $\theta$ is given by
\begin{equation*}
\begin{aligned}
    \phi_{\mathrm{quad}}(\theta';\theta, z)\coloneqq& \phi(\theta;\theta, z)
    + (\theta'-\theta)^\top \nabla \log \pi_{\Theta}(\theta) \\
    &-\frac{1}{2}(\theta'-\theta)^\top\Big(\mathrm{FIM}(\theta, z)-\nabla^2 \log \pi_{\Theta}(\theta)\Big)(\theta'-\theta),
\end{aligned}
\end{equation*}
where the matrix $\mathrm{FIM}(\theta, z)-\nabla^2 \log \pi_{\Theta}(\theta)$ is the negative Hessian of $\phi$ at $\theta$. We define the corresponding local Gaussian approximation
\begin{equation*}
    \pi_{\mathrm{loc}}(\theta';\theta, z)\propto \exp\big(\phi_{\mathrm{quad}}(\theta';\theta, z)\big).
\end{equation*}
Completing the square for $\phi_{\mathrm{quad}}$ gives us the local Gaussian approximation
\begin{subequations}\label{eq:local_gauss_approx_genprior}
    \begin{gather}
    \pi_{\mathrm{loc}}(\cdot;\theta, z)=\mathcal{N}(m_{\mathrm{loc}}(\theta, z),\Gamma_{\mathrm{loc}}(\theta, z)),\\
    m_{\mathrm{loc}}(\theta, z)
    =\theta+\Gamma_{\mathrm{loc}}(\theta, z)\,\nabla \log \pi_{\Theta}(\theta),\quad\Gamma_{\mathrm{loc}}(\theta, z)
    =\Big(\mathrm{FIM}(\theta, z)-\nabla^2 \log \pi_{\Theta}(\theta)\Big)^{-1}.
\end{gather}
\end{subequations}

\subsubsection{Applying the Gaussian Approximation} \label{sssec:apply_gauss}
The approximate IG using \cref{eq:local_gauss_approx_genprior}, denoted as $\widetilde{\textrm{IG}}$, is given by
\begin{equation}\label{eq:ig_local_approx}
     \textrm{IG}(y, z)\approx\widetilde{\textrm{IG}}(\theta, z) \coloneqq \mathcal{H}(\pi_{\Theta}) - \mathcal{H}(\pi_{\textrm{loc}}(\cdot;\theta, z)) ,\quad y\sim \pi_{Y(z)|\Theta}(\cdot\mid \theta),
\end{equation}
where $\theta$ is the generating parameter of the data $y$. Since the information entropy of Gaussian distributions has a closed-form expression, we have the following formula for the approximate IG in \cref{eq:ig_local_approx}: 
\begin{subequations}\label{eq:approximate_ig}
\begin{align}
    \widetilde{\textrm{IG}}(\theta, z)
    & = \frac{1}{2} \log \textrm{det} \left(\mathrm{FIM}(\theta, z)-\nabla^2 \log \pi_{\Theta}(\theta)\right) + \mathcal{H}(\pi_{\Theta}),
\end{align}
\end{subequations}
where the first term measures the volume of the posterior confidence ellipsoid.

Using the approximate IG, we obtain an approximation to the EIG:
\begin{gather}\label{eq:eig_global_approx}
     \textrm{EIG}(z)= \mathbb{E}_{\theta\sim\pi_{\Theta}}\left[\mathbb{E}_{y\sim \pi_{Y(z)\mid\Theta}(\cdot\mid \theta)}\left[\textrm{IG}(y, z)\right]\right]
     \approx \mathbb{E}_{\theta\sim\pi_{\Theta}}\left[\widetilde{\textrm{IG}}(\theta, z)\right], 
\end{gather}
where the expectation with respect to the data $y$ collapses after the approximation. The EIG approximation in \cref{eq:eig_global_approx} accounts for global parametric uncertainty by prior expectation and is often referred to as the utility function for \textit{Bayesian D-optimal design} \cite{chaloner1995bayesian}. This utility can be estimated using the sample average approximation, which leads to the \textit{estimator based on Gaussian approximation} (EGA):
\begin{equation}\label{eq:eig_gauss}
    \begin{gathered}
    \theta^{(i)}\iid \pi_{\Theta},\quad 1\leq i\leq N_{P},
    \\ \textrm{EGA}(z; N_{P})\coloneqq \frac{1}{2N_{P}}\sum_{i=1}^{N_{P}}\log\textrm{det}\left(\mathrm{FIM}(\theta^{(i)}, z)-\nabla^2 \log\pi_{\Theta}(\theta^{(i)})\right) + \mathcal{H}(\pi_{\Theta}).
\end{gathered}
\end{equation}

\begin{remark}[Gaussian Prior as a Special Case]
    When the prior is Gaussian with mean $m_0\in\R^{d_{\theta}}$ and covariance $\Gamma_0\in\R^{d_{\theta}\times d_{\theta}}$, we have
    \begin{equation*}
        m_{\mathrm{loc}}=\theta-\Gamma_{\mathrm{loc}}\Gamma_0^{-1}(\theta-m_0),\quad\Gamma_{\mathrm{loc}}=(\mathrm{FIM}(\theta, z)+\Gamma_0^{-1})^{-1}.
    \end{equation*}
    Consequently, the EGA becomes:
    \begin{equation*}
        \mathrm{EGA}(z; N_{P})\coloneqq \frac{1}{2N_{P}}\sum_{i=1}^{N_{P}}\log\mathrm{det}\left(\mathrm{FIM}(\theta^{(i)}, z)+\Gamma_0^{-1}\right) + \frac{1}{2} \log\mathrm{det}(\Gamma_0).
    \end{equation*}
    This formula is used in our numerical case studies.
\end{remark}

\begin{remark}[Gaussian Approximation Breaks Form Equivalence]
Under a Gaussian posterior approximation, the two equivalent definitions in \cref{lem:eig_equivalence} diverge. When the prior is also Gaussian, the KL divergence from the posterior to the prior has a closed-form expression that includes a log-determinant term, along with additional terms accounting for the averaged posterior variance and the mean shift. While this approximation is widely used \cite{long2013fast, Long2015Laplace, wu2023fast}, it differs from standard Bayesian D-optimal design and does not generally apply to non-Gaussian priors. Furthermore, this approach underestimates the EIG as evident by the Barber--Agakov bound \cite{Barber2003IM}. Indeed, this specific property makes it useful for variational approximation of the EIG; see, e.g., \cite{foster2019variational, baptista2024bayesian, dong2025variational, li2025expected}.
\end{remark}

\subsubsection{Numerical Evaluation}\label{ssec:numerical_evalaution}

Here, we discuss the numerical evaluation of the EGA. First, the FIM is equivalent to the Gauss--Newton Hessian for likelihood based on certain exponential families of distributions. When the equivalence holds, the FIM can be evaluated directly from the forward or adjoint sensitivity of the forward model. For example, when the noise model is additive and Gaussian, i.e., $y = \calF_{\textrm{pto}}(\theta, z) + \eta$ with $\eta\sim \mathcal{N}(0, \Gamma_{\eta})$, we have
\begin{equation*}
    \textrm{FIM}(\theta, z) = \partial_{\theta}\calF_{\textrm{pto}}(\theta, z)^{\top}\Gamma_{\eta}^{-1}\partial_{\theta}\calF_{\textrm{pto}}(\theta, z),
\end{equation*}
where $\partial_{\theta}\calF_{\textrm{pto}}(\theta, z)\in\R^{d_y\times d_{\theta}}$ is the partial derivative of the observables, i.e., noise-free simulated data, with respect to the parameters. In this case, evaluating the FIM requires $\min(d_y, d_\theta)$ evaluations of the direct or adjoint sensitivity of the forward problem, where each sensitivity computation represents a matrix-vector multiplication or transpose multiplication by $\partial_{\theta}\calF_{\textrm{pto}}(\theta, z)$. Second, the log-determinant of the posterior covariance matrices in \cref{eq:approximate_ig} can be evaluated using the eigenvalues of an eigendecomposition of the FIM. This means that $\widetilde{\textrm{IG}}$ can be evaluated without forming FIM using matrix-free eigendecomposition algorithms. One can exploit the fast eigenvalue decay of the FIM to remove the dependence of the number of forward and/or adjoint sensitivity computations on the parameter dimension \cite{saibaba2017randomized}. It has been applied for D- and A-optimal design for parameters with arbitrarily high dimensions \cite{alexanderian2014aoptimal, alexanderian2018efficient, wu2023fast, alexanderian2021optimal}.

\subsubsection{Limitations and Advantages}\label{sssec:limitation} The Bayesian D-optimal design utility approximates the EIG effectively when posteriors are unimodal and locally Gaussian, as supported by the Bernstein–von Mises theorem \cite[Chapter 4]{Nickl2023Bayesian} in the large-data or small noise limit. Leveraging the asymptotic consistency of the MLE, we construct our local approximation via a quadratic expansion of the log-posterior around the data-generating parameter $\theta$. However, this approximation requires that the FIM be non-singular and the likelihood be unimodal. It breaks down due to unidentifiability or multimodality arising from under-constrained models (e.g., interchangeable viscous branches) or from designs that produce uninformative data.

Despite the EGA's limited use cases, it often strikes a superior balance between design improvements and cost, as discussed in \cref{subsec:discussion_estimators}. Additionally, it enables interpretation of the experimental design via EGAs for marginal parameters, as extracting the marginal posteriors for the Gaussian approximation is straightforward. In our numerical examples, we use the EGA to interpret how experimental designs affect the identifiability of subsets of the constitutive parameters associated with different material behaviors.

\subsection{Estimator Based on Surrogate Fisher Information Matrix}\label{subsec:ESFIM}

The computational cost of EGA-based design optimization can be overwhelming when designing multiple experiments at once (see \cref{subsec:multiple_experiments}), where EGA becomes
\begin{equation}\label{eq:multiple_experiments_local_gaussian}
    \begin{gathered}
    \theta^{(i)}\iid \pi_{\Theta},\quad 1\leq i\leq N_{P},\\ \textrm{EGA}(\{z_j\}; N_{P})\coloneqq \frac{1}{2N_{P}}\sum_{i=1}^{N_{P}}\log \textrm{det} \left(\sum_{j=1}^{N_{E}}\textrm{FIM}(\theta^{(i)}, z_j) - \nabla^2\log\pi_{\Theta}(\theta^{(i)})\right) + \calH(\pi_{\Theta}).
\end{gathered}
\end{equation}
Each evaluation of the EGA requires $N_{P}\times N_E$ evaluations of the forward model and $N_{P}\times N_E\times \min(d_{\theta}, d_{y})$ linearized PDE solutions corresponding to the direct or adjoint sensitivity of the forward model. Repeated evaluations of FIM at different constitutive parameters and design variables become a bottleneck in the design optimization. To remove this bottleneck, we use a fast-to-evaluate neural network surrogate of the FIM, denoted as $\widetilde{\textrm{FIM}}(\theta, z; w)$ with trainable weights $w\in\R^{d_{w}}$, to replace the FIM evaluations in EGA. We refer to the resulting surrogate approximation of the EGA as the \textit{estimator based on surrogate FIM} (ESFIM).

The surrogate is formulated to accommodate key features of FIMs, namely symmetric positive semi-definiteness and potentially extreme spectrum spreads. We consider using a neural network to emulate the lower-triangular components of the logarithmic regularized FIM:
\begin{gather*}
      B(\theta, z) \coloneqq \textrm{logm}(\textrm{FIM}(\theta, z) + \epsilon I),\quad \epsilon>0,\\
      \textrm{NN}(\theta, z;w) \approx \begin{bmatrix}
         B_{11} & B_{22} & \dots &
         B_{d_{\theta} d_{\theta}} &
         \sqrt{2}B_{12}& \sqrt{2}B_{13} & \cdots
    \end{bmatrix}^{\top}, 
\end{gather*}
where logm$(\cdot)$ is the matrix logarithm, $\epsilon$ is a small regularization constant, and $\textrm{NN}:\R^{d_{\theta}}\times \R^{d_{z}}\to\R^{d_{\theta}(d_{\theta}+1)/2}$ is the neural network predictions of the components of $B(\theta, z)$. The scaling by $\sqrt{2}$ of the off-diagonal term is introduced to be consistent with the scaling in the Frobenius matrix norm. The neural network can be trained to improve its FIM prediction quality, for example, by using the mean relative Frobenius norm of the prediction error as the loss function. The training data can be generated by evaluating the FIM at samples of constitutive parameters and design variables. Once the neural network is trained, it can be used to predict FIMs:
\begin{gather*}
    \begin{bmatrix}
         \widetilde{B}_{11} & \widetilde{B}_{22} & \dots &
         \widetilde{B}_{d_{\theta} d_{\theta}} &
         \sqrt{2}\widetilde{B}_{12}& \sqrt{2}\widetilde{B}_{13} & \cdots
    \end{bmatrix}\coloneqq \textrm{NN}(\theta, z;w)^{\top},\\
    \widetilde{\textrm{FIM}}(\theta, z; w) \coloneqq \textrm{expm}\left(\widetilde{B}(\theta, z;w)\right),
\end{gather*}
where expm$(\cdot)$ is the matrix exponential. The surrogate predictions of the FIM are then used in the evaluations of ESFIM by replacing \textrm{FIM} with $\widetilde{\textrm{FIM}}$ in \cref{eq:multiple_experiments_local_gaussian}, and subsequently in \cref{eq:approximate_ig,eq:eig_gauss}.

Since the surrogate $\textrm{FIM}$ is trained on random designs, the optimized designs found by ESFIM are likely to outperform random designs and perform similarly or worse than those found by EGA. The dominant computational cost of ESFIM-based optimal design arises from surrogate construction, which is independent of the number of experiments, $N_E$. Consequently, ESFIM enables optimization of experimental design for a large number of experiments, achieving a greater reduction in parametric uncertainty than random designs at a much lower computational cost than EGA. This advantage of ESFIM is demonstrated in our numerical examples.

\subsection{Nested Monte Carlo Estimator}\label{subsec:NMCE}

Lastly, we introduce the baseline method for EIG estimation: the nested Monte Carlo estimator. When the likelihood $\pi_{Y(z)\mid\Theta}(\cdot\mid\theta)$ is tractable to evaluate and generative, i.e., sampling $y\sim\pi_{Y(z)\mid\Theta}(\cdot\mid\theta)$ is straightforward, one often resorts to the likelihood-based mutual information form of the EIG as in \cref{lem:eig_equivalence}:
\begin{equation}\label{eq:eig_nmc}
    \textrm{EIG}(z)=\mathbb{E}_{\theta\sim\pi_{\Theta}}\left[\mathbb{E}_{y\sim \pi_{Y(z)\mid\Theta}(\cdot\mid\theta)}\left[\log\left(\frac{\pi_{Y(z)\mid\Theta}(y\mid \theta)}{ \pi_{Y(z)}(y)}\right)\right]\right].
\end{equation}
Here, the data marginal distribution $\pi_{Y(z)}$ is also defined through an expectation as in \cref{eq:data_marginal}. Using the Monte Carlo (MC) method to estimate the expectations in \cref{eq:eig_nmc}, we arrive at a nested Monte Carlo estimator (NMCE) \cite{Ryan2003Estimating} given by:
\begin{equation}\label{eq:eig_nmc_estimator}
    \begin{gathered}
    \theta^{(i)} \iid \pi_{\Theta},\quad y^{(i)}\iid \pi_{Y(z)|\Theta}(\cdot\mid \theta^{(i)}),\quad \theta^{(ij)}\iid \pi_{\Theta},\quad 1\leq i\leq N_O,\quad 1\leq j\leq N_I,\\
    \textrm{NMCE}(z; N_O, N_I)\coloneqq \frac{1}{N_O}\sum_{i=1}^{N_O}\log \left(\frac{\pi_{Y(z)\mid\Theta}(y^{(i)}\mid \theta^{(i)})}{ \frac{1}{N_I}\sum_{j=1}^{N_I}\pi_{Y(z)|\Theta}\left(y^{(i)}\mid\theta^{(ij)}\right)}\right).
    \end{gathered}
\end{equation}

The NMCE is biased for a finite $N_I$ but is asymptotically unbiased and consistent \cite[Theorem 1, Appendix B]{Rainforth2018Nested}. The computational effort required to achieve a mean squared error of $\epsilon^2$ is shown to be $O(\epsilon^{-3})$ for continuously differentiable log-likelihood, which is slower than the standard MC rate of $O(\epsilon^{-2})$ \cite[Appendix G]{Rainforth2018Nested}. Each evaluation of the NMCE requires $N_O(N_I+1)$ solutions of the balance equation in \cref{subsec:material_response}, and achieving a desirable level of error tolerance in the EIG estimation can be prohibitively costly. Typically, the number of inner MC samples, $N_I$, needs to be large because the likelihood can be concentrated when the data are informative. Additional techniques, such as importance sampling, are often necessary when evaluating the NMCE; see \cite[Section 3.1]{Huan2024Optimal}.

\subsection{Discussion}\label{subsec:discussion_estimators}

We close this section by highlighting two central aspects of BOED, for learning history-dependent constitutive models, that make the EGA and ESFIM more robust and practical than the NMCE baseline in design optimization.

\paragraph{Cost-Accuracy Trade Off}
For history-dependent constitutive models, the dominant cost is the repeated solution of the balance equation over the loading history. In this setting, NMCE can be prohibitively expensive because it places this cost inside a nested Monte Carlo loop. EGA removes this bottleneck by replacing this nested loop with a single loop of FIM evaluations, and ESFIM bypasses this cost entirely after a surrogate FIM is trained. It remains to be shown that EGA and ESFIM produce good designs despite the bias they introduce into the EIG estimation, and this is addressed empirically through numerical examples in the next two sections.

\paragraph{High-Dimensional Data}
High-resolution images can extract large amounts of information relevant to parameter identification, provided that the design excites the relevant heterogeneous responses. EGA and ESFIM remain viable in this regime because their dominant costs, namely linearized PDE solves and surrogate training, are independent of the data dimension. By contrast, \cref{ssec:ve_nmce} of our numerical example shows that the NMCE baseline is intractable to evaluate even with importance sampling due to extreme likelihood concentration induced by the image data.

\section{Application: Learning Linear Viscoelasticity}\label{sec:results_linear}

In this section, we present a numerical study on designing uniaxial tests, as described in \cref{subsec:material_response}, to learn the constitutive parameters of the linear viscoelastic model described in \cref{subsec:viscoelastic}. First, in \cref{subsec:viscoelastic_setup} we specify the design problem's settings, including the prior and observation operators. We then present results on design optimization for a single experiment and for a batch of experiments in \cref{subsec:ve_single_experiment,subsec:ve_batched_design}, respectively, which are overviewed as follows.
\begin{enumerate}[label=(\roman*)]
    \item Surrogate FIM sample complexity that reaches below 3\% relative error in log-FIM.
    \item Visualizations of the EGA- and ESFIM-optimized designs with load/unload patterns and tilted elliptical holes that agree with physical intuition about designs that maximize information.
    \item Quantitative interpretations of the optimized designs compared to random designs through credible interval size (averaged 47\% uncertainty reduction) and marginal EGA values (10\% improvement in learning relaxation times).
    \item A Spearman correlation of $0.67$ between EGA and stress-state entropy, indicating that designs favored by EGA tend to induce more diverse stress states, but not strongly enough for the latter to serve as a surrogate utility.
\end{enumerate}
Finally, in \cref{ssec:ve_nmce} we demonstrate the intractability of NMCE and the suitability of EGA for this design problem. The implementation details are provided in \cref{app:numerics}.

\subsection{Settings}\label{subsec:viscoelastic_setup}

\paragraph{Experimental design}

\begin{figure}
    \centering
    \includegraphics[width=0.58\linewidth]{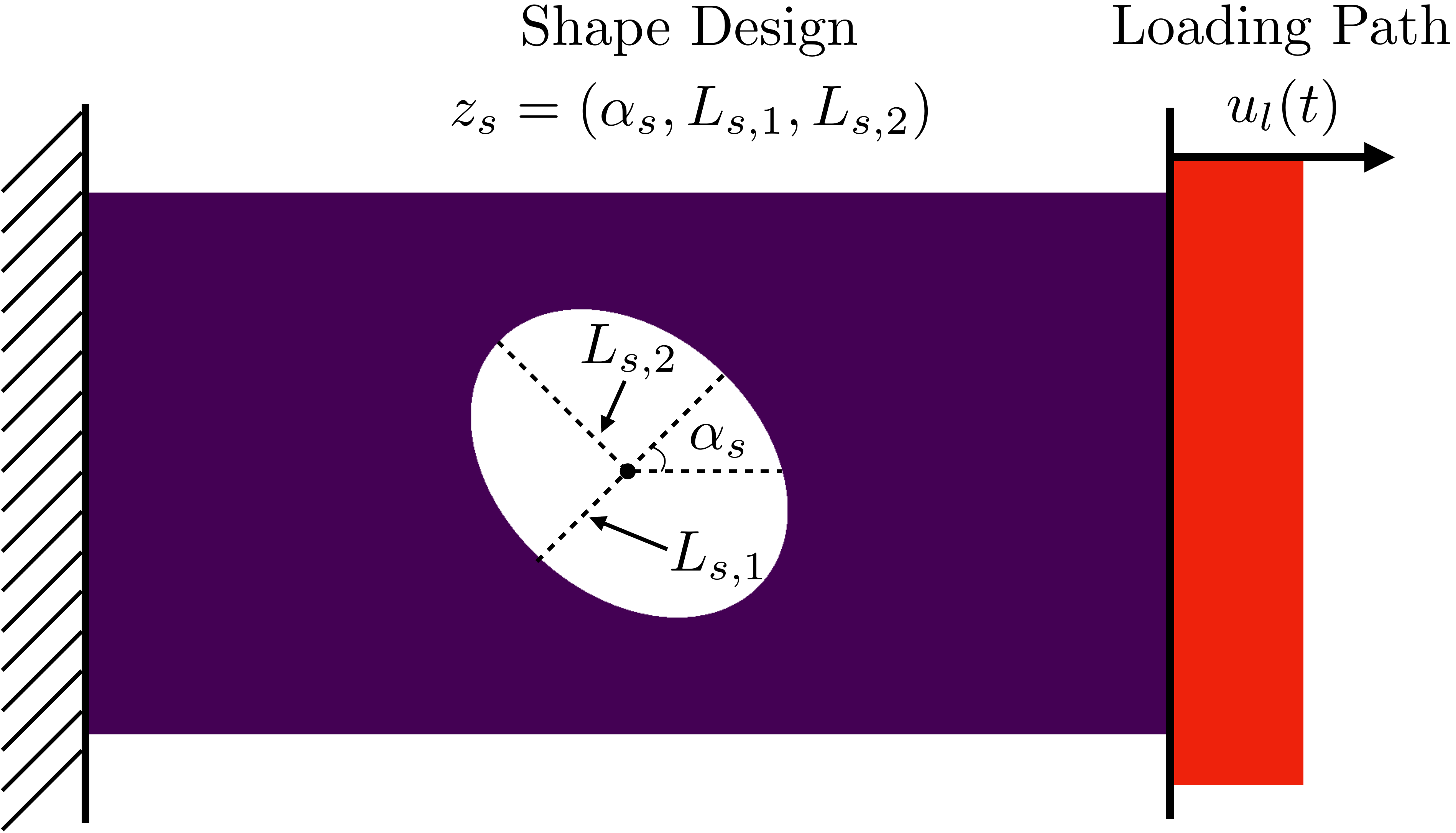}\hspace{0.05\linewidth}\includegraphics[width=0.27\linewidth]{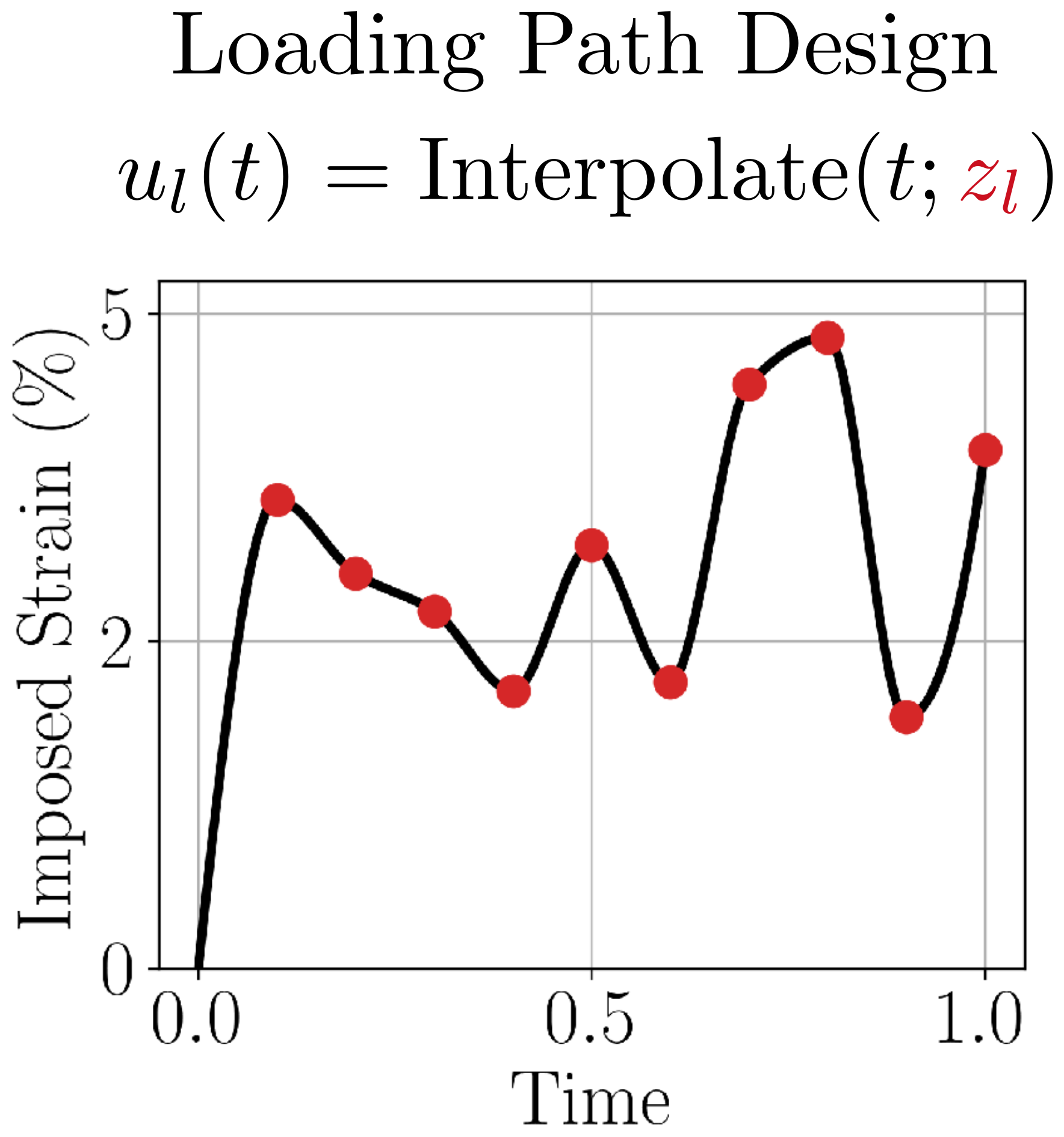}
    \caption{An example of the uniaxial testing setup and the design variables. (\emph{Left}) The setup and the specimen geometry design. (\emph{Right}) The loading path design.}
    \label{fig:viscoelastic_setup}
\end{figure}

The uniaxial testing setup described in \cref{subsec:material_response} is used with the following specifications shown in \cref{fig:viscoelastic_setup}. For shape design, we consider a specimen with normalized widths $L_1 = 2$ and $L_2 = 1$. The hole $\Omega_{H}$ in the specimen has an elliptical shape, parametrized by one of the principal angles $\alpha_s\in[0, \pi/2]$. The principal axis lengths, denoted as $L_{s, 1}$ and $L_{s, 2}$, are each chosen in the interval $[0.1, 0.35]$. Together, there are three shape design variables $z_s=(\alpha_s, L_{s, 1}, L_{s, 2})$. For the loading path design, a normalized total test time of $T=1$ is considered, and $10$ loading path control values are used. Specifically, the $10$ components of $z_{l}\in[0, 0.1]^{10}$ are taken as loading values on an equally spaced grid of $10$ points in time-interval $[0.1,1.0]$, and interpolated, starting from zero loading at time $t=0.$ The interpolation is performed using a monotonic piecewise cubic Hermite interpolating polynomial (PCHIP). The choice of range of $z_{l}$ leads to a maximum imposed strain of $5\%$. In \cref{fig:viscoelastic_setup}, this experimental setup and the design variables are visualized.\hfill$\diamond$

\paragraph{Reparameterization And The Prior Distribution}

We consider the linear viscoelastic constitutive model in \cref{subsec:viscoelastic}, simplified under the orthotropic and plane-stress assumptions with two viscous branches. This reparametrization consists of two steps. First, a standard normal prior distribution, i.e., $\pi_{\Theta} = \mathcal{N}(0, I)$, is prescribed. Then, the physical parameters, denoted as $\vartheta$, are obtained via the following transformation 
\begin{equation}\label{eq:uniform_transform}
    \vartheta_i = a_i + \frac{(b_i-a_i)}{2}\left(1+\textrm{erf}\left(\frac{\theta_i}{\sqrt{2}}\right)\right),\quad \vartheta_i\sim \mathcal{U}(a_i, b_i).
\end{equation}
where $\{a_i\}$ and $\{b_i\}$ are the lower and upper bounds of the uniform distributions of the physical parameters. The physical parameters and their ranges are specified in \cref{tab:prior_bound_viscoelastic}.

Here, we consider a parametrization for the elastic and viscous tensors $\{\mathbb{C}_i\}$ and justify the associated ranges. Throughout, Voigt notation uses the in-plane ordering $[11,22,12]$, and a single material orientation is shared across all branches:
\begin{equation*}
    \mathbb{C}_i = \mathbb{T}(\alpha_c)\mathbb{S}_i^{-1}\mathbb{T}(\alpha_c)^{\top},
    \quad
    \mathbb{S}_i =
    \begin{bmatrix}
        1/E^{(i)}_{1} & \nu_{21}^{(i)}/E^{(i)}_{2} & 0\\
        \nu_{12}^{(i)}/E^{(i)}_{1} & 1/E_2^{(i)} & 0\\
        0 & 0 & 1/G_{12}^{(i)}
    \end{bmatrix},
    \quad 0\le i\le 2,
\end{equation*}
where $\mathbb{T}(\alpha_c)$ is the in-plane rotation by angle $\alpha_c$, and $\mathbb{S}_i$ denotes the material-frame plane-stress orthotropic compliance with four independent components per branch among $E^{(i)}_{1}$, $E^{(i)}_{2}$, $G_{12}^{(i)}$, $\nu_{12}^{(i)}$ and $\nu_{21}^{(i)}$. The material axis orientation is uniformly distributed as $\alpha_c \sim \mathcal{U}(-\pi/4,\pi/4)$. This range of orientations is chosen because we hypothesize that the optimal ellipse is likely tilted, which creates an interesting experimental design problem.

We set uniform priors for the logarithm of the modulus $E^{(0)}_{1}$ in the equilibrium branch and the ratio between $E^{(0)}_{2}$ and $E^{(0)}_{1}$,
\begin{equation*}
    \log E^{(0)}_{1} \sim \mathcal{U}(2, 3.5),\quad r_E\sim \mathcal{U}(0.2, 1),\quad E^{(0)}_{2}\coloneqq r_EE^{(0)}_{1}. 
\end{equation*}
The range of $E_1^{(0)}$ is chosen so that the reaction force in \cref{eq:force_data} is, on average, of order 1. The viscous modulus $E^{(i)}_1$ and $E^{(i)}_2$ for $i=1,2$ are related to the equilibrium ones via the total fractions $f_1, f_2$:
\begin{equation*}
    f_j\sim \mathcal{U}(0.2, 0.8),\quad E^{(i)}_j = f_j w_j^{(i)} E^{(0)}_j,\quad 1\leq i,j\leq 2,
\end{equation*}
where the weights are given by
\begin{gather*}
    w_j^{(1)}\sim \calU(0, 1),\quad w_j^{(2)}\coloneqq 1- w_j^{(1)},\quad 1\leq j\leq 2.
\end{gather*}
For all branches, the shear moduli $\{G^{(i)}\}$ are defined through a single ratio $r_G$
\begin{equation*}
    r_G\coloneqq G_{12}^{(i)}/\sqrt{E^{(i)}_{1}E^{(i)}_{2}}\sim \mathcal{U}(0.2, 0.8).
\end{equation*}
The Poisson's ratios $\{\nu_{12}^{(i)}\}$ and $\{\nu_{21}^{(i)}\}$ are determined by a single ratio $r_{\nu}$:
\begin{gather*}
    r_{\nu} \coloneqq \nu_{12}^{(i)}\nu_{21}^{(i)}\sim \mathcal{U}(0, 0.2), \quad \nu_{12}^{(i)} \coloneqq \sqrt{r_{\nu}E_1^{(i)}/E_2^{(i)}},\quad \nu_{21}^{(i)}\coloneqq \sqrt{r_{\nu}E_2^{(i)}/E_1^{(i)}}.
\end{gather*}
For this case study, the relaxation times are set to distinct scales, i.e., the time between control points ($\sim 0.1$) and the experiment time window ($\sim 1$), to ensure identifiability of the branches:
\begin{equation*}
    \log \tau_1\sim\mathcal{U}(-3.4, -1.2),\quad \log \tau_2\sim\mathcal{U}(-1.2, 1).
\end{equation*}
\hfill$\diamond$

\begin{table}[htb]
\centering
\renewcommand{\arraystretch}{1.3}
\begin{tabular}{|c|c|c|}\hline
Physical Parameter & Description & Bound \\\hline\hline 
$\log E_1^{(0)}$ & Log.\ of equilibrium modulus in $1$-dir.& (2.0, 3.5) \\
$r_E $ & Equilibrium modulus anisotropy, $E_2/E_1$ & (0.2, 1) \\
$r_{G}$ & Normalized shear, $G/\sqrt{E_1E_2}$ & (0.2, 0.8) \\
$r_{\nu}$ & Poisson's ratio product, $\nu_{12}\nu_{21}$ & (0.0, 0.2) \\
$\alpha_c$ & Material axis angle & (-$\pi$/4, $\pi$/4) \\
$f_i$, $i=1,2$ & Total relaxed fraction in dir.\ $i$ & (0.2, 0.8) \\
$w^{(1)}_j$, $j=1,2$ & Branch-1 share of $f_i$ & (0, 1)\\
$\log \tau_1$ & Log.\ of relaxation time, fast & (-3.4, -1.2) \\
$\log \tau_2$ & Log.\ of relaxation time, slow & (-1.2, 1) \\\hline
\end{tabular}
\renewcommand{\arraystretch}{1.0}
\caption{The bounds for the uniform distributions of the physical parameters in the re-parameterized linear viscoelastic constitutive model.}
\label{tab:prior_bound_viscoelastic}
\end{table}

\begin{figure}
    \centering
    \begin{minipage}{0.3\linewidth}
        \includegraphics[width=\linewidth]{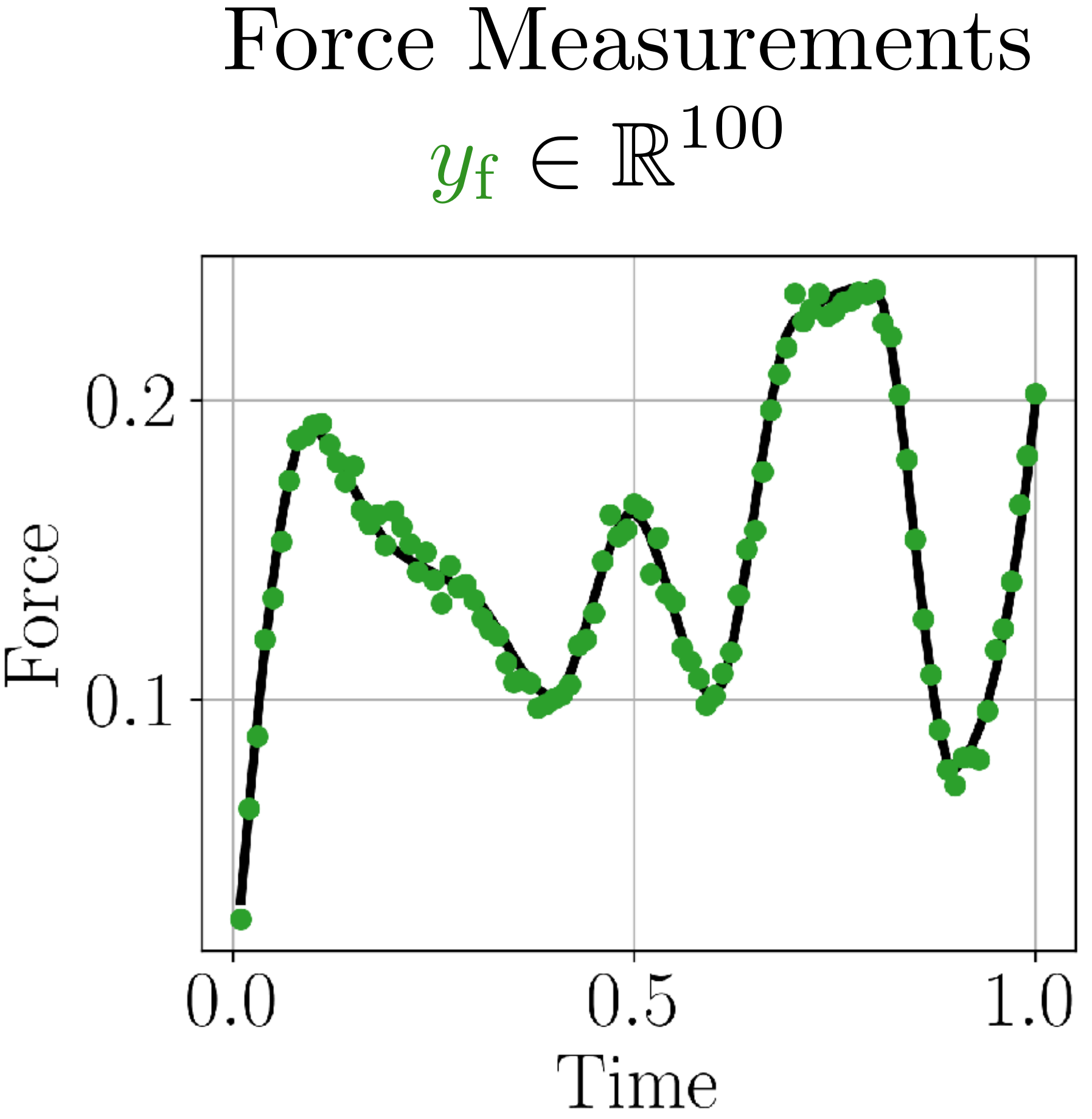}
    \end{minipage}
    \hspace{0.05\linewidth}
    \begin{minipage}{0.62\linewidth}
        \includegraphics[width=\linewidth]{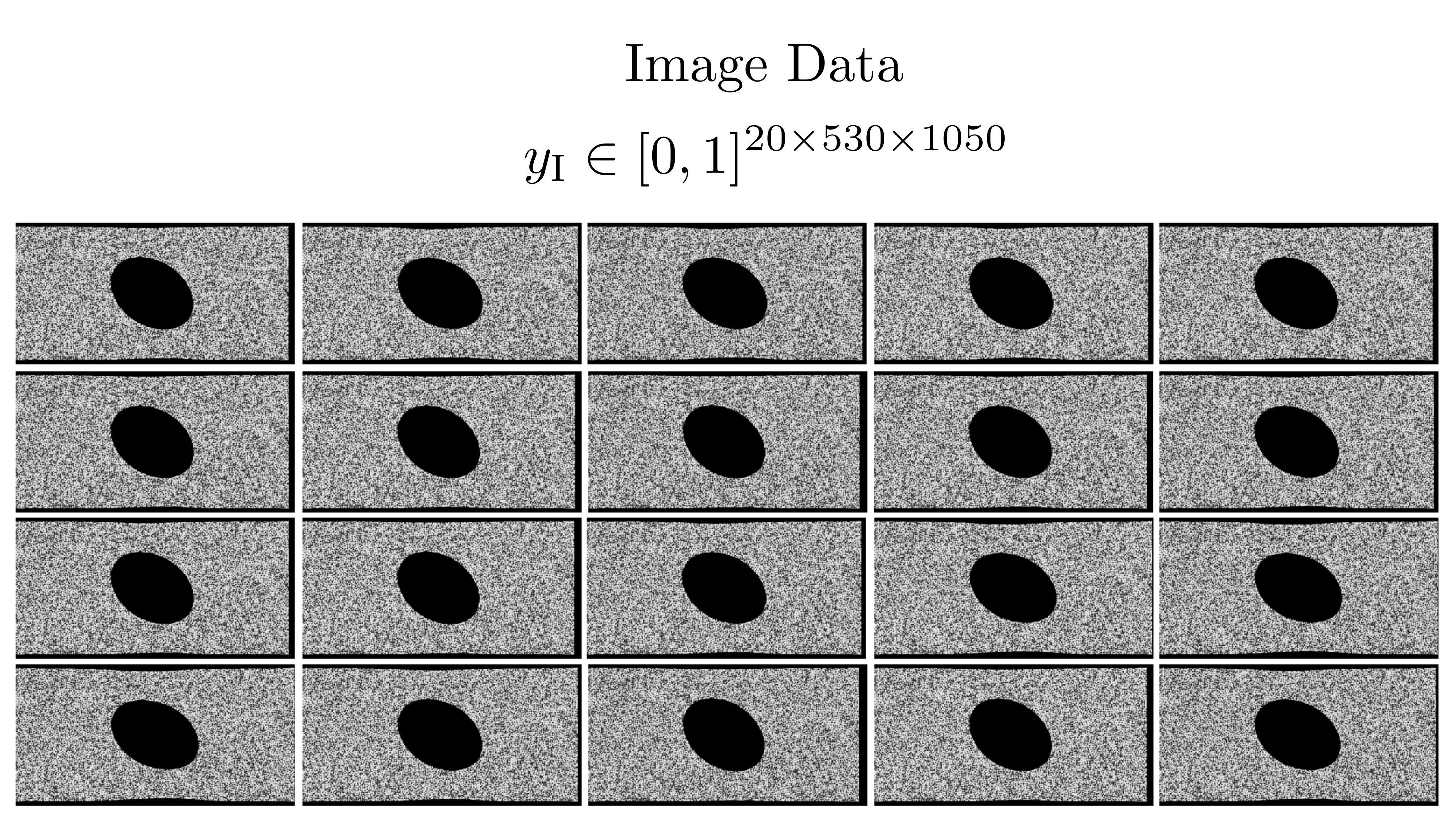}
    \end{minipage}
    
    \caption{An example of simulated force and image data for learning linear viscoelasticity. This simulated dataset is created via the experimental setup in \cref{subsec:viscoelastic_setup,fig:viscoelastic_setup}, and the constitutive parameters are drawn from the prior. Note that the resolution of the image data visualized here is much lower than that of the actual data used in the numerical study; see an image snapshot with high resolution in \cref{fig:high_res_images}. The material deformation is discernible yet small due to the narrow range of the imposed strain ($<5\%$) suitable for linear viscoelasticity.}
    \label{fig:linear_data}
\end{figure}

\paragraph{Observation Operator}
We use a combination of image data and force data for our numerical study. For image data, 20 snapshots are taken, equally spaced in time. We assume a pixel density of 500, with a field of view of height 1.06 and width 2.1, resulting in an image size of $530\times1050$ (height $\times$ width). A fixed speckle pattern is used across different experiments, with an averaged speckle radius of $6\times 10^{-3}$ (6 pixels across a speckle on average), a contrast of 0.8, and coverage of approximately 50\%. We assume $\sigma_{\eta}=0.02$. The image observation operator is defined in \cref{subsec:observation} and detailed in \cref{app:dic}, and the image generation procedure is detailed in \cref{app:image_generation}. For the force data, 100 values are collected, equally spaced in time. The force data at time $t_i$ is modeled as the outward force at the right boundary where the loading is applied:
\begin{equation} \label{eq:force_data}
    y_{{\textrm{f}},i} = \int_{\{2\}\times [0, 1]} [P(X, t_i)]_{11} \dd X + \eta_{\textrm{f}},\quad \eta_{\textrm{f}}\sim \mathcal{N}(0, 2.5^{-5}),
\end{equation}
where a signal-to-noise ratio of about 0.5\% is used. An example of simulated experimental data is shown in \cref{fig:linear_data}.\hfill$\diamond$

\subsection{Designing a Single Experiment}\label{subsec:ve_single_experiment}

We maximize EGA and ESFIM to obtain designs of a single experiment. Here, we briefly describe the procedures for EGA evaluation, surrogate FIM training, and utility maximization. Implementation details are provided in \cref{app:numerics}

\paragraph{EGA evaluation} At each parameter sample and design candidate, we compute the Jacobian of the parameter-to-observable map, $\partial_{\theta}\mathcal{F}_{\textrm{pto}}$ using forward sensitivity solves. This requires 11 linear, time-dependent PDE solves per parameter sample at each design candidate, in addition to the one required for forward-model evaluations.

\begin{figure}[htb]
    \centering
    \scalebox{0.9}{
    \begin{tabular}{c c}
        \hspace{0.09\linewidth}\makecell{Surrogate FIM Sample Complexity\\
        for Linear Viscoelasticity} & \hspace{0.09\linewidth}\makecell{Surrogate FIM Sample Complexity \\ for Nonlinear Viscoelasticity}  \\
    \includegraphics[width=0.42\linewidth]{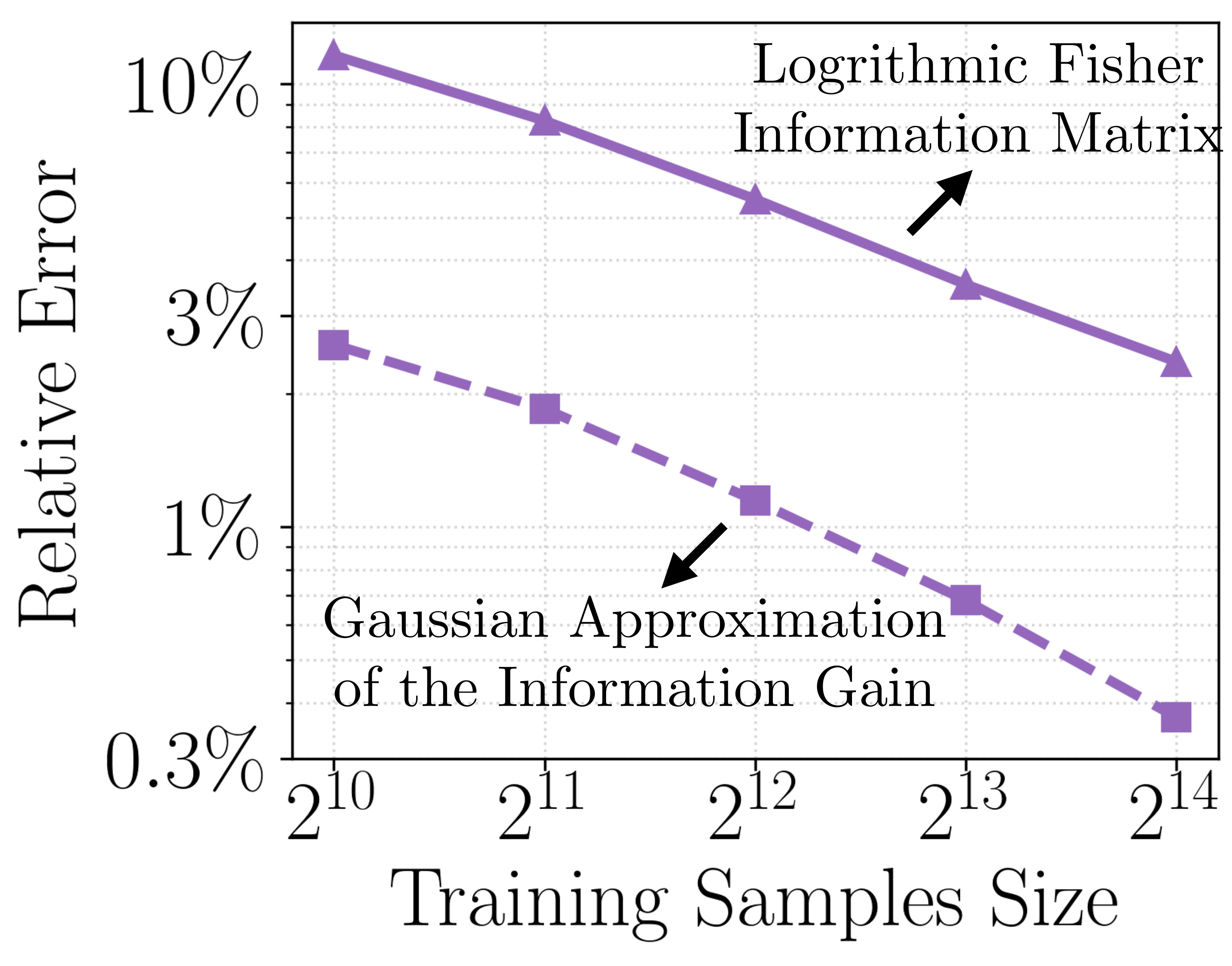} & \includegraphics[width=0.42\linewidth]{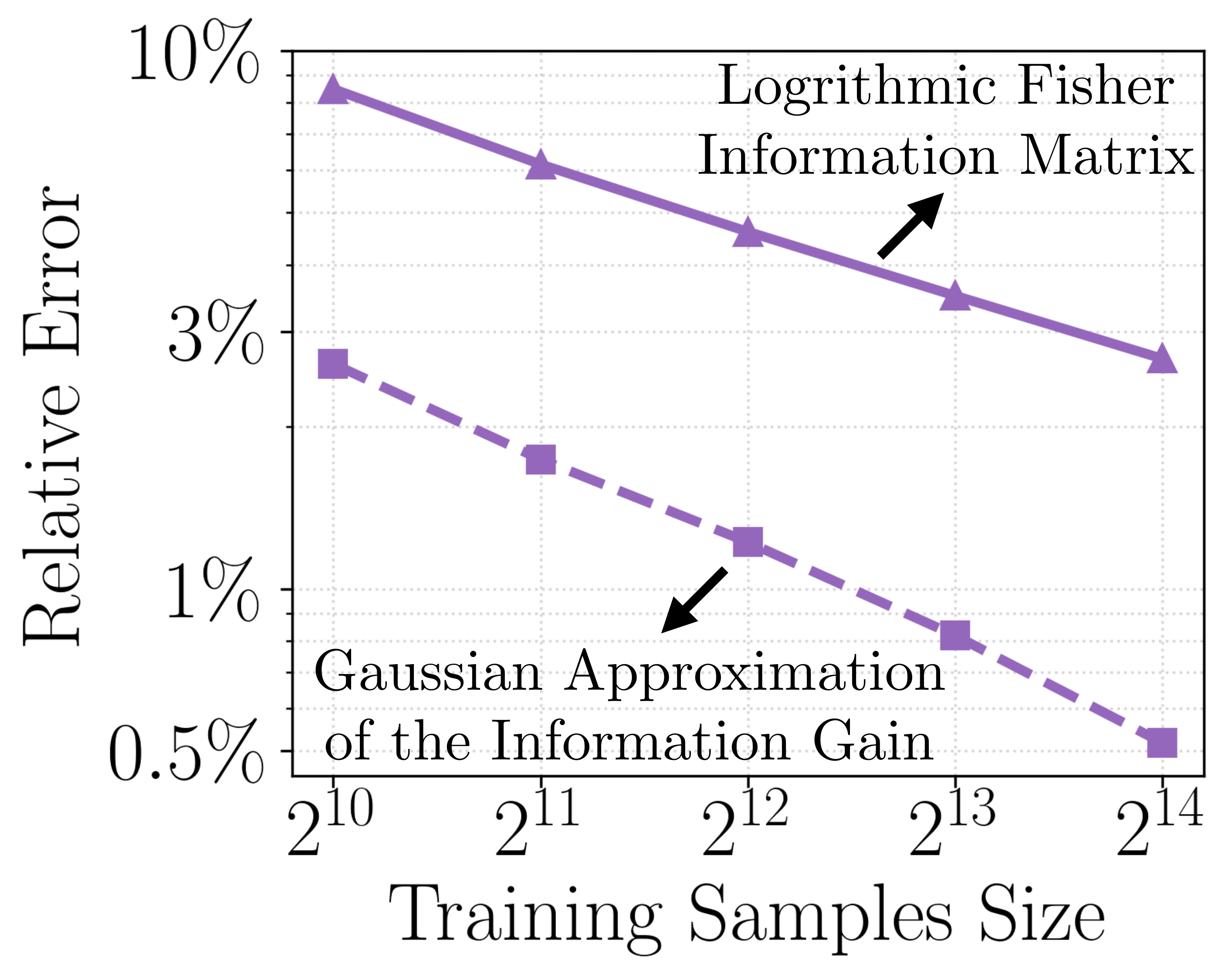}
    \end{tabular}
    }
    \caption{The surrogate FIM testing accuracy as a function of training sample size for designing uniaxial tests of linear and nonlinear viscoelastic materials in \cref{sec:results_linear,sec:results_nonlinear}, respectively. The solid and dashed lines indicate the mean relative errors for predicting the log.\ FIM (measured in the Frobenius norm) and the approximate information gain, respectively. The surrogate prediction of the approximate information gain is obtained from the surrogate FIM predictions using \cref{eq:approximate_ig}.}
    \label{fig:linear_testing_accuracy}
\end{figure}

\paragraph{Surrogate FIM} The surrogate model is a residual network with 6 hidden layers and Gaussian Error Linear Unit (GELU) activation, with a layer width of 200. The surrogate is trained on data consisting of FIMs computed at 16,384 parameter and design variable samples generated via QMC. The surrogate achieved a mean relative accuracy of over 97\% in predicting the log.\ FIM on a testing dataset of 4,096 samples. Moreover, the surrogate achieves a mean relative accuracy of over 99.6\% in predicting the approximate information gain in \cref{eq:approximate_ig} on the test set. On the left panel of \cref{fig:linear_testing_accuracy}, we visualize the surrogate FIM sample complexity, i.e., testing error as a function of training sample size. A standard MC rate of convergence at $N_{\textrm{train}}^{-1/2}$ is observed.\hfill$\diamond$

\paragraph{Design Optimization} We use EGA and ESFIM to approximate the EIG in the design optimization of \cref{eq:eig_maximization}, subject to a bound constraint on the design variables and without additional regularization. The EGA is maximized using the Bayesian optimization method with $N_P=128$. The ESFIM is maximized using L-BFGS-B with an ensemble of initial starts.  We note that the data generation and training costs for constructing the surrogate FIM model are similar to the total costs of Bayesian optimization for EGA maximization. The ESFIM maximization costs on a Graphics Processing Unit (GPU) are relatively negligible compared to those of EGA maximization.\hfill$\diamond$

Now we present and analyze the EGA- and ESFIM-optimized designs, both qualitatively and quantitatively.

\subsubsection{Optimized Designs}

\begin{figure}[htb]
    \centering
    \begin{minipage}{0.42\linewidth}
    \centering
    \includegraphics[width=\linewidth]{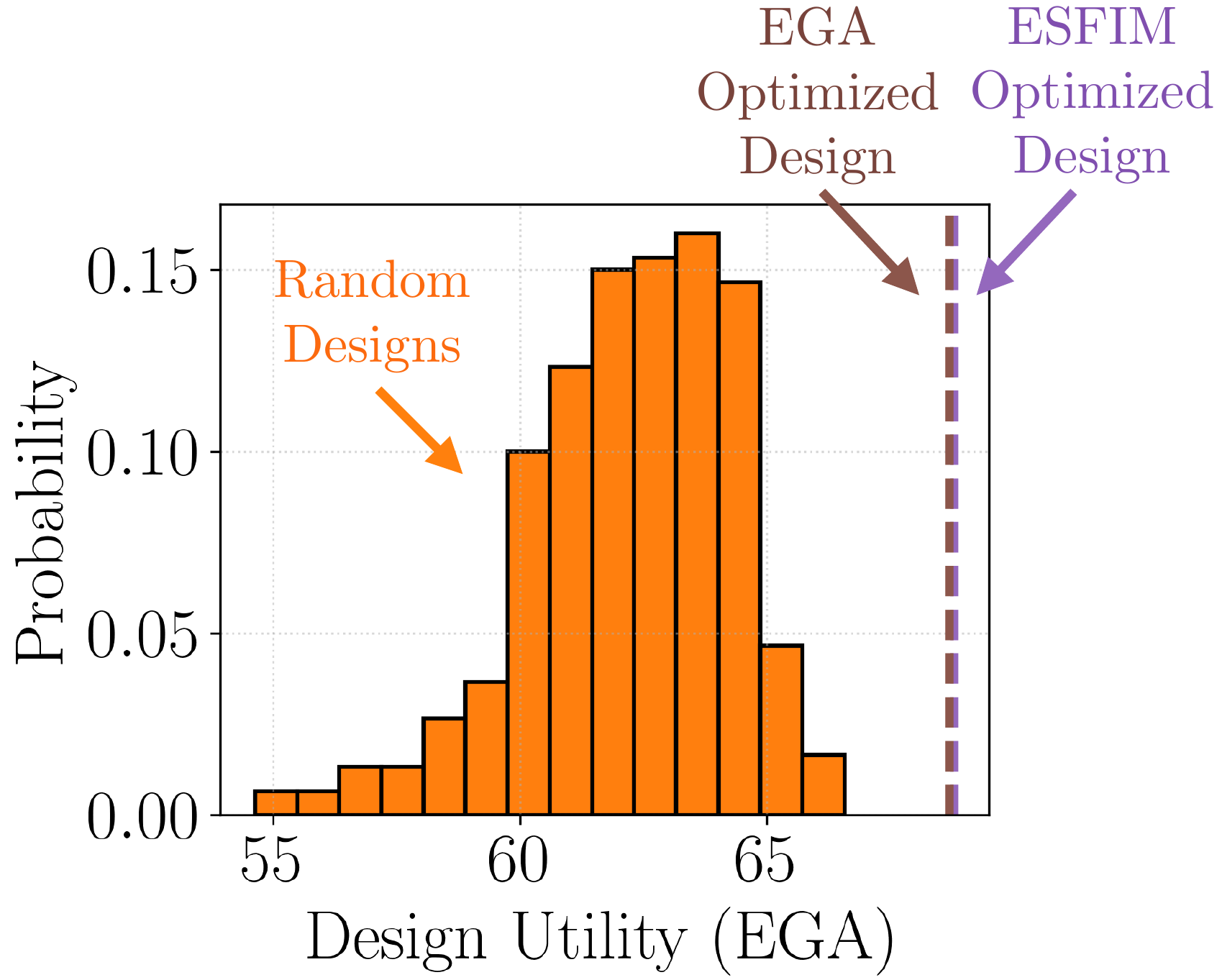}
    \end{minipage}
    \begin{minipage}{0.27\linewidth}
    \centering
        \begin{tabular}{c}
           \makecell{\small EGA-Optimized Design} \\ \includegraphics[width=\linewidth]{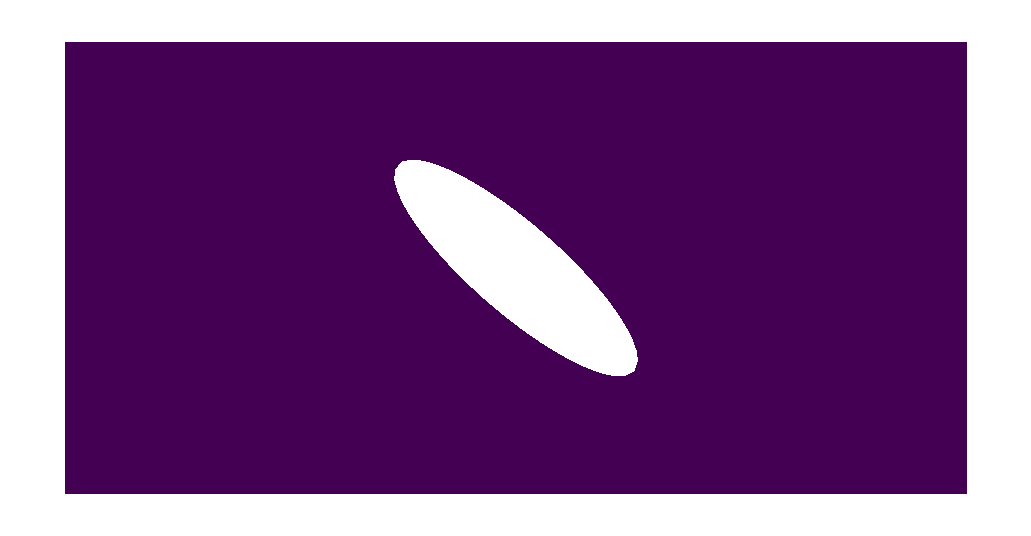}\\
            \includegraphics[width=0.9\linewidth]{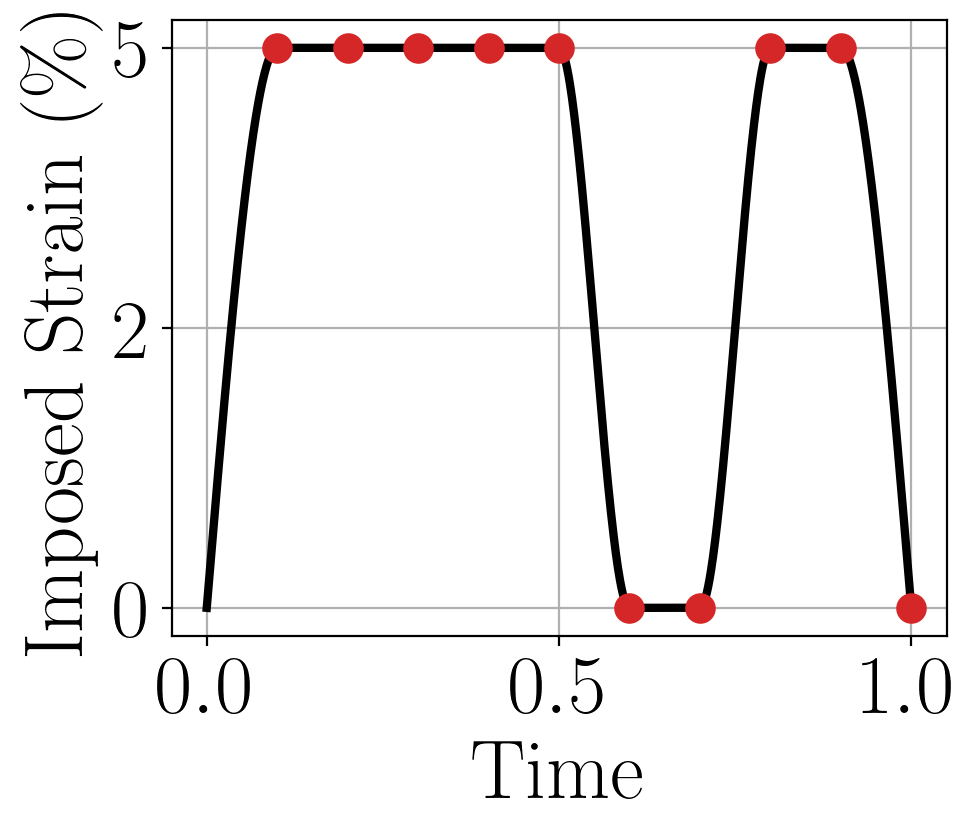} 
        \end{tabular}
    \end{minipage}
    \begin{minipage}{0.27\linewidth}
    \centering
        \begin{tabular}{c}
           \makecell{\small ESFIM-Optimized Design} \\ \includegraphics[width=\linewidth]{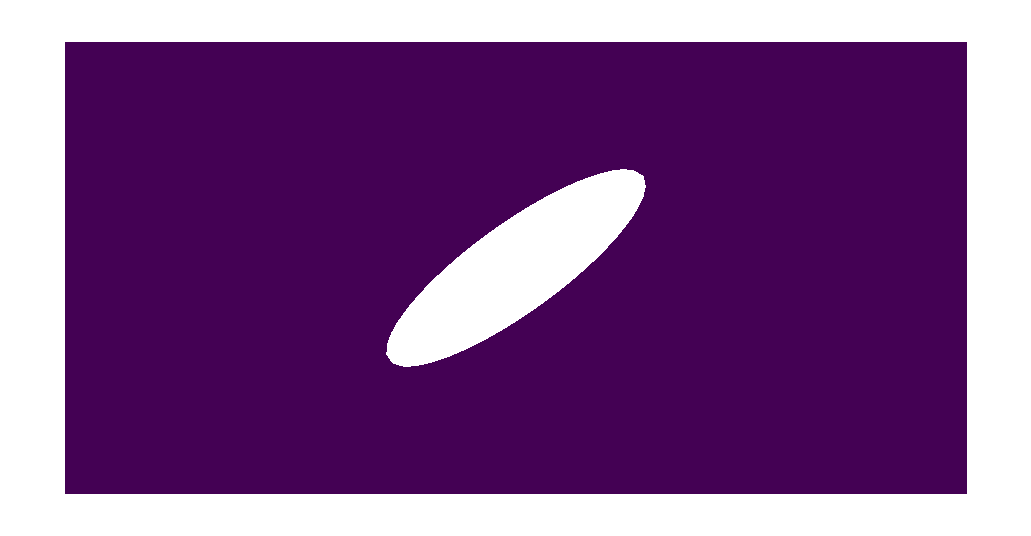}\\
            \includegraphics[width=0.9\linewidth]{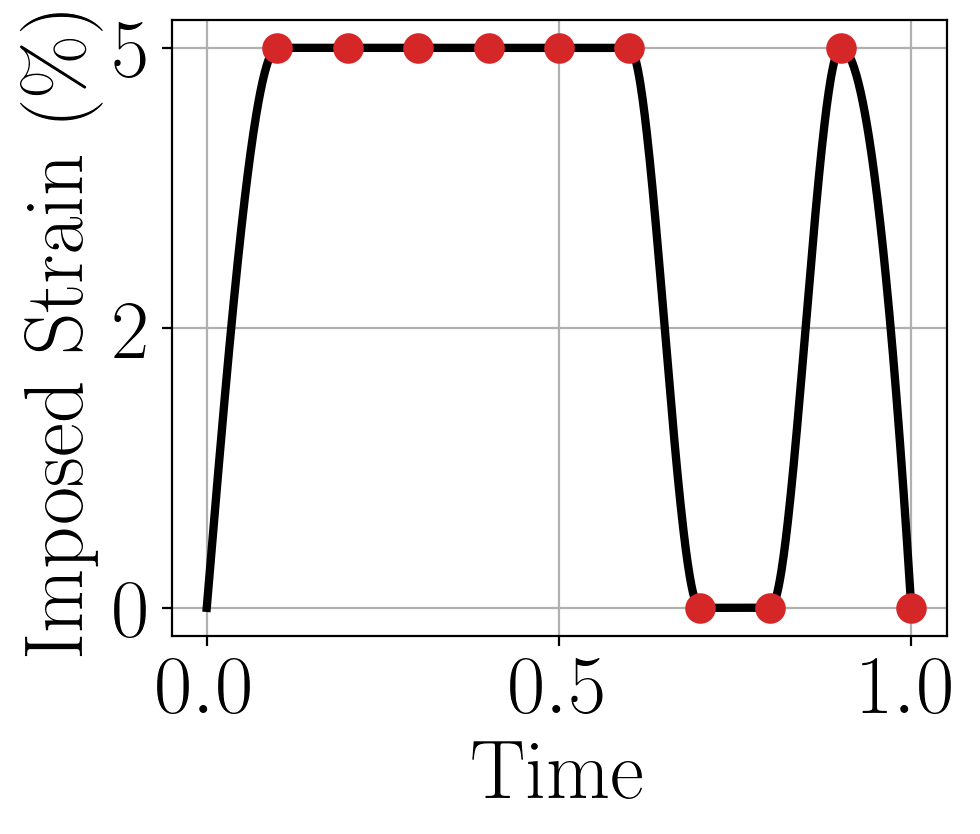} 
        \end{tabular}
    \end{minipage}

    \caption{The optimized designs of uniaxial test for reliable learning of linear viscoelasticity found by EGA and ESFIM maximization. (\emph{Left}) Comparison of the EGA values at random, EGA-optimized, and ESFIM-optimized designs. (\emph{Right}) Visualization of the optimized designs found via EGA maximization and ESFIM maximization.}
    \label{fig:linear_designs}
\end{figure}

The EGA-optimized design is shown in \cref{fig:linear_designs}. The loading path increases to the maximum strain at the first control point, holds for 6 consecutive control points, then rapidly unloads and holds for 2 control points, and then rapidly loads to the maximum strain and holds for 2 control points. The rapid loading and unloading, along with long holding periods, can be interpreted as maintaining a balance between learning the two relaxation times. The optimized shape of the elliptical hole has a maximum aspect ratio with $L_{s,1}=0.1$ and $L_{s,2}=0.35$, and an angle of $\alpha_s=0.27\pi$ (48.9\degree). This tilted shape is expected due to the prior on the material axis angle specified in \cref{tab:prior_bound_viscoelastic}. 

The ESFIM-optimized design is also shown in \cref{fig:linear_designs}. The loading path increases to the maximum strain at the first control point, holds for 6 consecutive control points, then rapidly unloads and holds for 2 consecutive control points, and then rapidly loads and unloads for the final two control points. The optimized shape of the ellipse hole has $L_{s,1}=0.35$ and $L_{s,2}=0.1$, and an angle of $\alpha_s=0.2\pi$ (35.9\degree). The ESFIM-optimized design can be interpreted similarly to the EGA-optimized design, with notable variations in hole orientation and holding duration. However, the resulting discrepancy between the two designs has a negligible effect on parameter identification, as indicated by their EGA values, which differ by less than 0.2\%.

We compare optimized designs with an ensemble of 300 designs randomly sampled from the uniform distribution within the design variable's bounds. The distribution of the EGA utility values for random designs is shown on the left panel of \cref{fig:linear_designs}, along with the utility values of the EGA- and ESFIM-optimized designs. Compared to random designs, both the EGA- and ESFIM-optimized designs yield, on average, 10\% higher EGA utility values. We note that the same set of random designs is used for the subsequent results in this section.

\subsubsection{Quantitative Interpretation: Credible Intervals of the Posteriors}

\begin{figure}[htb]
    \centering
    \begin{minipage}{0.8\linewidth}
        \includegraphics[width=\linewidth]{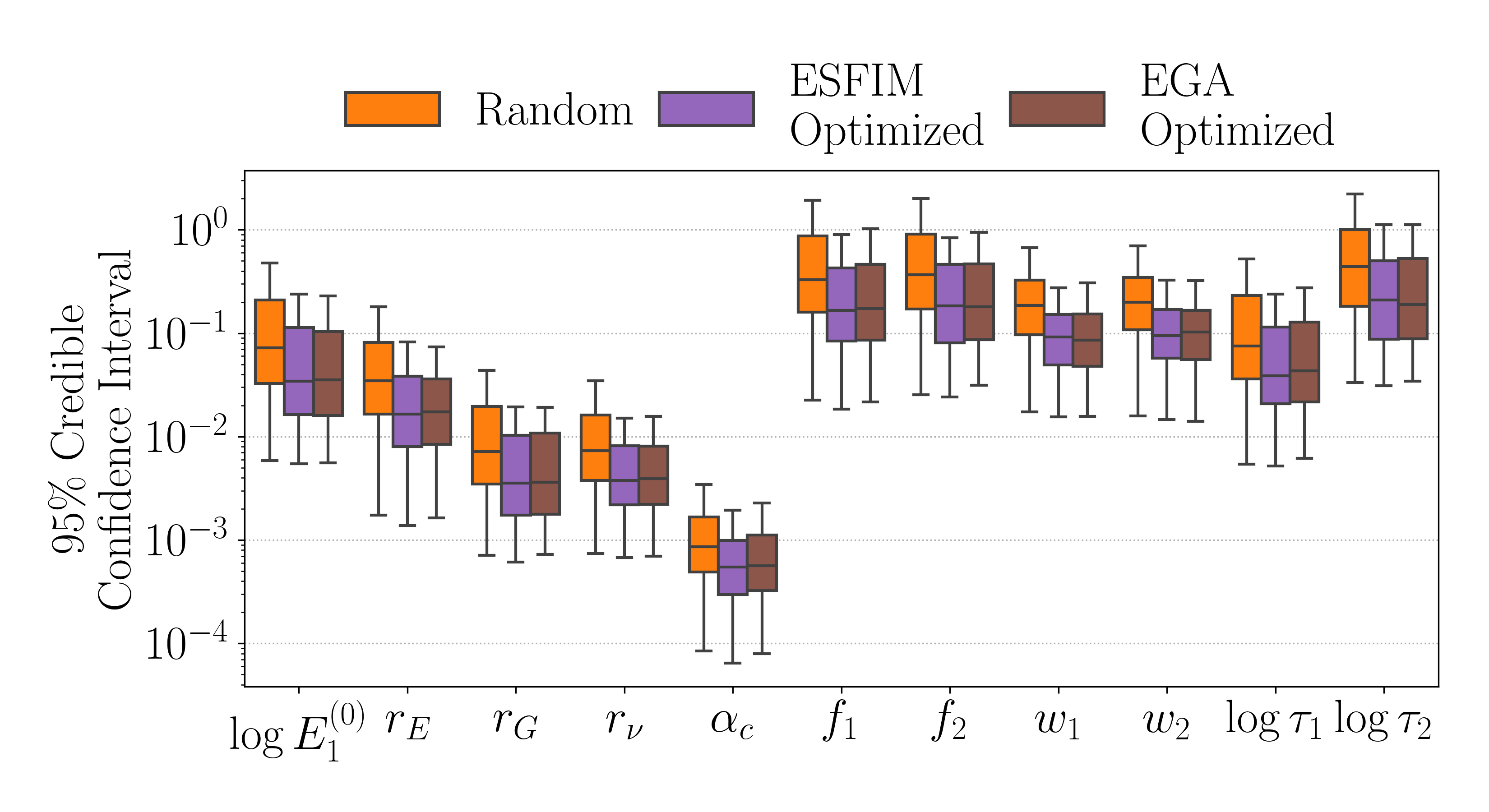}
    \end{minipage}
    \scalebox{0.9}
    {\renewcommand{\arraystretch}{2}
    \begin{tabular}{|c | c | c |c | c |c |c |c |c| c| c| c| c|}\hline
    \multicolumn{13}{|c|}{\makecell{Expected Reduction of the $95\%$ Credible Intervals \\ By Optimized Designs Relative  to Random Designs}}\\\hline\hline
    & $\log E_1^{(0)}$ & $r_E$ & $r_G$ & $r_{\nu}$ & $\alpha_c$ & $f_1$ & $f_2$ & $w_1$ & $w_2$ & $\log\tau_1$ & $\log\tau_2$ & Averaged\\\hline
    ESFIM & 55\% & 50\% & 48\% & 33\% & 48\% & 47\% & 44\% & 55\% & 51\% & 48\% & 42\% & 47\%\\\hline
    EGA & 54\% & 50\% & 50\% & 34\% & 52\% & 47\% & 44\% & 54\% & 51\% & 49\% & 44\% & 48\% \\\hline
    \end{tabular}
    \renewcommand{\arraystretch}{1.0}
    }
    \caption{Visualizations and statistics of the 95\% CIs for constitutive parameters of linear viscoelasticity inferred from experiments with random, ESFIM-optimized, and EGA-optimized designs. (\emph{Top}) The box plots of the 95\% CI sizes. The size distributions arise from the prior-based variation in the experimental outcomes. For random designs, the distributions also account for design variations. (\emph{Bottom}) The relative expected reduction of the 95\% CI sizes by optimized designs relative to random designs. The 95\% CI sizes are estimated for $\theta$, which is transformed as in \cref{subsec:viscoelastic_setup} from the physical parameters used as labels.}
    \label{fig:linear_ci}
\end{figure}

In \cref{fig:linear_ci}, we present the visualizations and statistics of the 95\% credible intervals (CIs) for the posteriors. The credible intervals are computed using the Gaussian approximation in \cref{subsubsec:gaussia_approx}. For each experimental design, we use 128 samples of the data-generating parameters, drawn via QMC, to capture the variation in the experimental outcomes. In the top panel of \cref{fig:linear_ci}, we visualize the distribution of 95\% CI sizes across prior-based variation in experimental outcomes. For random designs, the size distribution also accounts for design variations. We observe that the optimized designs consistently shift the distribution of the 95\% CI sizes toward lower values. In the bottom panel of \cref{fig:linear_ci}, we provide the relative expected reduction of the 95\% CIs by optimized designs compared to random designs. The formula is given as follows:

\begin{equation*}
    \frac{\sum_{i,j}\left(\textrm{CI}\left(\theta^{(i)}, z^{(j)}\right) - \textrm{CI}\left(\theta^{(i)}, z^{\dagger}\right)\right)}{\sum_{i,j}\textrm{CI}\left(\theta^{(i)}, z^{(j)}\right)}\times 100\%,
\end{equation*}
where CI corresponds to the estimated 95\% CI size for the parameter of interests, $\{\theta^{(i)}\}$ are samples of the data-generating parameters, $\{z^{(j)}\}$ are random design samples, and $z^{\dagger}$ is the optimized design. The table shows that the design optimization significantly reduces the 95\% CI sizes, with an average reduction of 47\%. These results demonstrate that our optimized designs yield more reliable parameter estimation across a range of materials whose constitutive parameters are modeled by the prior.

\subsubsection{Quantitative Interpretation: Design Utility for Learning Different Behaviors}

\begin{figure}[htb]
    \centering
    \begin{minipage}{0.55\linewidth}
        \includegraphics[width=\linewidth]{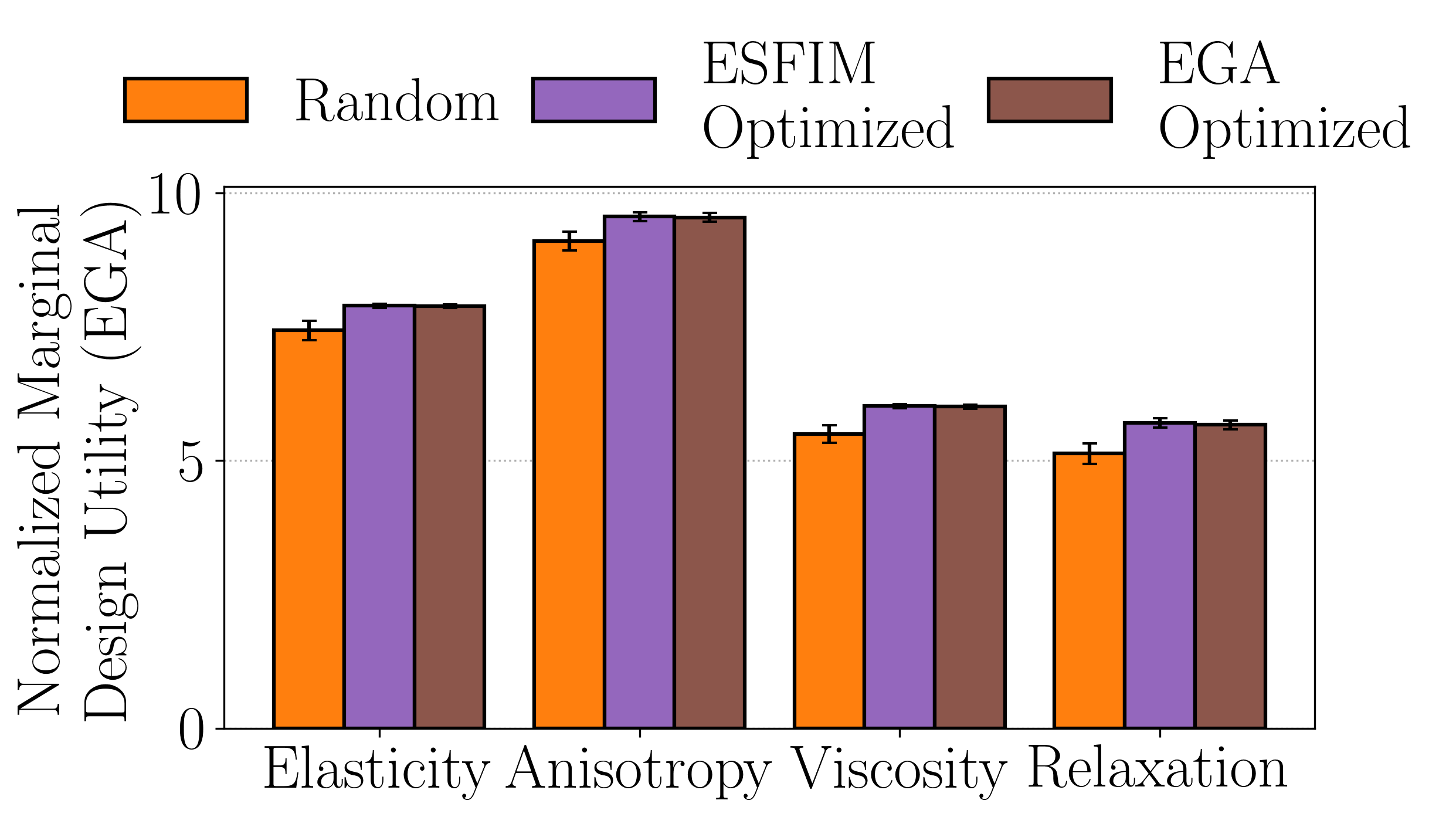} 
    \end{minipage}
    \begin{minipage}{0.39\linewidth}
    \scalebox{0.9}{
    \renewcommand{\arraystretch}{2}
        \begin{tabular}{|c |c |c|}\hline
        \multicolumn{3}{|c|}{\makecell{Expected Improvement in \\
        the EGA Design Utility \\
        of Optimized Designs \\
        Relative to Random Designs}}\\\hline\hline
             & ESFIM & EGA  \\\hline
            Elasticity & 6.2\%  & 6.1\%  \\\hline
            Anisotropy & 5.0\% & 4.8\%  \\\hline
            Viscosity & 9.6\% & 9.4\% \\\hline
            Relaxation & 11.2\% & 10.4\% \\\hline
        \end{tabular}
    \renewcommand{\arraystretch}{1}
    }
    \end{minipage}
        
    \caption{The normalized EGA design utility for learning marginal parameters of linear viscoelasticity grouped by material behaviors. (\emph{Left}) The bar plots of the utility of random, ESFIM-optimized, and EGA-optimized designs for learning elasticity, anisotropy, viscosity, and relaxation. The error bars for the random designs represent the standard deviations of EGA with respect to the design variables. In contrast, the error bars for the optimized designs represent the standard errors of EGA attributed to finite parameter sample sizes. (\emph{Right}) The expected improvement in the utility of optimized design relative to random designs for learning different behaviors.}
    \label{fig:linear_marginal}
\end{figure}

In \cref{fig:linear_marginal}, we provide the EGA for inferring four disjoint subsets of parameters. The four subsets corresponding to different modes of material response: (i) elasticity, ($\log E_1^{(0)}$, $r_E$, $r_G$, $r_{\nu}$), (ii) anisotropy, $\alpha_c$, (iii) viscosity, ($f_1$, $f_2$, $w_1$, $w_2$), and (iv) stress relaxation, ($\log \tau_1$, $\log \tau_2$). The relative expected improvements in the EGA for marginal parameters are computed as follows:
\begin{equation*}
    \frac{\sum_{j}\left(\textrm{EGA}_{\textrm{marg}}\left(z^{(j)}\right) - \textrm{EGA}_{\textrm{marg}}(z^{\dagger})\right)}{\sum_{j}\textrm{EGA}_{\textrm{marg}}\left(z^{(j)}\right)}\times 100\%,
\end{equation*}
where $\textrm{EGA}_{\textrm{marg}}$ corresponds to the EGA for marginal parameters, $\{z^{(j)}\}$ are random design samples, and $z^{\dagger}$ is the optimized design. The results show that the optimized designs consistently improve the identifiability of these parameter subsets compared to random designs. The relative improvements are most significant for the relaxation times at 11\%, whereas anisotropy is the least significant, yet still at 5\%.

\begin{remark}
    Although the marginal EGA improvements in \cref{fig:linear_marginal} are only 4--12\%, they are computed based on the log-determinant of the marginal posterior precision matrix, so even a modest improvement corresponds to a substantial reduction in marginal posterior uncertainty. For example, in \Cref{fig:linear_ci}, the optimized designs reduce the average 95\% CI width of the orthotropic rotation parameter ($\alpha_c$) by 52\%, whereas in \Cref{fig:linear_marginal}, the relative improvement, measured by the marginal EGA, is 4.8\%. The improvements are not larger because the baseline EGA value is already high, which we attribute to the high-dimensional data with a low noise level.
\end{remark}

\subsubsection{Relation Between Information Entropy and Stress-State Entropy}

\begin{figure}[!htp]
        \centering
        \includegraphics[width=0.4\linewidth]{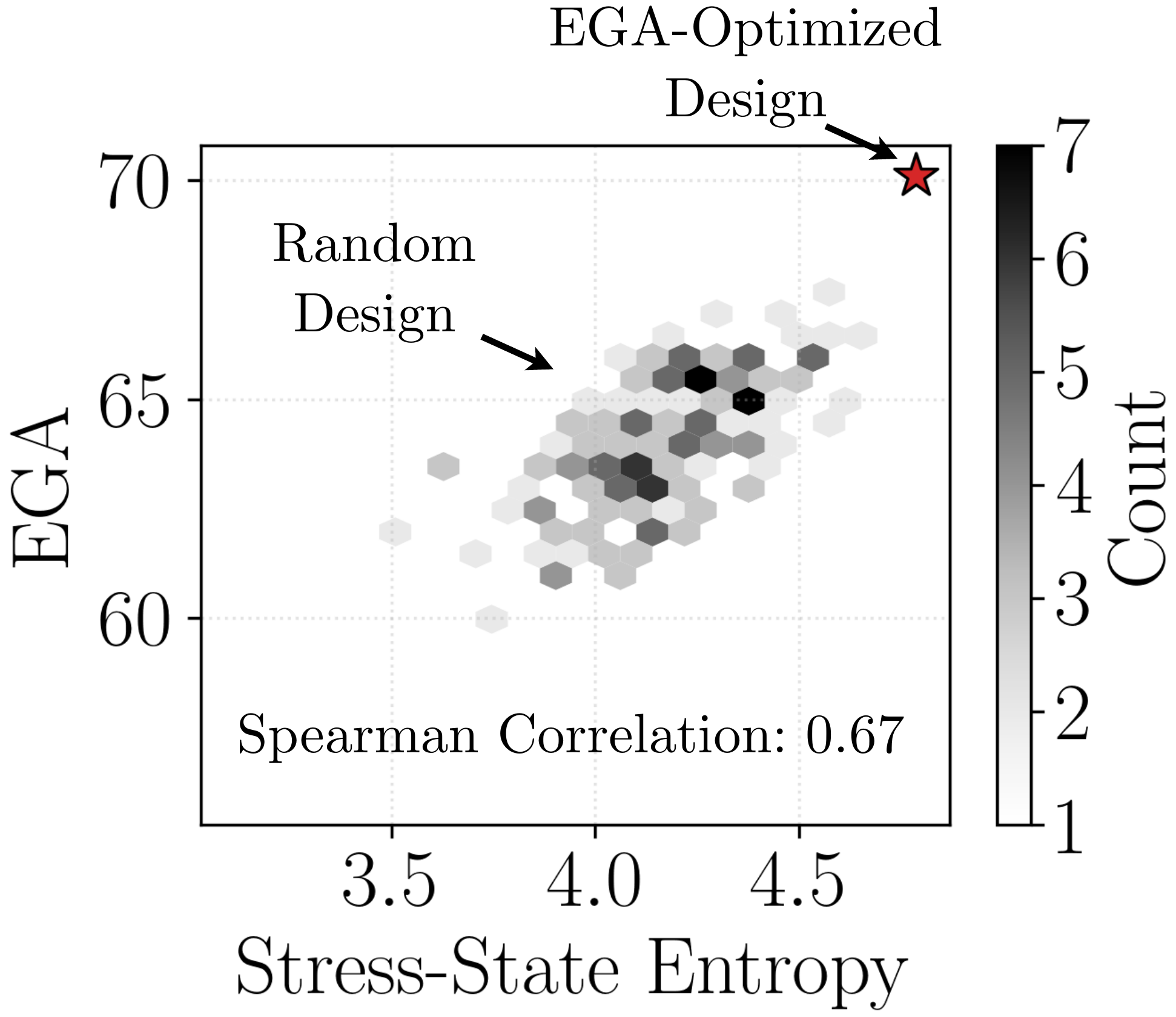}
        \caption{Relation between principal-stress-based stress-state entropy and EGA for randomly selected designs and the EGA-optimized design. The EGA-optimized design also has the largest stress-state entropy, with an overall rank correlation of $0.67$.}
        \label{fig:sse}
    \end{figure}

The utility of a mechanical testing design is often tied to the diversity of local strain histories and the stresses it induces, since parameter identifiability can be improved by richer mechanical responses. To examine this connection in our setting, we consider a principal-stress-based stress-state entropy \cite{ihuaenyi2024seeking, ihuaenyi2025mechanics}; its definition is given in \cref{app:sse}. \Cref{fig:sse} compares this mechanics-based quantity with EGA across random and optimized designs. The EGA-optimized design also attains the largest stress-state entropy, and the two quantities are positively correlated, with a Spearman rank correlation of 0.67. This indicates that designs favored by EGA tend to induce more diverse stress states, while also showing that stress-state entropy does not fully account for the information-based utility.

\subsection{Batched Experimental Design} \label{subsec:ve_batched_design}

We utilize ESFIM with bound constraints on the design variables for batched optimal experimental design. Note that the same surrogate FIM described in \cref{subsec:ve_single_experiment} is reused to find all the optimizers shown below.

\begin{figure}[htb]
    \centering
    \scalebox{0.9}{
    \renewcommand{\arraystretch}{1.5} 
        \begin{tabular}{|c c || c c c|}\hline
        \multicolumn{5}{|c|}{ESFIM-Optimized Batched Experimental Designs} \\\hline\hline
        \multicolumn{2}{|c||}{Batch Size of 2} & \multicolumn{3}{c|}{ Batch Size of 3}\\\hline
        \raisebox{-0.065\linewidth}{\includegraphics[width=0.18\linewidth]{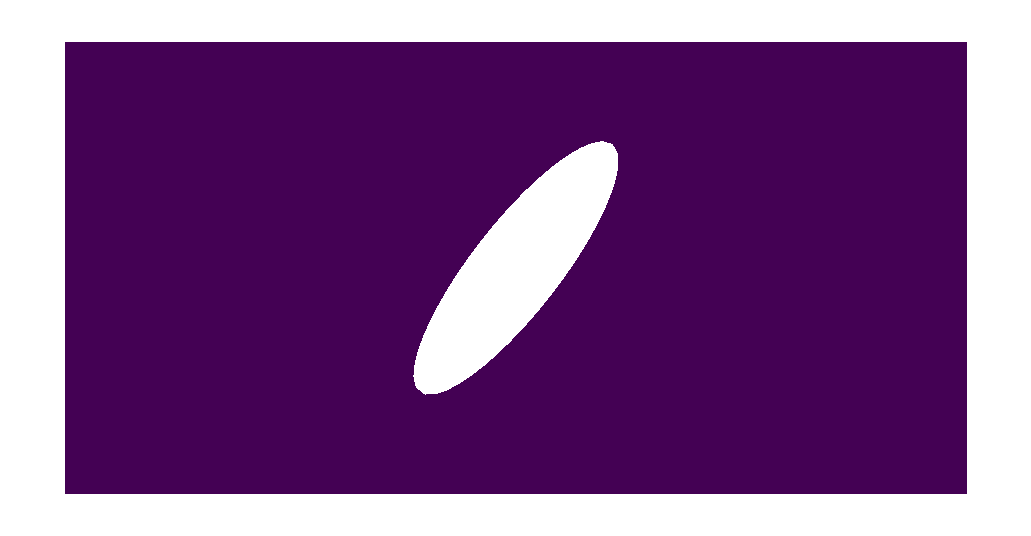}}  & \raisebox{-0.065\linewidth}{\includegraphics[width=0.18\linewidth]{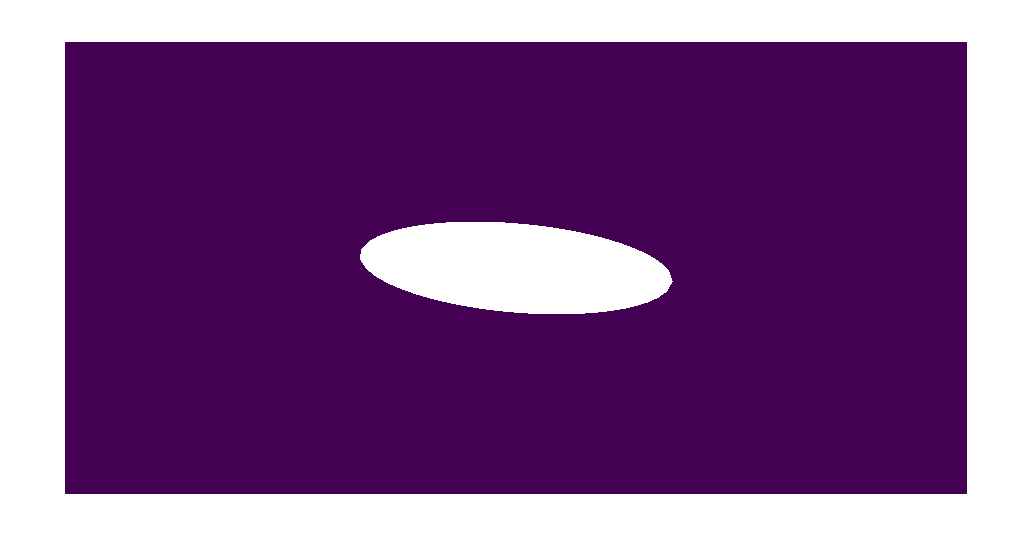}} & \raisebox{-0.065\linewidth}{\includegraphics[width=0.18\linewidth]{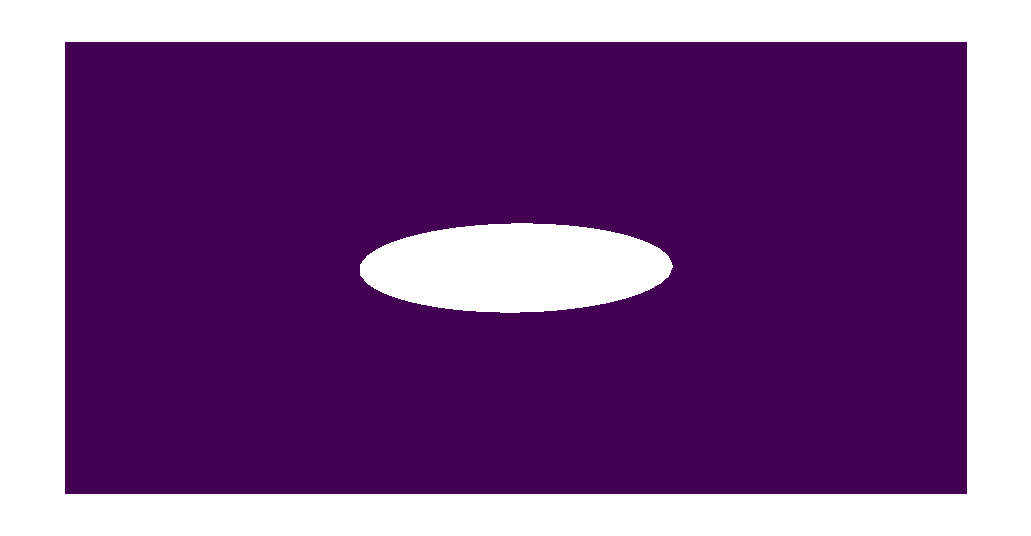}} & \raisebox{-0.065\linewidth}{\includegraphics[width=0.18\linewidth]{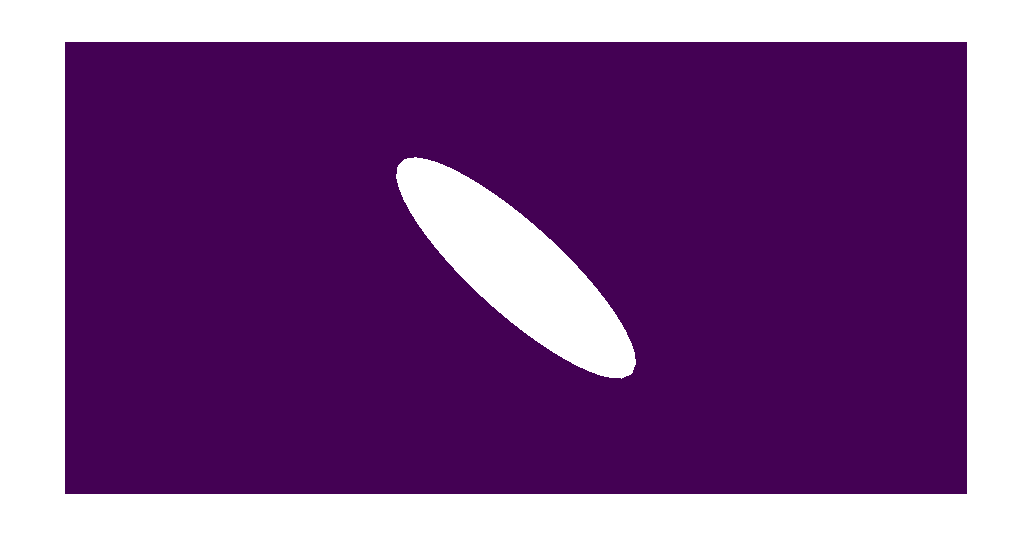}} & \raisebox{-0.065\linewidth}{\includegraphics[width=0.18\linewidth]{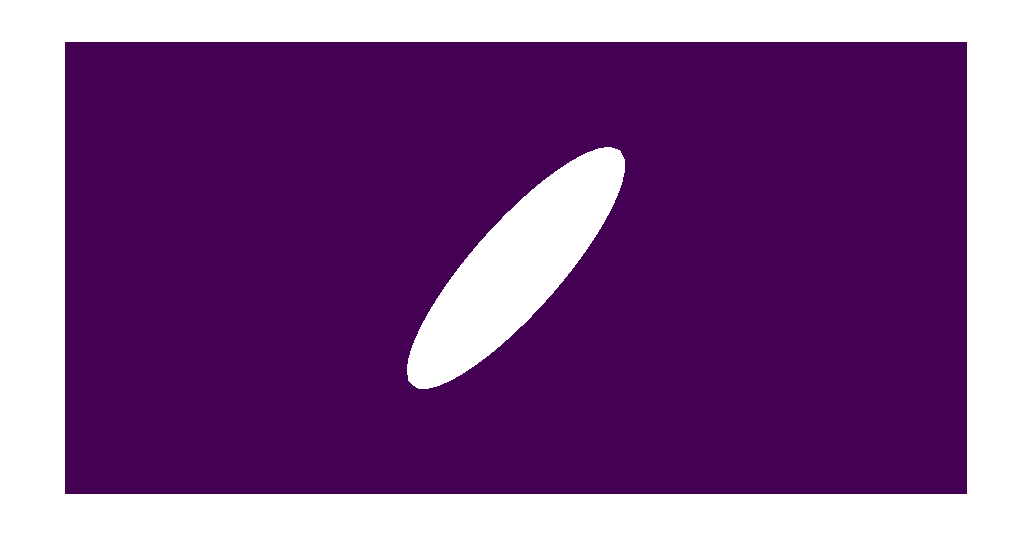}}\\
            \includegraphics[width=0.18\linewidth]{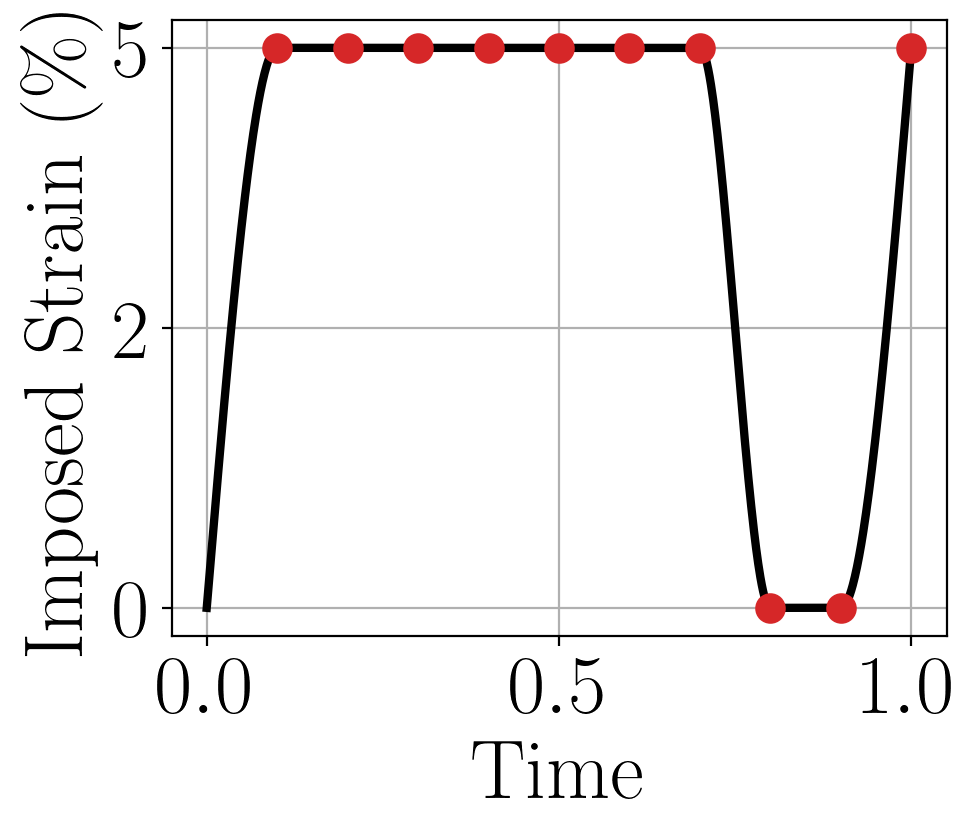} &  \includegraphics[width=0.18\linewidth]{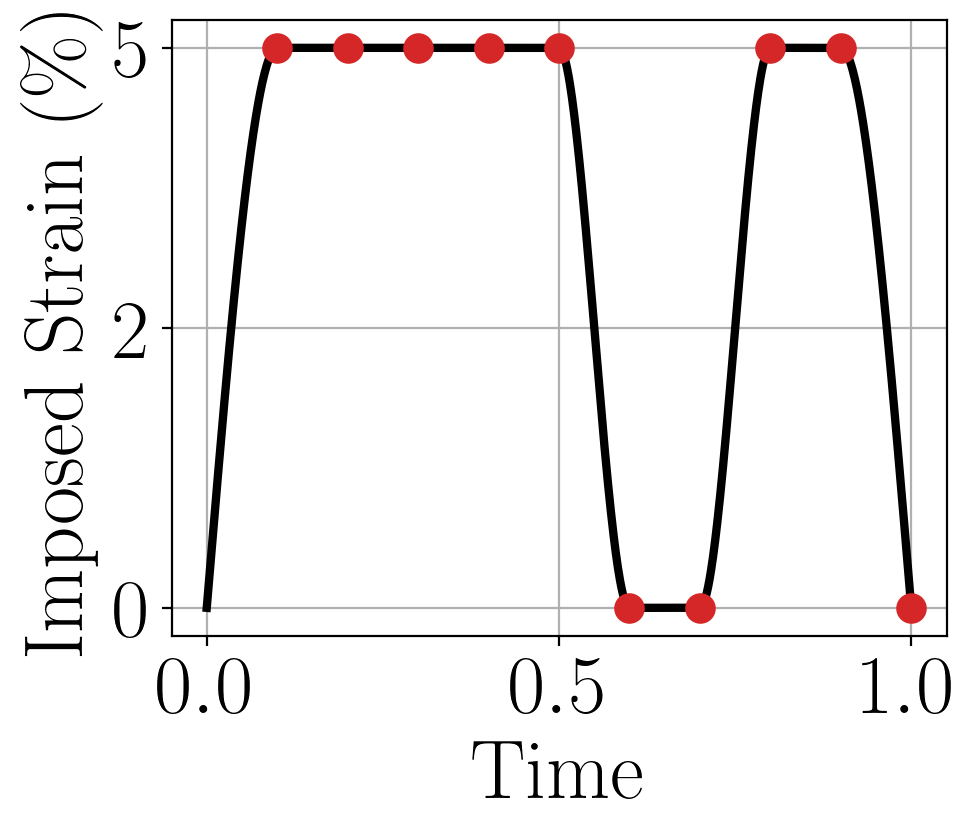} & \includegraphics[width=0.18\linewidth]{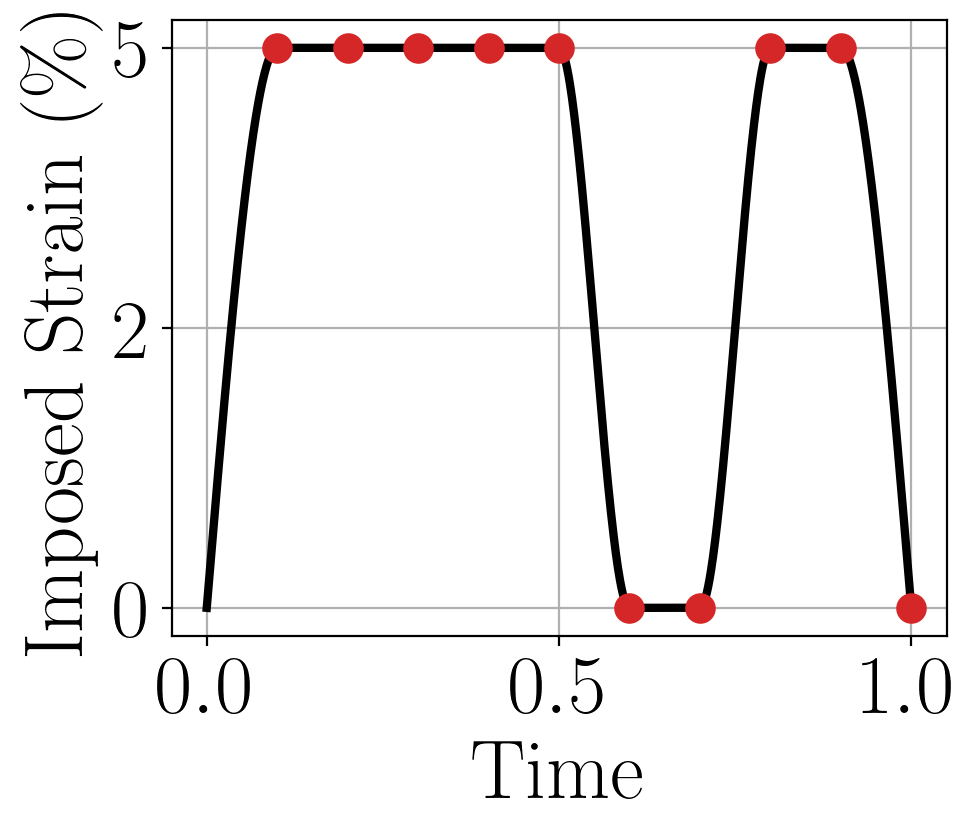} & \includegraphics[width=0.18\linewidth]{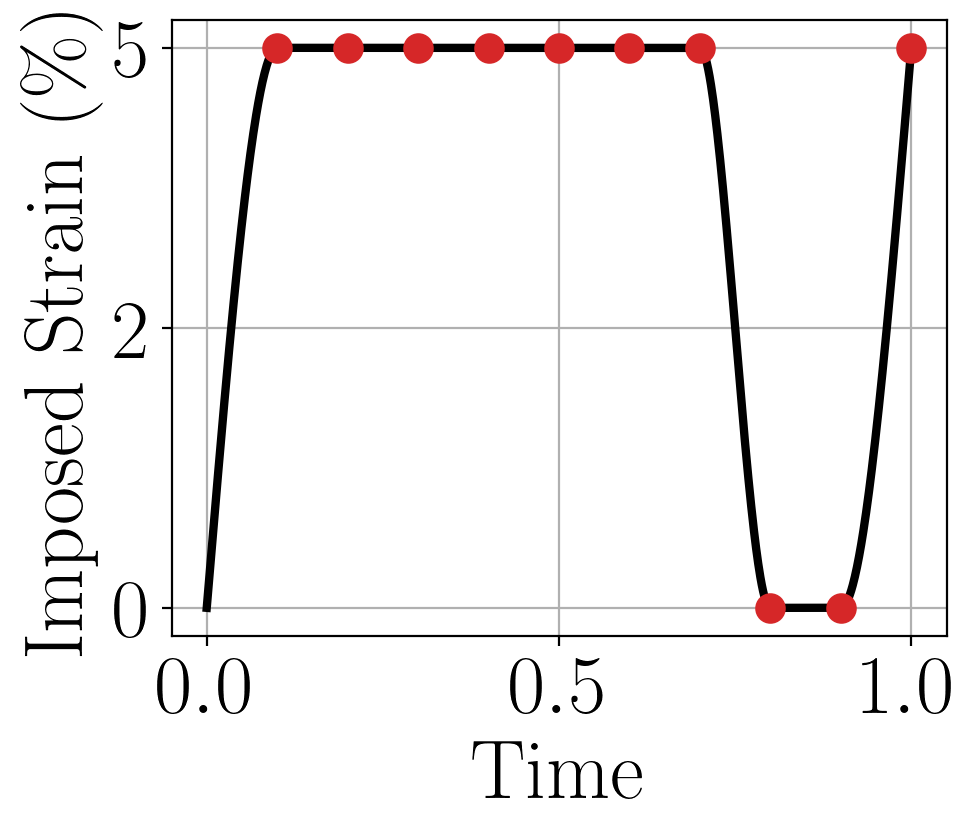} & \includegraphics[width=0.18\linewidth]{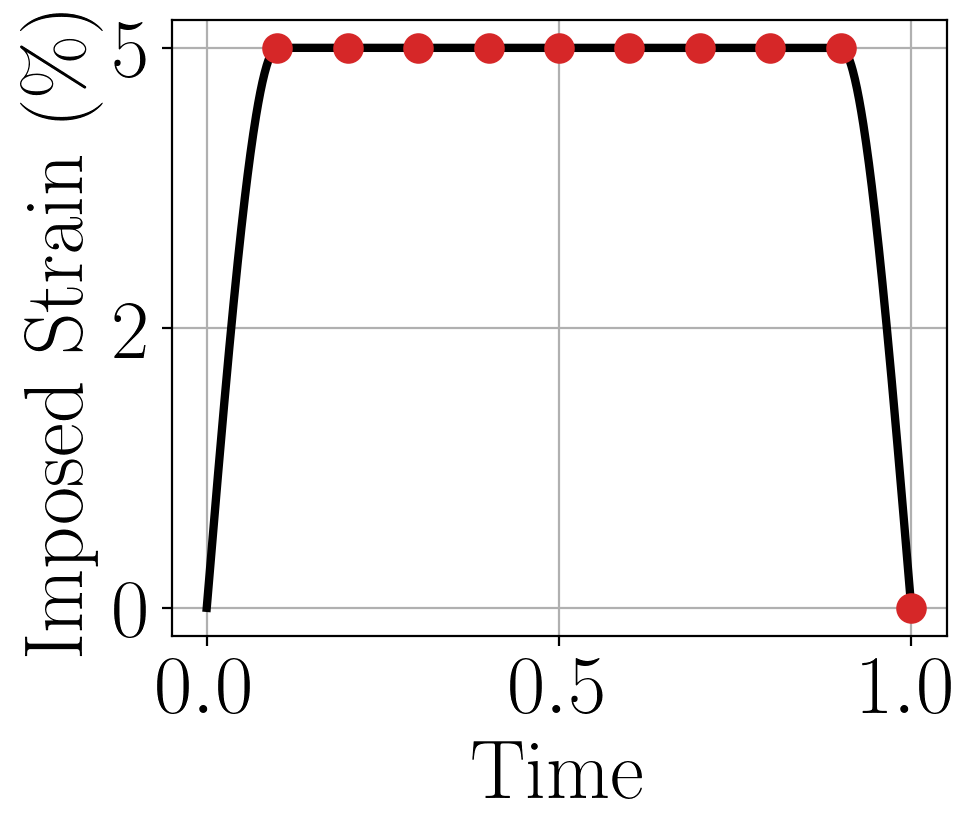}\\\hline
        \end{tabular}
    \renewcommand{\arraystretch}{1}
    }\\[5pt]
    \scalebox{0.9}{    \renewcommand{\arraystretch}{1.5} 
        \begin{tabular}{|c c || c c c|}\hline
       \multicolumn{5}{|c|}{ESFIM-Optimized Batched Experimental Designs with Shared Geometry} \\\hline\hline
        \multicolumn{2}{|c||}{Batch Size of 2} & \multicolumn{3}{c|}{ Batch Size of 3}\\\hline
        \multicolumn{2}{|c||}{\raisebox{-0.065\linewidth}{\includegraphics[width=0.18\linewidth]{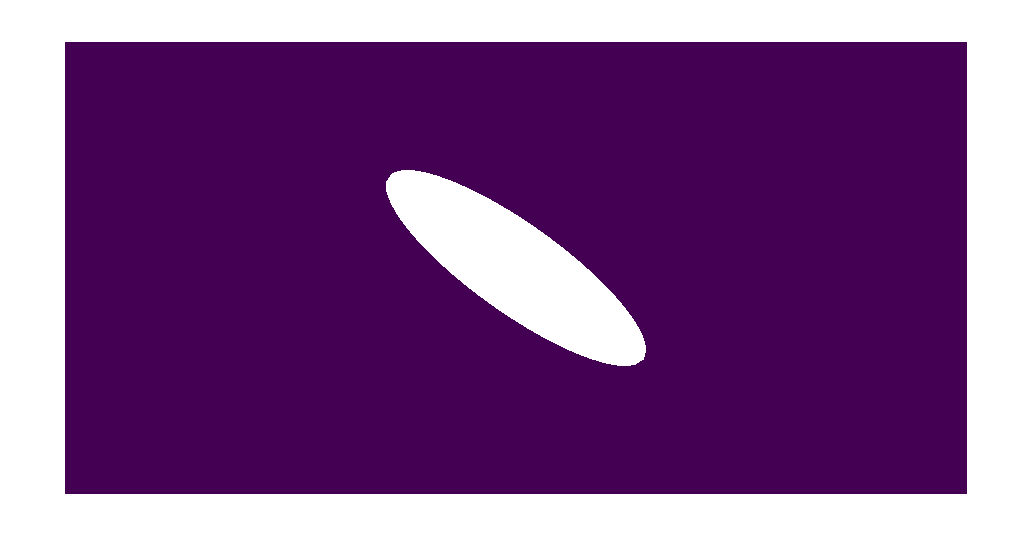}}} & \multicolumn{3}{c|}{\raisebox{-0.065\linewidth}{\includegraphics[width=0.18\linewidth]{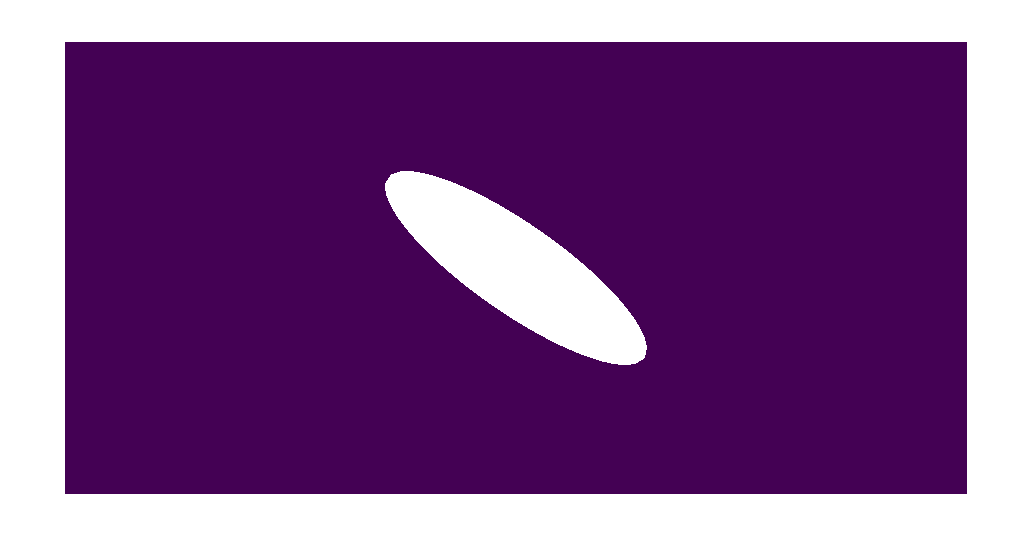}}}\\
            \includegraphics[width=0.18\linewidth]{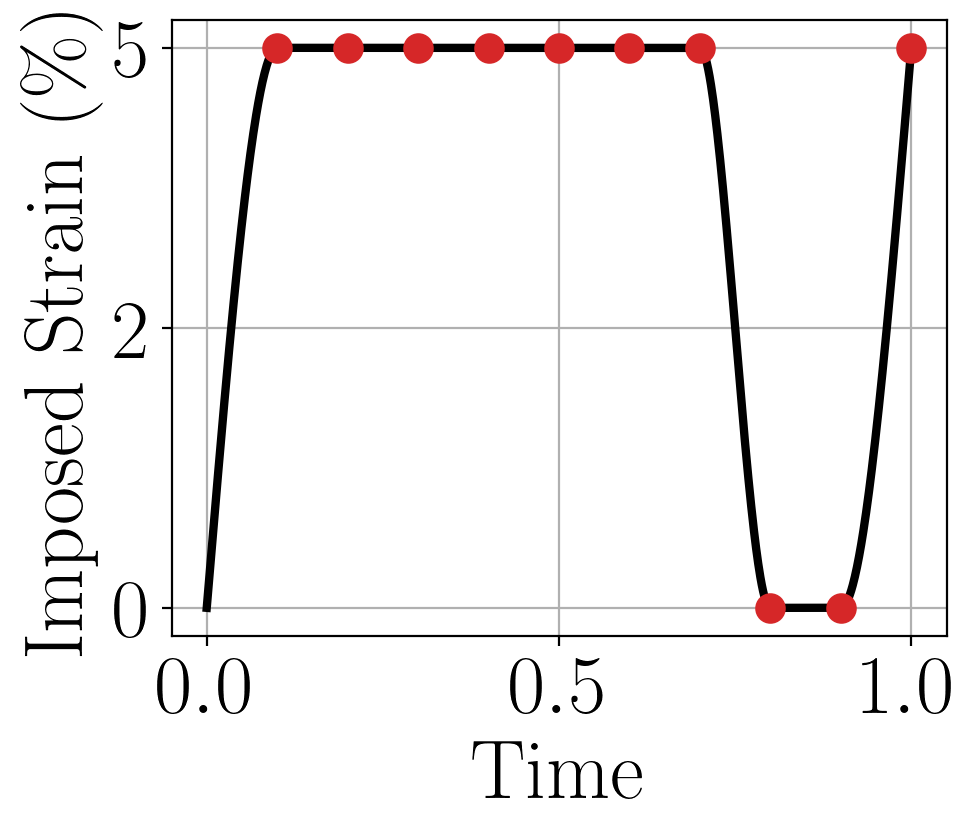} &  \includegraphics[width=0.18\linewidth]{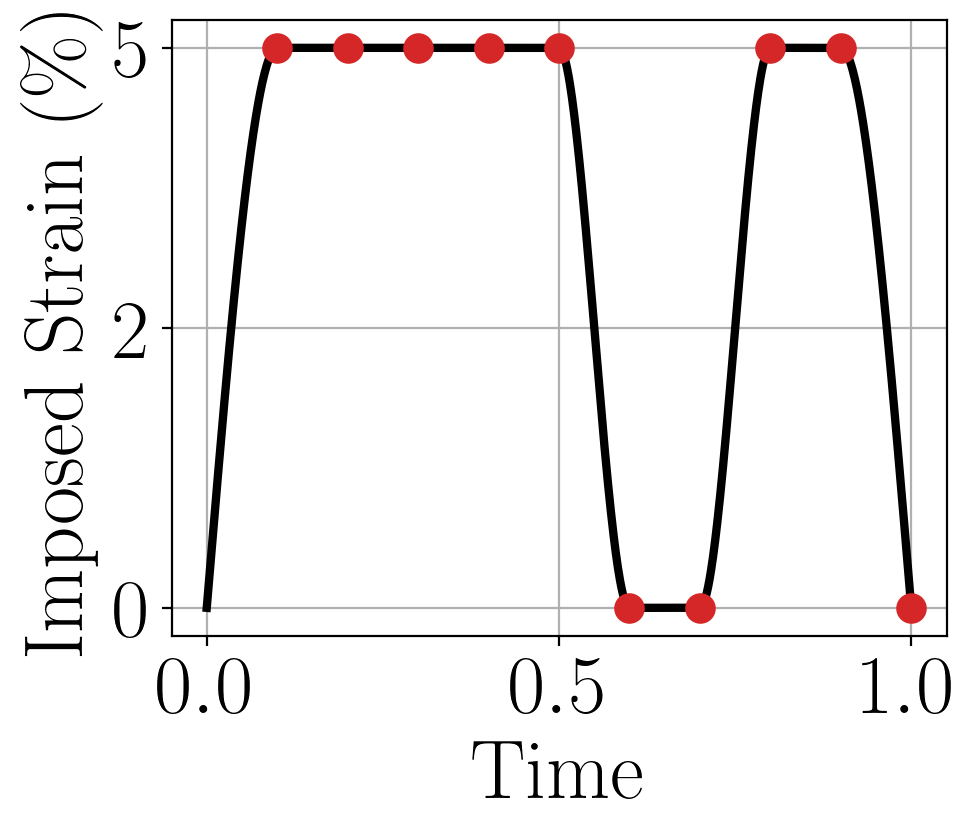} & \includegraphics[width=0.18\linewidth]{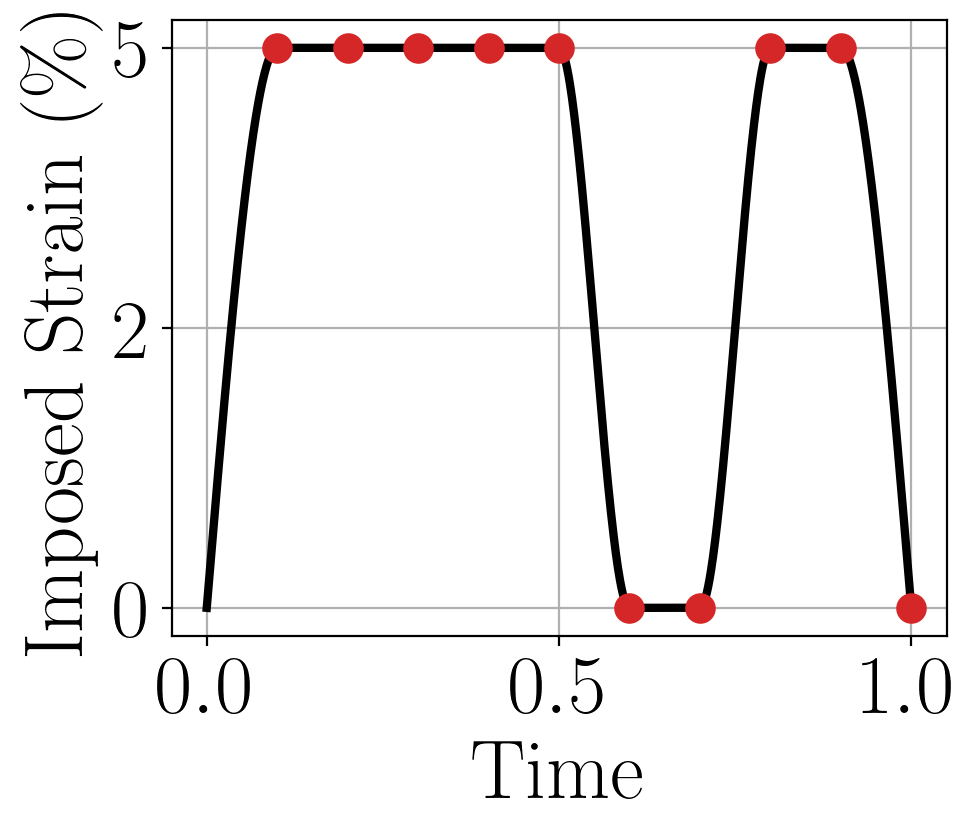} & \includegraphics[width=0.18\linewidth]{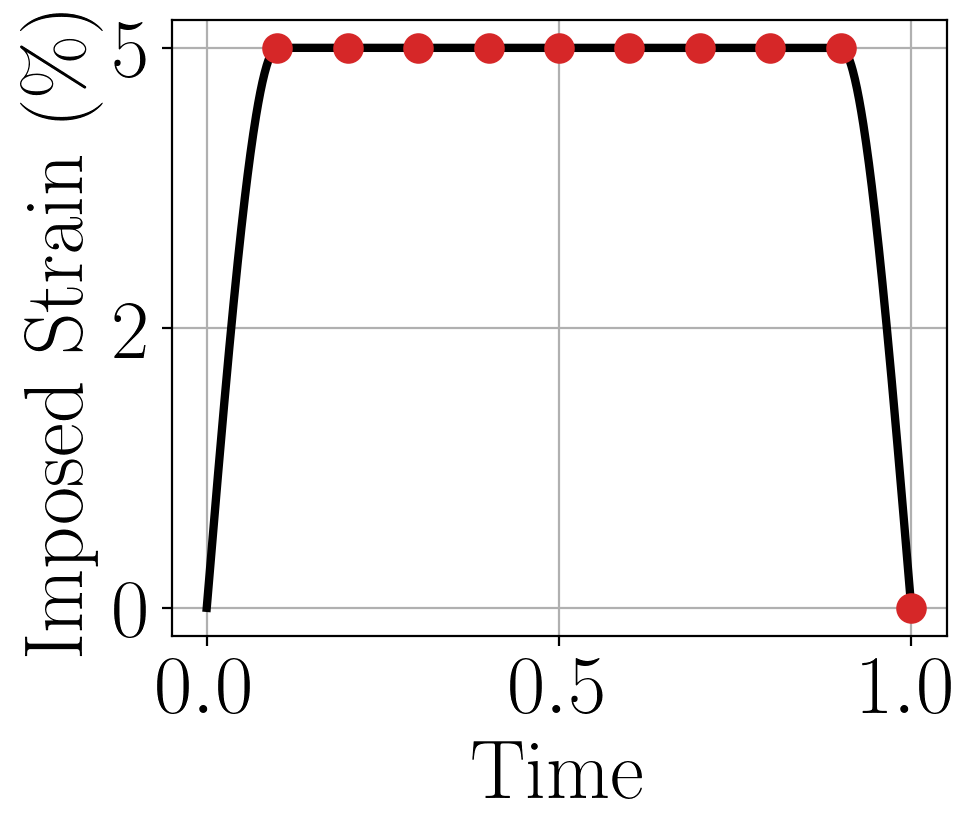} & \includegraphics[width=0.18\linewidth]{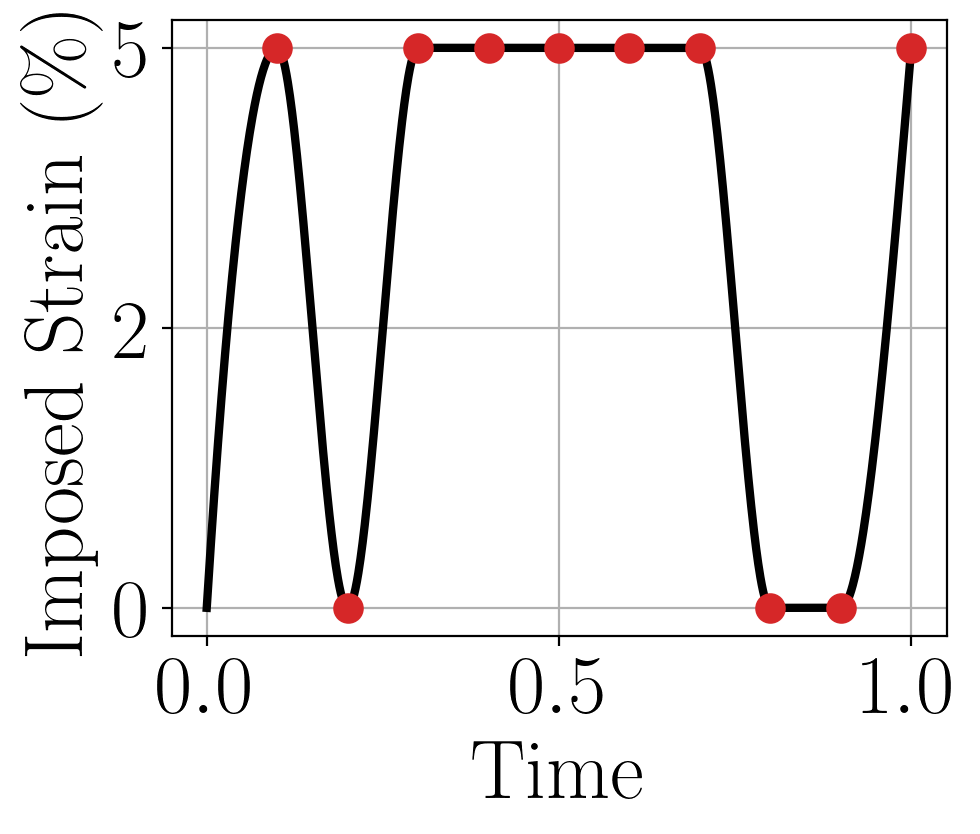}\\\hline
        \end{tabular}
    \renewcommand{\arraystretch}{1.0}
    }

    \caption{The optimal batched designs of uniaxial testing for learning linear viscoelasticity found by ESFIM maximization. The designs in the top panel vary in geometry within their batches, whereas those in the bottom panel share the same geometry.}
    \label{fig:linear_batched_design}
\end{figure}

The optimizers for experiment batch sizes of 2 and 3 are shown in the top panel of \cref{fig:linear_batched_design}. Within the same batch of experiments, the optimized shapes of the elliptical holes alternate in orientation, which is known to improve anisotropy identification. Loading paths with rapid loading and unloading, along with long holds, are still preferred by ESFIM, which is known to improve the identification of relaxation times. We observe an increase in the number of switches between no-strain and maximum-strain as the batch size increases, suggesting that improving the identifiability of the parameters associated with the instantaneous response becomes more rewarding.

In the bottom panel of \cref{fig:linear_batched_design}, we visualize the optimized designs for batch sizes of $2$ and $3$ when specimen geometry is shared. This corresponds to reusing the same specimen across multiple loading paths, which is more practical for our small-strain setting. The result shows that the shared geometry for both batch sizes of 2 and 3 is a tilted ellipse with maximal aspect ratio: $L_{s,1}=0.1$, $L_{s,2}=0.35$, and $\alpha_s=0.3\pi$ (54\degree). We again observe an increase in the number of switches between no strain and maximum strain as the batch size increases.

\begin{figure}[htb]
    \centering
    \begin{minipage}{0.42\linewidth}
    \centering
    \includegraphics[width=\linewidth]{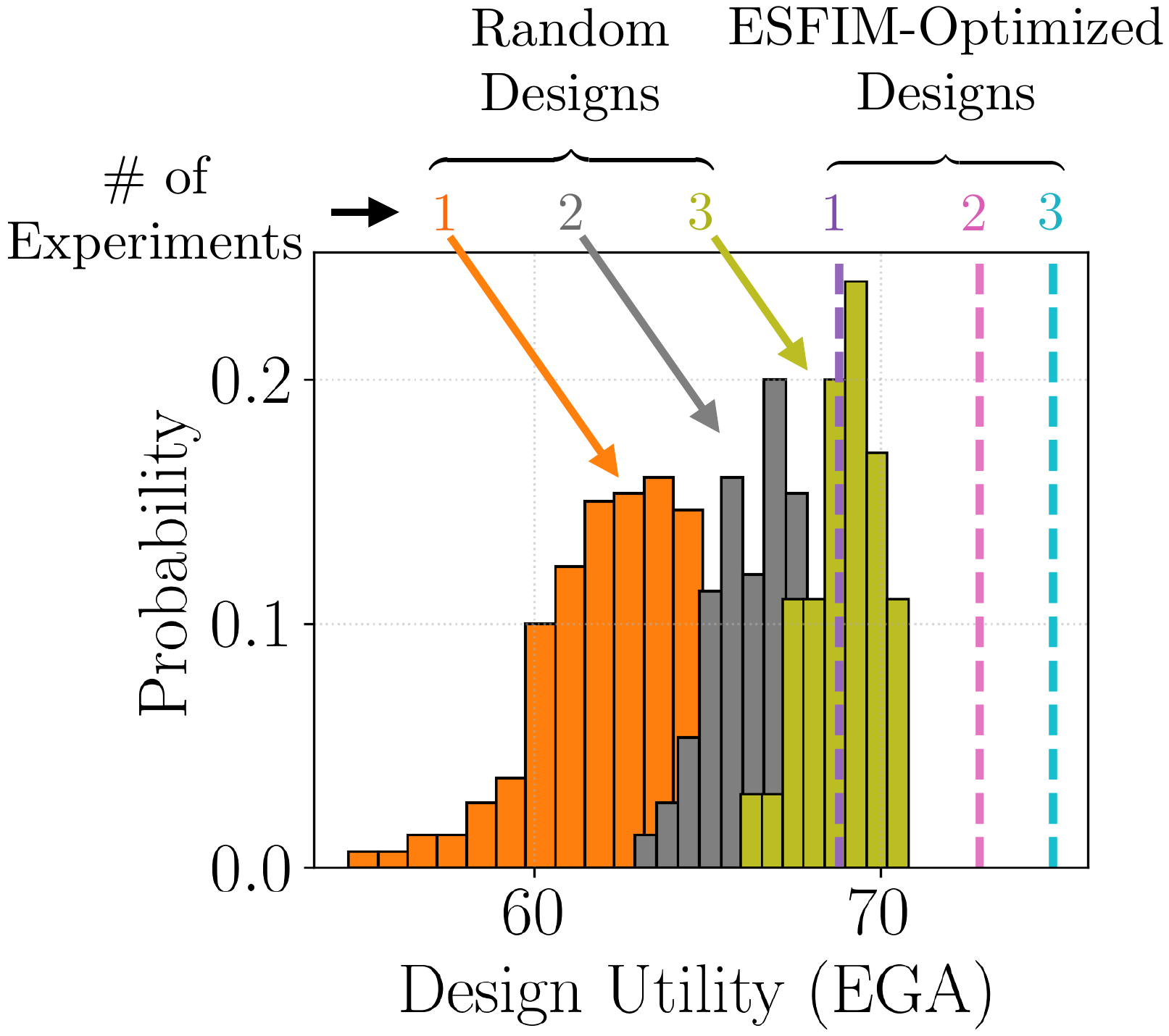}
    \end{minipage}
    \begin{minipage}{0.54\linewidth}
    \centering
    \scalebox{0.9}{
    \renewcommand{\arraystretch}{2}
    \begin{tabular}{|c|c|c|c|c|}\hline
    \multicolumn{5}{|c|}{\makecell{Expected Improvements in \\ the EGA Design Utility of Optimized Designs \\ Relative to Random Designs}}\\\hline\hline
    \multicolumn{2}{|c|}{\multirow{2}{*}{\backslashbox{Optimized}{Random}}} & \multicolumn{3}{c|}{Batch Size} \\\cline{3-5}
    \multicolumn{2}{|c|}{}& 1 & 2 & 3\\\hline
    \parbox[t]{6mm}{\multirow{3}{*}{\rotatebox[origin=c]{90}{Batch Size}}} & 1 & 10.5\% & 3.4\% & -0.2\%\\\cline{2-5}
     & 2 & 17.0\% & 9.5\% & 5.7\%\\\cline{2-5}
     & 3 & 20.4\% & 12.7\% & 8.8\%\\\hline
    \end{tabular}
    \renewcommand{\arraystretch}{1.0}
    }
    \end{minipage}

    \caption{ Visualization and statistics of the EGA design utility values of batched designs for uniaxial testing of linear viscoelastic materials. (\emph{Left}) The EGA values for random and ESFIM-optimized (batched) designs of 1--3 experiments.  (\emph{Right}) Relative expected improvements of the EGA for the optimized designs with batch sizes of 1--3 compared to random designs with batch sizes of 1--3. }
    \label{fig:linear_multiple_design_vs_random}
\end{figure}

In \cref{fig:linear_multiple_design_vs_random}, we visualize and quantify the improvements in the EGA for ESFIM-optimized designs shown in the top panel of \cref{fig:linear_batched_design} relative to random designs across different experiment batch sizes. For a single optimized experiment, the optimized design has a similar EGA utility value to that of three randomly designed experiments. We expect four randomly designed experiments to outperform one optimized experiment. On the other hand, at similar computational costs, ESFIM obtains optimized designs for experiment batch sizes of 1, 2, and 3, and each substantially outperforms the random designs of the same batch size by around 10\%. This demonstrates ESFIM's cost-amortization capability and its suitability for batched design.

\subsection{Nested Monte Carlo versus Gaussian Approximation}\label{ssec:ve_nmce}

We report and discuss our attempts to approximate the EIG using NMCE and EGA across three randomly sampled designs (see \cref{fig:nmce_design}). The outcomes suggest that NMCE, while asymptotically unbiased, is computationally intractable as a design utility due to the extreme likelihood concentration arising from high-resolution image data. The results motivate the use of alternative EIG approximation techniques, such as EGA, which strikes a superior balance between speed and accuracy compared to NMCE.

We use $N_{\textrm{out}}=64$ and $N_{\textrm{in}}=$16,384 for NMCE, with importance sampling using proposals given by the Gaussian approximation in \cref{eq:local_gauss_approx_genprior}. Both outer and inner sampling are performed with quasi-Monte Carlo (QMC) using scrambled Sobol sequences \cite{owen1998scramling, bartuska2025double}. We observe severe weight collapse in the importance sampling. The effective sample size for inner sampling is 1 per outer sample. A variance decomposition of the log-weights shows that variation in the log-likelihood is the dominant contributor to the weight collapse. In particular, the log-likelihood of the data-generating parameter exceeds that of the inner samples by a median of 159, indicating an extreme concentration of likelihood that can be attributed to the data's high dimensionality. A log-normal approximation to the importance weights suggests that approximately $10^{10^{7}}$ samples would be required for accurate inner sampling, rendering NMCE intractable. See a more detailed discussion of this estimation in \cref{app:nmc}.

In contrast, we observe that EGA evaluations, which share 64 outer samples with NMCE for each design, can be performed at a computational cost 4 orders of magnitude lower, while still yielding more reliable EIG approximations. Importantly, the costs of EGA evaluation are independent of the data dimension, whereas NMCE's efficiency and reliability degrade as the data dimension increases. Furthermore, this concentration of likelihood supports the validity of our Gaussian posterior approximation under the Bernstein-von Mises theorem; see \cref{sssec:limitation}. We expect this comparison to be more favorable for EGA as the data dimension increases.

\section{Application: Learning Nonlinear Viscoelasticity}\label{sec:results_nonlinear}

In this section, we present a numerical study on the design of uniaxial tests to determine the constitutive parameters of the anisotropic finite-strain viscoelasticity model in \cref{subsec:viscohyper}. The goal of this study is to demonstrate that the results in \cref{sec:results_linear} for the linear model can be generalized to more complicated constitutive models with large deformation. In \cref{subsec:nonlinear_viscoelastic_setup} we specify the settings of
the design problem. Next, we present the results of ESFIM-optimized design optimization in \cref{subsec:nonlinear_results}. Compared with the results in \cref{sec:results_linear}, we additionally visualize and quantitatively interpret designs that are optimized with regularization of the imposed strain rate, yielding numerically identified smooth loading protocols that balance utility maximization and design feasibility. The implementation details are in \cref{app:numerics}.

\subsection{Settings}\label{subsec:nonlinear_viscoelastic_setup}

\paragraph{Experimental design}
The uniaxial test has an almost identical setup to that in \cref{subsec:viscoelastic_setup}. Here, however, we consider a larger maximal imposed strain of $50\%$: $z_{l}\in[0, 1]^{10}$.\hfill$\diamond$

\paragraph{Reparameterization and The Prior Distribution}

We consider a reparameterization of the nonlinear viscoelastic constitutive model in \cref{subsec:viscohyper} with two viscous branches and 15 constitutive parameters. A two-step procedure, as in \cref{subsec:viscoelastic_setup}, is used. This yields a standard normal prior, and the physical parameters are uniformly distributed and defined via a transformation of the prior in \cref{eq:uniform_transform}.

We describe the physical parameters and justify the associated ranges; see \cref{tab:prior_bound} for a summary. First, we set the bound for the equilibrium shear modulus $\mu_{\infty}$ in the logarithmic scale, and the bulk modulus $\kappa$ is set using a distribution on the initial (small-strain) Poisson ratio $\nu$ to enforce near incompressibility:
\begin{equation*}
    \log \mu_{\infty}\sim \mathcal{U}(-1.5, -0.5), \quad \nu \sim \mathcal{U}(0.45, 0.49),\quad \kappa = \mu_{\infty}\frac{2(1+\nu)}{3(1-2\nu)}.
\end{equation*}
The range of the shear modulus is set to keep the overall reaction force of order 1. The equilibrium initial fiber stiffness is defined relative to the shear modulus through a logarithmic ratio $\phi_f$, and the stiffening rate is set to a uniform distribution in the logarithmic scale
\begin{equation*}
    \phi_f\sim \mathcal{U}(0, 3),\quad k_1 \coloneqq \mu_{\infty}\exp(\phi_f), \quad \log k_2 \sim\mathcal{U}(-1, 1).
\end{equation*}
The resulting maximum fiber stiffness is around 20 times the matrix stiffness. We set a uniform prior for fiber orientation angle, $\alpha_f\in\mathcal{U}(-\pi/4, \pi/4)$, where the fiber is assumed to be generally aligned with the uniaxial direction. For each viscous branch, we use logarithmic ratios to relate its parameters to the equilibrium parameters:
\begin{gather*}
    q_{\mu,i}, q_{k_1, i}, q_{k_2, i}\sim \mathcal{U}(-1.5, 0.7),\,\,\, \mu_{i} = \mu_{\infty}\exp(q_{\mu, i}),\,\,\, k_{1, i} = k_1 \exp(q_{k_1, i}),\,\,\, k_{2, i} = k_{2}\exp(q_{k_2, i}).
\end{gather*}
We assign a uniform prior on the rate sensitivity exponent $m_i\sim\mathcal{U}(1.5,3)$. Finally, to ensure the identifiability of the viscous branches, the logarithm of relaxation time for each branch, $\tau_i = (\gamma_i\mu_i)^{-1}$, is assigned to a distinct range,
\begin{equation*}
\log\tau_1\sim \mathcal{U}(-3.0, -1.5),\quad  \log\tau_2\sim \mathcal{U}(-1.5, 0).
\end{equation*}
Consequently, $\gamma_i$ is replaced by $\log \tau_i$ in the set of constitutive parameters. \hfill$\diamond$

\begin{table}[htb]
\centering
\renewcommand{\arraystretch}{1.3}
\begin{tabular}{|c||c|c|c|}\hline
& Physical Parameter & Description & Bound \\\hline\hline
\multirow{5}{*}{\rotatebox{90}{Equilibrium}} 
& $\log \mu_{\infty}$ & Log.\ of eq.\ shear modulus & (-1.5, -0.5) \\
& $\nu$ & Initial Poisson ratio & (0.45, 0.49) \\
& $\phi_{f}$ & Eq.\ fiber stiffness-to-shear log.\ ratio & (0, 3) \\
& $\log k_2$ & Log.\ of eq.\ fiber stiffening rate  & (-1, 1) \\
& $\alpha_f$ & Fiber orientation angle & (-$\pi$/4, $\pi$/4) \\\hline\hline
\multirow{7}{*}{\rotatebox{90}{Non-Equilibrium}} 
& $q_{\mu, i}$, $i=1,2$ & Viscous-to-eq.\ shear log.\ ratio & (-1.5, 0.7) \\
& $q_{k_1, i}$, $i=1,2$ & Viscous-to-eq.\ fiber stiffness log.\ ratio & (-1.5, 0.7) \\
& $q_{k_2, i}$, $i=1,2$ & Viscous-to-eq.\ fiber stiffening rate log.\ ratio & (-1.5, 0.7) \\
& $m_i$, $i=1,2$ & Rate-sensitivity exponent & (1.5, 3) \\
& $\log \tau_1$ & Log.\ of relaxation time (fast) & (-3.0, -1.5) \\
& $\log \tau_2$ & Log.\ of relaxation time (slow) & (-1.5, 0) \\\hline
\end{tabular}
\renewcommand{\arraystretch}{1.0}
\caption{The bounds for the uniform distributions of the physical parameters in the re-parameterized nonlinear viscoelastic constitutive model.}
\label{tab:prior_bound}
\end{table}

\begin{figure}[htb]
    \centering
    \begin{minipage}{0.3\linewidth}
        \includegraphics[width=\linewidth]{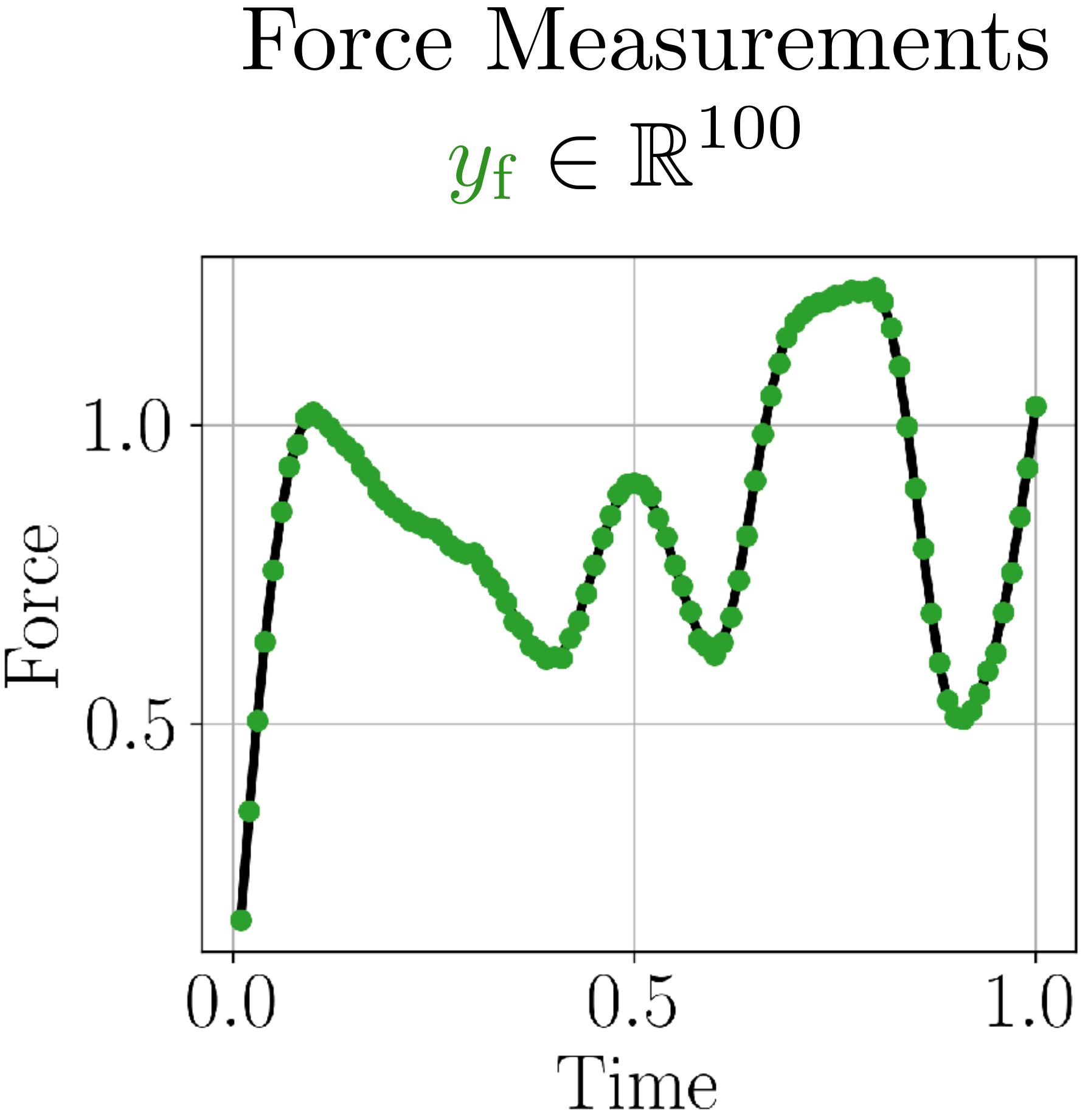}
    \end{minipage}
    \begin{minipage}{0.69\linewidth}
        \includegraphics[width=\linewidth]{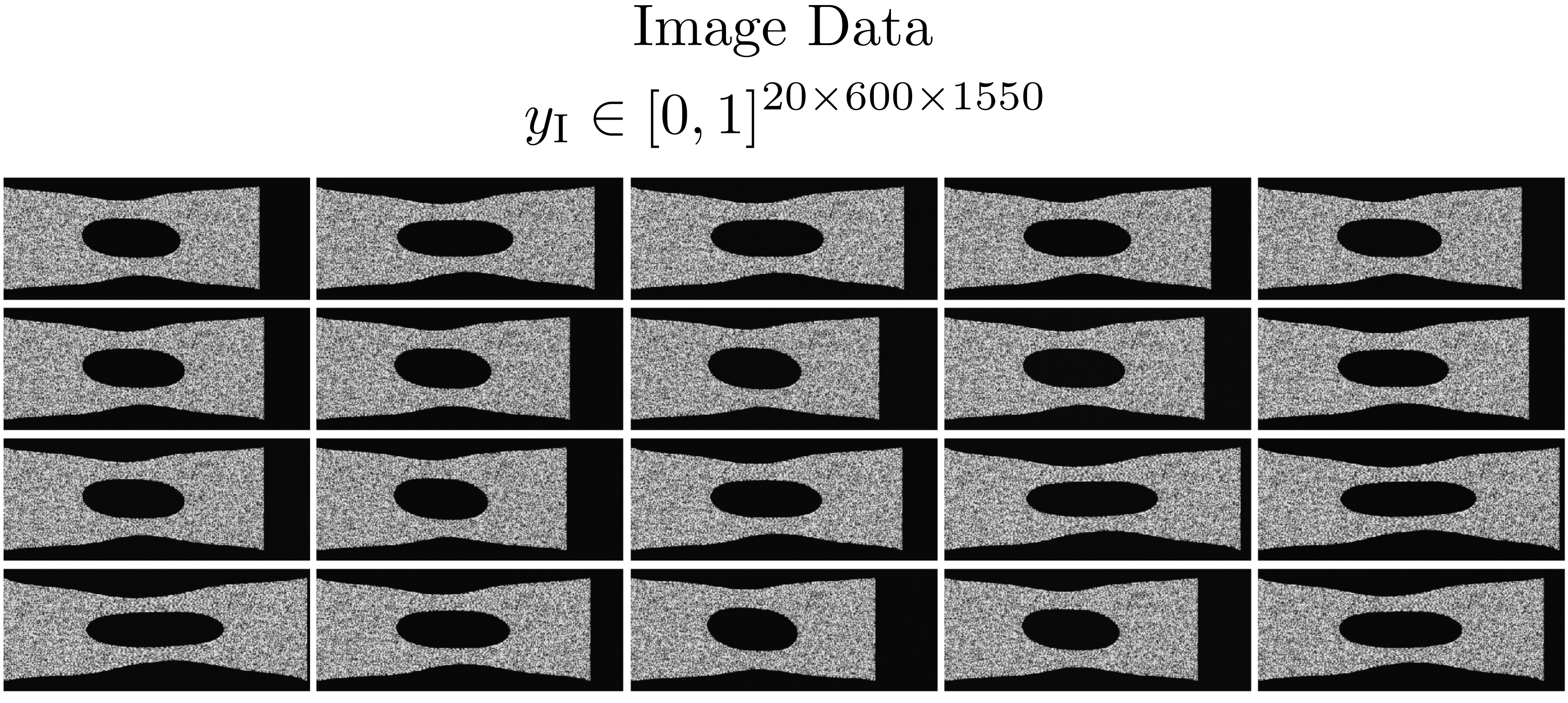}
    \end{minipage}
    \caption{An example of simulated force and image data for learning nonlinear viscoelasticity. This simulated dataset is generated using the experimental setup in \cref{subsec:viscoelastic_setup,fig:viscoelastic_setup}, with an order-of-magnitude higher imposed strain. The constitutive parameters are drawn from the prior. Note that the resolution of the image data visualized here is much lower than that of the actual data used in the numerical study; see an image snapshot with high resolution in \cref{fig:high_res_images}.}
    \label{fig:nonlinear_data}
\end{figure}

\paragraph{Observation Operator}
We use the same observation operators as in \cref{subsec:viscoelastic_setup}, except that (i) the field of view has a height of $1.2$ and a width of $3$ to accommodate the larger maximal prescribed strain, leading to an image size of $600\times 1550$ (height $\times$ width). (ii) the averaged speckle radius is $7\times 10^{-3}$. In \cref{fig:nonlinear_data}, we show an example of the simulated data generated from the forward model.\hfill$\diamond$

\subsection{Results}\label{subsec:nonlinear_results}
We utilize ESFIM for experimental design optimization. The surrogate FIM construction and the design optimization settings are given as follows.

\paragraph{Surrogate FIM} The surrogate is given by a residual network with 6 hidden layers and GELU activation, where each layer has a width of 240. The surrogate is trained on FIMs computed from 16,384 parameter and design-variable samples generated via QMC. The surrogate FIM achieves a mean relative accuracy of over 97\% in predicting the log.\ FIM on a testing dataset of 4,096 samples. Moreover, the surrogate achieves a mean relative accuracy of over 99.4\% in predicting the approximate information gain in \cref{eq:approximate_ig}. In the right panel of \cref{fig:linear_testing_accuracy}, we visualize the surrogate FIM sample complexity, i.e., testing error as a function of training sample size. A standard MC rate of convergence at $N_{\textrm{train}}^{-1/2}$ is observed.\hfill$\diamond$

\paragraph{Design Optimization}  Two sets of results on design optimization of batch sizes 1--3 are presented: (i) ESFIM maximization with bound constraints and no additional regularization, and (ii) ESFIM maximization with shared geometry, bound constraints, and Tikhonov regularization on the loading path. In particular, we consider the following regularization term for the control points $z_l$ of a single loading path
\begin{equation}\label{eq:strain_rate_reg}
    \mathcal{P}(z_l) = \lambda\int_{0}^1 |\dot{u}_l(s; z_l)|^2 \dd s + \chi_{[0, 1]^{10}}(z_l),
\end{equation}
where the $\lambda>0$ is a regularization constant, $u_l$ is the interpolated loading path, $\chi$ is the indicator function for enforcing bound constraints, which returns 0 if the control is inside the bounds and infinite if outside. For batched designs, the regularization terms for different experiments are averaged. This regularization term encourages the generated loading protocols to be smooth and physically realizable by eliminating step-like control inputs that require high actuation speeds.\hfill$\diamond$

Now we present and analyze the ESFIM-optimized designs.

\subsubsection{Optimized Designs}

\begin{figure}[htb]
    \centering
    \scalebox{0.9}{
    \renewcommand{\arraystretch}{1.5}
        \begin{tabular}{|c||c c || c c c|}\hline
        \multicolumn{6}{|c|}{ESFIM-Optimized Batched Experimental Designs With No Regularization ($\lambda=0$)} \\\hline\hline
        Batch Size of 1 &\multicolumn{2}{c||}{Batch Size of 2} & \multicolumn{3}{c|}{ Batch Size of 3}\\\hline
        \raisebox{-0.05\linewidth}{\includegraphics[width=0.15\linewidth]{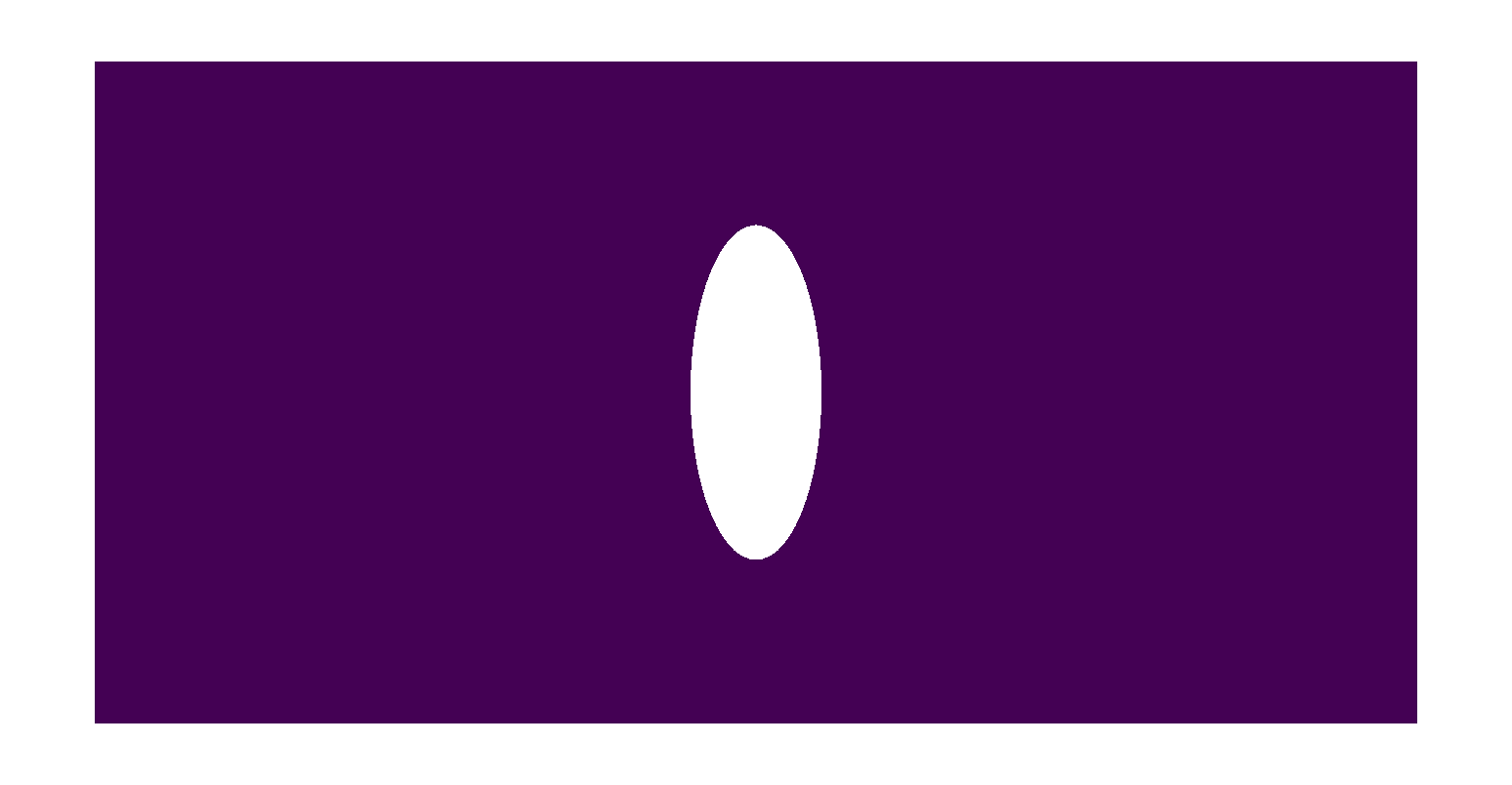}}&\raisebox{-0.05\linewidth}{\includegraphics[width=0.15\linewidth]{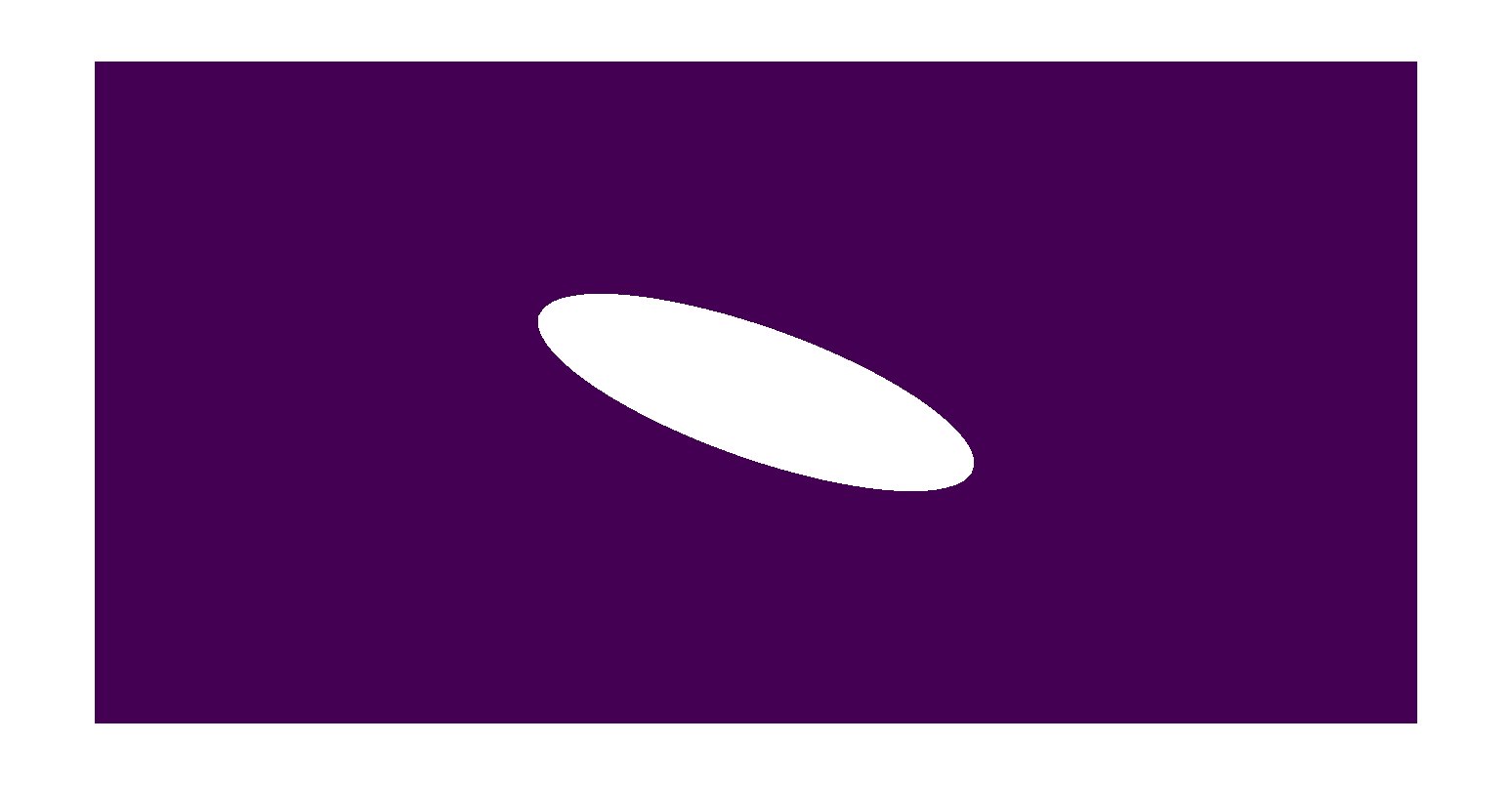}}  & \raisebox{-0.05\linewidth}{\includegraphics[width=0.15\linewidth]{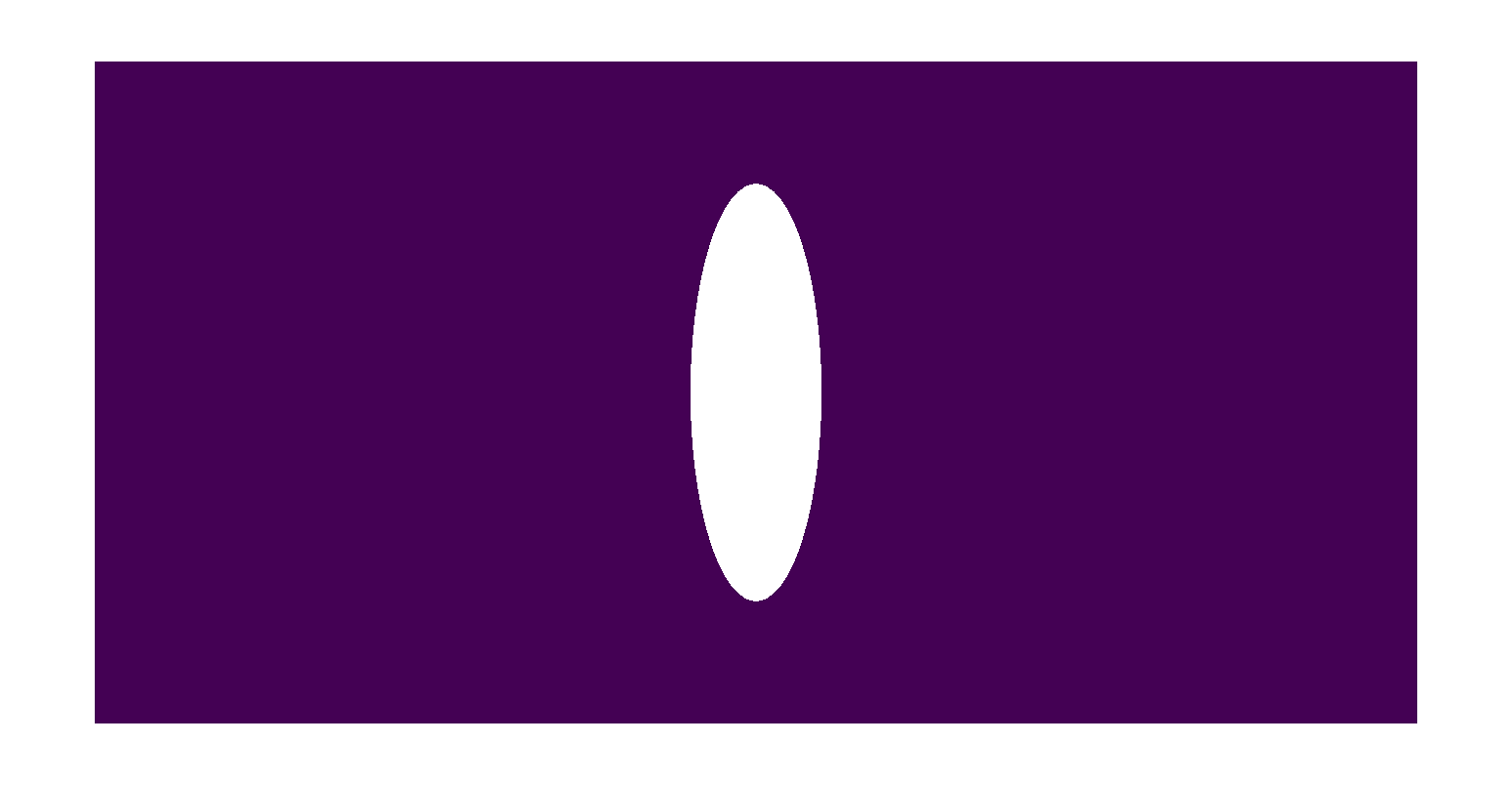}} & \raisebox{-0.05\linewidth}{\includegraphics[width=0.15\linewidth]{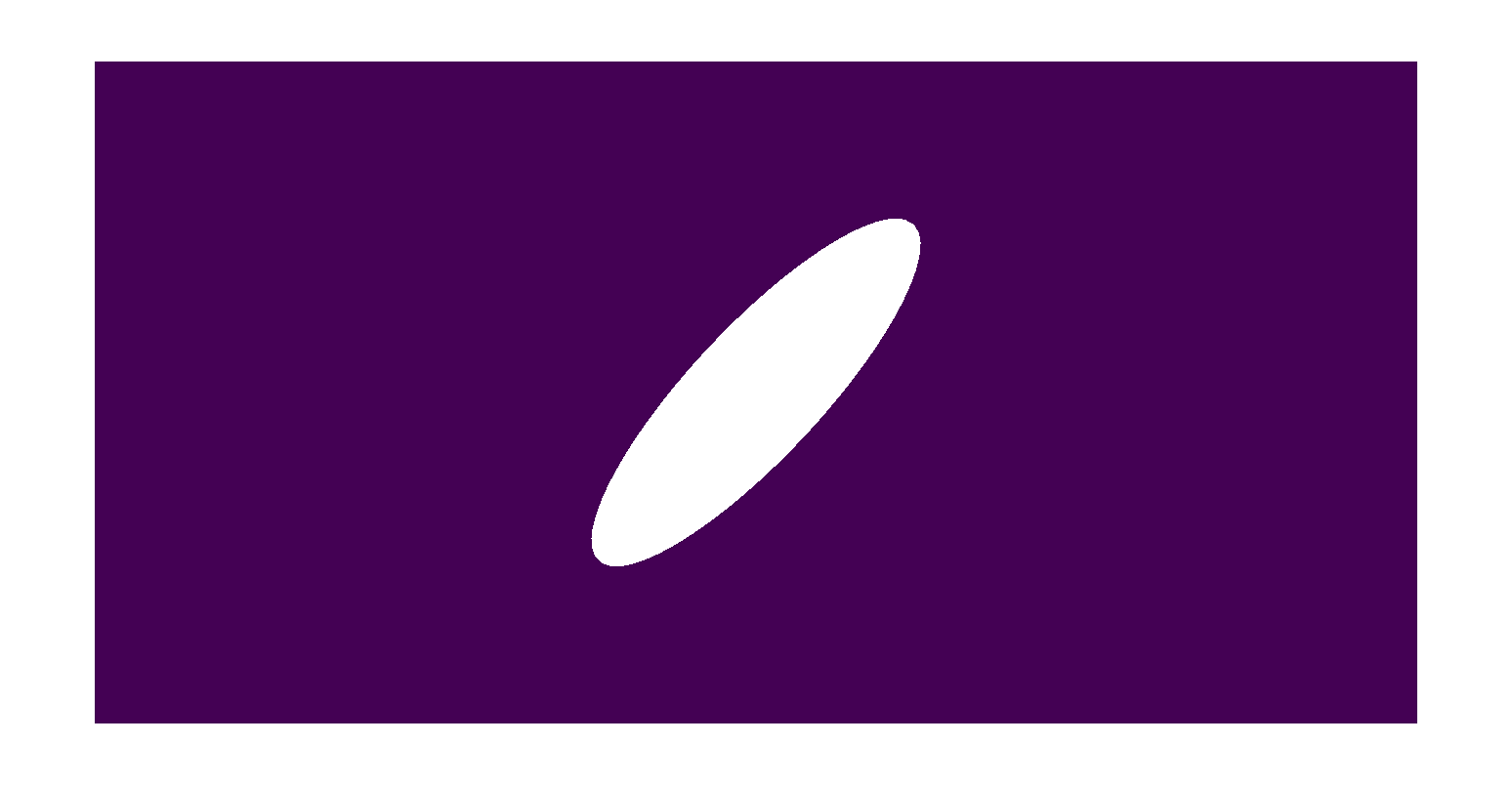}} & \raisebox{-0.05\linewidth}{\includegraphics[width=0.15\linewidth]{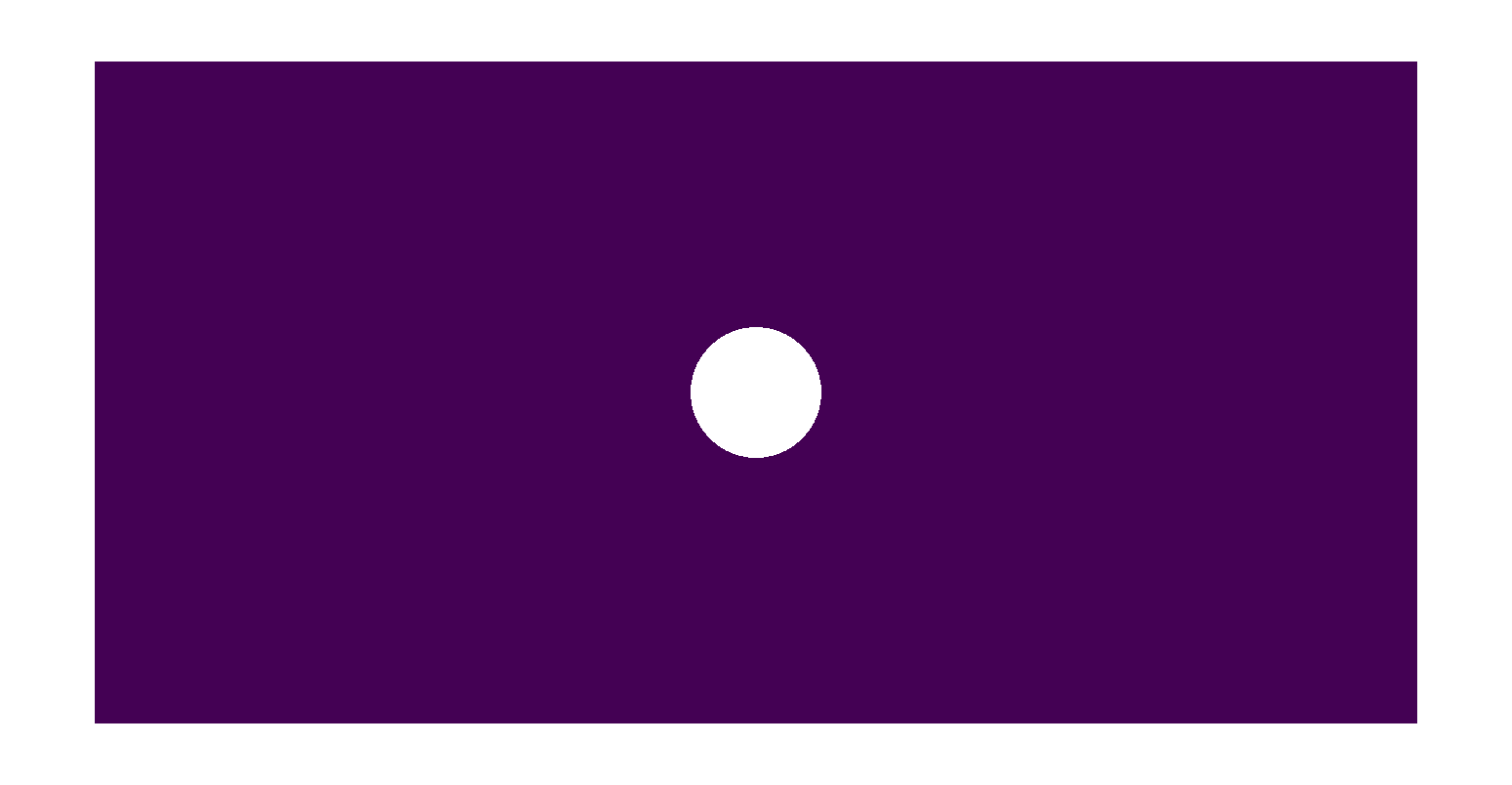}} & \raisebox{-0.05\linewidth}{\includegraphics[width=0.15\linewidth]{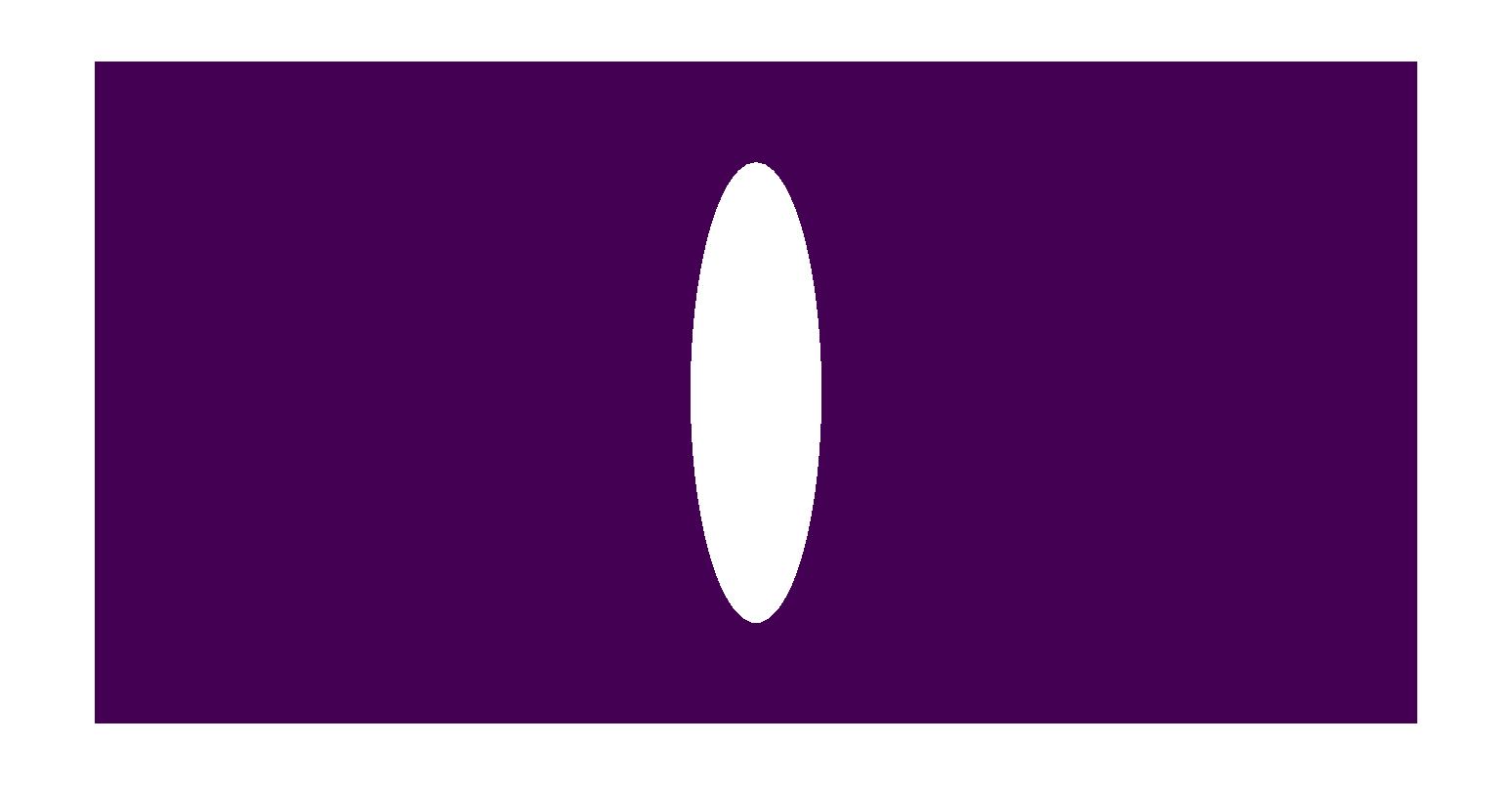}}\\
            \includegraphics[width=0.15\linewidth]{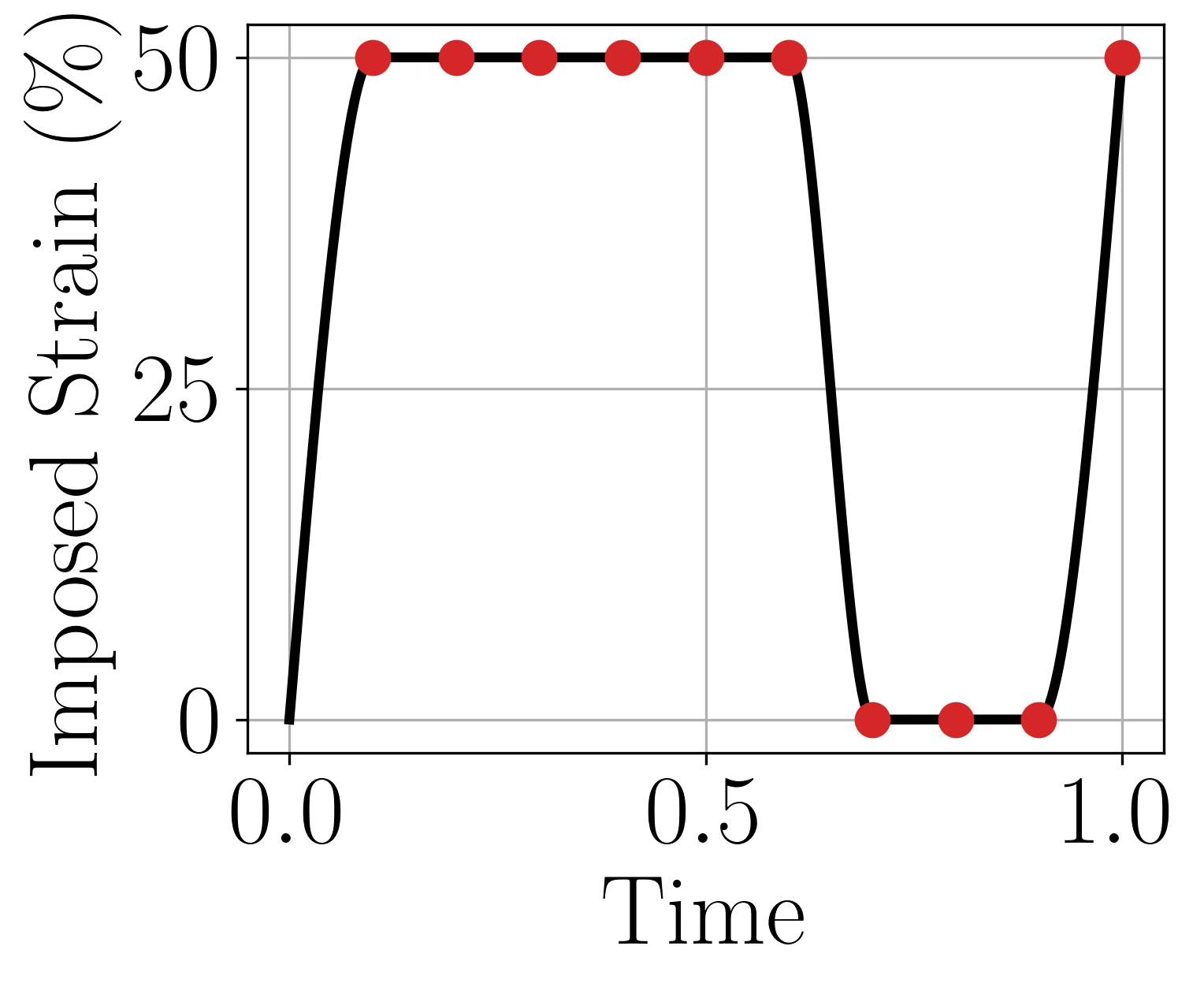}&\includegraphics[width=0.14\linewidth]{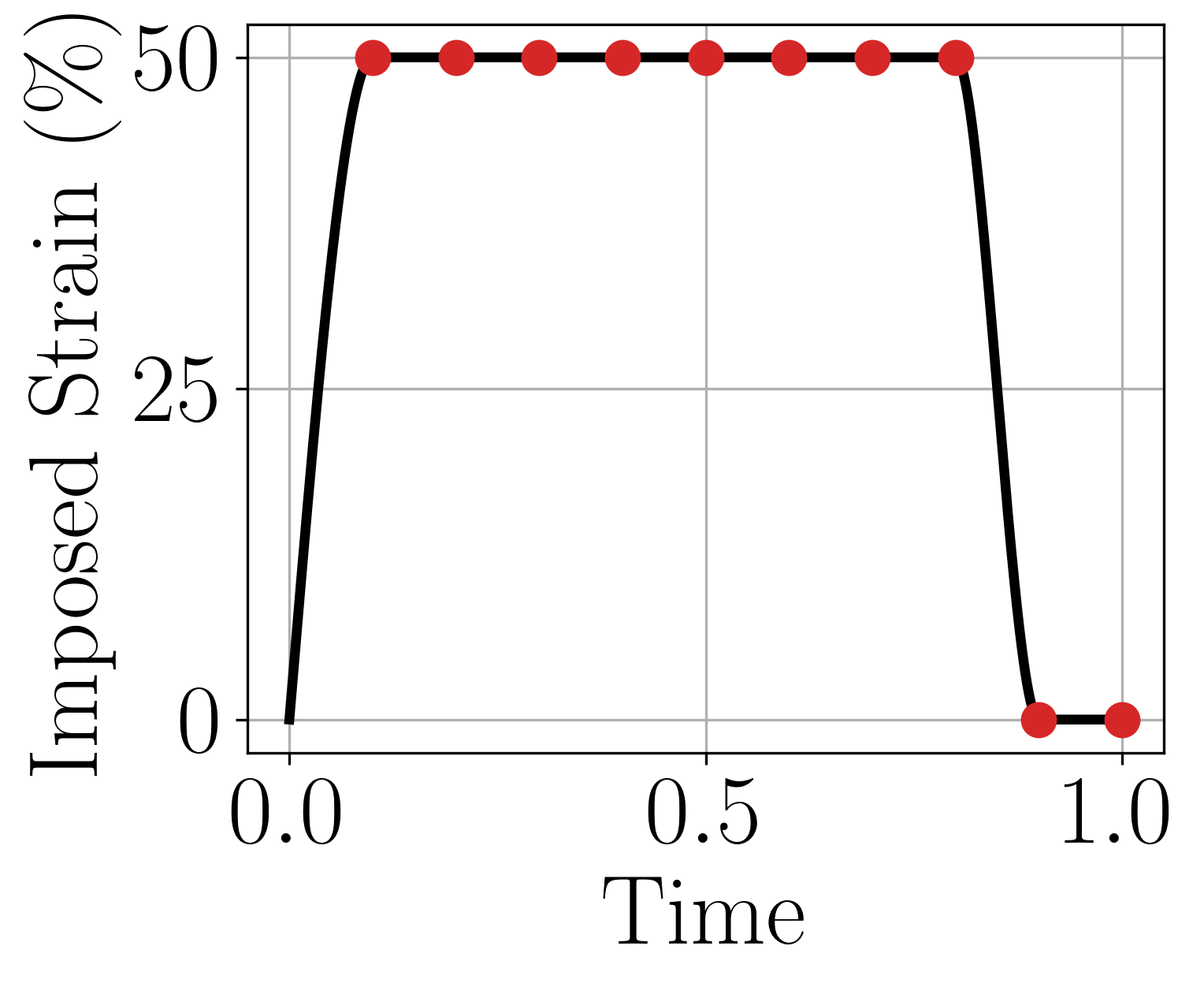} &  \includegraphics[width=0.15\linewidth]{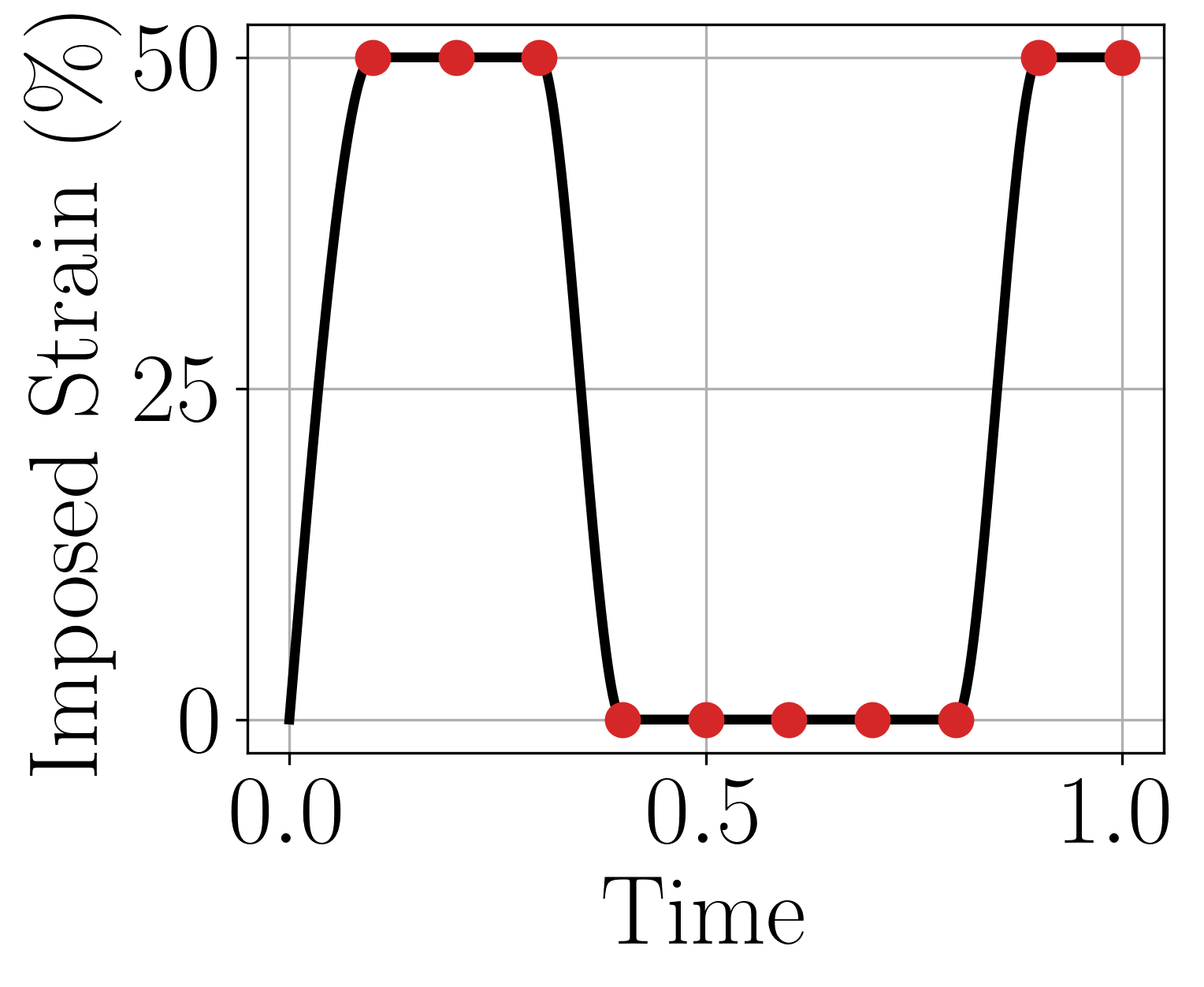} & \includegraphics[width=0.15\linewidth]{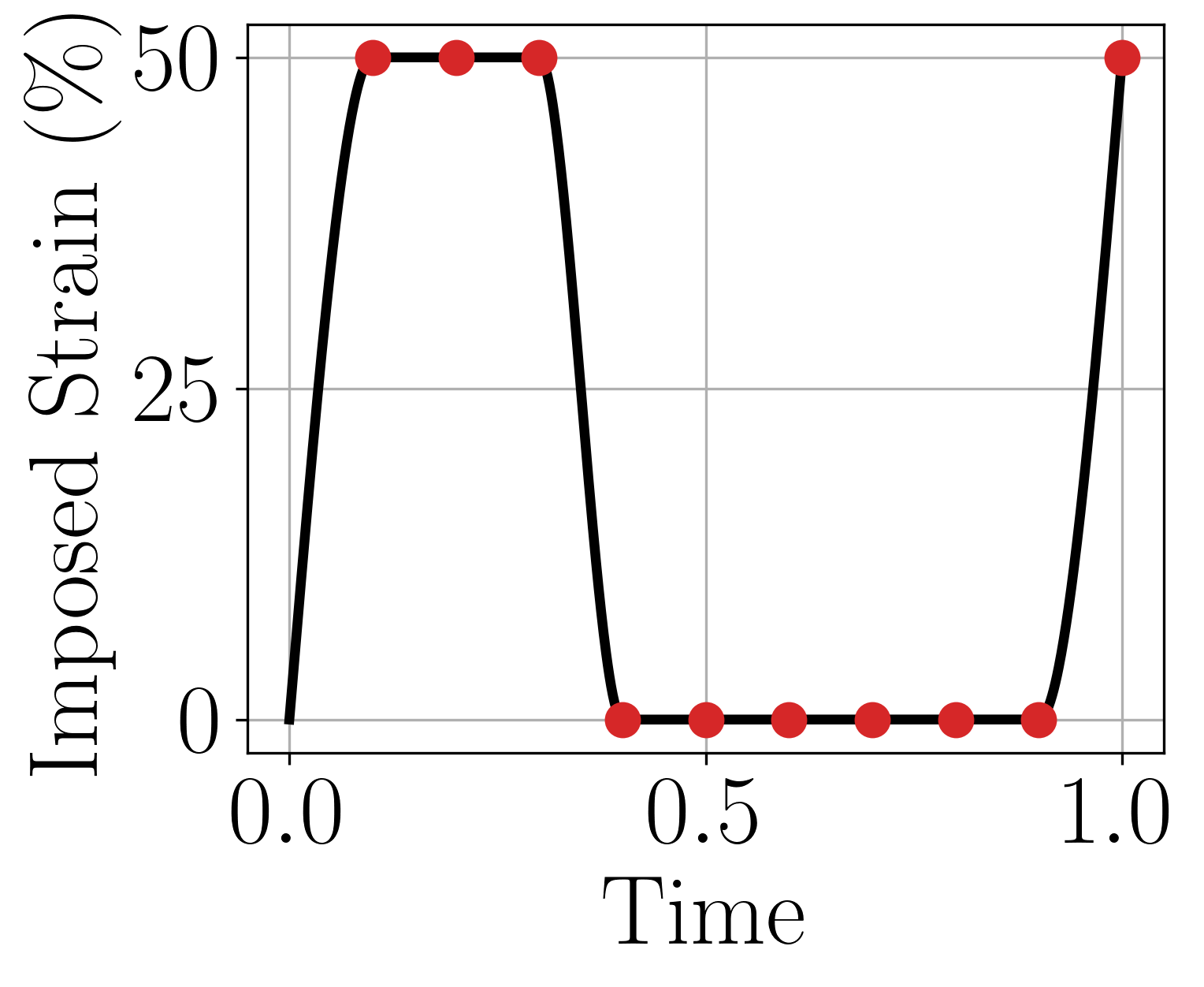} & \includegraphics[width=0.15\linewidth]{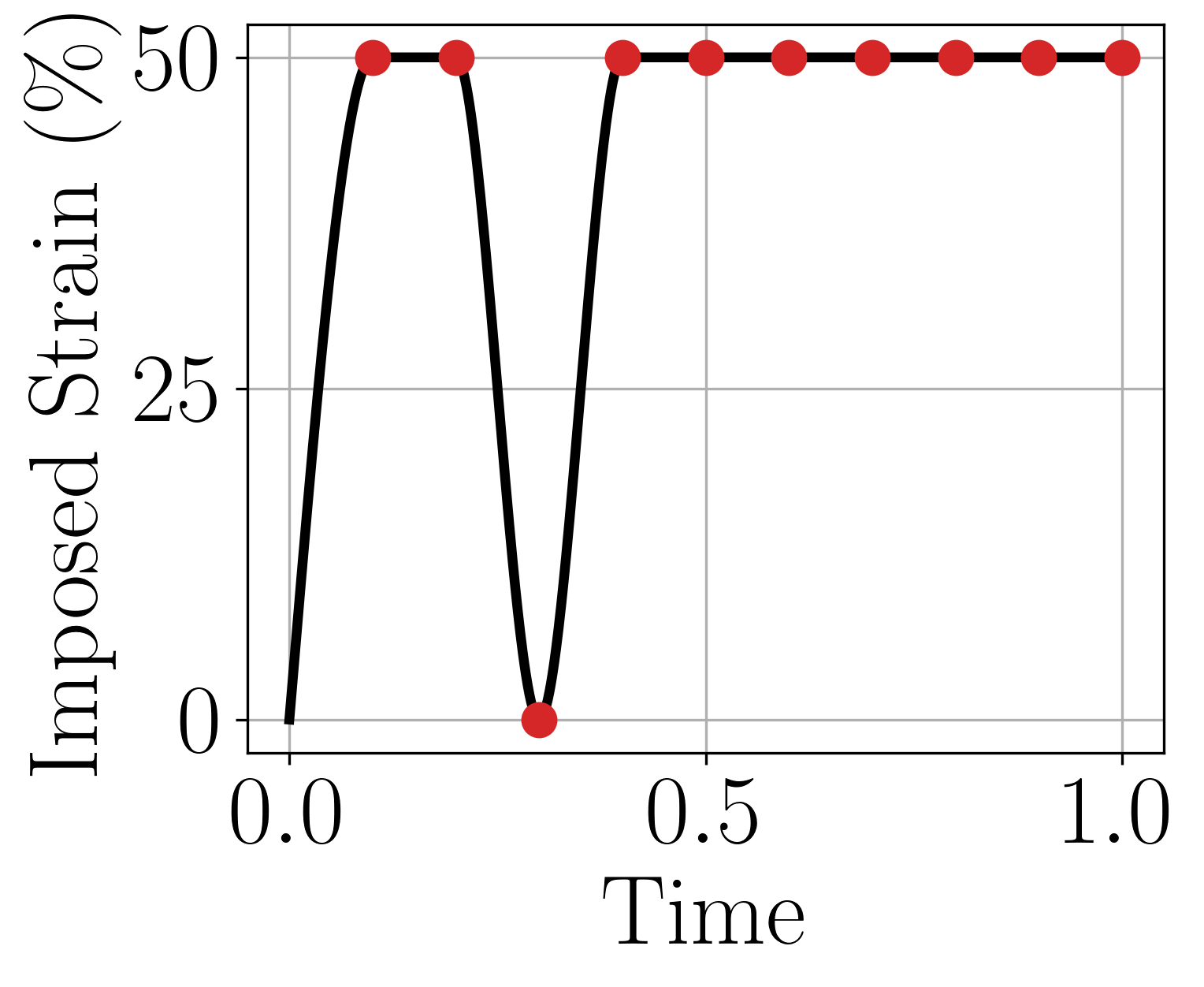} & \includegraphics[width=0.15\linewidth]{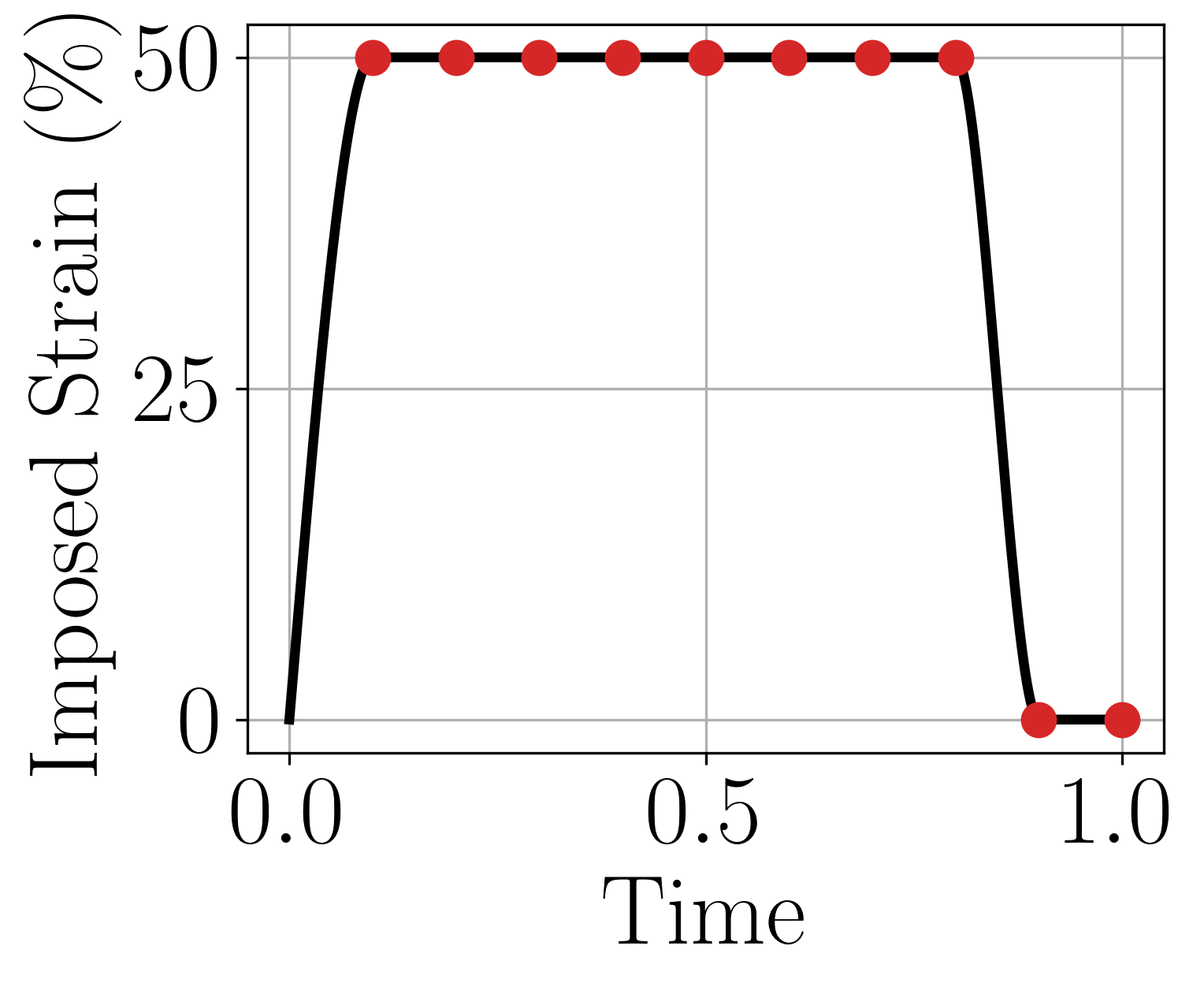}\\\hline
        \end{tabular}
    \renewcommand{\arraystretch}{1.0}
    }\\[5 pt]
    \begin{minipage}{0.42\linewidth}
    \centering
        \includegraphics[width=\linewidth]{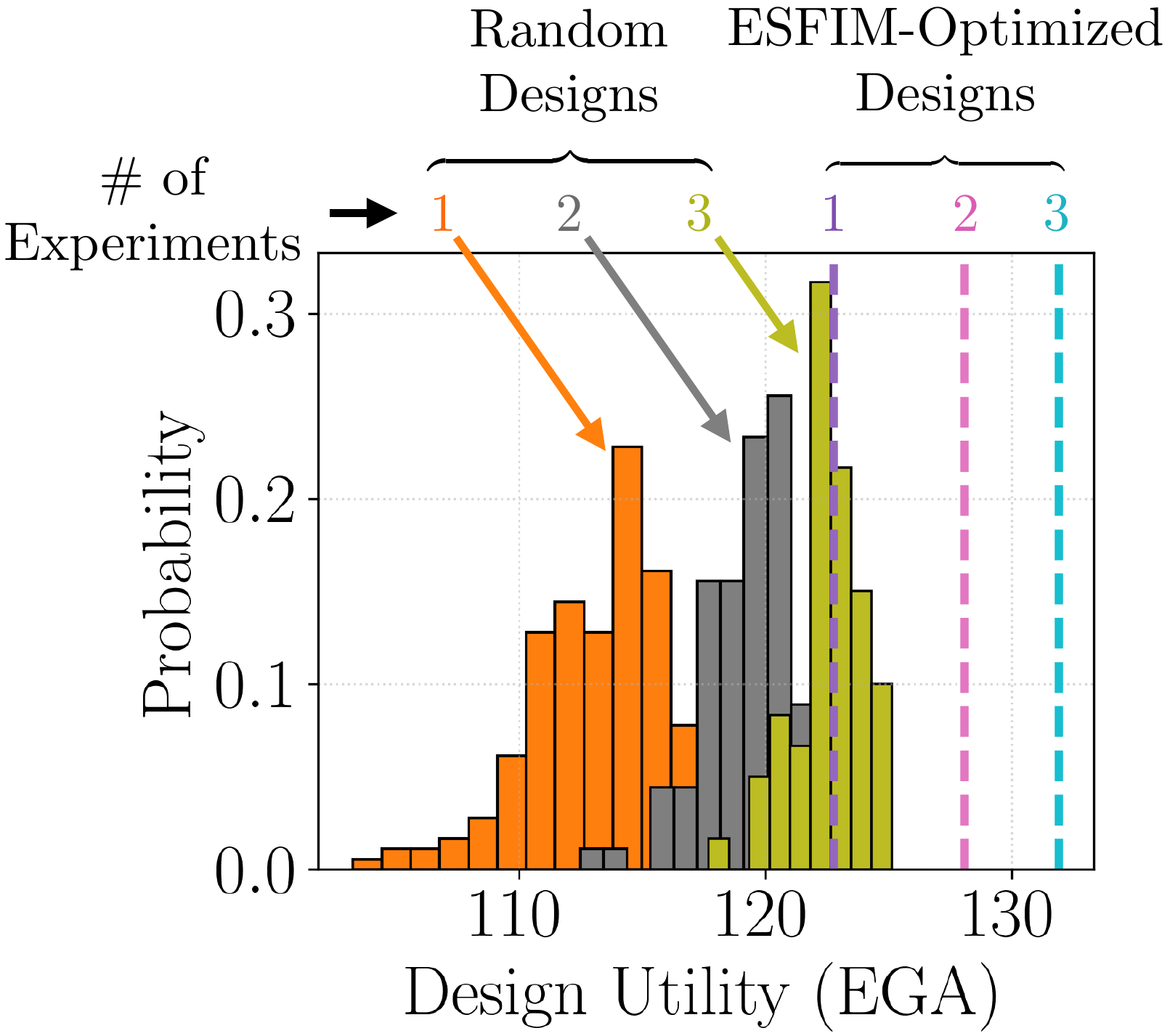}
    \end{minipage}
    \begin{minipage}{0.54\linewidth}
    \centering

    \scalebox{0.9}{
    \renewcommand{\arraystretch}{2}
    \begin{tabular}{|c|c|c|c|c|}\hline
    \multicolumn{5}{|c|}{\makecell{Expected Improvements in \\ the EGA Design Utility of Optimized Designs \\ Relative to Random Designs}}\\\hline\hline
    \multicolumn{2}{|c|}{\multirow{2}{*}{\backslashbox{Optimized}{Random}}} & \multicolumn{3}{c|}{Batch Size} \\\cline{3-5}
    \multicolumn{2}{|c|}{}& 1 & 2 & 3\\\hline
    \parbox[t]{5mm}{\multirow{3}{*}{\rotatebox[origin=c]{90}{Batch Size}}} & 1 & $9.1\%$ & $3.0\%$ & $0.2\%$\\\cline{2-5}
     & 2 & $13.3\%$ & $7.5\%$ & $4.5\%$\\\cline{2-5}
     & 3 & $16.7\%$ & $10.7\%$ & $7.7\%$\\\hline
    \end{tabular}
    \renewcommand{\arraystretch}{1.0}
    }
    \end{minipage}

    \caption{Visualization and analysis of ESFIM-optimized designs for uniaxial testing of nonlinear viscoelastic materials. (\emph{Top}) The designs obtained via ESFIM maximization with batch sizes of 1--3. (\emph{Bottom Left}) The EGA design utility values for random and ESFIM-optimized designs with batch sizes of 1--3. (\emph{Bottom Right}) The expected improvements in the EGA of the optimized designs with batch sizes of 1--3 relative to random designs with batch sizes of 1--3.}
    \label{fig:nonlinear_batched_design}
\end{figure}

The optimal designs for batch sizes of 1--3 with no regularization are shown in the top panel of \cref{fig:nonlinear_batched_design}. For a single experiment, the design optimization found a loading path that loads to maximum rapidly, holds for 8 control points, and rapidly unloads and holds for the last two control points. As the batch size increases, the additional loading path introduces transitions between maximal strain and no strain. This pattern is similar to the linear model---improving the identifiability of the parameters associated with the instantaneous response becomes more rewarding as the batch size increases. Regarding specimen geometry, a preference for large aspect ratios is observed across all batch sizes. Distinct geometries are introduced as batch size increases: tilted holes with large aspect ratios for batch sizes of 2 and 3, and a small hole with an equal aspect ratio for batch size of 3.

We compare the optimized designs with an ensemble of 160 designs randomly drawn from the uniform distribution within the design variable bounds. The distribution of the EGA values for random designs is shown on the bottom-left panel of \cref{fig:nonlinear_batched_design}, along with the EGA values of the ESFIM-optimized designs. The expected utility improvements in the bottom-right panel of \cref{fig:nonlinear_batched_design} indicate that the optimized designs increased confidence in parameter identification by approximately 7--9\% compared to random designs of the same batch size, demonstrating ESFIM's cost-amortization capability and suitability for batched design.

\begin{figure}[htb]
     \centering
    \scalebox{0.9}{
    \renewcommand{\arraystretch}{1.5}
        \begin{tabular}{|c||c||c c || c c c|}\hline
        \multicolumn{7}{|c|}{\makecell{ESFIM-Optimized Batched Experimental Designs\\ with Shared Geometry and Strain Rate Regularization}} \\\hline\hline
        & Batch Size of 1 &\multicolumn{2}{c||}{Batch Size of 2} & \multicolumn{3}{c|}{ Batch Size of 3}\\\hline
       \parbox[t]{2mm}{\multirow{3}{*}{\rotatebox[origin=c]{90}{$\lambda=0.25$}}}& \raisebox{-0.045\linewidth}{\includegraphics[width=0.14\linewidth]{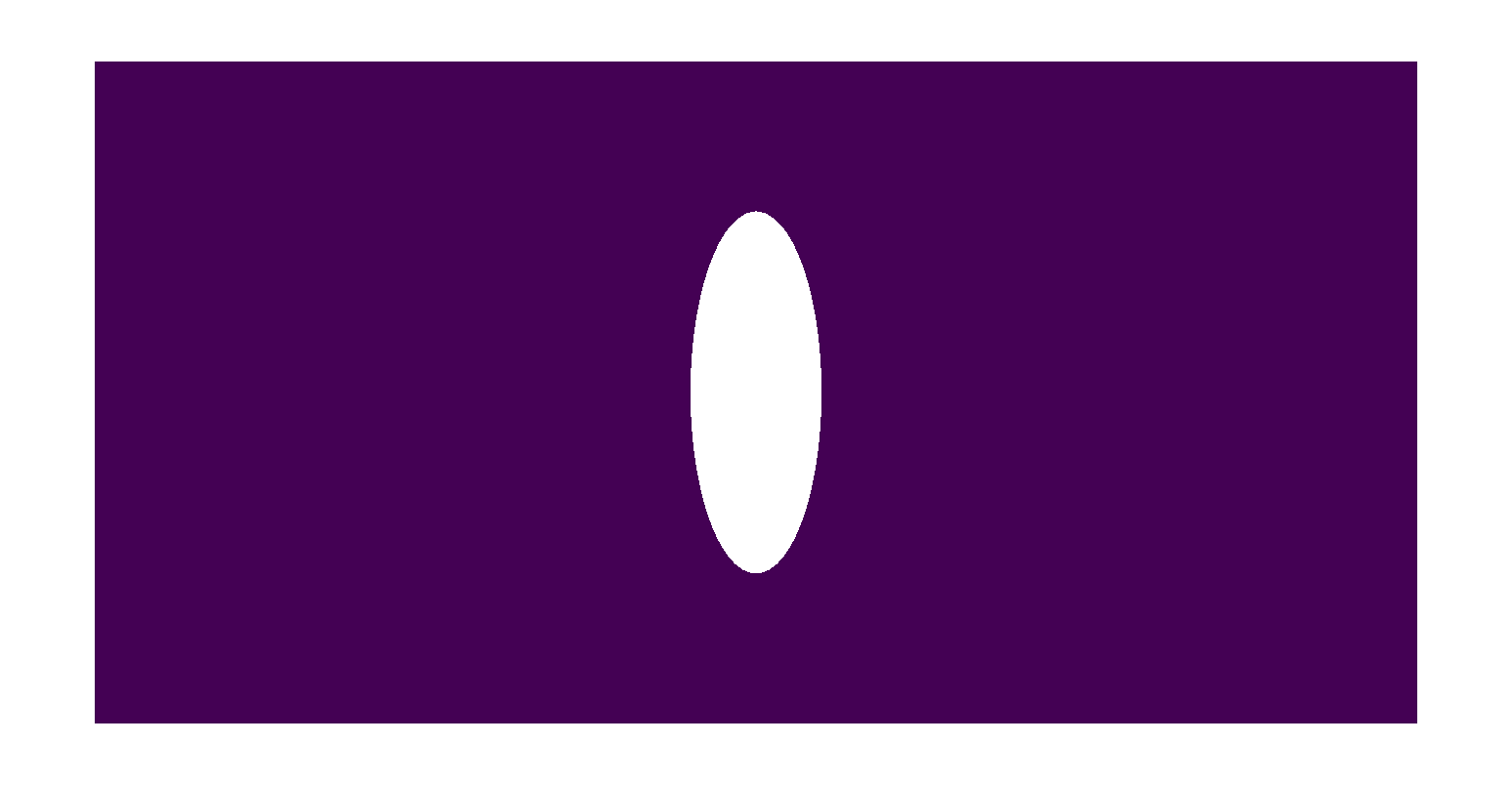}}&\multicolumn{2}{c||}{\raisebox{-0.045\linewidth}{\includegraphics[width=0.14\linewidth]{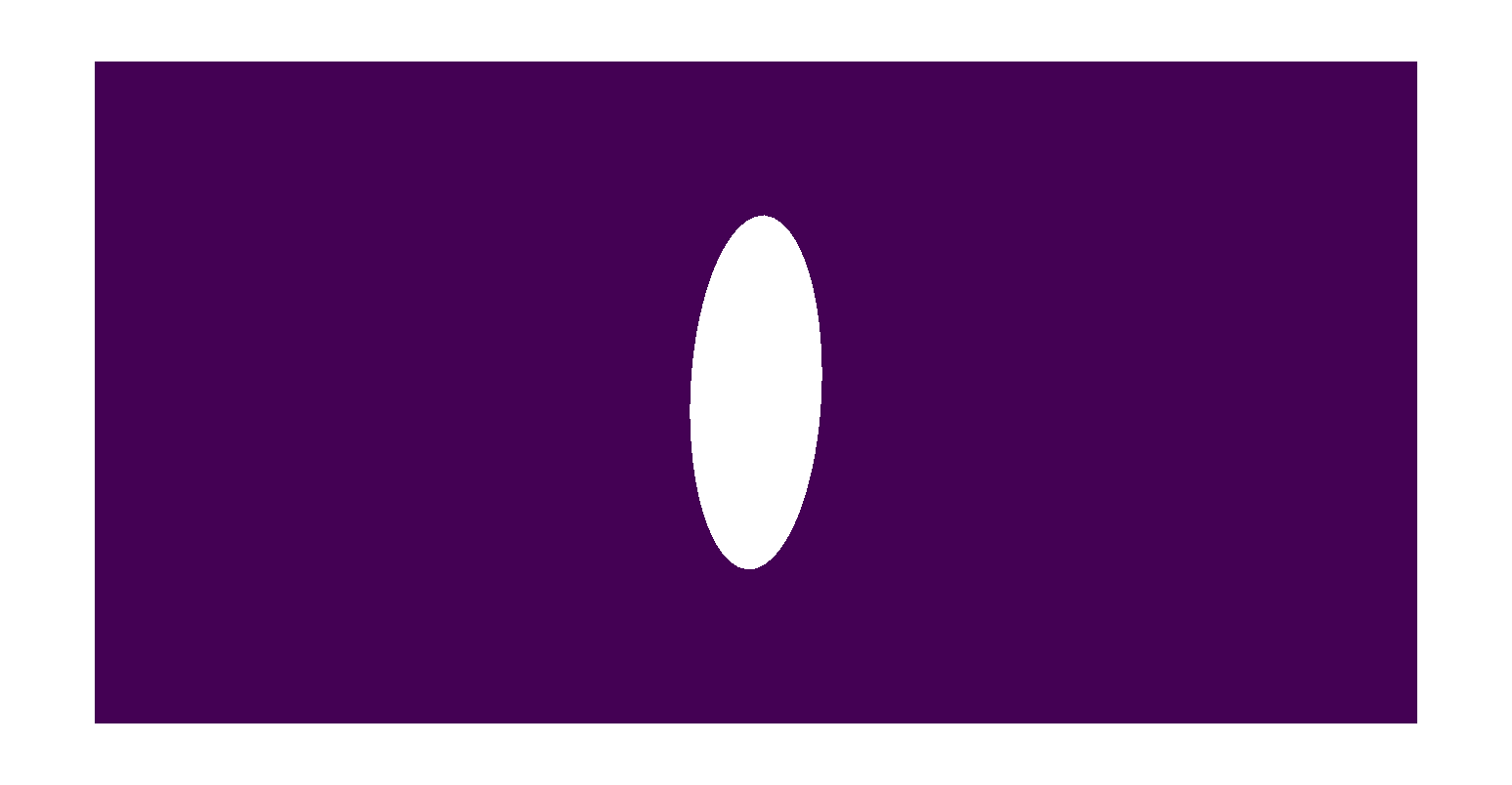}}} & \multicolumn{3}{c|}{\raisebox{-0.045\linewidth}{\includegraphics[width=0.14\linewidth]{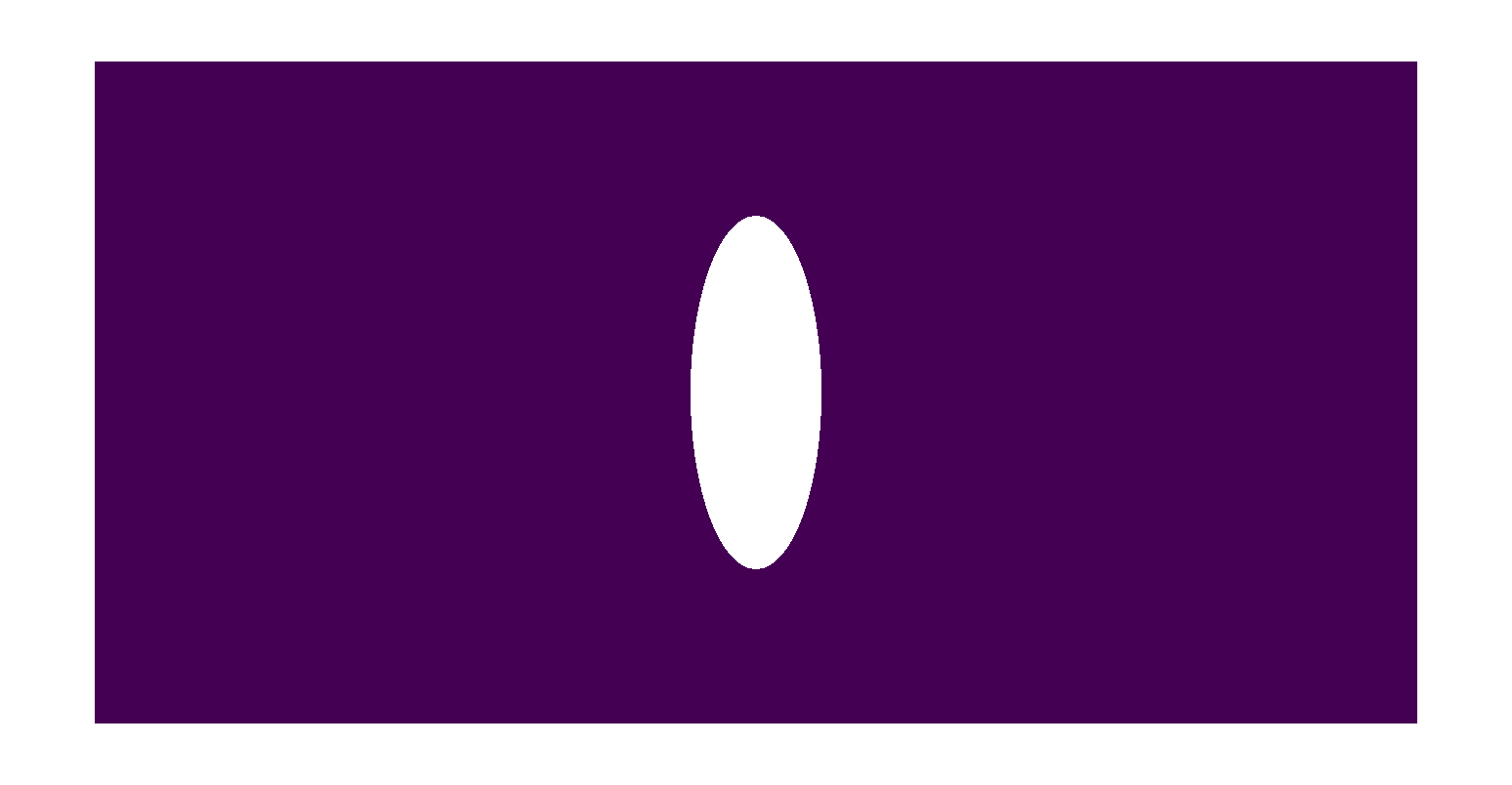}}} \\
        &    \includegraphics[width=0.14\linewidth]{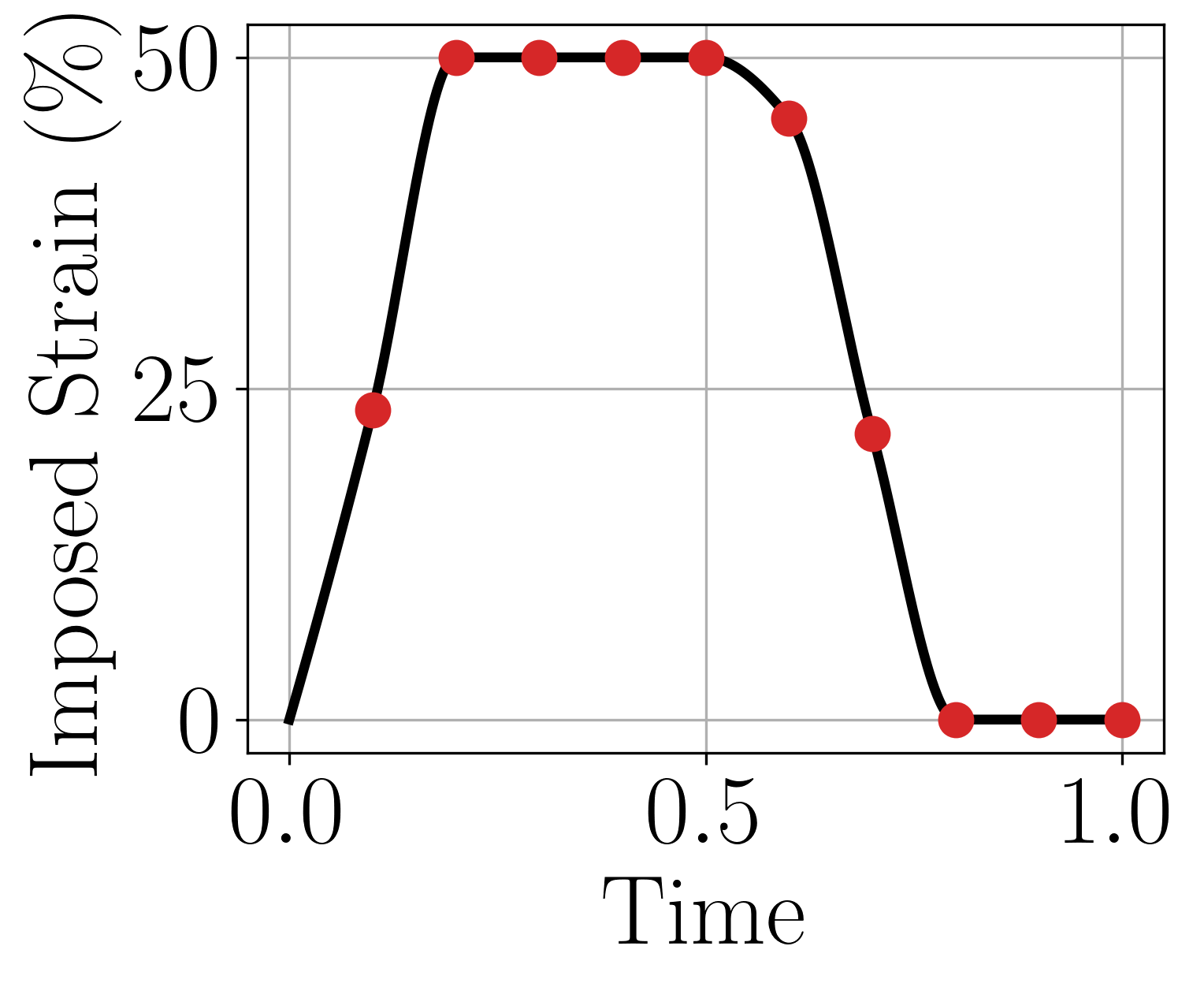}&\includegraphics[width=0.14\linewidth]{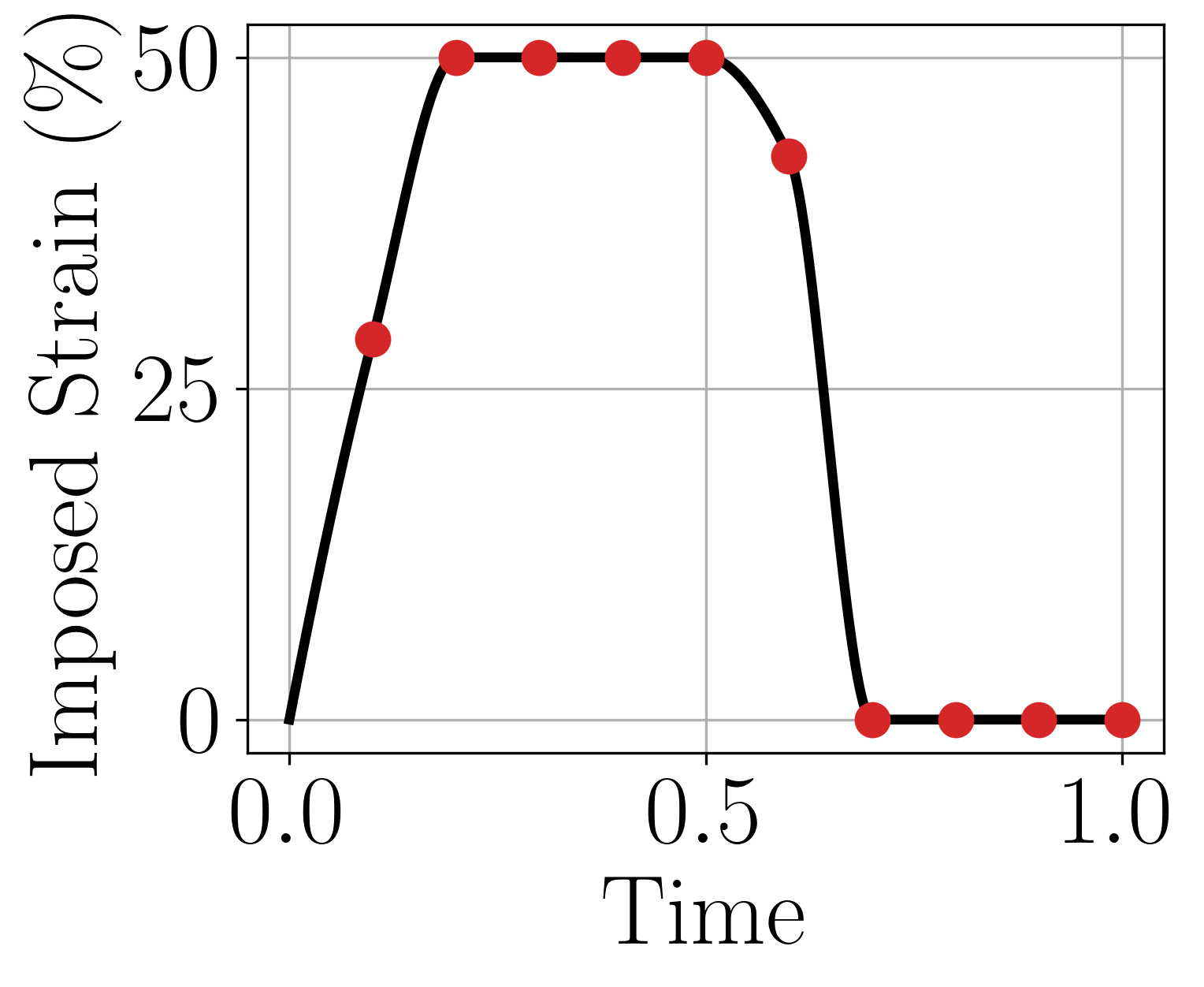} &  \includegraphics[width=0.14\linewidth]{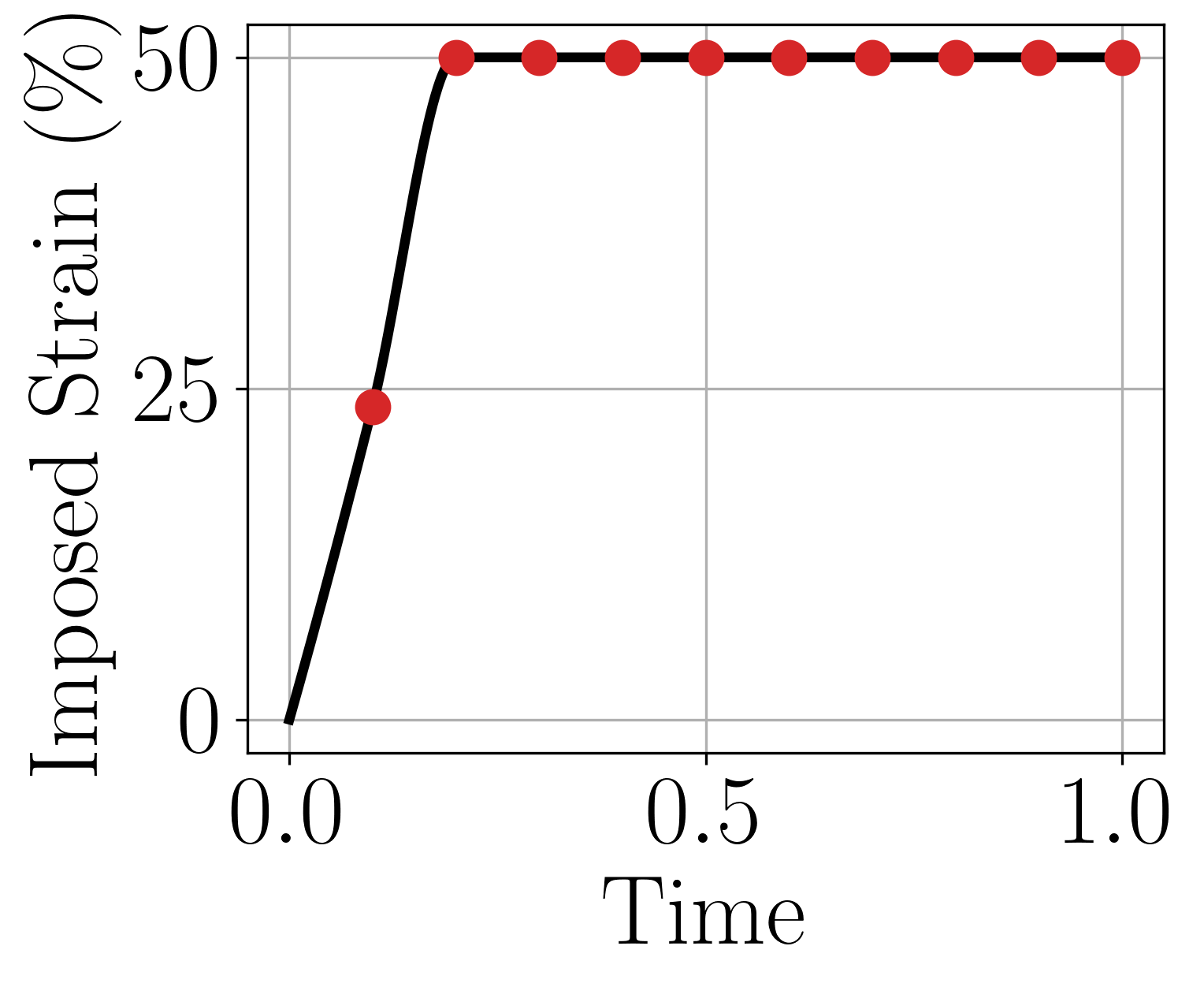} & \includegraphics[width=0.14\linewidth]{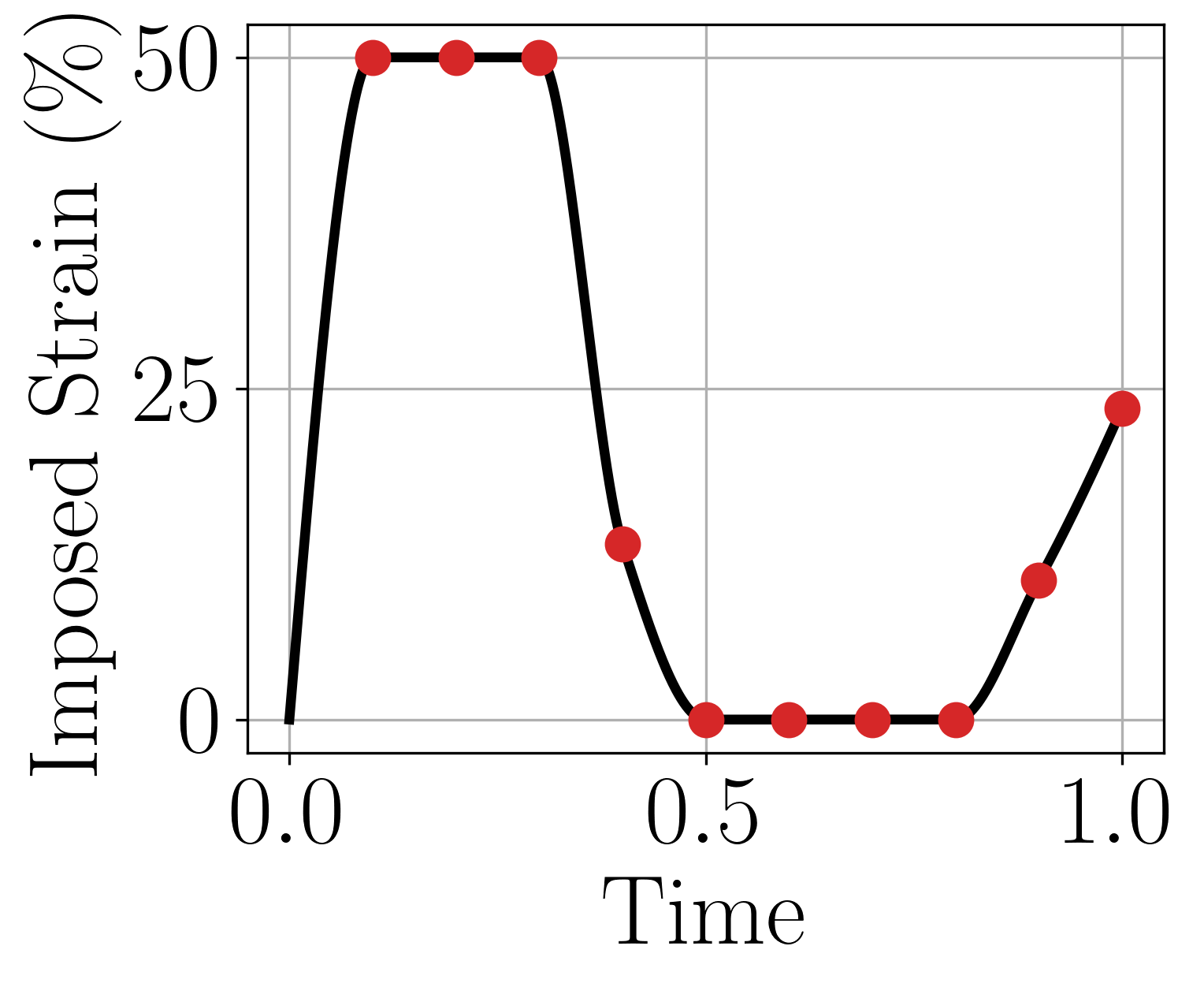} & \includegraphics[width=0.14\linewidth]{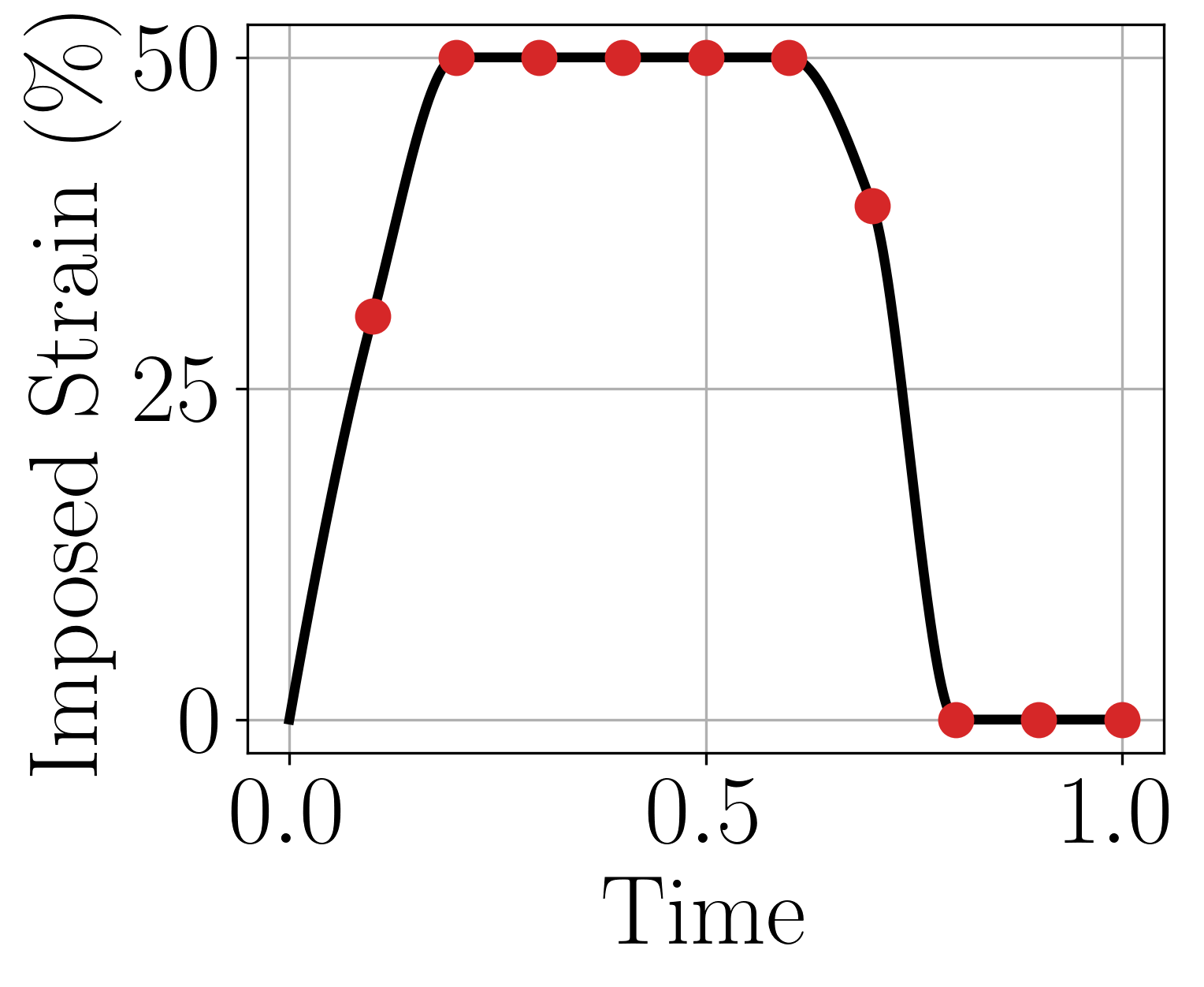} & \includegraphics[width=0.14\linewidth]{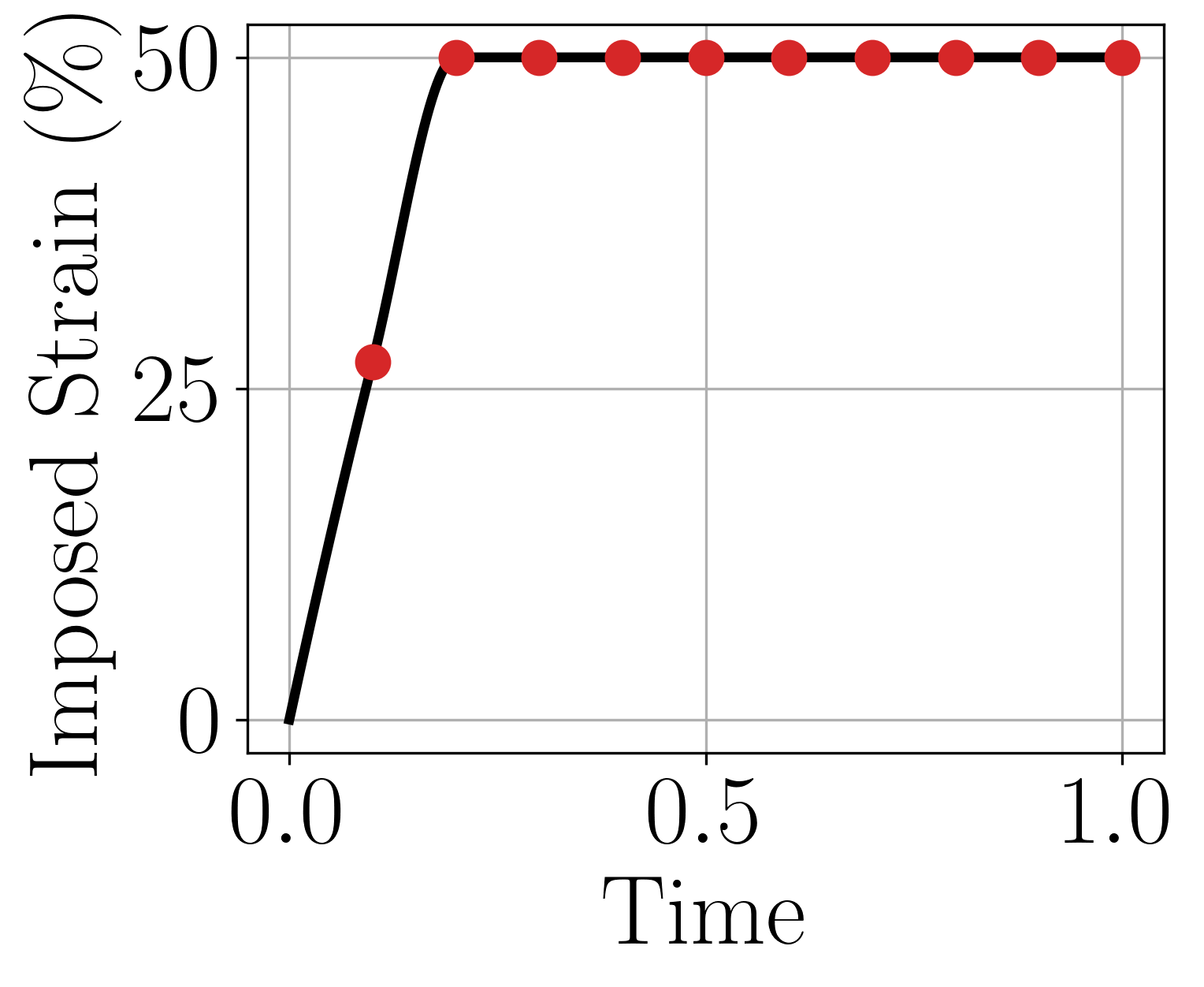}\\\hline
         \parbox[t]{2mm}{\multirow{3}{*}{\rotatebox[origin=c]{90}{$\lambda=1$}}} & \raisebox{-0.045\linewidth}{\includegraphics[width=0.14\linewidth]{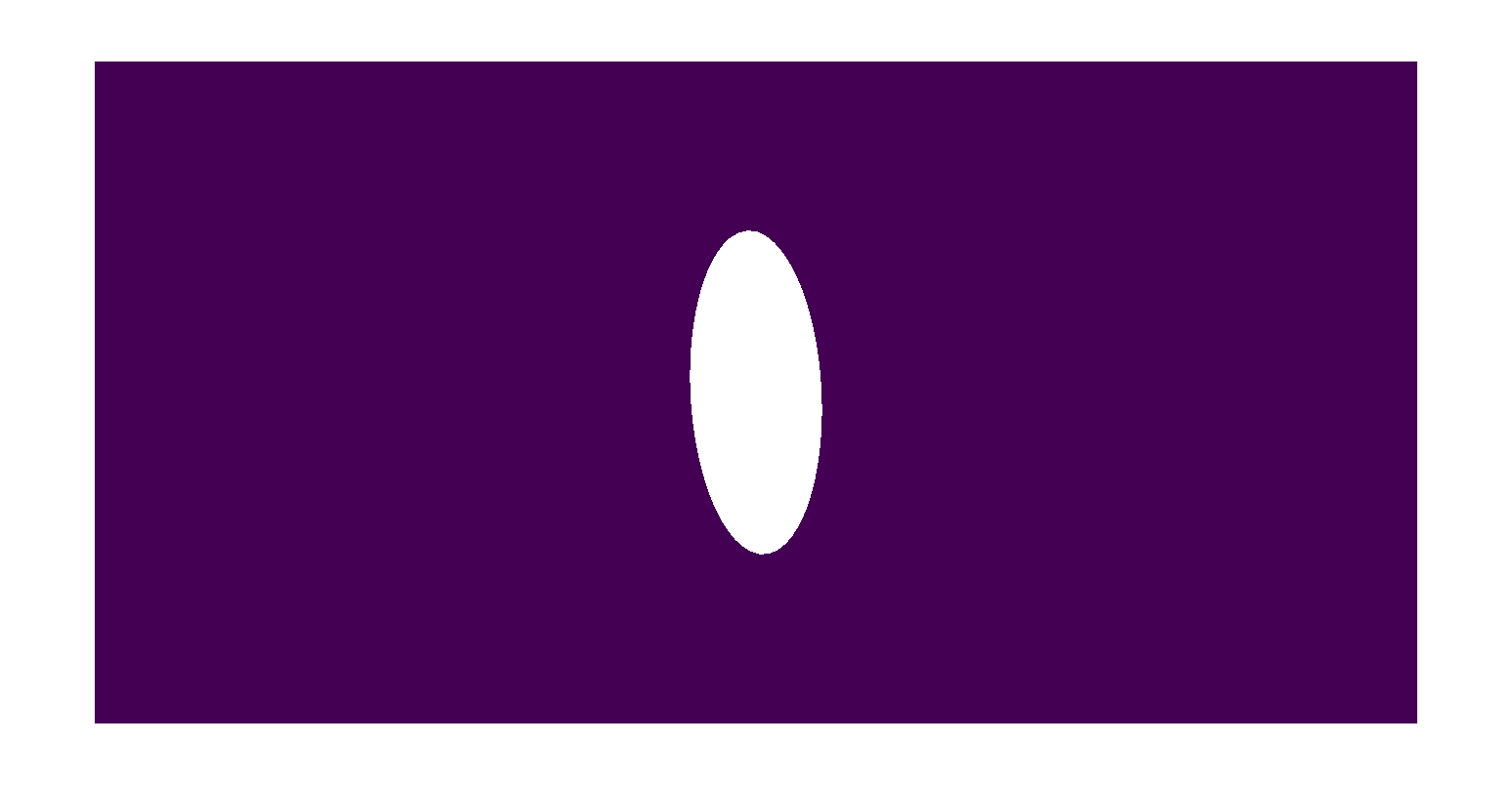}}&\multicolumn{2}{c||}{\raisebox{-0.045\linewidth}{\includegraphics[width=0.14\linewidth]{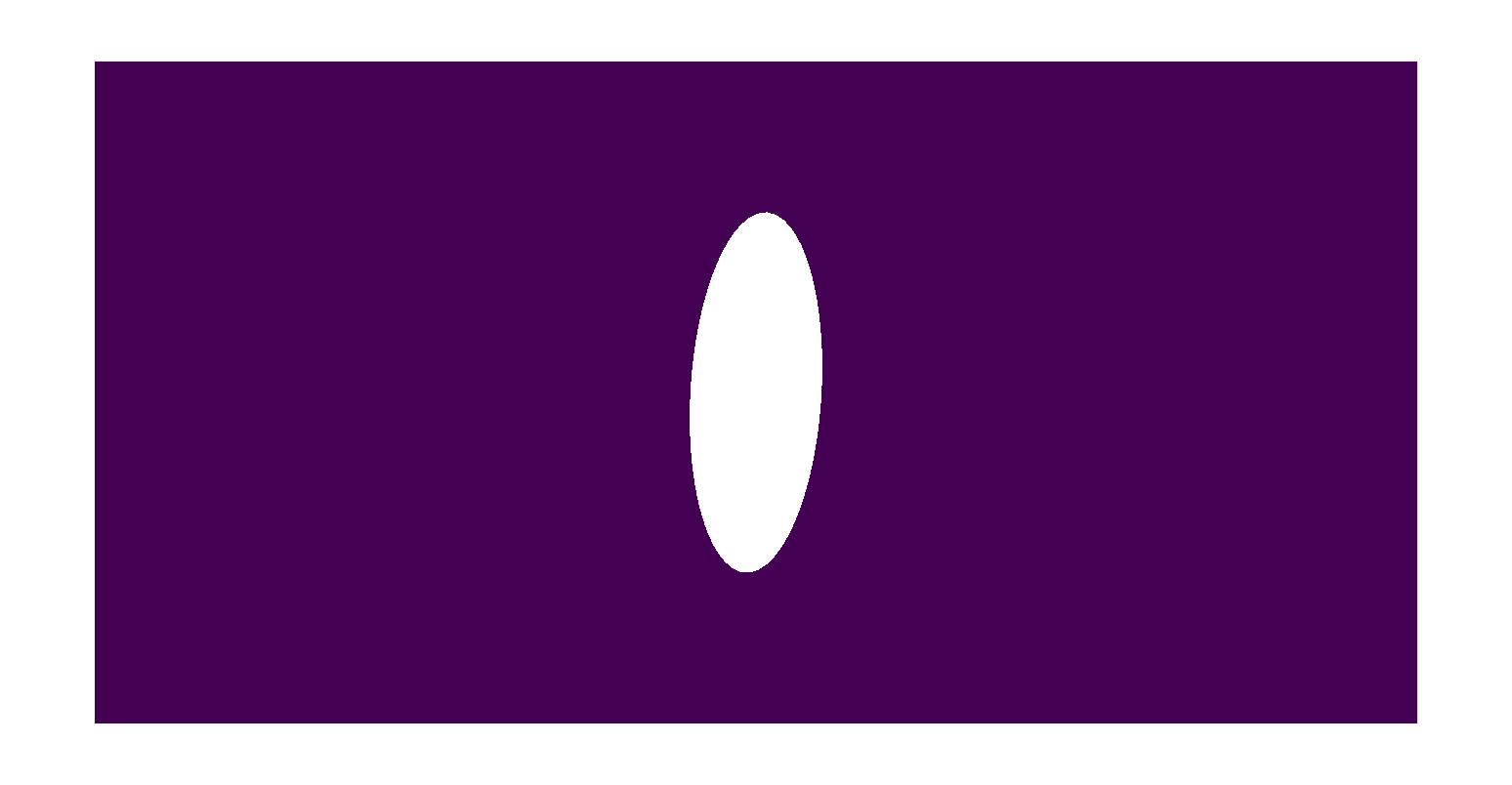}}} & \multicolumn{3}{c|}{\raisebox{-0.045\linewidth}{\includegraphics[width=0.14\linewidth]{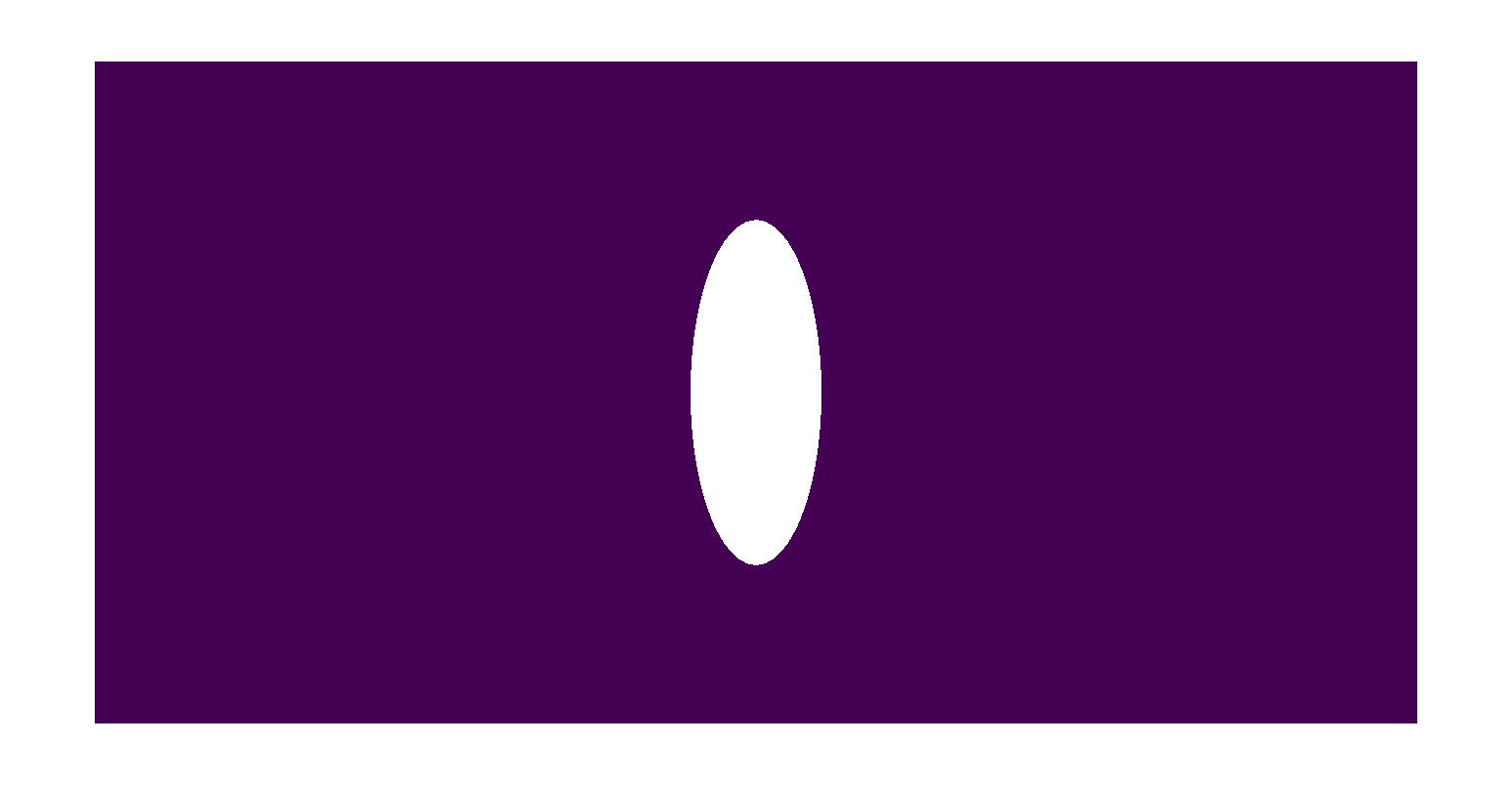}}} \\
        &    \includegraphics[width=0.14\linewidth]{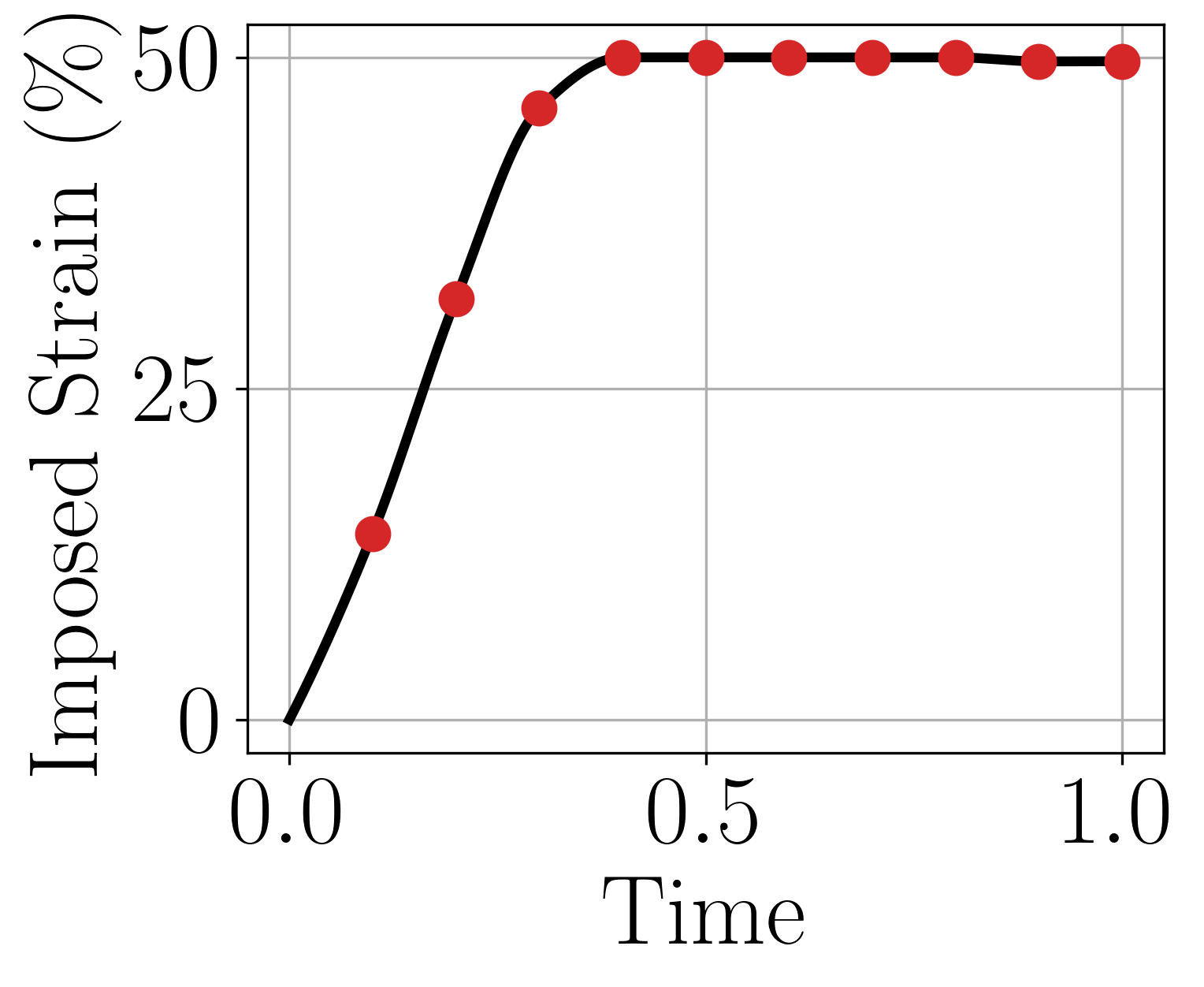}&\includegraphics[width=0.14\linewidth]{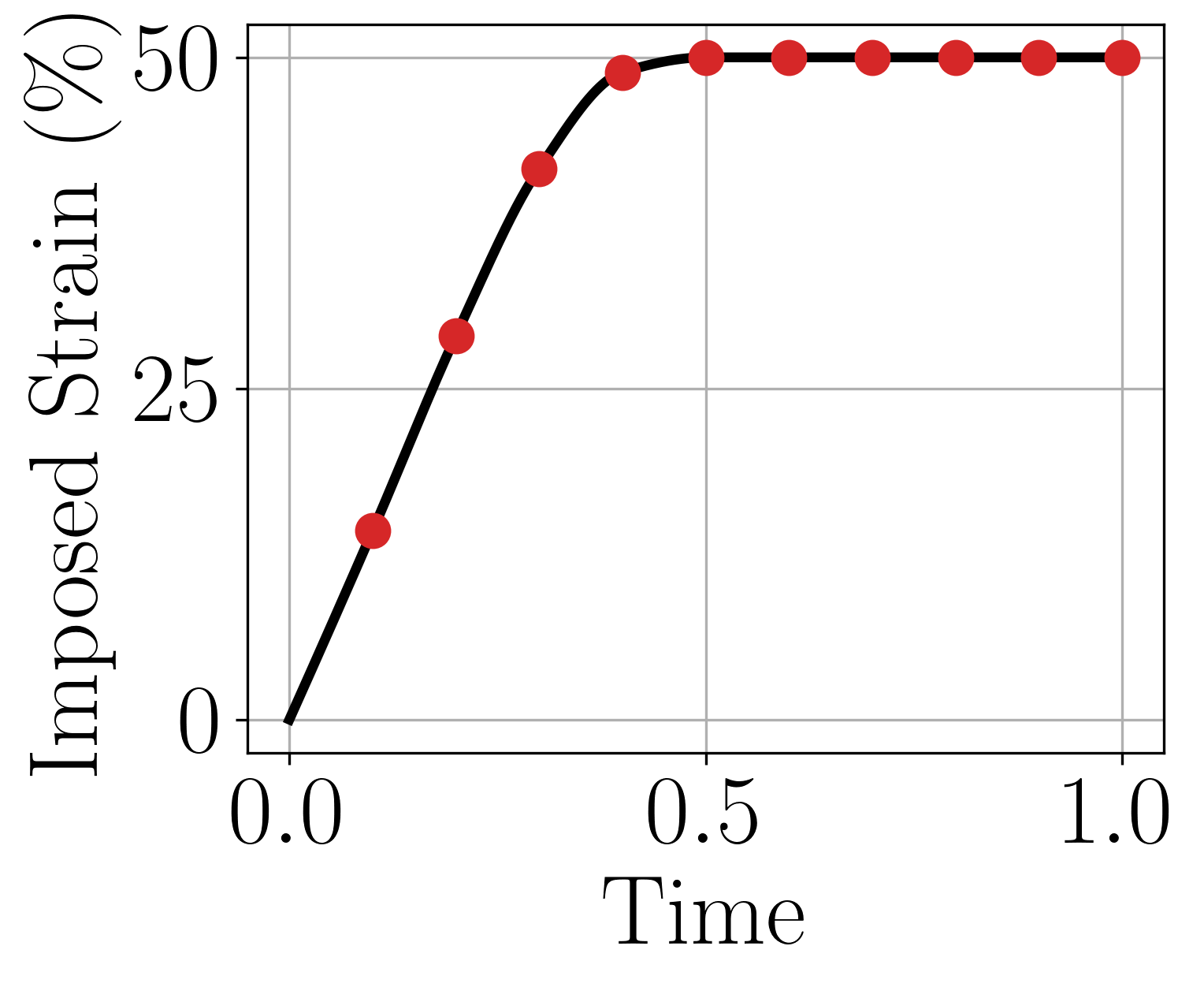} &  \includegraphics[width=0.14\linewidth]{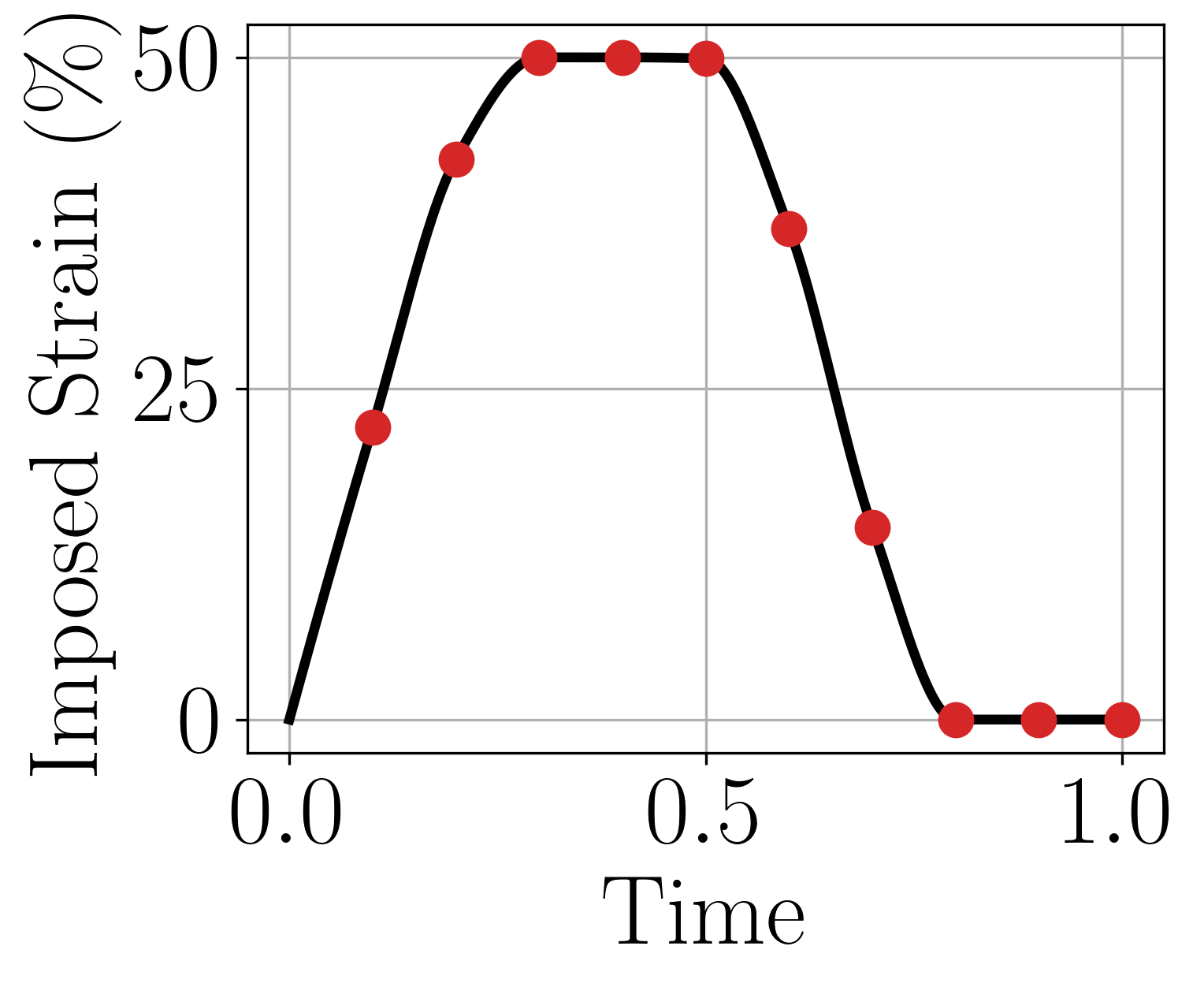} & \includegraphics[width=0.14\linewidth]{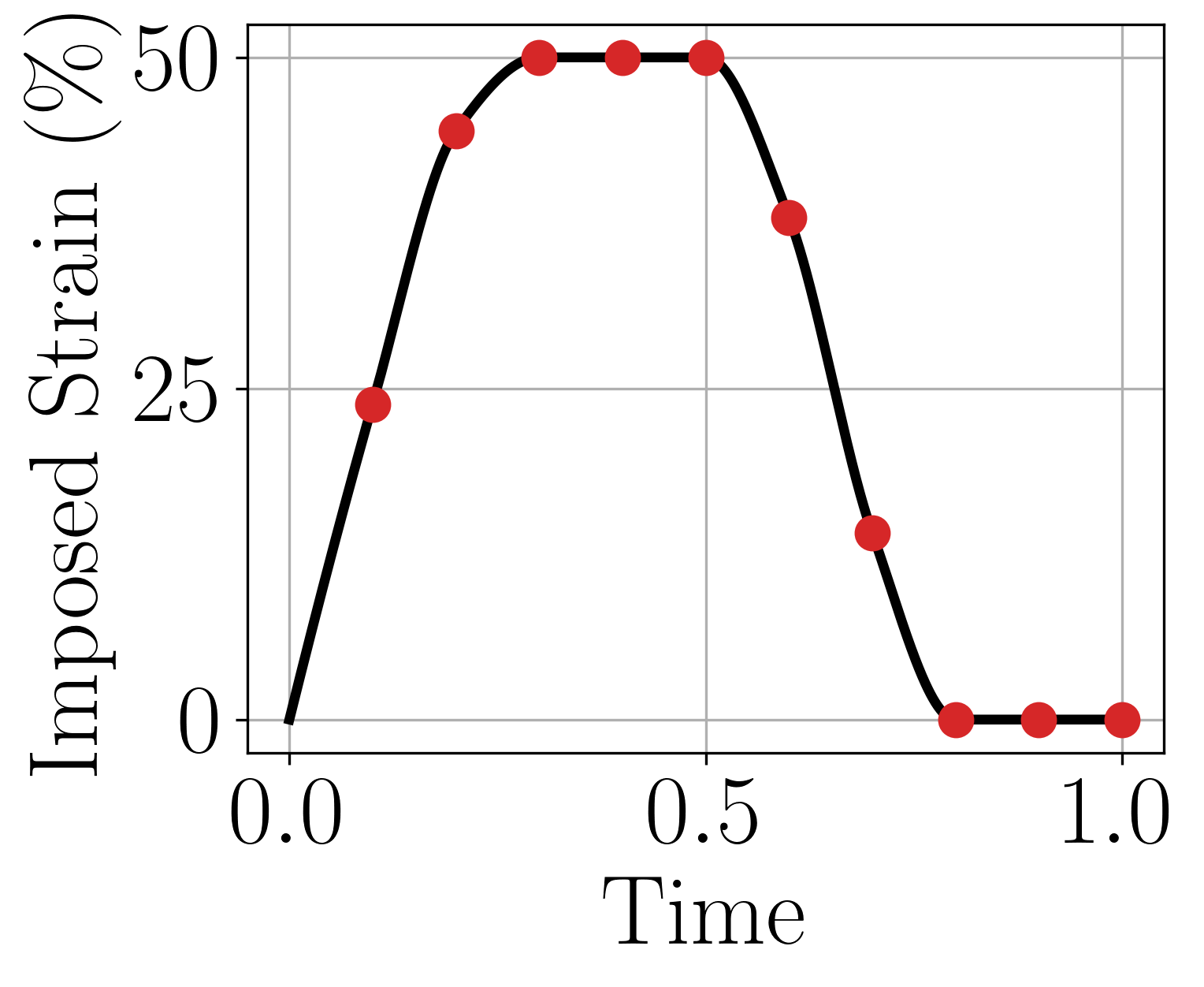} & \includegraphics[width=0.14\linewidth]{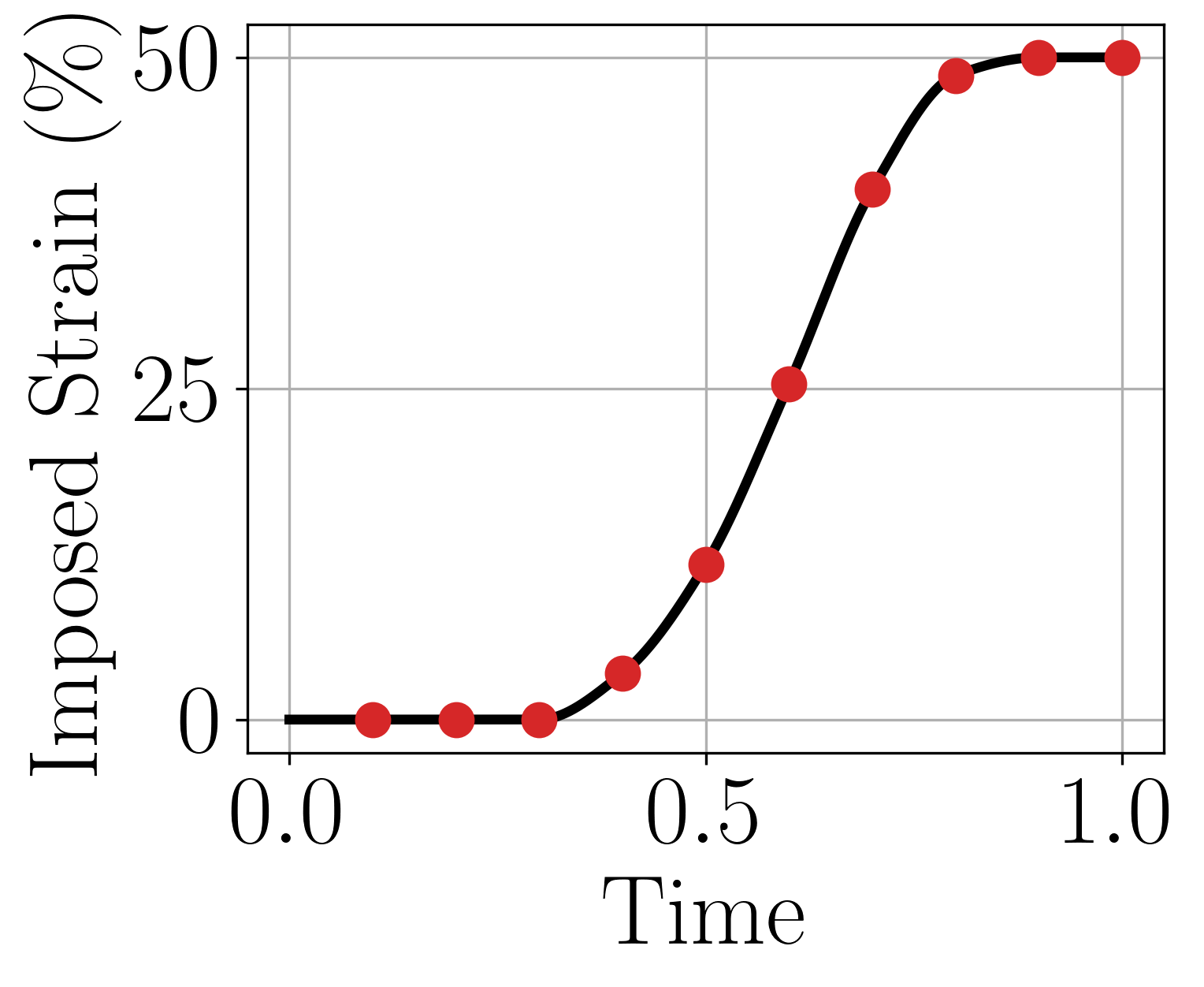} & \includegraphics[width=0.14\linewidth]{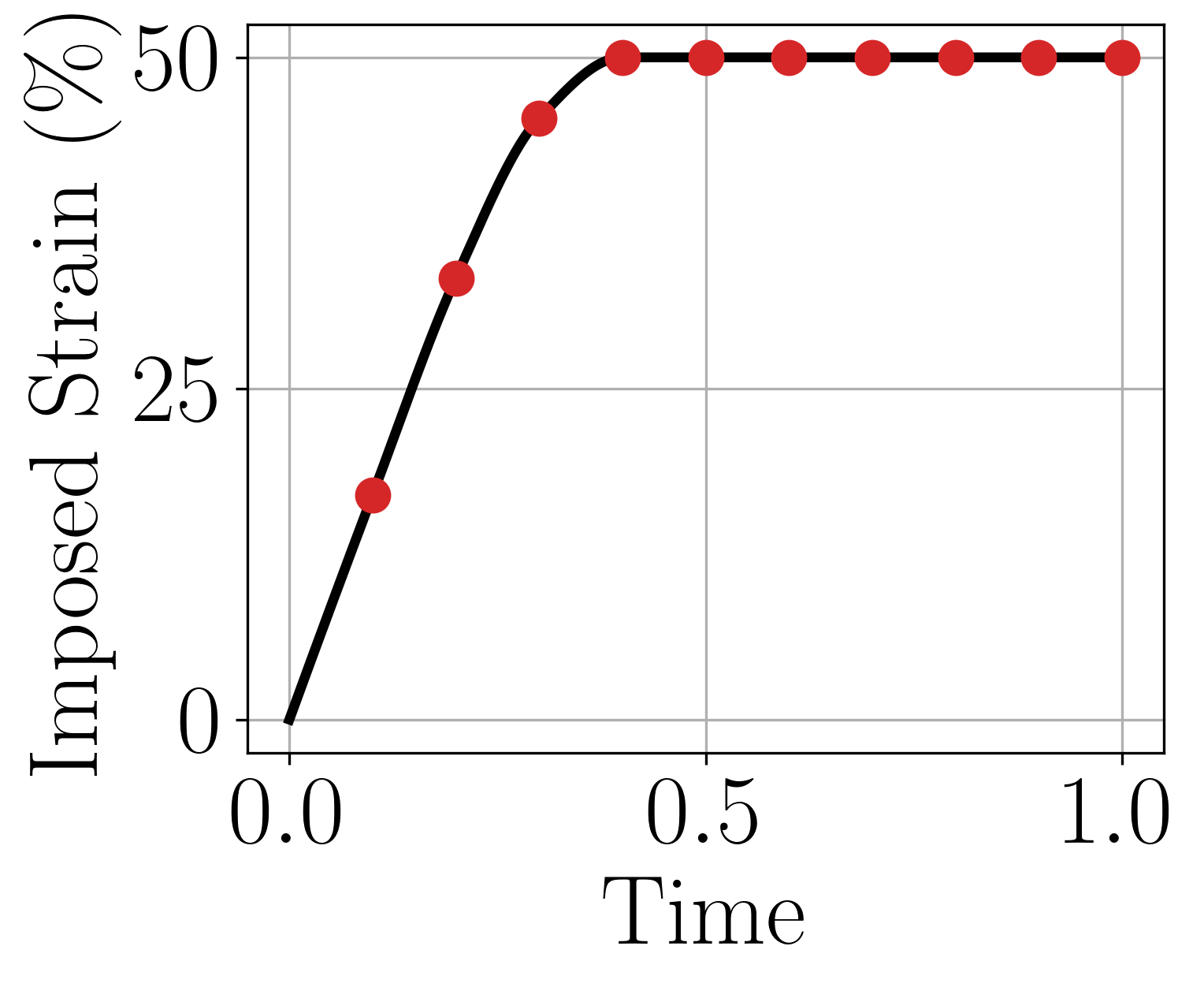}\\\hline
        \parbox[t]{2mm}{\multirow{3}{*}{\rotatebox[origin=c]{90}{$\lambda=4$}}} & \raisebox{-0.045\linewidth}{\includegraphics[width=0.14\linewidth]{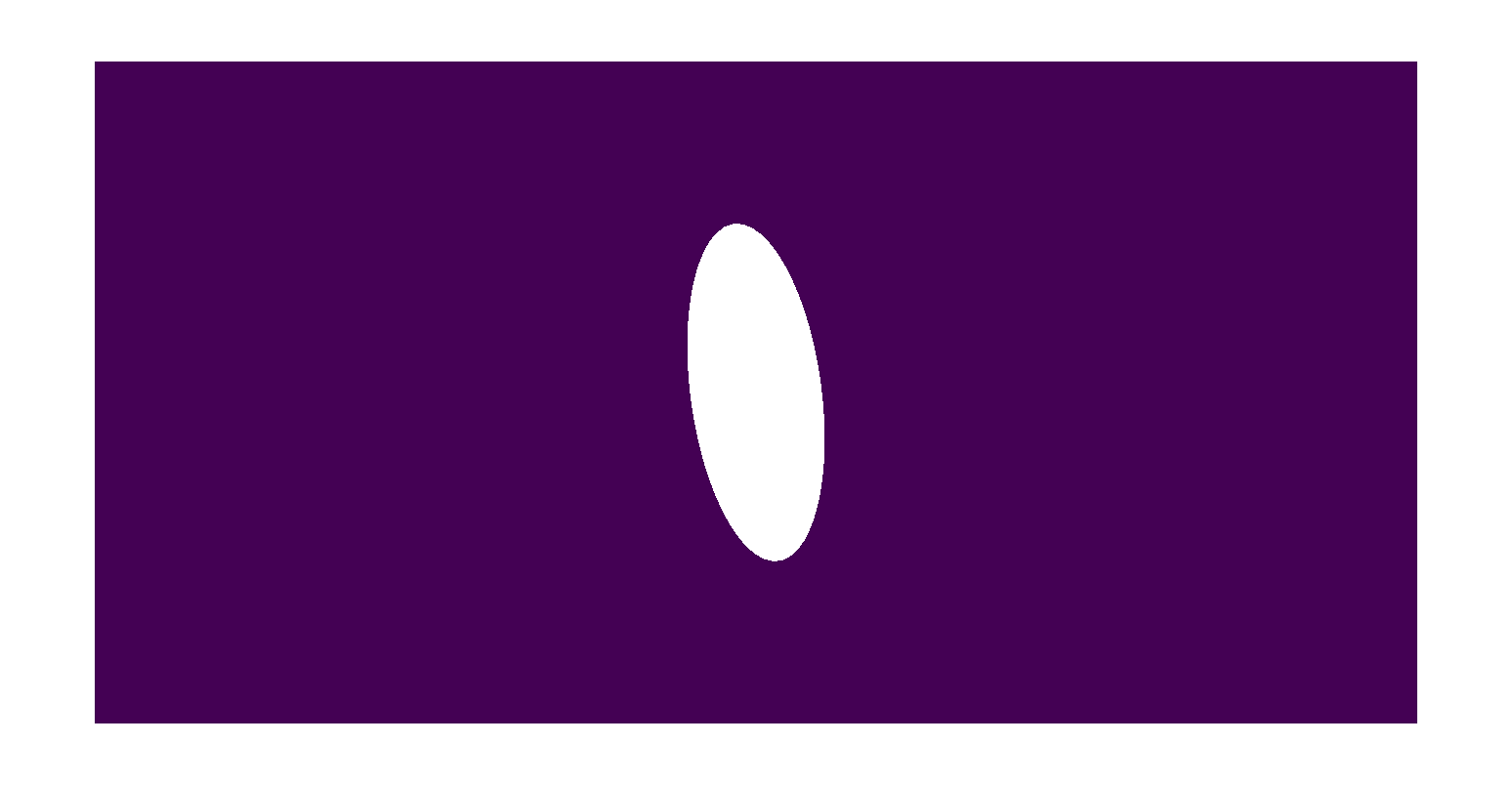}}&\multicolumn{2}{c||}{\raisebox{-0.045\linewidth}{\includegraphics[width=0.14\linewidth]{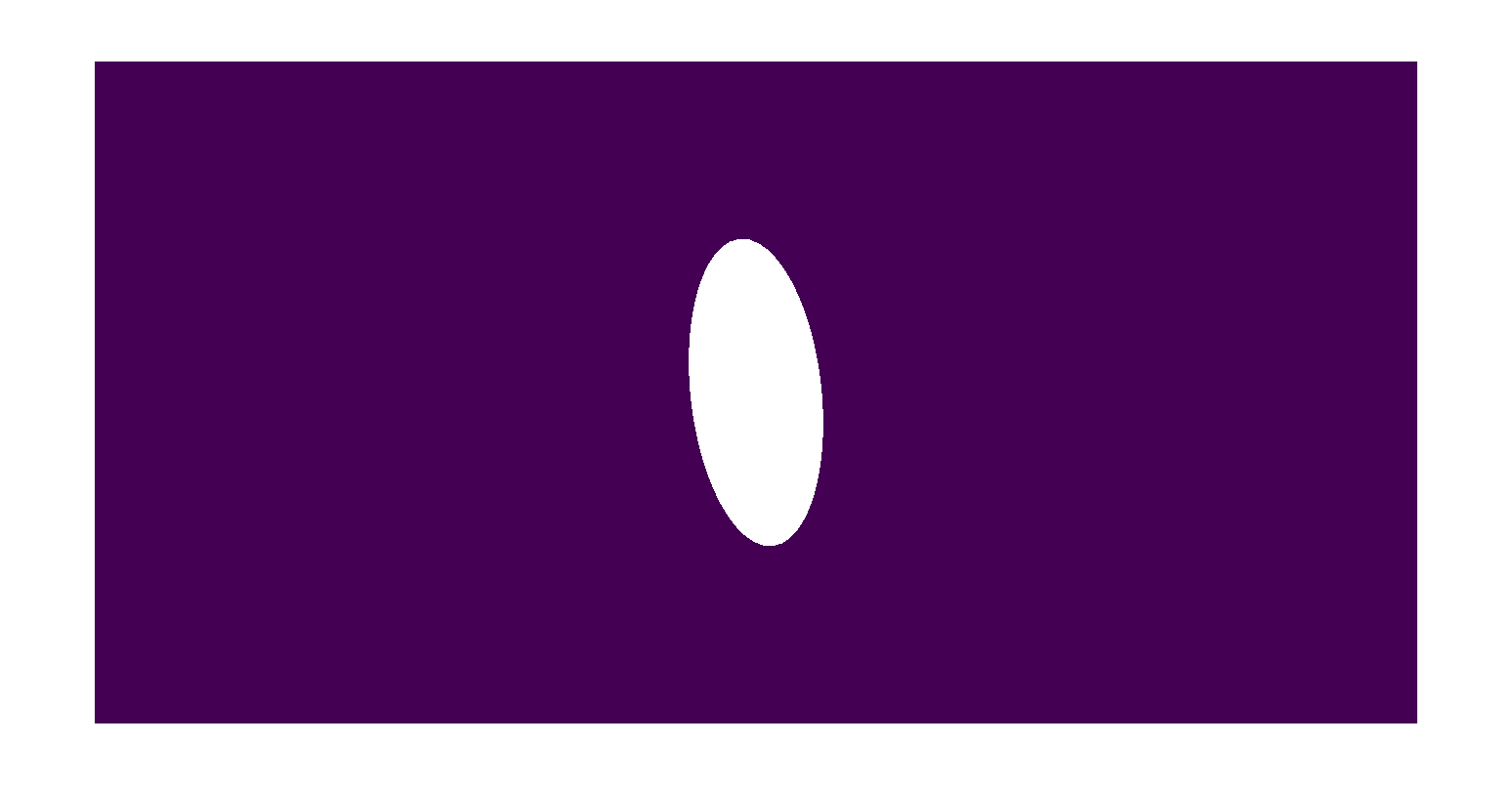}}} & \multicolumn{3}{c|}{\raisebox{-0.045\linewidth}{\includegraphics[width=0.14\linewidth]{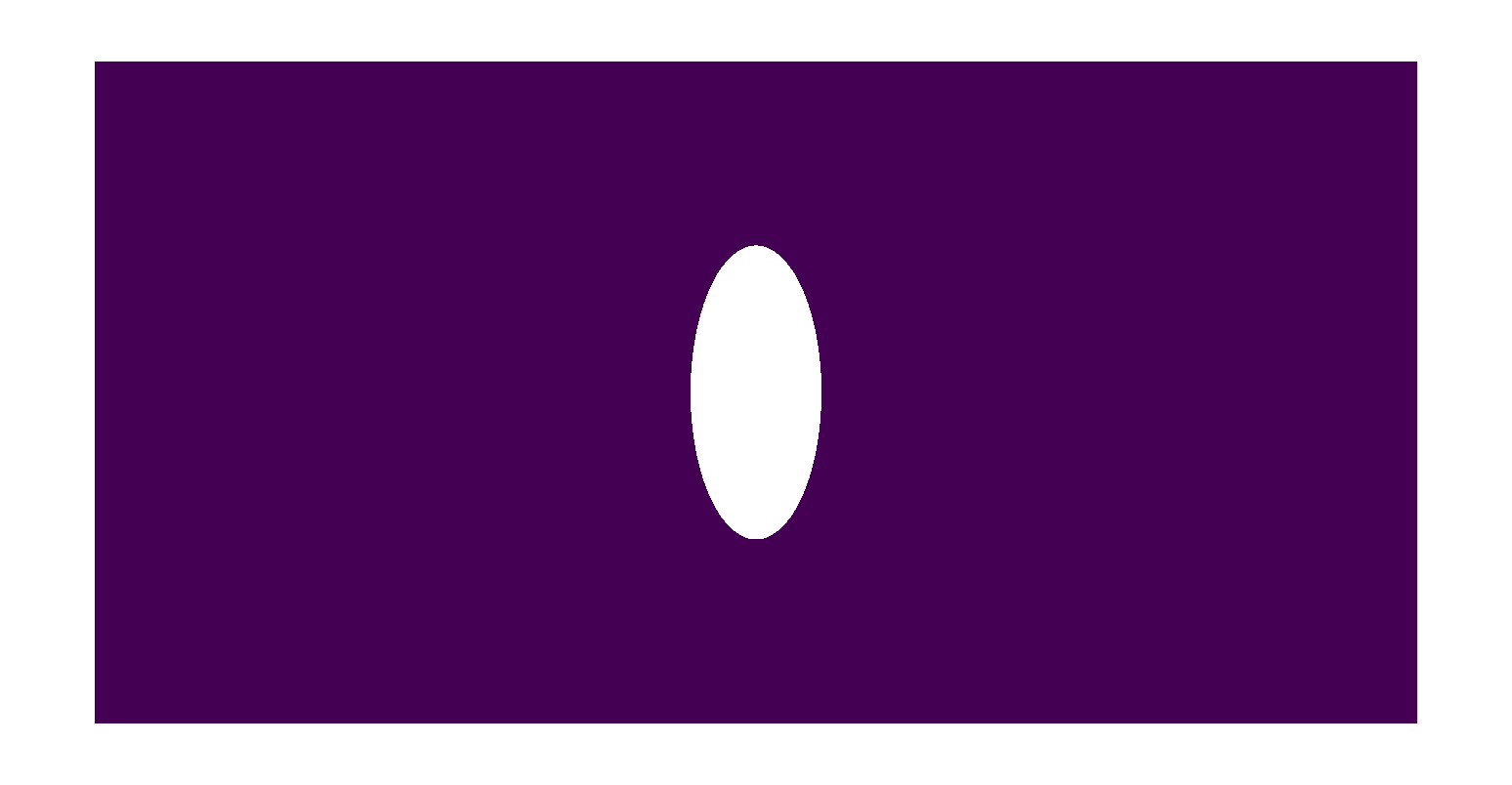}}} \\
        &    \includegraphics[width=0.14\linewidth]{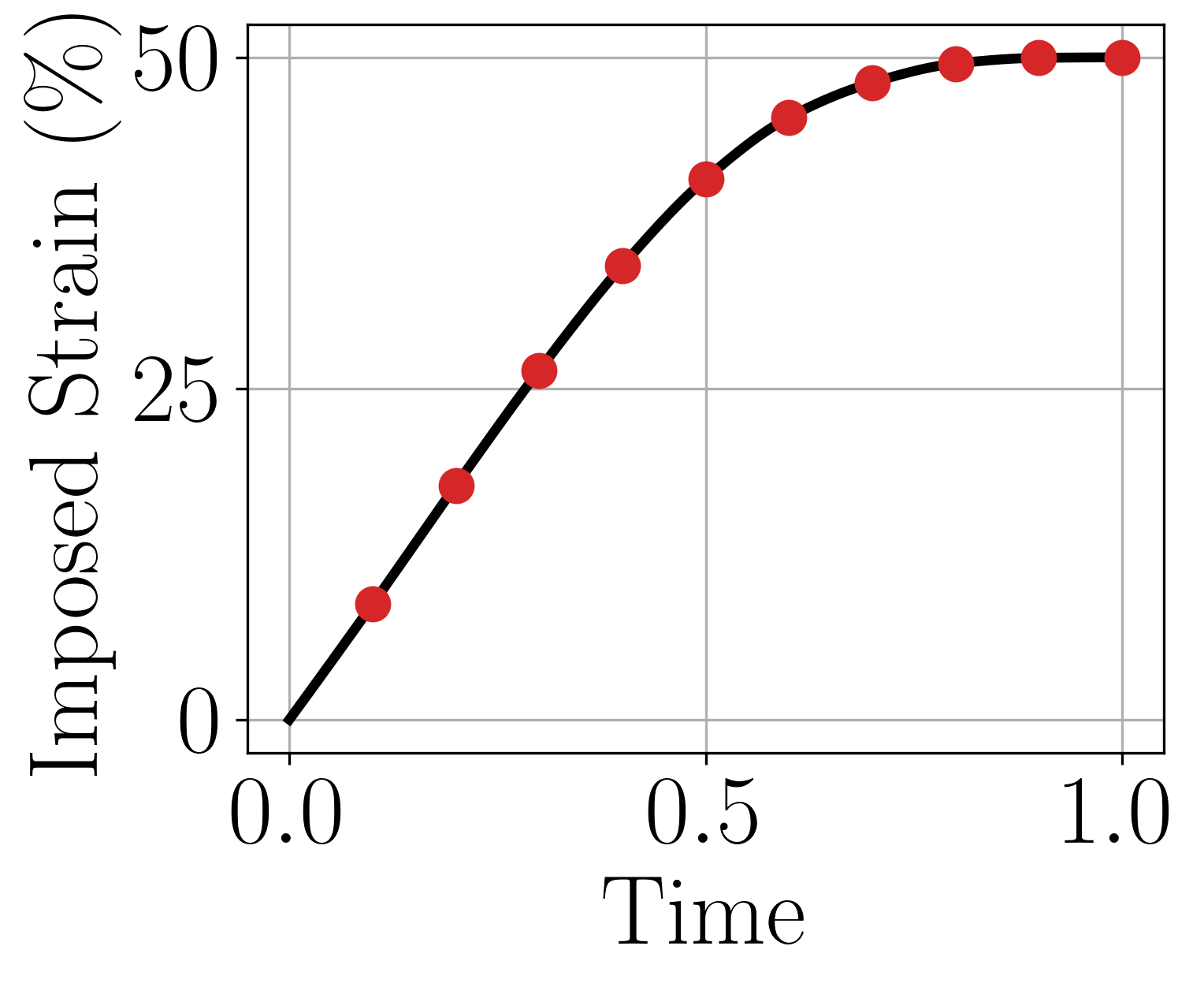}&\includegraphics[width=0.14\linewidth]{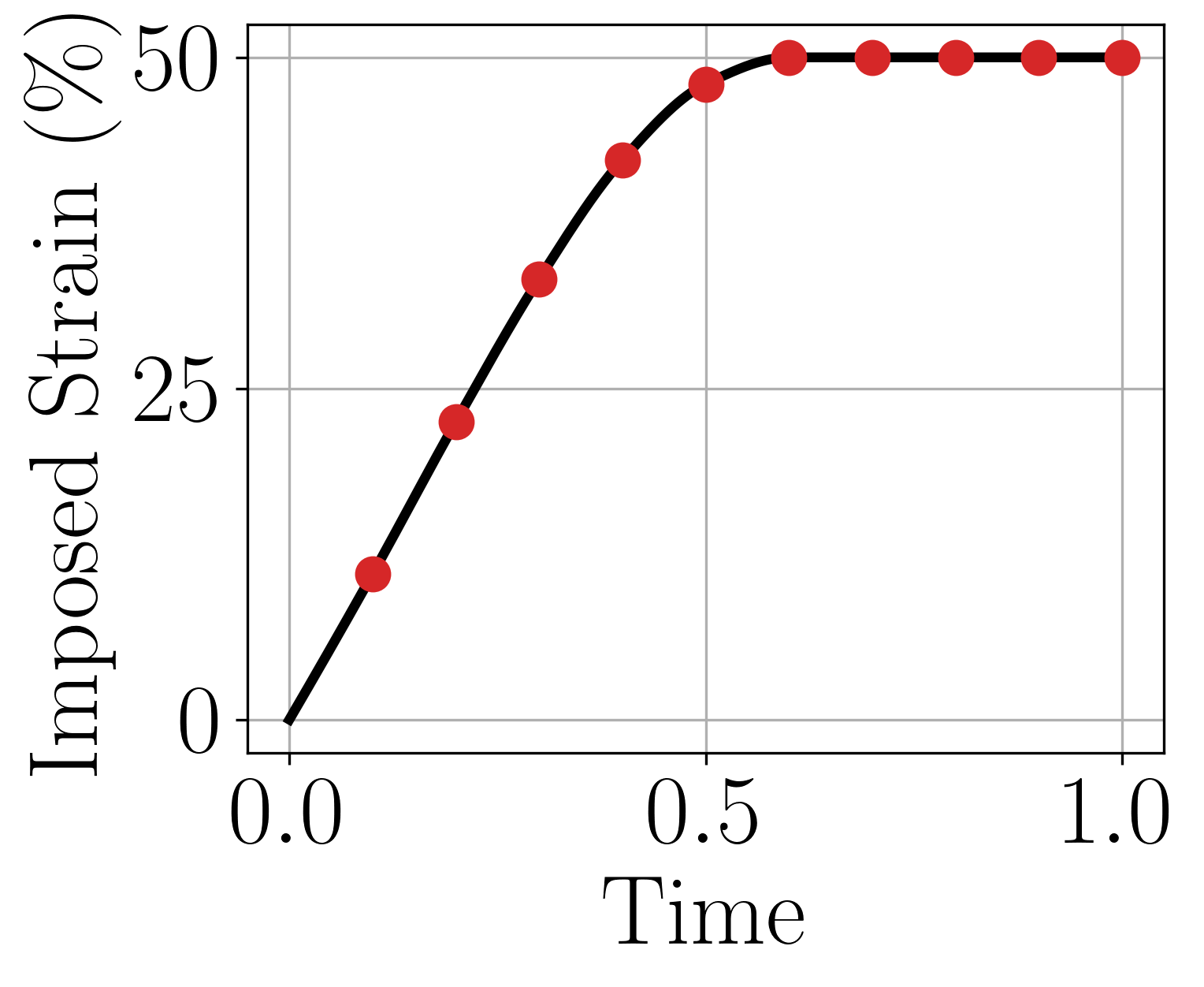} &  \includegraphics[width=0.14\linewidth]{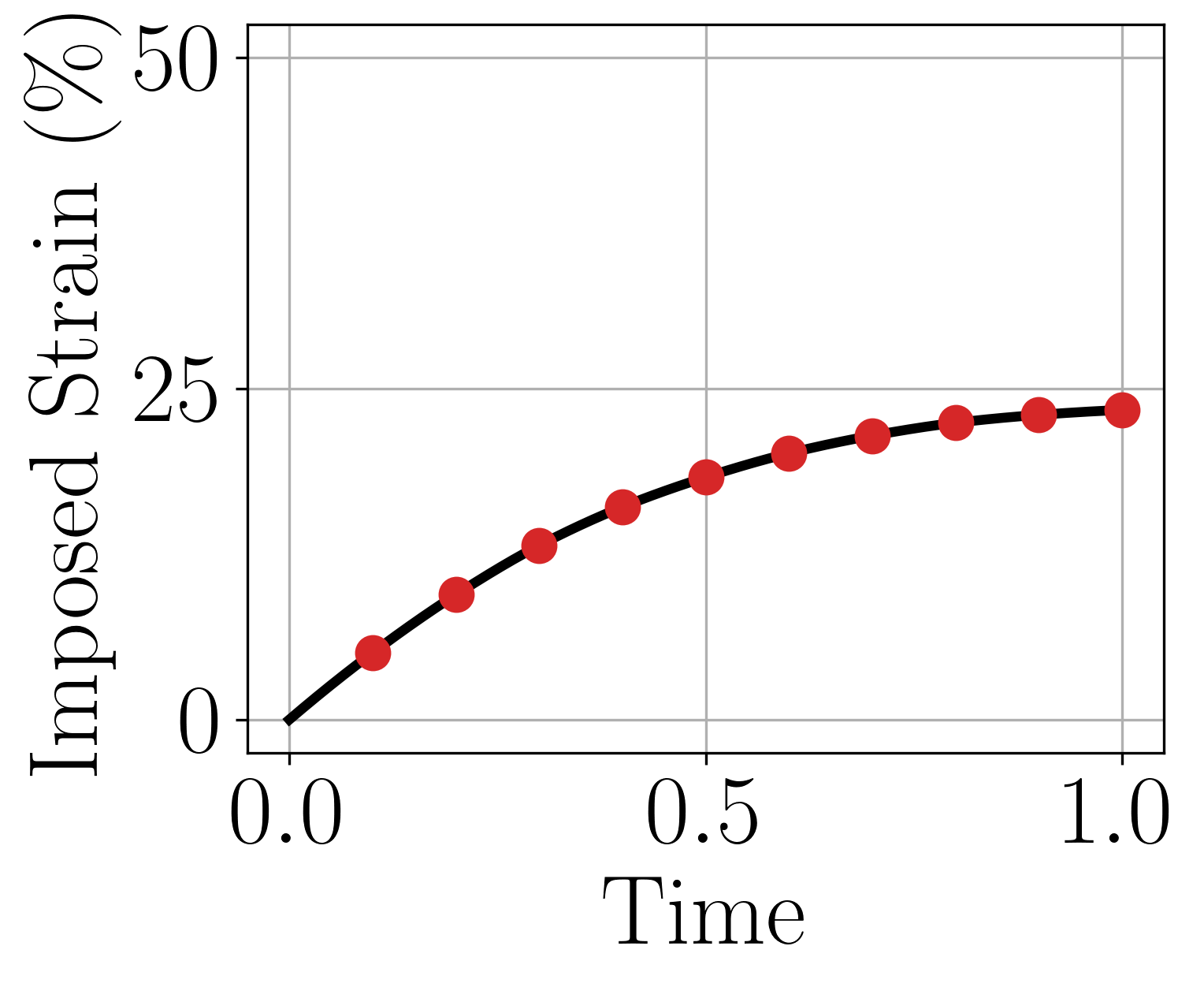} & \includegraphics[width=0.14\linewidth]{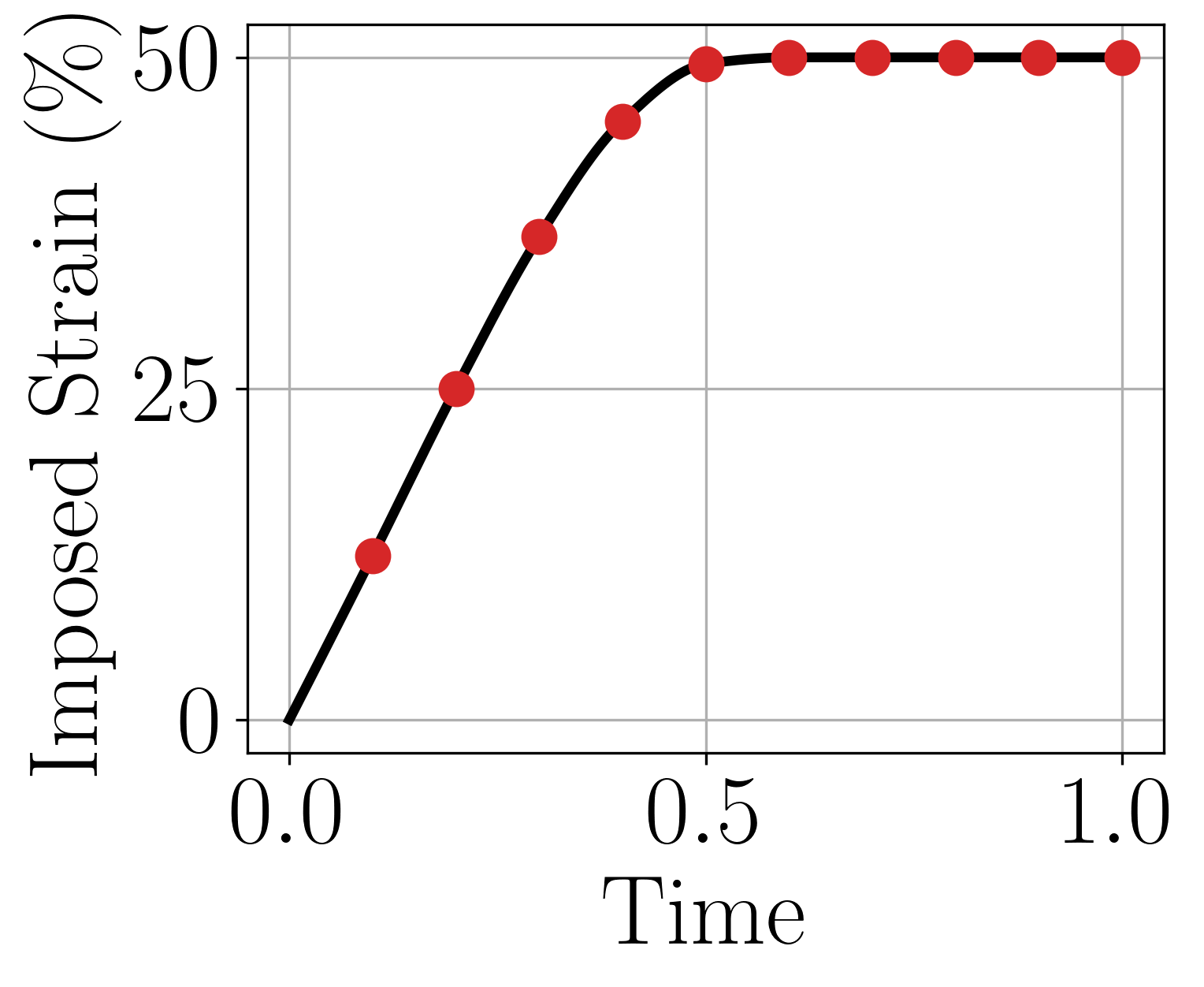} & \includegraphics[width=0.14\linewidth]{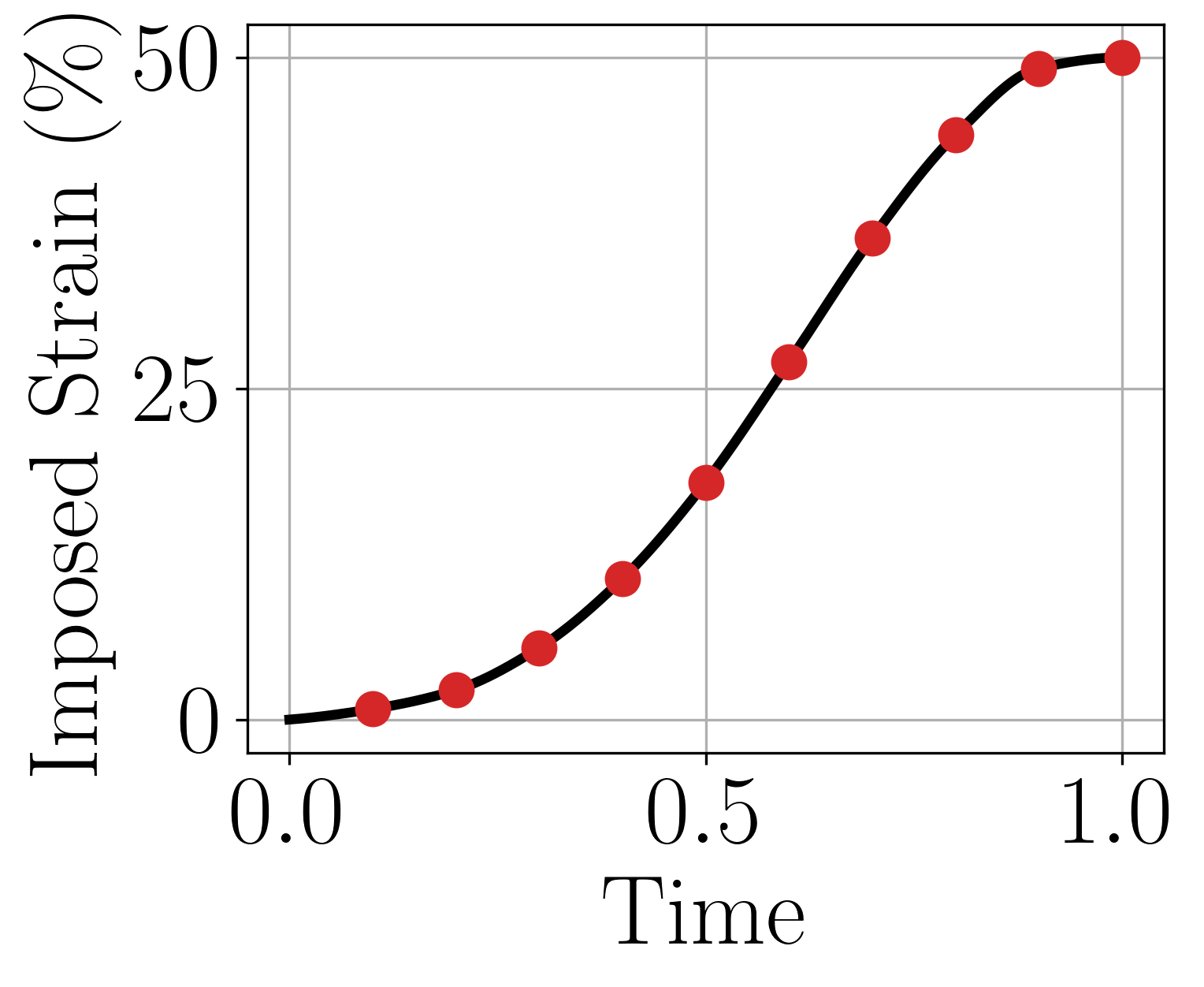} & \includegraphics[width=0.14\linewidth]{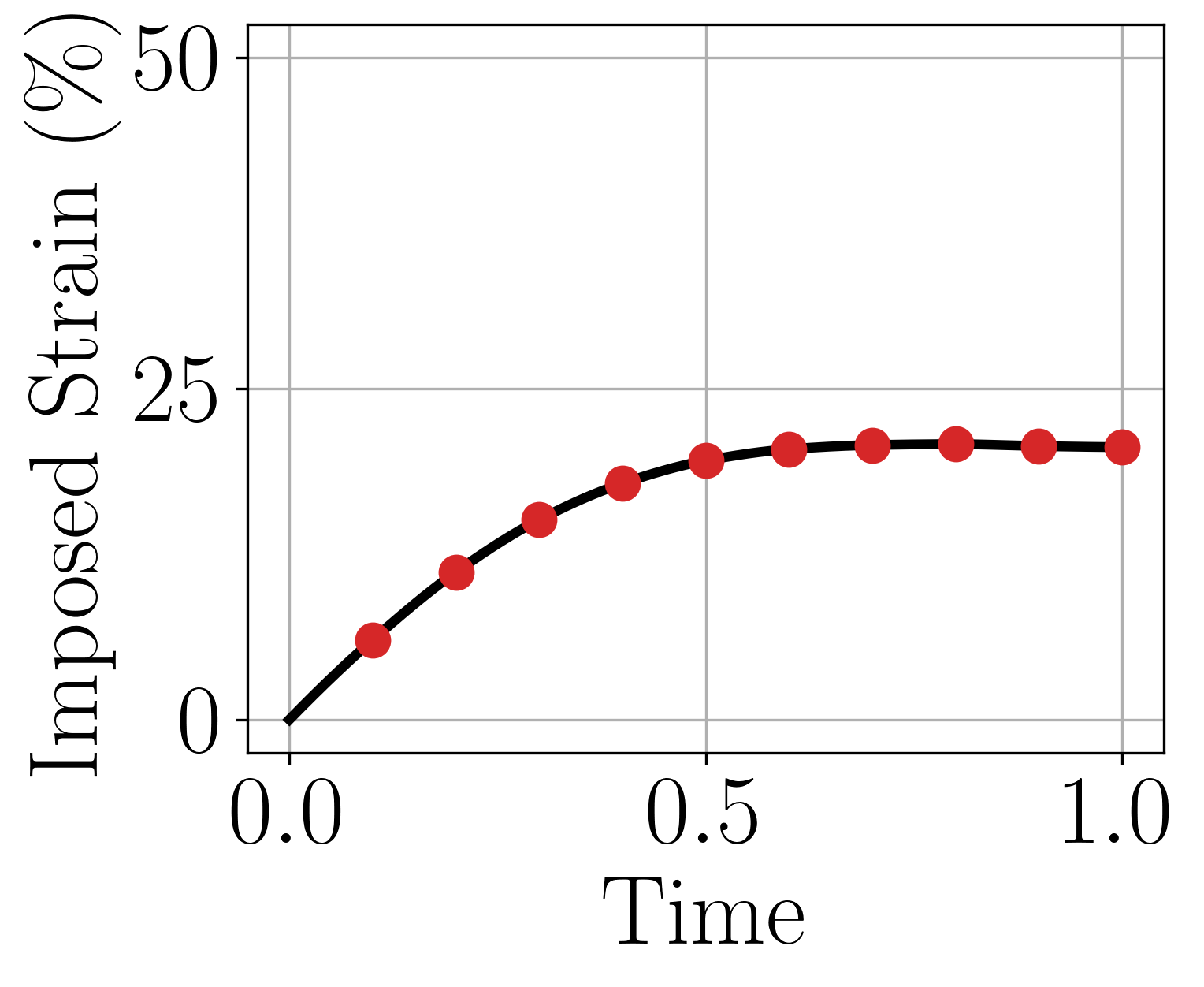}\\\hline
        \end{tabular}
    \renewcommand{\arraystretch}{1.0}
    }

    \caption{The experimental designs with batch sizes 1--3 found via ESFIM maximization for uniaxial testing of nonlinear viscoelastic materials. The designs within the same batch share a geometry, and the design optimization is subject to strain-rate regularization in \cref{eq:strain_rate_reg} with three regularization constants $\lambda\in\{0.25, 1, 4\}$.}
    \label{fig:nonlinear_batched_design_penalty}
\end{figure}

In \cref{fig:nonlinear_batched_design_penalty}, we visualize the optimized designs with shared geometry and strain rate regularization. We consider three regularization constants, $\lambda\in\{0.25, 1, 4\}$. The resulting optimized loading path is much smoother compared to those in \cref{fig:nonlinear_batched_design} with $\lambda=0$, suggesting that the design optimization finds a balance between ESFIM maximization and regularization minimization. While loading paths with switches between maximal and no-strain are still preferred, intermediate control values are introduced to reduce the imposed strain rate. For $\lambda=4$ and batch sizes of 2 and 3, one of the loading paths loads to half of the maximum strain. In contrast, the other loading paths in the same batch load to the maximum strain, indicating a further balance among the multiple regularization terms within the same batch. In general, these results demonstrate the benefit of the ESFIM in numerically discovering non-intuitive experimental designs, with cost amortization across batch sizes and design-optimization settings. In this case, ESFIM enables the study of how regularization constants affect design optimization across varying batch sizes at negligible computational cost beyond constructing the surrogate FIM.

\subsubsection{Quantitative Interpretations}

We provide quantitative interpretations of the ESFIM-optimized design of a single experiment without ($\lambda=0$, \cref{fig:nonlinear_batched_design}) and with ($\lambda=0.25$, \cref{fig:nonlinear_batched_design_penalty}) strain rate regularization.

\begin{figure}[htbp!]
    \centering
    \begin{minipage}{0.9\linewidth}
        \includegraphics[width=\linewidth]{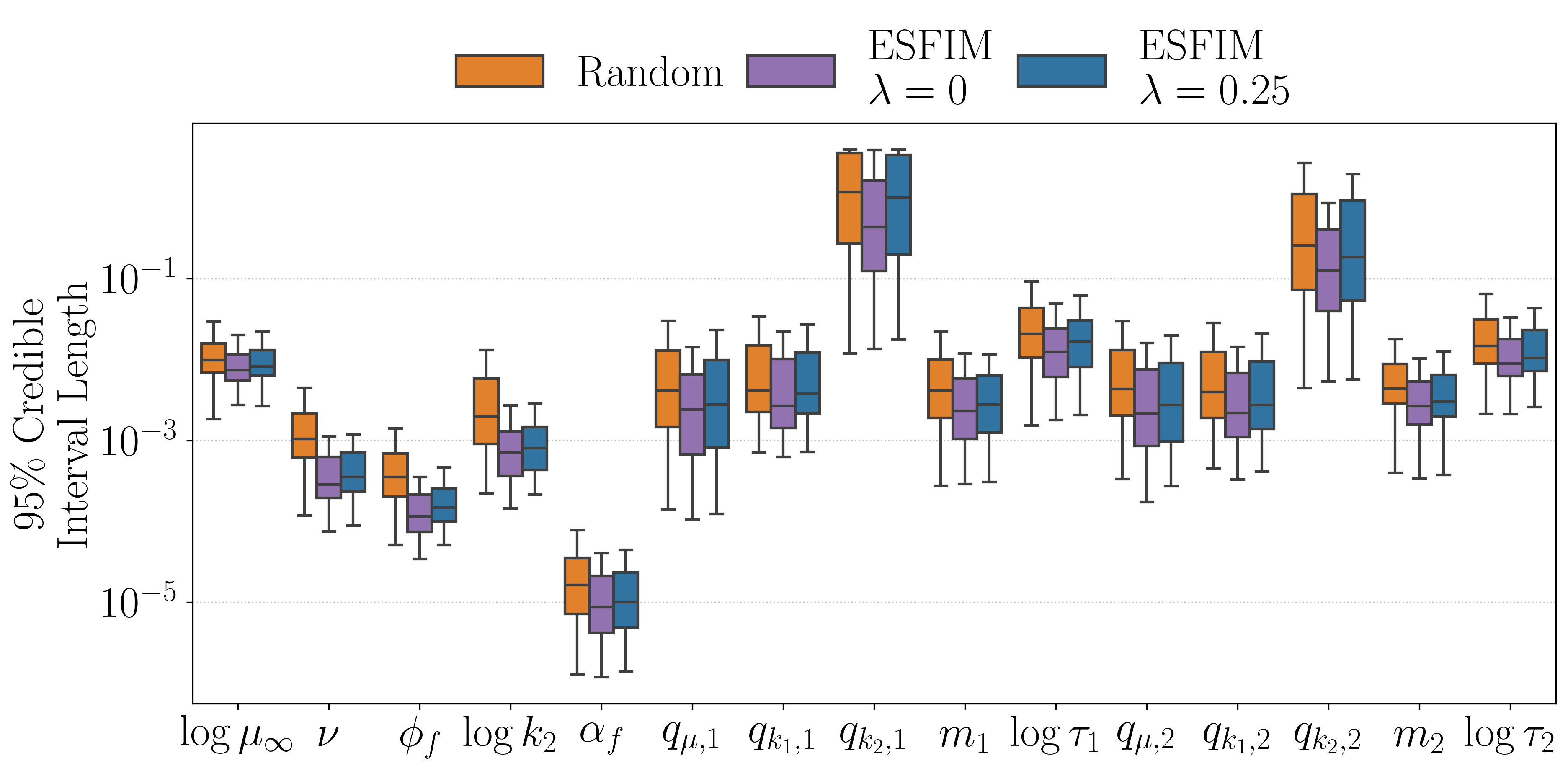}
    \end{minipage}
    \scalebox{0.9}
    {\renewcommand{\arraystretch}{2}
    \begin{tabular}{|c | c | c |c | c |c |c |c |c|}\hline
    \multicolumn{9}{|c|}{\makecell{Expected Reduction of the 95\% Credible Intervals \\ By Optimized Designs Relative  to Random Designs}}\\\hline\hline
    & $\log\mu_{\infty}$ & $\nu$ & $\phi_{f}$ & $\log k_2$ & $\alpha_f$ & $q_{\mu, 1}$ & $q_{k_1, 1}$ & $q_{k_2, 1}$\\\hline
    ESFIM ($\lambda=0$) & 25.5\% & 76.6\% & 68.3\% & 78.6\% & 40.3\% & 38.8\% & 42.8\% & 32.6\% \\\hline
    ESFIM ($\lambda=0.25$) & 16.4\% & 72.3\% & 61.3\% & 76.5\% & 34.4\% & 16.1\% & 10.4\% & 6.4\% \\\hline
     &$m_1$ &$\log\tau_1$ & $q_{\mu, 2}$ & $q_{k_1, 2}$ & $q_{k_2, 2}$ & $m_2$ & $\log\tau_2$ & Averaged \\\hline
 ESFIM ($\lambda=0$) & 42.3\% & 41.0\% & 36.8\% & 37.8\% & 33.2\% & 34.6\% & 33.2\% & 44.2\% \\\hline
    ESFIM ($\lambda=0.25$) & 14.6\% & 27.1\% & 19.9\% &  19.2\% & 10.8\% & 30.0\% & 22.9\% & 29.2\% \\\hline
    \end{tabular}
    \renewcommand{\arraystretch}{1.0}
    }
    \caption{Visualizations and statistics of the $95\%$ CIs for constitutive parameters of nonlinear viscoelasticity inferred from experiments with random and ESFIM-optimized designs. We consider design optimization without ($\lambda=0$) and with ($\lambda=0.25$) strain-rate regularization.  (\emph{Top}) The box plots of the 95\% CI sizes. The distributions arise from the prior-based variation in the experimental outcomes. For random designs, the distributions also account for design variations. (\emph{Bottom}) The expected reduction of the 95\% CI sizes by optimized designs relative to random designs. The sizes are estimated for $\theta$, which is transformed as in \cref{subsec:nonlinear_viscoelastic_setup} from the physical parameters used as labels.}
    \label{fig:nonlinear_ci}
\end{figure}

In \cref{fig:nonlinear_ci}, we present the visualizations and statistics of the $95\%$ credible intervals (CIs) for the posteriors. The credible intervals are again computed using the local Gaussian approximation in \cref{subsubsec:gaussia_approx}. In the top panel of \cref{fig:nonlinear_ci}, we visualize the distribution of $95\%$ CI sizes across prior-based variation in experimental outcomes. For random designs, the size distribution also accounts for design variations. The optimized designs consistently shift the distribution of 95\% CI lengths toward smaller values, with the design obtained without regularization being more effective. In the bottom panel of \cref{fig:nonlinear_ci}, we provide the relative expected reduction of the 95\% CIs by optimized designs compared to random designs. The design optimization significantly reduces the 95\% CI sizes, with average reductions of 44\% without regularization and 29\% with regularization. We observe that regularization substantially hinders parameter identifiability in the first viscous branch that captures fast stress relaxation.

\begin{figure}[htbp!]
    \centering
    \begin{minipage}{0.5\linewidth}
        \includegraphics[width=\linewidth]{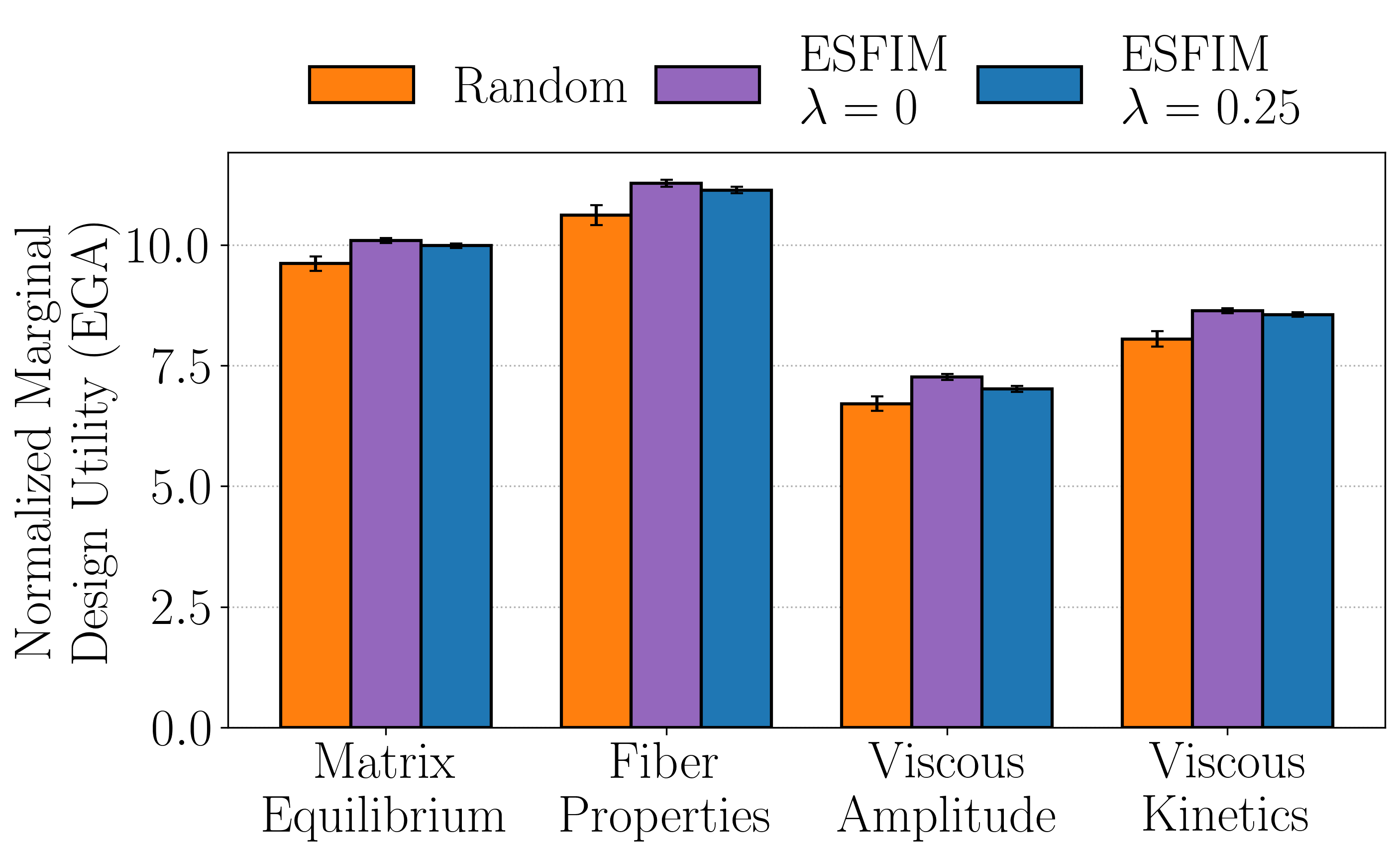} 
    \end{minipage}
    \begin{minipage}{0.48\linewidth}
    \scalebox{0.9}{
    \renewcommand{\arraystretch}{2}
        \begin{tabular}{|c |c|c|}\hline
        \multicolumn{3}{|c|}{\makecell{Expected Improvement in \\
        the EGA Design Utility \\
        of Optimized Designs \\
        Relative to Random Designs}}\\\hline\hline
        & \makecell{ESFIM \\$\lambda=0$} & \makecell{ESFIM\\ $\lambda=0.25$} \\ \hline
            Matrix Equilibrium & 5.0\% & 3.8\%    \\\hline
            Fiber Properties & 6.3\% &  4.9\% \\\hline
            Viscous Amplitude & 8.3\% & 4.5\% \\\hline
            Viscous Kinetics & 7.3\% & 6.3\% \\\hline
        \end{tabular}
    \renewcommand{\arraystretch}{1}
    }
    \end{minipage}
        
    \caption{The normalized EGA design utility for learning marginal parameters of nonlinear viscoelasticity grouped by material behaviors.  We consider design optimization without ($\lambda=0$) and with ($\lambda=0.25$) strain-rate regularization. (\emph{Left}) The bar plots of the utility values of random and ESFIM-optimized designs for learning matrix equilibrium, fiber, viscous amplitude, and viscous kinetics properties. The error bars for the random designs represent the standard deviations of the utility with respect to the design variables. In contrast, the error bars for the optimized designs represent the standard errors of EGA due to finite parameter sample sizes. (\emph{Right}) The expected improvement in the utility of optimized design relative to random designs for learning different properties.}
    \label{fig:nonlinear_marginal}
\end{figure}

In \cref{fig:nonlinear_marginal}, we provide the EGA for inferring four disjoint subsets of parameters. The four subsets corresponding to (i) matrix equilibrium properties, ($\log \mu_{\infty}$, $\nu$), (ii) fiber properties, ($\phi_f$, $\log k_2$, $\alpha_f$), (iii) viscous amplitude, ($q_{\mu, i}$, $q_{k_1, i}$, $q_{k_2, i}$, $i=1, 2$), and (iv) viscous kinetics, ($m_i$, $\log \tau_i$, $i=1, 2$). The optimized designs consistently improve the identifiability of these parameter subsets compared to random designs. For the design obtained without regularization, the relative utility improvement is greatest for the viscous amplitude (8.3\%), whereas the improvement in matrix equilibrium properties is the least significant, yet still 5\%. For the design obtained with regularization, the relative utility improvements are smaller, particularly for the matrix equilibrium behavior. The relative improvement of this design is most significant for viscous kinetics, at over 6\%, with matrix equilibrium properties the least significant, at 3\%.

\section{Conclusion and Outlook}\label{sec:conclusion}

In this work, we propose a simulation-based framework for quantifying, interpreting, and maximizing the utility of experimental design to reliably learn history-dependent constitutive models. Using a Bayesian formulation of the parameter identification problem, we assess the quality of experimental data by the expected information gain (EIG) to quantify the reduction in parametric uncertainty.

To address the computational challenges associated with EIG estimation, we consider the Bayesian D-optimal design based on a Gaussian approximation of the EIG. The resulting EGA utility not only significantly reduces computational complexity by replacing the nested Monte Carlo sampling of the forward model with a single-layer Monte Carlo sampling of the Fisher information matrix (FIM) but also enables insightful physical interpretations of design utility by analyzing the posterior marginals. Furthermore, we addressed the scalability issues in designing multiple experiments by introducing an estimator based on a surrogate FIM (ESFIM). This approach enables efficient optimization of batched experimental designs by amortizing the surrogate model's training costs across all batch sizes.

We demonstrate the efficacy of this framework through numerical case studies on uniaxial testing for both linear and nonlinear viscoelastic solids. The results indicate that optimized specimen geometries and loading paths yield high-quality force and image data, leading to significantly narrower credible intervals for constitutive parameters than those from random experimental designs. For simple models such as linear viscoelasticity, the framework identified designs that align with our expectations, including tilted elliptical holes and specific loading-holding or unloading-holding sequences that isolate and amplify the signals of anisotropy and stress relaxation. We also show that the framework can identify non-intuitive designs, such as those for nonlinear viscoelasticity with strain-rate regularization. Overall, this work provides a systematic tool for advanced material testing, enabling the automated discovery of experimental protocols that enrich the information content of experimental data and support reliable learning of constitutive models from designed experiments.

While we demonstrate the potential of the proposed framework to facilitate learning of constitutive models from experiments, several avenues for further improvement and extension remain. Methodologically, the Gaussian approximation to the posterior introduces bias in EIG estimation. Employing a more general non-parametric method for posterior density estimation can enhance the effectiveness of our proposed framework; see, e.g., \cite{foster2019variational, baptista2024bayesian, koval2024tractable, li2025expected, dong2025variational,cao2026lazydino}. There are density estimation methods that can readily incorporate complex, high-dimensional nuisance uncertainty, e.g., uncertain boundary and initial conditions \cite{baptista2024bayesian}, into design utility evaluations to enhance the robustness of parameter identification. In terms of material testing, current numerical studies have primarily focused on uniaxial loading of planar specimens under simplified conditions. Extending this framework to complex three-dimensional geometries and loading scenarios is essential for calibrating more general constitutive models. Such an expansion necessitates integrated studies that simultaneously optimize specimen topology and multi-axial loading protocols, potentially utilizing 3D volumetric data.

\section*{Acknowledgment}
This work is supported by the ONR MURI on Data-Driven Closure Relations (N00014-23-1-2654). AMS is also supported by a Department of Defense Vannevar Bush Faculty Fellowship (N00014-22-1-2790).

\bibliographystyle{elsarticle-num-names}
\bibliography{main}

\appendix
\crefalias{section}{appendix}
\crefalias{subsection}{appendix}
\crefalias{subsubsection}{appendix}

\section{The Image Observation Model}\label{app:dic}

In this appendix, we detail the observation model used for the image data.

\subsection{Image Deformation via Splatting}
The image formation operator $\mathcal{W}$ returns the predicted deformed image $I_{\text{pred}}\in[0,1]^{n_W\times n_{H}}$ given the reference image $I_0\in [0,1]^{n_W\times n_{H}}$ and the 2D displacement field $u: \partial\Omega_I\to\R^2$. Note that here we take $u$ directly as the surface displacement at time $t$, i.e., $u(\cdot, t)|_{\partial\Omega_I}$, for notational convenience.  This operator is implemented using a pushforward algorithm that maps each pixel's intensity in the reference configuration to its new location in the deformed configuration.

Let $\Lambda\subset \Z^2$ denote the pixel indices and $\calD\subset\R^2$ denote the physical coordinates of pixels, with $\phi:\Lambda\to\calD$ denoting the index to coordinate map. For each source pixel $p_s\in \Lambda$ of the reference image located in the material domain, i.e., $\phi(p_s)\in \calD\cap \partial\Omega_I$, its displacement value $(u\circ \phi)(p_s)$ yields a destination coordinate $x_d = \phi(p_s) + (u\circ \phi)(p_s)$. Since typically $x_d\not\in\calD$, the source intensity at $p_s$, denoted by $[I_0]_{p_s}$, is distributed to each of the target pixels located near $x_d$ using a Gaussian kernel $K:\R\to\R$.

We define an accumulated weight $W\in\R^{n_W\times n_{H}}$ and an accumulated intensity $V\in \R^{n_W\times n_{H}}$. For a target pixel coordinate $p\in\Lambda$, they are computed by summing contributions from all source pixels:
\begin{align}
    [W]_{p} &= \sum_{p_s \in \Lambda} [M_0]_{p_s} \, K\left( \| \phi(p) - \phi(p_s) - (u\circ \phi)(p_s) \|;\sigma_s \right), \\
    [V]_{p} &= \sum_{p_s \in \Lambda} [M_0]_{p_s} \, [I_0]_{p_s} \, K\left( \| \phi(p) - \phi(p_s) - (u\circ \phi)(p_s)\|; \sigma_s \right),
\end{align}
where $M_0 \in \{0, 1\}^{n_W \times n_H}$ indicates the pixels of the reference image in the material domain, ensuring that the weights and intensities do not include the image intensities of the background. We use a kernel defined as $K(r;\sigma_s) = \exp(-r^2 / 2\sigma_s^2)$, where $\sigma_s$ is the diagonal length of a single pixel. In practice, the kernel is localized via truncation and simplified to a weighted average over a $3 \times 3$ pixel window for small deformations and to a $5 \times 5$ or larger window for large deformations. The predicted deformed image is obtained by normalizing the accumulated intensity by the accumulated weights:
\begin{equation}
    [I_{\text{pred}}]_p = \begin{cases} 
        \frac{[V]_p}{[W]_p} & \text{if } [W]_p > \epsilon, \\
        0 & \text{otherwise},
    \end{cases}
\end{equation}
where $\epsilon$ is a small threshold to avoid division by zero in regions where no material is present (i.e., the background). This push-forward approach, known as splatting in computer graphics \cite{westover1990footprint}, reconstructs the deformed image via kernel regression \cite{Nadaraya1964estimating, watson1964smooth}, distinct from the pullback interpolation typically used in DIC \cite{sutton2009image}.

\subsection{A Heteroscedastic Noise Model with Soft Masking}
We consider an additive noise model for the deformed image, with soft masking to eliminate noise in background pixels. This masking function $\calA:[0,1]\to [0, 1]$ generates a near-binary mask indicating the region occupied by the deformed material. The mask is computed using a sigmoid function:
\begin{equation}
    \mathcal{A}(x) = \left(1 + \exp\left(-\beta(x-\gamma)\right)\right)^{-1},
\end{equation}
where $\gamma$ represents a normalized intensity threshold, and $\beta$ controls the steepness of the transition between the material and the background. We set $\gamma=0.05$ and $\beta=10$. The image data $y$ is modeled as the masked prediction perturbed by additive white Gaussian noise $\eta$:
\begin{equation}
    y = \mathcal{A}(I_{\text{pred}}) \odot (I_{\text{pred}} + \sigma_\eta\eta), \quad [\eta]_{ij} \sim \mathcal{N}(0, 1).
\end{equation}
where $\mathcal{A}$ is applied pixel-wise and $\odot$ is pixel-wise multiplication. The corresponding image misfit function (negative log-likelihood) is given by:
\begin{equation}
    \Phi(I_{\text{pred}}) = \frac{1}{2\sigma_\eta^2} \| y \oslash \mathcal{A}(I_{\text{pred}})  - I_{\text{pred}}\|_F^2 - \sum_{p\in \Lambda} \log \mathcal{A}([I_{\text{pred}}]_p),
\end{equation}
where $\oslash$ denotes pixel-wise division.

We contrast our heteroscedastic noise model with two alternatives: hard thresholding and homoscedastic modeling. Hard thresholding discards background pixels in the predicted image, introducing a non-differentiable cutoff that reduces information relevant to boundary features. Homoscedastic modeling assumes that background and material pixels have the same noise level and share the same residual weighting constant. This is an unnatural assumption for the probabilistic modeling of masked data, where background pixels are obtained from post-processing and are noise-free. Our approach resolves these issues by deriving a probabilistic model for masked data. This naturally leads to a negative log-likelihood that strongly penalizes mismatches at material boundaries.

\section{Proof of the EIG Equivalence Lemma}\label{app:proof}

We consider two factorizations of the joint density of the parameters and data:
\begin{subequations}\label{eqns:factorization}
\begin{align}
    \pi(\theta, y) &= \pi_{\Theta\mid Y(z)}(\theta\mid y)\pi_{Y(z)}(y)\label{eq:factorization_posterior}\\
    &= \pi_{Y(z)\mid\Theta}(y\mid\theta)\pi_{\Theta}(\theta).\label{eq:factorization_likelihood}
\end{align}
\end{subequations}
Expanding the entropy terms in the EIG and applying these factorizations, we have:
\begin{align*}
\textrm{EIG}(z) &= -\int \pi_{\Theta}(\theta)\log\pi_{\Theta}(\theta)\,\mathrm{d}\theta + \iint \pi(\theta, y)\log\pi_{\Theta\mid Y(z)}(\theta\mid y)\,\mathrm{d}\theta\,\mathrm{d}y && \left(\textrm{Apply }\cref{eq:factorization_posterior}\right) \\
&= \iint \pi(\theta, y)\log \frac{\pi_{\Theta\mid Y(z)}(\theta\mid y)}{\pi_{\Theta}(\theta)} \,\mathrm{d}\theta\,\mathrm{d}y &&\left(\pi_{\Theta} = \int\pi(\cdot, y)\,\mathrm{d} y\right)\\
&= \iint \pi(\theta, y)\log \frac{\pi_{Y(z)\mid\Theta}(y\mid\theta)}{\pi_{Y(z)}(y)} \,\mathrm{d}\theta\,\mathrm{d}y. && \left(\textrm{Apply }\cref{eqns:factorization}\right)
\end{align*}
By factoring the joint density through the posterior, we recover the first form:
\begin{align*}
    \textrm{EIG}(z) &= \int \left( \int \log\frac{ \pi_{\Theta\mid Y(z)}(\theta\mid y)}{ \pi_{\Theta}(\theta)}\pi_{\Theta\mid Y(z)}(\theta\mid y)\,\mathrm{d}\theta\right)\pi_{Y(z)}(y)\,\mathrm{d}y && \left(\textrm{Apply } \cref{eq:factorization_posterior}\right) \\
    &=\mathbb{E}_{y\sim\pi_{Y(z)}}\left[D_{\mathrm{KL}}\left( \pi_{\Theta\mid Y(z)}(\cdot\mid y) \,\|\, \pi_{\Theta}\right)\right].
\end{align*}
Similarly, by factoring the joint density through the likelihood, we recover the second form:
\begin{align*}
    \textrm{EIG}(z) &= \int  \left(\int \log\frac{ \pi_{Y(z)\mid \Theta}(y\mid\theta)}{\pi_{Y(z)}(y)}\pi_{Y(z)\mid\Theta}( y\mid\theta)\,\mathrm{d}y\right)\pi_{\Theta}(\theta)\,\mathrm{d}\theta && \left(\textrm{Apply } \cref{eq:factorization_likelihood}\right) \\
    &=\mathbb{E}_{\theta \sim\pi_{\Theta}}\left[D_{\mathrm{KL}}\left( \pi_{Y(z)\mid\Theta}(\cdot\mid \theta) \,\|\, \pi_{Y(z)}\right)\right].
\end{align*}

\section{Numerical Implementation Details}\label{app:numerics}

\subsection{Numerical Solver}\label{app:details_solver}

The balance equation and its sensitivity equations are solved using the finite element method implemented with the FEniCS \cite{logg2012automated, alnaes2015fenics} library, with inverse-problem functionality (including the prior, the likelihood, and the symbolic derivation of the sensitivity equations) provided by the hIPPYlib \cite{villa2021hippylib} and Geometric MCMC \cite{cao2025derivative} library. In particular, we use linear triangular elements for the displacement field and assign internal variable values at quadrature points, with approximately 17,000 and 30,000 degrees of freedom for the linear and nonlinear models, respectively. In particular, the mesh is refined by a factor of 2.5 and 4 near the elliptical holes for the linear and nonlinear models, respectively. The implicit Euler scheme is used for time evolution. A sparse direct solver is used to solve systems of equations. The Newton method with backtracking line search is used for nonlinear solves. We use 100 time steps for the linear model and 200 time steps for the nonlinear model with adaptive refinement to 300 or more when the Newton method fails to converge. The linear model takes, on average, 10 seconds to solve, while the nonlinear model takes, on average, 3.2 minutes. Evaluating the FIM takes, on average, 3.5 minutes for linear models and 26 minutes for nonlinear models. The timings are recorded on an Intel Cascadelake 8276 2.1 GHz CPU.

\subsection{Image Data Generation}\label{app:image_generation}

The synthetic speckle pattern is generated in two stages. First, speckle centers are sampled. To obtain a spatially uniform yet non-periodic distribution, we use a Poisson disk sampling algorithm based on Bridson’s method \cite{bridson2007fast}, which enforces a minimum distance between any two candidate centers to avoid clustering. In the second stage, we adjust the pattern to match a prescribed area coverage by estimating the number of speckles needed to cover the desired fraction of the domain and randomly subsampling that number of centers. Each retained center is then assigned a radius drawn from a normal distribution around the base radius. The resulting pattern consists of circular speckles with slightly varying size and controlled overall density, providing a realistic, high-contrast pattern.

\begin{figure}[htb]
    \centering
    \begin{minipage}{0.41\linewidth}
        \includegraphics[width=\linewidth]{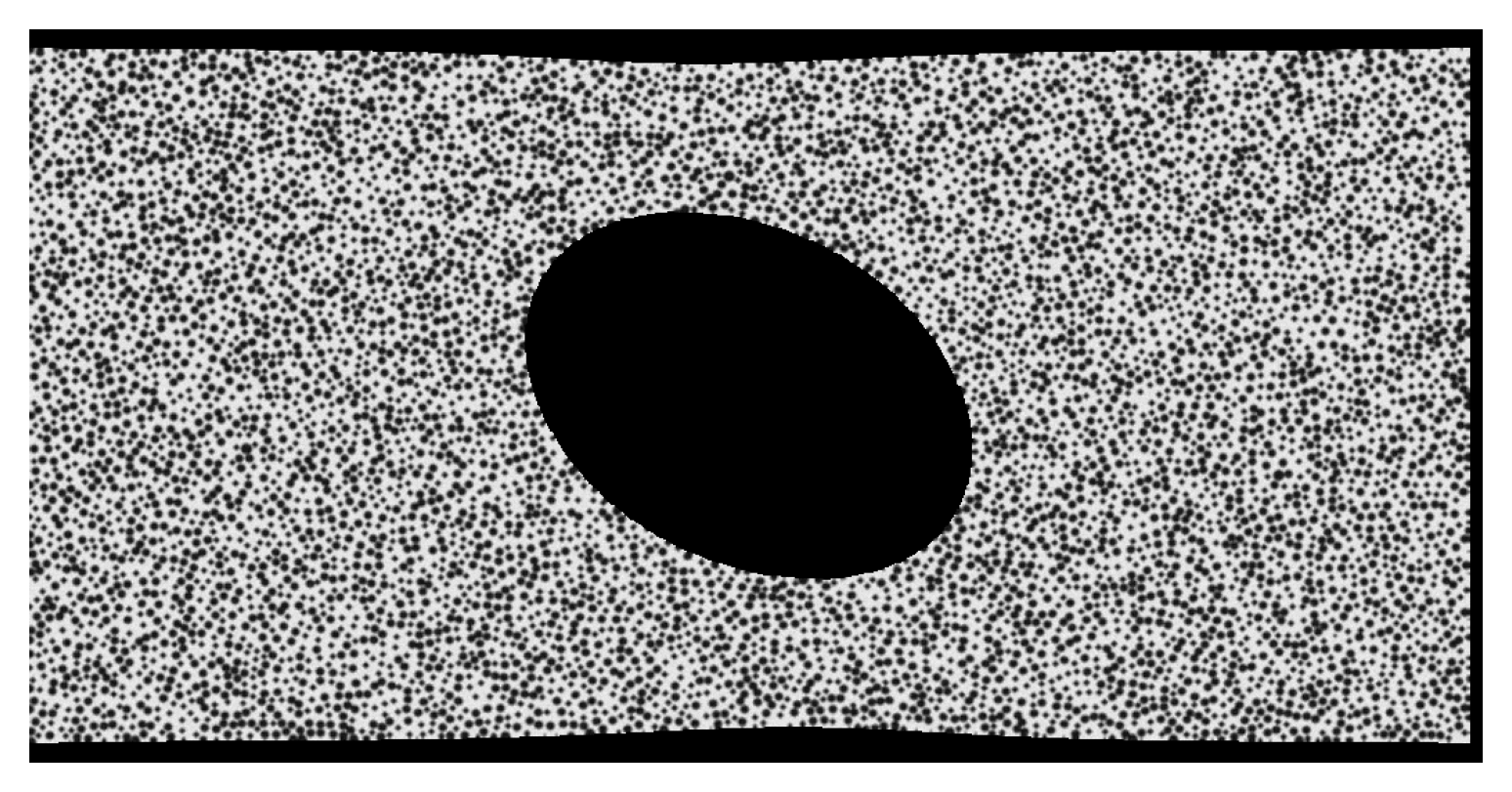}
    \end{minipage}
    \begin{minipage}{0.58\linewidth}
        \includegraphics[width=\linewidth]{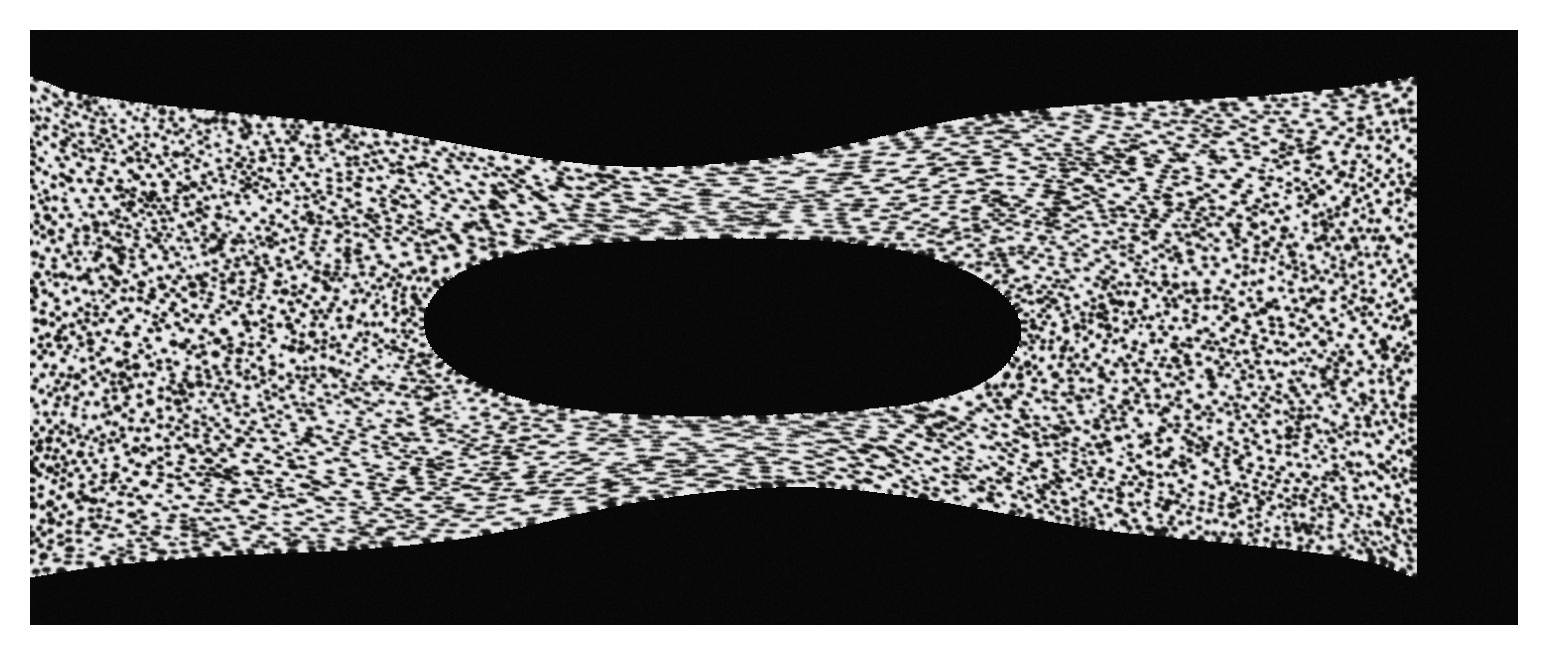}
    \end{minipage}
    
    \caption{High resolution snapshots of image data. (\emph{Left}) The last snapshot in \cref{fig:linear_data} for uniaxial testing of linear viscoelastic materials. (\emph{Right}) The last snapshot in \cref{fig:nonlinear_data} for uniaxial testing of nonlinear viscoelastic materials.}
    \label{fig:high_res_images}
\end{figure}

The image observation model is implemented using JAX \cite{jax2018github} to enable automatic differentiation of the image formation operator $\mathcal{W}$ defined in \cref{app:dic}. It takes approximately 7 seconds and 22 seconds to create the full image data for linear and nonlinear models, respectively. The timings are recorded on an Intel Cascadelake 8276 2.1 GHz CPU. We visualize two high-resolution snapshots of image data in \cref{fig:high_res_images}.

\subsection{Principal-Stress-Based Stress-State Entropy}\label{app:sse}

We consider the principal-stress-based stress-state entropy following \cite{ihuaenyi2024seeking, ihuaenyi2025mechanics}. For a given design $z$, this quantity is evaluated using a fixed set of $N_P=300$ prior samples $\{\theta^{(i)}\}$.

For each prior sample $\theta^{(i)}$ and design $z$, we solve the forward model and evaluate the stress tensor at the times where the image snapshots are taken. At each spatial quadrature point and time point, the stress tensor is mapped to its principal stresses, yielding points in a two-dimensional principal-stress space, which produces a stress-state distribution associated with $(\theta^{(i)}, z)$. To quantify the diversity of the stress-state distribution, we discretize the principal-stress space using a rectangular grid shared across all designs and prior samples. Let $p_j(\theta^{(i)}, z)$ denote the empirical probability associated with bin $j$. The corresponding stress-state entropy is given by
\begin{equation*}
    \calH_{ss}(\theta^{(i)}, z)
    =
    -
    \sum_{j:\,p_j(\theta^{(i)}, z)>0}
    p_j(\theta^{(i)}, z)\log p_j(\theta^{(i)}, z).
\end{equation*}
We then define the stress-state entropy of the design $z$ by averaging over the prior samples:
\begin{equation*}
    \overline{\calH_{ss}}(z)
    =
    \frac{1}{N_P}
    \sum_{i=1}^{N_P}
    \calH_{ss}(\theta^{(i)}, z).
\end{equation*}
This quantity is used to produce \cref{fig:sse}.

\subsection{Nested Monte Carlo Estimator Evaluation}\label{app:nmc}

To quantify the intractability of the NMC estimator, we estimate the number of inner samples $N_{\mathrm{in}}^*$ required to achieve a minimal effective sample size of $\mathrm{ESS} \approx 2$. Under a log-normal approximation for the importance weights, one has $\mathrm{ESS} \approx N_{\mathrm{in}} \exp(-V)$, where $V = \mathrm{Var}[\log w]$ denotes the variance of the log-weights across inner samples. Solving for the sample size yields $N_{\mathrm{in}}^* = 2\exp(V)$, or equivalently $\log_{10} N_{\mathrm{in}}^* = (\ln 2 + V)/\ln 10$. 

We observe a median log-weight variance of $V \approx 5.9 \times 10^{7}$, driven almost entirely by the likelihood term rather than the prior-to-proposal reweighting. Substituting $V$, we obtain $\log_{10} N_{\mathrm{in}}^* \approx 2.6 \times 10^{7}$, i.e., $N_{\mathrm{in}}^* \approx 10^{2.6 \times 10^{7}}$. This sample requirement, which far exceeds any conceivable computational budget, demonstrates that the NMC estimator is fundamentally intractable for this problem, a consequence of the likelihood's extreme sensitivity to perturbations in the model parameters.

We note that the estimate $N_{\mathrm{in}}^* \approx 10^{2.6 \times 10^7}$ should be interpreted as an order-of-magnitude diagnostic rather than a precise prediction. The log-normal approximation $\mathrm{ESS} \approx N_{\mathrm{in}} \exp(-V)$ is exact only when the log-weights are Gaussian. Moreover, since we observed $\mathrm{ESS} = 1$ with $N_{\mathrm{in}}=$16,384, the empirical variance $V$ may over- or under-estimate the true log-weight variance. However, because $\log_{10} N_{\mathrm{in}}^*$ is proportional to $V$, even orders-of-magnitude reduction in $V$ (e.g., a better proposal distribution) would only rescale the exponent of $N_{\mathrm{in}}^*$. For instance, a $10^6$ times reduction of the empirical variance $V$ would still yield $N_{\mathrm{in}}^* \approx 10^{26}$, which remains intractable.

\begin{figure}[htb]
    \centering
    \scalebox{0.9}{
    \renewcommand{\arraystretch}{2}
    \begin{tabular}{cc c}
        Design \# 1 & Design \#2 & Design \#3 \\
        \raisebox{-0.04\linewidth}{\includegraphics[width=0.18\linewidth]{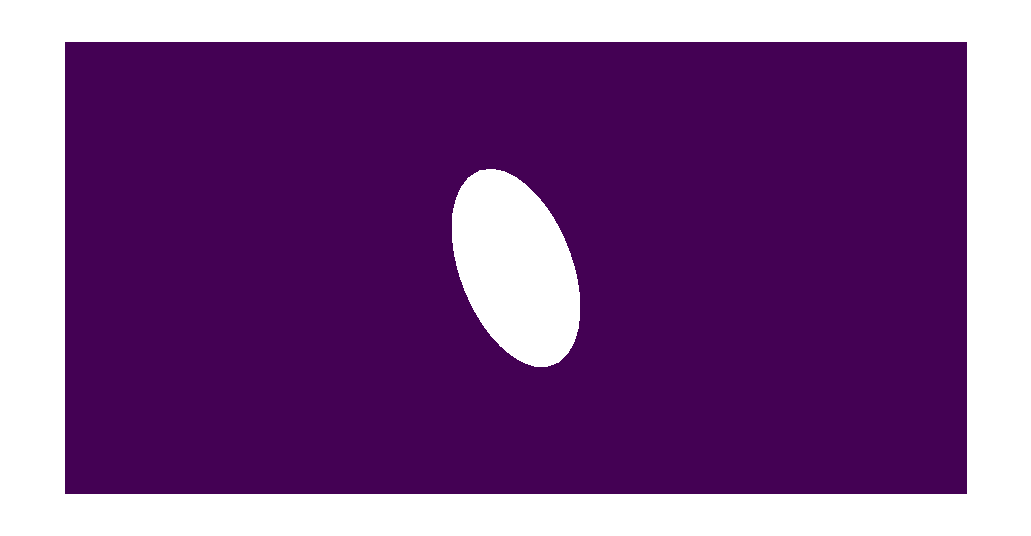}} & \raisebox{-0.04\linewidth}{\includegraphics[width=0.18\linewidth]{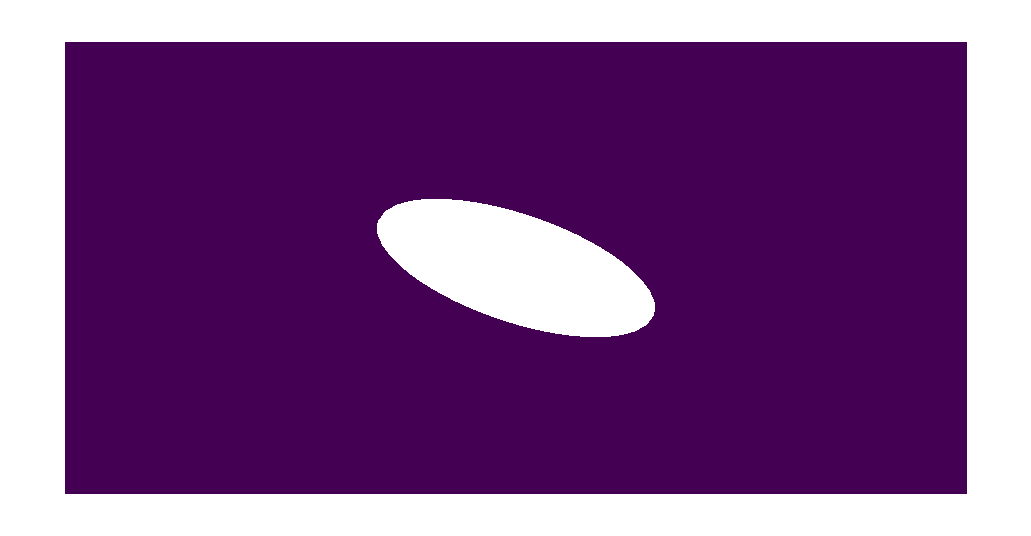}} & \raisebox{-0.04\linewidth}{\includegraphics[width=0.18\linewidth]{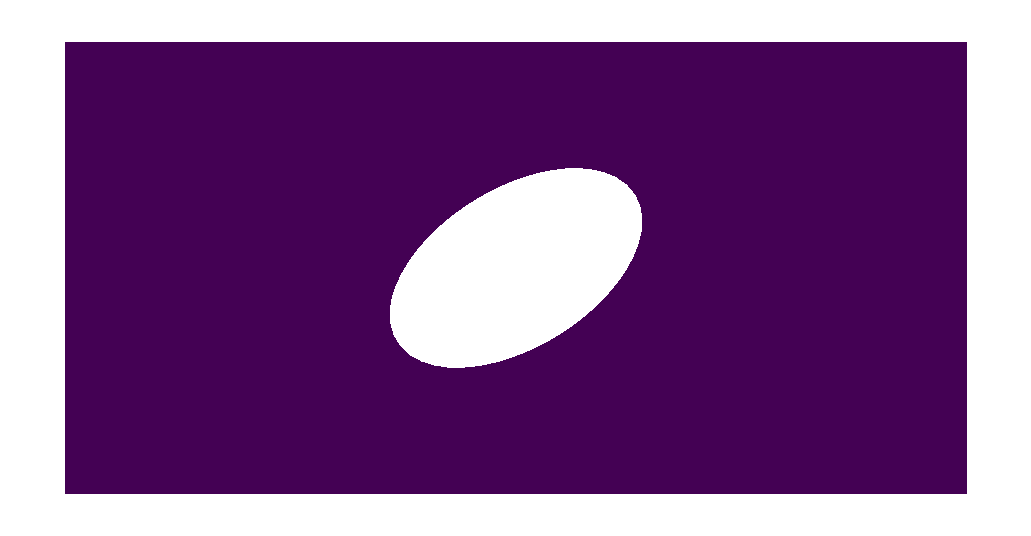}} \\
        
        \includegraphics[width=0.16\linewidth]{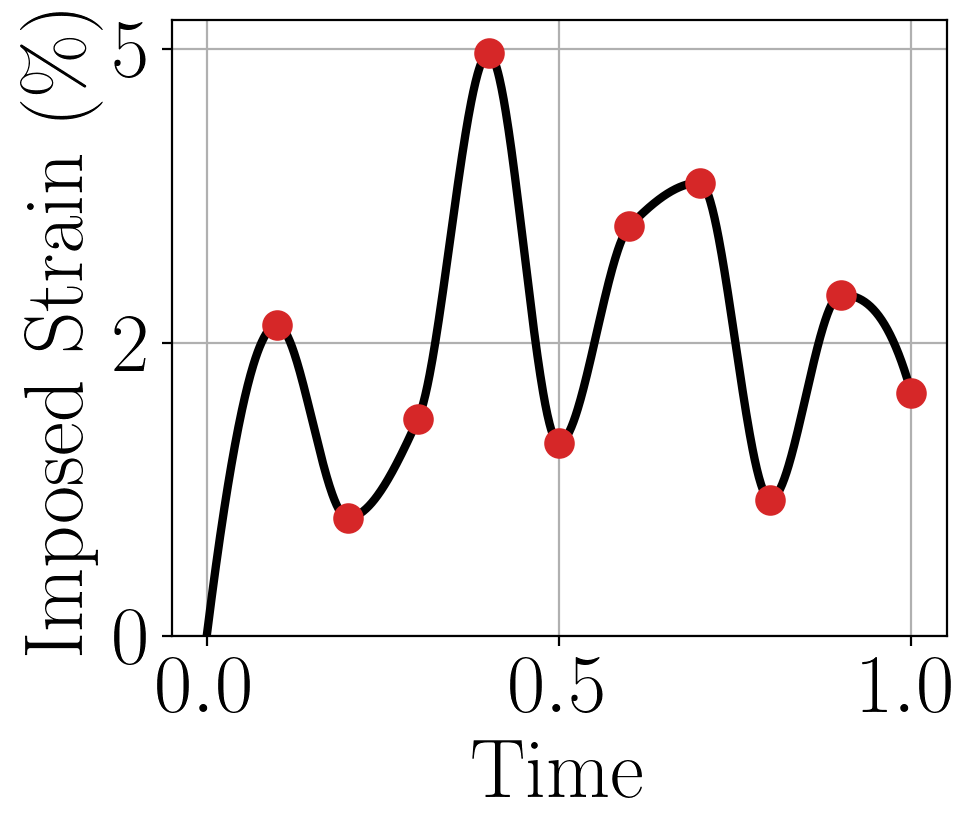} & \includegraphics[width=0.16\linewidth]{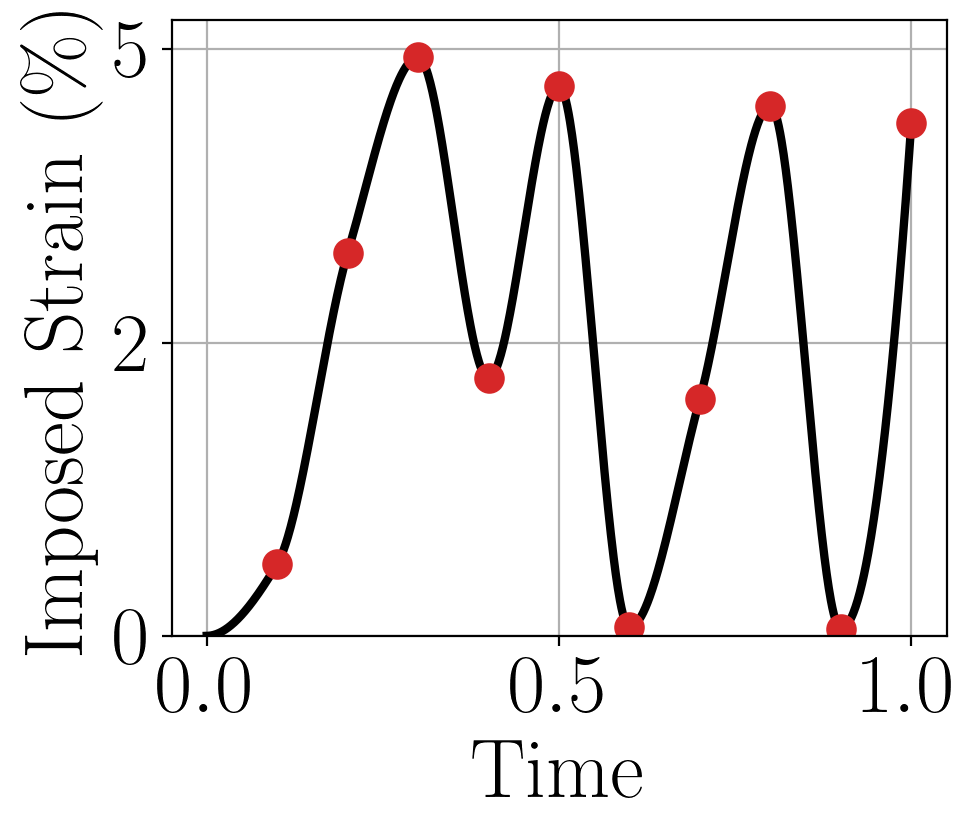} & \includegraphics[width=0.16\linewidth]{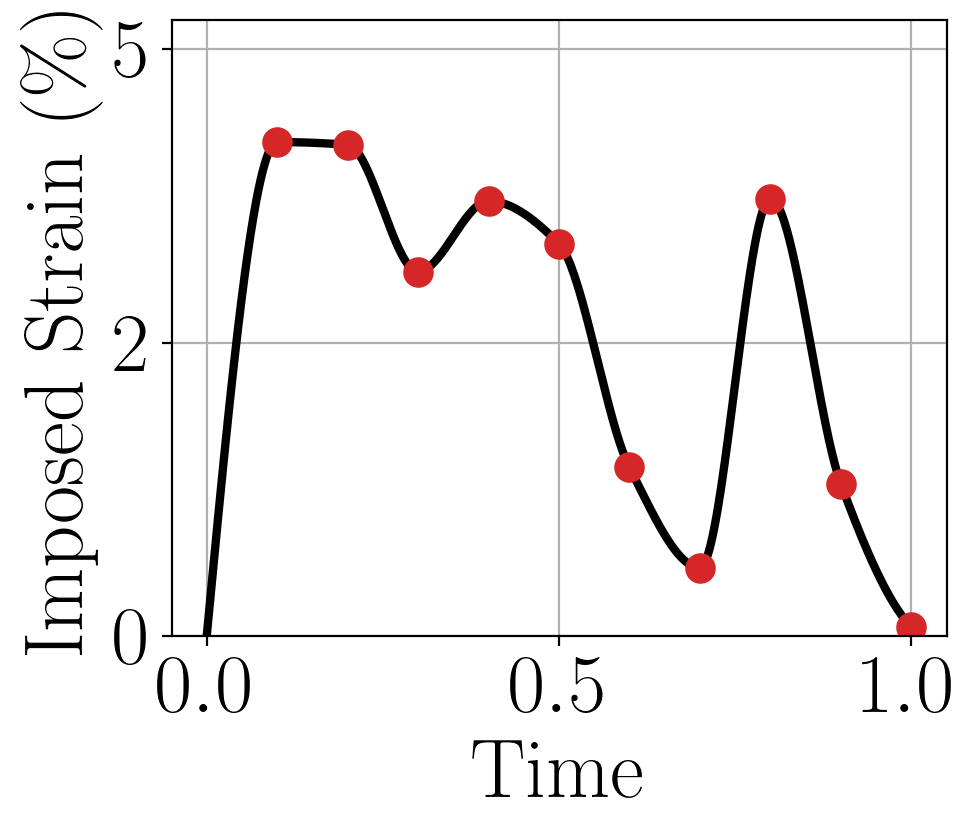}
    \end{tabular}
    \renewcommand{\arraystretch}{1.0}
    }
    \caption{Randomly sampled designs used for NMCE evaluations.}
    \label{fig:nmce_design}
\end{figure}

In \cref{fig:nmce_design}, we visualize the randomly sampled designs for NMCE in \cref{ssec:ve_nmce}. 

\subsection{Surrogate Training} \label{app:details_surrogate}
Additional details of the surrogate training are provided as follows. The mean relative error in predicting the log-FIM is used as the loss function, where the Frobenius norm for matrices is employed. We use an AdamW optimizer with an initial learning rate of $10^{-3}$ and a weight decay of $10^{-3}$. We use cosine learning rate annealing to decay the learning rate to $10^{-5}$ at the end of training. We consider 250 total epochs with a batch size of 32. A validation dataset of size 4096 is used to tune the architecture parameters and select the best-performing model during training. The surrogate training is implemented in PyTorch \cite{paszke2019pytorch}.

\subsection{Design Utility Maximization} \label{app:details_maximization}

The EGA maximization is performed using Bayesian optimization implemented through the software library in \cite{nogueira1024bayesian}. The design space is first normalized to the unit hypercube. We employ a Gaussian process surrogate of the EGA utility function with a Mat\'ern kernel ($\nu = 5/2$) and Automatic Relevance Determination \cite{snoek2012practical}:
\begin{equation}
    K(z, z') = \sigma_f^2 \left( 1 + \sqrt{5}r + \frac{5}{3}r^2 \right) \exp\left(-\sqrt{5}r\right), \quad r = \sqrt{\sum_{j=1}^{d_z} \frac{\left([z]_j - [z']_j\right)^2}{[l]_j^2}},
\end{equation}
where $l_j$ are the length-scale hyperparameters optimized for each design dimension and $\sigma_f^2$ is the signal variance. The search strategy proceeds in two phases: (i) an initialization phase with $N_{\text{init}} = 64$ design samples drawn via QMC, and (ii) an adaptive phase of $N_{\text{iter}} = 268$ iterations. In the adaptive phase, new designs are selected by maximizing the expected improvement acquisition function. To balance exploration and exploitation, the exploration parameter is set to $0.05$ and decays by 5\% per iteration. The sample-averaged approximation in \cref{eq:eig_gauss} is parallelized, where the FIM evaluations at 128 parameter samples occur concurrently. Instead of iid parameter samples in \cref{eq:eig_gauss}, we consider correlated samples drawn via QMC.

The ESFIM maximization is performed by first evaluating the objective over a large number of randomly sampled design variables (e.g., $10^{6}$), which can be done efficiently on a GPU. Then, a smaller pool of 256 samples with high ESFIM values is selected as initial guesses for ESFIM maximization using the L-BFGS-B algorithm. The final optimal design is selected as the one with the maximum ESFIM across all optimization runs.

\end{document}